\documentclass[aps,prx,reprint,twocolumn,footinbib,footinbib,longbibliography,superscriptaddress]{revtex4-2}
\usepackage{amsmath}
\usepackage{amsfonts}
\usepackage{amssymb}
\usepackage{times}
\usepackage{graphicx}
\usepackage[usenames,dvipsnames]{xcolor}
\usepackage{bm}
\usepackage{amsthm}
\usepackage{mathtools}
\usepackage{physics}
\usepackage{mathrsfs}
\usepackage{hyperref}

\begin{document}
\title{Speed-accuracy relations for diffusion models: Wisdom from nonequilibrium thermodynamics and optimal transport}
\date{\today}
\author{Kotaro Ikeda}
\email{kotaro-ikeda@g.ecc.u-tokyo.ac.jp}
\affiliation{Department of Mathematical Engineering and Information Physics, School of Engineering, The University of Tokyo, 7-3-1 Hongo, Bunkyo-ku, Tokyo 113-0033, Japan}
\author{Tomoya Uda}
\affiliation{Department of Earth and Planetary Physics, School of Science, The University of Tokyo, 7-3-1 Hongo, Bunkyo-ku, Tokyo 113-0033, Japan}
\author{Daisuke Okanohara}
\affiliation{Preferred Networks Inc., 1-6-1, Otemachi, Chiyoda-ku, Tokyo 100-0004, Japan}
\author{Sosuke Ito}
\email{sosuke.ito@ubi.s.u-tokyo.ac.jp}
\affiliation{Universal Biology Institute, Graduate School of Science, The University of Tokyo, 7-3-1 Hongo, Bunkyo-ku, Tokyo 113-0033, Japan}
\begin{abstract}
   We discuss a connection between a generative model, called the diffusion model, and nonequilibrium thermodynamics for the Fokker-Planck equation, called stochastic thermodynamics. Using techniques from stochastic thermodynamics, we derive the speed-accuracy relations for diffusion models, which are inequalities that relate the accuracy of data generation to the entropy production rate. This relation can be interpreted as the speed of the diffusion dynamics in the absence of the non-conservative force. From a stochastic thermodynamic perspective, our results provide quantitative insight into how best to generate data in diffusion models. The optimal learning protocol is introduced by the geodesic of space of the 2-Wasserstein distance in optimal transport theory. We numerically illustrate the validity of the speed-accuracy relations for diffusion models with different noise schedules and different data. We numerically discuss our results for optimal and suboptimal learning protocols. We also demonstrate the applicability of our results to data generation from the real-world image datasets.
\end{abstract}
\maketitle  

\section{Introduction}
Diffusion processes are irreversible phenomena that cause thermodynamic dissipation. Diffusion processes are described by stochastic processes such as Brownian motion~\cite{van1992stochastic}, and thermodynamic irreversibility is quantified by the entropy production in stochastic thermodynamics~\cite{sekimoto2010stochastic, seifert2012stochastic}. In stochastic thermodynamics, there have been various discussions about the relationship between information processing and thermodynamic dissipation for diffusion processes~\cite{sekimoto1998langevin,kurchan1998fluctuation,hatano2001steady,seifert2005entropy,chernyak2006path,schmiedl2007efficiency,allahverdyan2009thermodynamic,van2010three,sagawa2012nonequilibrium,ito2013information,horowitz2014second, ito2015maxwell,gingrich2017inferring, dechant2018entropic,li2019quantifying, hasegawa2019uncertainty, ito2020stochastic, otsubo2020estimating, dechant2021continuous, otsubo2022estimating, koyuk2020thermodynamic}. Based on optimal transport theory~\cite{villani2009optimal}, inevitable thermodynamic dissipation for diffusion processes has been discussed in stochastic thermodynamics~\cite{jordan1998variational,aurell2011optimal, chen2019stochastic,nakazato2021geometrical}, and the thermodynamic trade-off relations among speed, accuracy, and dissipation for diffusion processes have been discussed as a generalization of the second law of thermodynamics~\cite{aurell2012refined,nakazato2021geometrical, dechant2022geometric2,ito2023geometric,nagayama2023geometric}.

Diffusion processes have recently been discussed in the context of statistical machine learning models called generative models~\cite{tomczak2022deep}. The diffusion-based generative models called diffusion models~\cite{sohl2015deep, song2020scorebased} were originally inspired by nonequilibrium thermodynamics~\cite{sohl2015deep}.
By introducing time-reversed dynamics, which are well studied in the context of the fluctuation theorem~\cite{evans2002fluctuation,crooks1999entropy, kurchan1998fluctuation} and Jarzynski equality~\cite{jarzynski1997nonequilibrium}, diffusion models generate data with spatial structure from noisy data without spatial structure. Several improvements have been made from different perspectives~\cite{ho2020denoising,kingma2021variational,song2019generative,nichol2021improved,rombach2022high, song2020improved,song2023consistency,dhariwal2021diffusion,song2020denoising,karras2022elucidating,chen2021likelihood,ramesh2022hierarchical,ho2022classifier,nichol2021glide}, and the improved diffusion models have reached the state of the art in an image generation task~\cite{song2019generative, ho2020denoising}. 

The proposed diffusion models~\cite{song2019generative,rombach2022high,ramesh2022hierarchical,sauer2024fast,podell2023sdxl} are currently understandable without nonequilibrium thermodynamics, and the improved diffusion models are regarded as variants of other statistical machine learning models. For example, diffusion models~\cite{song2019generative} are related to the score estimation methods including the score matching~\cite{hyvarinen2005estimation,vincent2011connection, kingma2010regularized}. By incorporating the existing flow-based method~\cite{dinh2014nice,rezende2015variational,chen2018neural,song2022applying}, another improved method called the flow matching method~\cite{lipman2022flow} has also been proposed in the context of diffusion models. 

\begin{figure*}
    \centering
    \includegraphics[width=\linewidth]{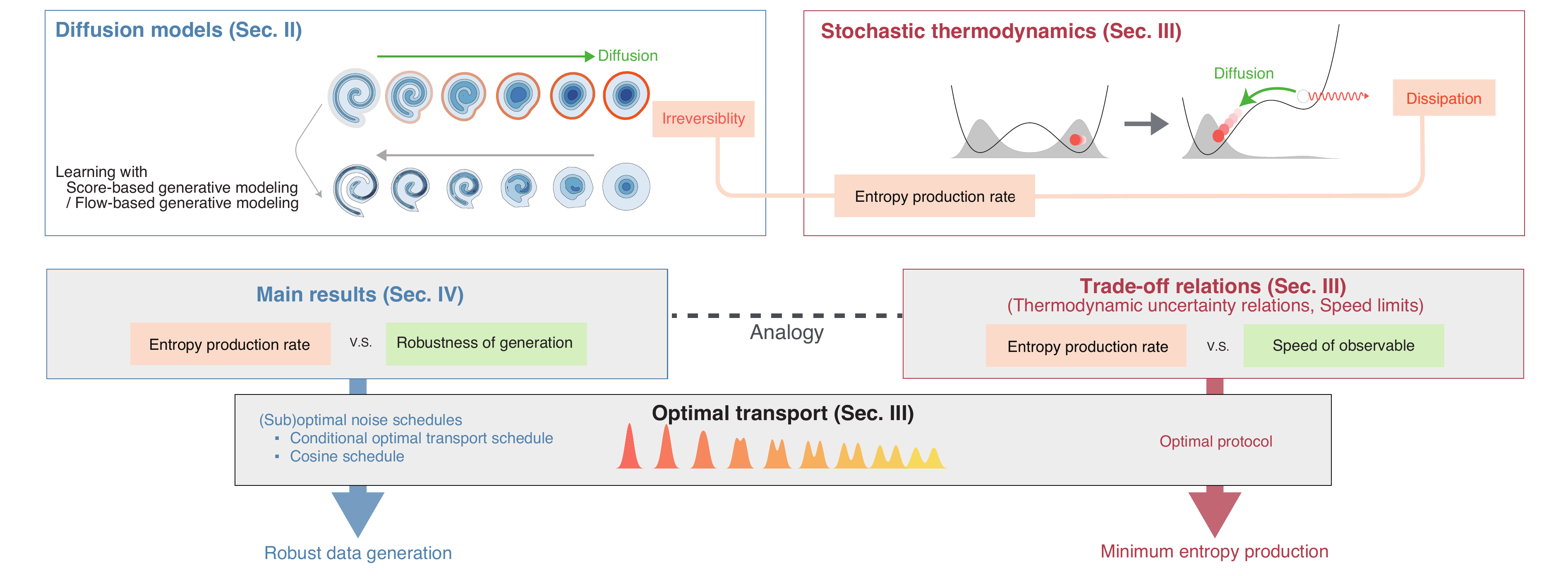}
    \caption{The organization of this paper. In Sec.~\ref{Generative-models&diffusion-models}, we explain the basic concepts of diffusion models and discuss their methods, including score-based generative modeling and flow-based generative modeling. In Sec.~\ref{Stochastic-thermo&diffusion}, we discuss stochastic thermodynamics and optimal transport theory, and introduce the entropy production rate. We also introduce several thermodynamic trade-off relations, such as thermodynamic uncertainty relations and speed limits, which provide the minimum entropy production in a finite time via optimal transport. In Sec.~\ref{Main-result}, We derive our main results in analogy to thermodynamic trade-off relations. Our main results reveal the essential role of the entropy production rate in diffusion models. Noise schedules based on optimal transport will lead to the most robust generation in diffusion models, and existing noise schedules such as the conditional optimal transport schedule and the cosine schedule are regarded as suboptimal noise schedules.}
    \label{fig:basic-concepts}
\end{figure*}
Because diffusion models have been improved outside the context of nonequilibrium thermodynamics, the insights of stochastic thermodynamics are not fully exploited in diffusion models. For example, the discussion of minimum entropy production in stochastic thermodynamics is related to the Wasserstein distance~\cite{villani2009optimal} in optimal transport theory~\cite{aurell2012refined,nakazato2021geometrical}, which is well used in generative models such as the Wasserstein generative adversarial network~\cite{arjovsky2017wasserstein} and diffusion models~\cite{kwon2022score,oko2023diffusion,de2022convergence}.  Discussions on the relationship between optimal transport and dynamics in the diffusion model have only recently begun~\cite{kornilov2024optimal,shaul2023kinetic,tong2023improving,shi2022conditional, liu2022flow,esser2024scaling}. In these discussions, various insights from stochastic thermodynamics, such as the thermodynamic trade-off relations for diffusion processes based on optimal transport theory~\cite{aurell2012refined,nakazato2021geometrical, dechant2022geometric2,ito2023geometric}, are not well used. There is only one paper that discusses the flow matching method from the perspective of both optimal transport theory and stochastic thermodynamics~\cite{klinger2024universal}. 

In this paper, we reconsider diffusion models in terms of stochastic thermodynamics, and derive speed-accuracy relations for diffusion models, which are analogous to the thermodynamic trade-off relations based on optimal transport theory. The speed-accuracy relations explain that the theoretical limits of accurate data generation are generally bounded by the entropy production rate and the temperature, which can be interpreted as the speed of the diffusion dynamics in the absence of the non-conservative force.
Furthermore, the speed-accuracy relations explain that the optimal diffusion dynamics for learning is given by the optimal transport. The results provide theoretical support for the trial-and-error protocols of diffusion dynamics, which have been characterized by noise schedules such as the cosine schedule~\cite{nichol2021improved} and the conditional optimal transport schedule~\cite{lipman2022flow}. We numerically illustrate the speed-accuracy relations for diffusion models using simple one-dimensional diffusion processes, two-dimensional diffusion processes, and diffusion processes on latent space~\cite{rombach2022high}. We show that accurate data generation can be discussed from a quantitative comparison of the speed-accuracy relations for different diffusion processes. We numerically illustrate that the optimal transport provides the most accurate data generation. We also investigate the situation where the inaccurate data generation is achieved due to a non-conservative force and a non-optimal protocol as predicted by the speed-accuracy relations. Furthermore, we demonstrate the applicability of the speed-accuracy relations to practical diffusion models trained on real-world image datasets.

This paper is organized as follows (see also Fig.~\ref{fig:basic-concepts}). In Sec.~\ref{Generative-models&diffusion-models}, we discuss the generative models and diffusion models. We first explain the generative models [Sec.~\ref{Generative-models}] and then move on to the basic concepts of diffusion models [Sec.~\ref{Diffusion-models}]. We then present the mathematical details of the diffusion model [Sec.~\ref{Forward-process}, \ref{Reverse&Estimated}]. We also explain a practical formulation based on the conditional Gaussian probabilities [Sec.~\ref{sec.conditionalGauss}], as well as some examples [Sec.~\ref{sec.concreteSystem}]. In Sec.~\ref{Stochastic-thermo&diffusion}, we explain stochastic thermodynamics, especially from the perspective of its relationship to diffusion models and optimal transport theory. In Sec~\ref{sec.OverDampedThermodynamics}, we introduce the entropy production rate in stochastic thermodynamics. In Sec.~\ref{sec.OverDampedSystemAndOptimalTransport}, we explain applications of optimal transport theory in stochastic thermodynamics such as the thermodynamic trade-off relations. In Sec.~\ref{Main-result}, we derive the main results, which are the speed-accuracy relations for diffusion models. The main results imply that the entropy production rate introduced in Sec.~\ref{sec.OverDampedThermodynamics} provides a fundamental limit on the accuracy of data generation in the diffusion model, and the main result is analogous to the thermodynamic trade-off relations in Sec.~\ref{sec.OverDampedSystemAndOptimalTransport}.  Furthermore, we discuss the optimal protocol based on the main results in Sec.~\ref{sec.optimal_noise_sce}, and confirm the main results by numerical calculations of diffusion models with three different situations in Sec.~\ref{sec.numericalCalculation}. Finally, we conclude with the main results and discuss different perspectives on the results, a generalization of the main results, and future studies in Sec.~\ref{discussion}.

\section{Generative models and diffusion models}
\label{Generative-models&diffusion-models}
\subsection{Generative models}\label{Generative-models}
In this paper, we discuss a class of statistical machine learning models known as generative models~\cite{tomczak2022deep}. A generative model is a class of statistical models capable of generating a new dataset that resembles the input dataset. The data generation in generative models can be considered as an estimation of the input data distribution, while the generated dataset is a sample from the estimated distribution.

More specifically, generative models can be described as follows. Let $d_{\mathrm{in}}=\{\bm{x}^{(1)},\ldots, \bm{x}^{(N_{\rm D})}\}$ be the set of input dataset, where each element is a member of the $n_{\rm d}$-dimensional Euclidean space $\bm{x}^{(n)}\in \mathbb{R}^{n_{\rm d}}$. The data $\bm{x}^{(n)}$ are assumed to be independent and identically distributed, and are sampled from a probability density function $q(\bm{x})$ that satisfies $q(\bm{x}) > 0$ and $\int d\bm{x} q(\bm{x}) =1$. Here, the positivity of the probability density function $q(\bm{x}) > 0$ is required for quantities such as $\ln q(\bm{x})$ or $1/q(\bm{x})$ not to diverge, and we will implicitly assume that any probability density functions are positive throughout this paper. We consider the problem of constructing a new probability density function $p(\bm{x})$ that satisfies $p(\bm{x}) > 0$ and $\int d\bm{x} p(\bm{x}) =1$, based on the estimation of the probability density function $q(\bm{x})$. Statistical machine learning models that generate samples from the probability density function $p(\bm{x})$ are called generative models.

Several studies have shown that generative models can be applied to practical datasets such as images~\cite{brock2018large}, natural language~\cite{brown2020language}, and audio~\cite{oord2016wavenet}. In these applications, generative models are constructed by solving optimization problems using machine learning methods such as deep learning~\cite{goodfellow2016deep,tomczak2022deep}. Machine learning optimization problems are solved by numerically minimizing or maximizing some objective function~\cite{goodfellow2016deep,tomczak2022deep}. To judge whether learning was successful, we sometimes discuss whether a distance or a pseudo-distance between the distribution of the input data $q(\bm{x})$ and the distribution of the generated data $p(\bm{x})$ becomes smaller~\cite{theis2015note,betzalel2022study,bischoff2024practical,heusel2017gans}. We call the value of this (pseudo-) distance the estimation error. Typical examples of the estimation error are the Kullback-Leibler divergence~\cite{amari2016information}, which is a pseudo-distance in information theory and information geometry, and the Wasserstein distance~\cite{villani2009optimal}, which is a distance in optimal transport theory.

\subsection{Diffusion models}\label{Diffusion-models}
\begin{center}
\begin{figure}
    \centering
    \includegraphics[width=\linewidth]{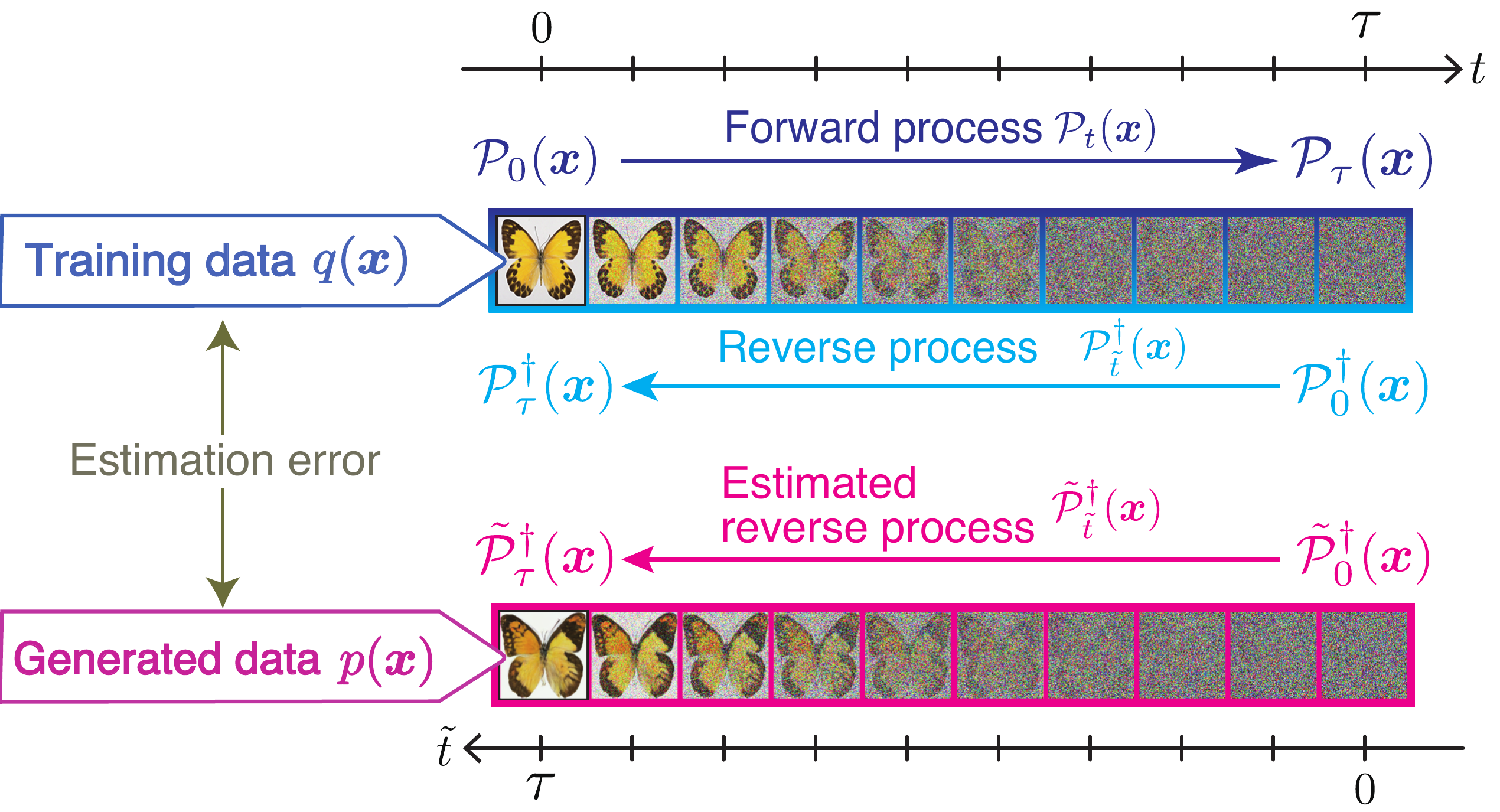}
    \caption{Illustration of diffusion models in the case of image data. The forward process, denoted by $\mathcal{P}_t(\bm{x})$, is a stochastic process which adds noise to the training data sampled from $q(\bm{x})$. By inverting this forward process in time, we prepare the reverse process $\mathcal{P}^{\dagger}_{\tilde{t}}(\bm{x})$. The generated data sampled from $p(\bm{x})$ is constructed by an estimated reverse process $\tilde{\mathcal{P}}^{\dagger}_{\tilde{t}}(\bm{x})$ that mimics the reverse process. The image in the figure is taken from the dataset~\cite{huggan/smithsonian_butterflies_subset}.}
    \label{fig:diffusionConcept}
\end{figure}
\end{center}

Among these machine learning based generative models, a method called diffusion models~\cite{sohl2015deep,ho2020denoising,song2019generative,song2020denoising,song2020scorebased} achieves lower estimation errors rather than previous generative models for various data~\cite{poole2022dreamfusion,xu2022geodiff, abramson2024accurate,song2020scorebased,dhariwal2021diffusion,esser2024scaling,ho2022imagen,videoworldsimulators2024,chen2020wavegrad,kong2020diffwave}, including image data~\cite{song2020scorebased, dhariwal2021diffusion,esser2024scaling}, video~\cite{ho2022imagen,videoworldsimulators2024}, and audio~\cite{chen2020wavegrad,kong2020diffwave}. Diffusion models generate samples from the probability density function $p(\bm{x})$ by estimating the time-reversed diffusion process that generates the data structure, while the corresponding forward process destroys the data structure by adding stochastic noise to the input dataset $d_{\mathrm{in}}$ [see Fig.~\ref{fig:diffusionConcept}]~\cite{song2020scorebased,sohl2015deep}. In practice, we sometimes convert the raw data into a low-dimensional latent space and discuss a diffusion process in the latent space to extract the features of the data and reduce the computational cost~\cite{rombach2022high}. In this case, the input dataset $d_{\rm in}$ can be regarded as a dataset of latent variables, which is created by converting the raw dataset into the latent space.

Specifically, the input probability density function $\mathcal{P}_{0}(\bm{x}) = q(\bm{x})$ is transformed into a known noise distribution $\mathcal{P}_{\tau}(\bm{x})$ by adding noise to the input data, where $\mathcal{P}_{t}(\bm{x})$ is the probability density function at time $t \in [0, \tau]$. The time evolution of $\mathcal{P}_{t}(\bm{x})$ is given by diffusion processes described by the Fokker-Planck equation~\cite{risken1996fokker} or the master equation~\cite{van1976stochastic} in diffusion models~\cite{song2020scorebased,sohl2015deep}. Sometimes we consider the stochastic dynamics of the data~\cite{song2020scorebased} from the initial state $\bm{x}_{t=0}$ to the final state $\bm{x}_{t=\tau}$, where $\bm{x}_{t}$ is the state of the data at time $t$ instead of the time evolution of the probability density function $\mathcal{P}_{t}(\bm{x})$. In this case, the stochastic time evolution of $\bm{x}_{t}$ can be described by the Langevin equation~\cite{van1992stochastic} or the Markov chain Monte Carlo (MCMC) method~\cite{brooks1998markov,andrieu2003introduction,welling2011bayesian}. The diffusion process from $t=0$ to $t=\tau$ is called the forward process~\cite{sohl2015deep}, and the dependence of the noise intensity at time $t$ in the forward process is called the noise schedule, which characterizes the time evolution of the diffusion dynamics~\cite{sohl2015deep,nichol2021improved,chen2023importance}.

Next, we will explain the reverse process, which is the time-reversed dynamics of the forward process. Here, we consider the forward process described by the time evolution of the probability density function $\mathcal{P}_{t}(\bm{x})$. We write the reverse process as $\mathcal{P}^{\dagger}_{\tilde{t}}(\bm{x})$, defined as $\mathcal{P}^{\dagger}_{\tilde{t}}(\bm{x})\coloneqq \mathcal{P}_{\tau-\tilde{t}}(\bm{x})$. In the reverse process, we consider the time evolution from $\tilde{t}=0$ to $\tilde{t}=\tau$. If $\tau$ is large enough, the final probability of the forward process $\mathcal{P}_{\tau}(\bm{x})$ can be considered as a noisy state where the data structure is well broken. Thus, the reverse process means that this noisy state $\mathcal{P}_{0}^{\dagger}(\bm{x})=\mathcal{P}_{\tau}(\bm{x})$ is the initial probability density function at time $\tilde{t}=0$ in the reverse process, and the time evolution in the reverse process allows the reproduction of the input data distribution $\mathcal{P}^{\dagger}_{\tau}(\bm{x})=\mathcal{P}_{0}(\bm{x})=q(\bm{x})$ containing the data structure at the final time $\tilde{t}= \tau$.

We introduce the stochastic process for generating the data from $p(\bm{x})$, namely the estimated reverse process, as an imitation of the reverse process. This estimated reverse process may differ from the reverse process in general, and so we write it as $\tilde{\mathcal{P}}^{\dagger}_{\tilde{t}}(\bm{x})$ and distinguish it from the reverse process $\mathcal{P}_{\tilde{t}}^{\dagger}(\bm{x}) ( \neq\tilde{\mathcal{P}}_{\tilde{t}}^{\dagger}(\bm{x}))$. Since the initial probability density function of the estimated reverse process $\tilde{\mathcal{P}}_{0}^{\dagger}(\bm{x})$ can be different from the initial probability density function of the reverse process $\mathcal{P}_{0}^{\dagger}(\bm{x}) (\neq \tilde{\mathcal{P}}_{0}^{\dagger}(\bm{x}))$, the final probability density function of the estimated reverse process $\tilde{\mathcal{P}}^{\dagger}_{\tau}(\bm{x})$ can be different from the input distribution $q(\bm{x}) (\neq\tilde{\mathcal{P}}^{\dagger}_{\tau}(\bm{x}))$. Here, the idea of diffusion models is to set this probability density function $\tilde{\mathcal{P}}_{\tau}^{\dagger}(\bm{x})$, which is the output of the estimated reverse process, as the generated data distribution $p(\bm{x})$~\cite{sohl2015deep}. In practical diffusion models, we do not necessarily compute the time evolution of the probability density function. For example, there are several methods to generate the data from the estimated reverse process using the Langevin equation~\cite{song2019generative,song2020scorebased} or a deterministic time evolution via flow~\cite{lipman2022flow}.

\subsection{Fokker--Planck equation for the forward process}
\label{Forward-process}
We discuss the time evolution of $\mathcal{P}_{t}(\bm{x})$ for the forward process in diffusion models~\cite{song2020scorebased} based on the Fokker--Planck equation~\cite{risken1996fokker}. The time evolution of the probability density function $\mathcal{P}_t(\bm{x})$ is described by the Fokker--Planck equation as follows.
\begin{align}
    \partial_t \mathcal{P}_t(\bm{x})&=-\nabla\cdot\left(\bm{\nu}_t^{\mathcal{P}}(\bm{x}) \mathcal{P}_t(\bm{x})\right), \nonumber\\
    \bm{\nu}_t^{\mathcal{P}}(\bm{x})&\coloneqq\bm{f}_t(\bm{x})-T_t\nabla\ln \mathcal{P}_t(\bm{x}),
\label{eq.FP}
\end{align}
where $\partial_t\coloneqq\partial /\partial t$ is the partial differential operator and $\nabla$ is the del operator. $\bm{\nu}^{\mathcal{P}}_t(\bm{x})\in\mathbb{R}^{n_{\rm d}}$ is the velocity field in a continuous equation. The parameters $T_t\in\mathbb{R}_{\geq0}$ and $\bm{f}_t(\bm{x})\in\mathbb{R}^{n_{\rm d}}$ are functions of the time $t$ in general. The forward process can be regarded as the time evolution of $\mathcal{P}_t(\bm{x})$ by the Fokker--Planck equation [Eq.~(\ref{eq.FP})] under the initial condition $\mathcal{P}_{0}(\bm{x})=q(\bm{x})$ with  fixed time dependence of the parameters $T_t$ and $\bm{f}_t(\bm{x})$. The time dependence of $T_t$ characterizes the time variation of the noise intensity in the diffusion process, which is known as the noise schedule in diffusion models.

The Fokker--Planck equation [Eq.~(\ref{eq.FP})] is statistically equivalent to the stochastic differential equation, namely the Langevin equation~\cite{risken1996fokker,van1976stochastic} for the state $\bm{x}_t$,
\begin{align}
    d\bm{x}_t=\bm{f}_t(\bm{x}_t)dt + \sqrt{2T_t}dB_t,\label{eq.Langevin}
\end{align}
where $dB_t=B_{t+dt}-B_t$ is the random white Gaussian noise, mathematically defined as the difference of the Wiener processes $B_t$. If the Langevin equation is considered as Brownian dynamics for the position of a Brownian particle, the parameters $T_t$ and $\bm{f}_t(\bm{x}_t)$ correspond to the temperature and the external force, respectively~\cite{risken1996fokker}. In the context of diffusion models, the forward process described by the Langevin equation can be viewed as a time-evolving process in which the random noise is sequentially added to the data. Using the Langevin equation, the time evolution of the forward process is computed using the input data $\boldsymbol{x}^{(n)} \in d_{\mathrm{in}}$ $(n=1, \dots, N_{\rm D})$ for fixed time dependence of the parameters $T_t$ and $\bm{f}_t(\bm{x}_t)$. The time evolution of the probability density function $\mathcal{P}_{t}(\bm{x})$ can be computed numerically by Monte Carlo sampling using Langevin dynamics~\cite{welling2011bayesian}.

\subsection{Reverse and estimated reverse process}\label{Reverse&Estimated}
Next, we discuss the reverse process by considering the time-reversed dynamics of the Fokker-Planck equation for the forward process [Eq.~(\ref{eq.FP})]. Using the reversed time $\tilde{t}=\tau -t$, the time evolution of $\mathcal{P}^{\dagger}_{\tilde{t}}(\bm{x})=\mathcal{P}_{\tau-\tilde
t}(\bm{x})=\mathcal{P}_{t}(\bm{x})$ is given by the Fokker--Planck equation for the forward process [Eq.~(\ref{eq.FP})] as follows,
\begin{align}
    \partial_{\tilde{t}} \mathcal{P}_{\tilde{t}}^{\dagger}(\bm{x})&= -\partial_{t} \mathcal{P}_{t}(\bm{x}) \nonumber\\
    &= 
    \nabla\cdot\left(\bm{\nu}_{t}^{\mathcal{P}}(\bm{x}) \mathcal{P}_{t}(\bm{x})\right) \nonumber\\
    &= 
    - \nabla\cdot\left(\bm{\nu}_{\tilde{t}}^{\dagger}(\bm{x}) \mathcal{P}_{\tilde{t}}^{\dagger}(\bm{x})\right),\label{eq.reverseFP}
\end{align}
where the velocity field for the reverse process $\bm{\nu}_{\tilde{t}}^{\dagger}(\bm{x})$ is defined as ${\bm{\nu}_{\tilde{t}}^{\dagger}(\bm{x})\coloneqq - \bm{\nu}^{\mathcal{P}}_{\tau - \tilde{t}}}(\bm{x})$. 

Diffusion models are introduced by the estimated reverse process that mimics the reverse process [Eq.~(\ref{eq.reverseFP})]. We denote the probability density function of the data $\bm{x}$ in the estimated reverse process by $\tilde{\mathcal{P}}^{\dagger}_{\tilde{t}}(\bm{x})$. We assume that the time evolution of this probability density function $\tilde{\mathcal{P}}^{\dagger}_{\tilde{t}}(\bm{x})$ can be written by the continuity equation,
\begin{align}
    \partial_{\tilde{t}}\tilde{\mathcal{P}}^{\dagger}_{\tilde{t}}(\bm{x})=-\nabla \cdot \left(\tilde{\bm{\nu}}_{\tilde{t}}^{\dagger}(\bm {x}) \tilde{\mathcal{P}}_{\tilde{t}}^{\dagger}(\bm{x})\right). \label{eq.inferedFP}
\end{align}
Here, $\tilde{\bm{\nu}}_{\tilde{t}}^{\dagger}(\bm{x})$ represents the velocity field in the estimated reverse process, which may differ from that of the reverse process ${\bm{\nu}}_{\tilde{t}}^{\dagger}(\bm{x})$. The velocity field $\tilde{\bm{\nu}}_{\tilde{t}}^{\dagger}(\bm{x})$ is estimated numerically from the dynamics of the forward process. 

In the following section, we will explain two methods for estimating the velocity field in the reverse process and constructing the velocity field of the estimated reverse process $\tilde{\bm{\nu}}_{\tilde{t}}^{\dagger}(\bm{x})$ for data generation, called the score-based generative modeling~\cite{song2019generative} and the flow-based generative modeling~\cite{lipman2022flow}. 

\subsubsection*{Example 1: Score-based generative modeling}
 First, we explain a method called score-based generative modeling~\cite{song2019generative}. This method consists of two components, score matching~\cite{hyvarinen2005estimation,vincent2011connection,kingma2010regularized,song2019generative} and data generation. Score matching is a method that estimates the score function $\nabla \ln \mathcal{P}_t(\bm{x})$ in the forward process [Eq.~(\ref{eq.FP})] in order to arrange the estimated reverse process~\cite{song2019generative,song2020scorebased}. The score function is typically estimated using neural networks. Specifically, we consider a situation where the score function $\nabla \ln \mathcal{P}_t(\bm{x})$ is modeled by a neural network denoted by $\bm{s}^{\theta}_t(\bm{x})$, where $\theta$ represents the set of the network's parameters. We then numerically solve an optimization problem to estimate the score function. The optimized parameters, denoted by $\theta^*_{\rm{SM}}$, are employed to generate the velocity field of the estimated reverse process $\tilde{\bm{\nu}}_{\tilde{t}}^{\dagger}(\bm{x})$.

The optimization problem is constructed as follows. The optimization problem to estimate the score function $\nabla \ln \mathcal{P}_t(\bm{x})$ is formulated as a minimization problem of the following score matching objective function~\footnote{The score matching objective function allows for arbitrary time-dependent weights~\cite{song2020scorebased}; in this study, we have adopted $T_t^2$ for the consistency with Eq.~\eqref{eq:underestimateSADM}.}.
\begin{align}
    L_{\rm SM}(\theta)=\mathbb{E}_{\mathcal{P}_t,\mathcal{U}}\left[T_t^2\|\bm{s}^{\theta}_t(\bm{x})-\nabla\ln \mathcal{P}_t(\bm{x})\|^2\right].\label{eq.SMLoss}
\end{align}
Here, the uniform distribution for $t \in [0,\tau]$ is given by $\mathcal{U}(t)=1/\tau$. The expected value with respect to this uniform distribution and the distribution $\mathcal{P}_t(\bm{x})$ is defined as $\mathbb{E}_{\mathcal{P}_t,\mathcal{U}} \left[ \cdots \right] = \int_0^{\tau} dt \mathcal{U}(t) \int d\bm{x} \mathcal{P}_t(\bm{x}) 
 \cdots$. Then, the optimal set of the parameters $\theta^*_{\rm{SM}}$ is obtained as~\cite{vincent2011connection,song2020scorebased},
\begin{align}
    \theta^*_{\rm SM} \in {\rm argmin}_{\theta} L_{\rm SM}(\theta).
    \label{opt.SMLoss}
\end{align}
In this optimization problem,  $\bm{s}^{\theta^*_{\rm SM}}_t(\bm{x})$, which reaches the minimum value $L_{\rm SM}(\theta^*_{\rm SM})=0$, can be the score function $\nabla\ln \mathcal{P}_t(\bm{x})$.

\subsubsection*{Example 1-1: Stochastic differential equation}
For the data generation by the estimated reverse process, we first explain the method based on the stochastic differential equations~\cite{song2020scorebased,song2019generative,ho2020denoising}.
The estimated score function, $\bm{s}^{\theta^*_{\rm SM}}_t(\bm{x})$, is employed to reconstruct the velocity field of the forward process $\bm{\nu}_t^{\mathcal{P}}(\bm{x})$ as,
\begin{align}
    \hat{\bm{\nu}}_t(\bm{x})\coloneqq\bm{f}_t(\bm{x})-T_t{\bm{s}}^{\theta^*_{\rm SM}}_t(\bm{x}).\label{eq.SMestimatedvelocity}
\end{align}
With this estimated velocity field $ \hat{\bm{\nu}}_t(\bm{x})$, we construct the velocity field of the estimated reverse process $\tilde{\bm{\nu}}_{\tilde{t}}^{\dagger}(\bm{x})$ as 
\begin{align}
    \tilde{\bm{\nu}}_{\tilde{t}}^{\dagger}(\bm{x})&=[\bm{\nu}_{\tilde{t}}^{\tilde{\mathcal{P}}^{\dagger}}(\bm{x})]^{\dagger} -2\hat{\bm{\nu}}_{\tau -\tilde{t}}(\bm{x}) \nonumber  \\
    &=-\bm{f}_{\tilde{t}}^{\dagger}(\bm{x})+2 T_{\tilde{t}}^{\dagger}{\bm{s}}^{\theta^*_{\rm SM} \dagger}_{\tilde{t}}(\bm{x}) -T_{\tilde{t}}^{\dagger}\nabla\ln \tilde{\mathcal{P}}_{\tilde{t}}^{\dagger}(\bm{x}),
    \label{eq.SMVelField}
\end{align}
so that the estimated reverse process [Eq.~(\ref{eq.inferedFP})] mimics the reverse process [Eq.~(\ref{eq.reverseFP})], where $[\bm{\nu}_{\tilde{t}}^{\tilde{\mathcal{P}}^{\dagger}}(\bm{x})]^{\dagger}\coloneqq\bm{f}_{\tilde{t}}^{\dagger}(\bm{x})-T_{\tilde{t}}^{\dagger}\nabla\ln \tilde{\mathcal{P}}_{\tilde{t}}^{\dagger}(\bm{x})$, $\bm{f}_{\tilde{t}}^{\dagger}(\bm{x})\coloneqq \bm{f}_{\tau-\tilde{t}}(\bm{x})$, $T_{\tilde{t}}^{\dagger}\coloneqq T_{\tau-\tilde{t}}$, and ${\bm{s}}^{\theta^*_{\rm SM} \dagger}_{\tilde{t}}(\bm{x})\coloneqq{\bm{s}}^{\theta^*_{\rm SM}}_{\tau-\tilde{t}}(\bm{x})$. If ${\bm{s}}^{\theta^*_{\rm SM}}_t(\bm{x})$ is exactly equal to $\nabla \ln \mathcal{P}_t(\bm{x})$ and the initial condition for the estimated reverse process is equivalent to the initial condition for the reverse process ($\tilde{\mathcal{P}}_{0}^{\dagger}(\bm{x})=\mathcal{P}_{0}^{\dagger}(\bm{x})$), the equation $\tilde{\bm{\nu}}_{\tilde{t}}^{\dagger}(\bm{x}) = \boldsymbol{\nu}_{\tilde{t}}^{\dagger} (\bm{x})$ holds and the reverse process [Eq.~(\ref{eq.reverseFP})] is consistent with the estimated reverse process [Eq.~(\ref{eq.inferedFP})].

The continuity equation of the estimated reverse process $\partial_{\tilde{t}}\tilde{\mathcal{P}}^{\dagger}_{\tilde{t}}(\bm{x})=-\nabla \cdot \left(\tilde{\bm{\nu}}_{\tilde{t}}^{\dagger}(\bm{x}) \tilde{\mathcal{P}}_{\tilde{t}}^{\dagger}(\bm{x})\right)$ using the estimated velocity field $\tilde{\bm{\nu}}^{\dagger}_{\tilde{t}}(\bm{x})$ can be regarded as the Fokker--Planck equation with the external force $-\bm{f}^{\dagger}_{\tilde{t}}(\bm{x})+2 T^{\dagger}_{\tilde{t}}{\bm{s}}^{\theta^*_{\rm SM} \dagger}_{\tilde{t}}(\bm{x})$. The stochastic differential equation, (i.e., the Langevin equation) corresponding to this Fokker--Planck equation~\cite{song2020scorebased,anderson1982reverse} is
\begin{align}
d\tilde{\bm{x}}^{\dagger}_{\tilde{t}}=\left(-\bm{f}_{\tilde{t}}^{\dagger}(\tilde{\bm{x}}^{\dagger}_{\tilde{t}})+2T^{\dagger}_{\tilde{t}}{\bm{s}}^{\theta^*_{\rm SM} \dagger}_{\tilde{t}}(\tilde{\bm{x}}^{\dagger}_{\tilde{t}})\right)d\tilde{t} + \sqrt{2T_{\tilde{t}}^{\dagger}}d{B}_{\tilde{t}},\label{eq.SMLangevin}
\end{align}
where $d{B}_{\tilde{t}}={B}_{\tilde{t}+d\tilde{t}} -{B}_{\tilde{t}}$ is the difference of the Wiener process $B_{\tilde{t}}$, and $\tilde{\bm{x}}^{\dagger}_{\tilde{t}}$ is the state of the data in the estimated reverse process. By simulating this stochastic differential equation numerically, we can generate the data in score-based generative modeling~\cite{song2020scorebased,song2019generative,ho2020denoising}.

\subsubsection*{Example 1-2: Probability flow ordinary differential equation}
Next, we discuss an alternative method for data generation in score-based generative modeling, namely the probability flow ordinary differential equation (ODE)~\cite{song2020scorebased}. The probability flow ODE is a method in which we use an ODE instead of the stochastic differential equation [Eq.~(\ref{eq.SMLangevin})] for data generation under the same objective function [Eq.~(\ref{eq.SMLoss})]. At present, this probability flow ODE is well used instead of the stochastic differential equation because it is possible to generate data faster and more accurately by performing time evolution based on numerical ODE solvers~\cite{lu2022dpm,lu2022maximum,zhang2022fast}.

Specifically, using the velocity field $\hat{\bm{\nu}}_t(\bm{x})$ [Eq.~(\ref{eq.SMestimatedvelocity})] estimated through the objective function [Eq.~(\ref{eq.SMLoss})], the velocity field of the estimated reverse process $\tilde{\bm{\nu }}^{\dagger}_{\tilde{t}}(\bm{x})$ is arranged as follows.
\begin{align}
    \tilde{\bm{\nu}}^{\dagger}_{\tilde{t}}(\bm{x})=-\hat{\bm{\nu}}_{\tau-\tilde{t}}(\bm{x})\label{eq.FMVelField}.
\end{align}
This expression is parallel to the expression of the velocity field in the reverse process $\bm{\nu}_{\tilde{t}}^{\dagger}(\bm{x})=-\bm{\nu}_{\tau-\tilde{t}}^{\mathcal{P}}(\bm{x})$. In this case, the continuity equation [Eq.~(\ref{eq.inferedFP})] corresponds to the following ODE~\cite{song2020scorebased,chen2018neural}.
\begin{align}
    \frac{d\tilde{\bm{x}}_{\tilde{t}}^{\dagger}}{d\tilde{t}}=-\hat{\bm{\nu}}_{\tau-\tilde{t}}(\tilde{\bm{x}}_{\tilde{t}}^{\dagger})=-\bm{f}^{\dagger}_{\tilde{t}}(\tilde{\bm{x}}^{\dagger}_{\tilde{t}})+T^{\dagger}_{\tilde{t}}{\bm{s}}^{\theta^*_{\rm SM} \dagger}_{\tilde{t}}(\tilde{\bm{x}}^{\dagger}_{\tilde{t}}),\label{eq.deterministicDiffusion}
\end{align}
where $\tilde{\bm{x}}^{\dagger}_{\tilde{t}}$ denotes the sample from $\tilde{\mathcal{P}}^{\dagger}_{\tilde{t}}(\bm{x})$. The probability flow ODE method generates data by numerically simulating this ODE~\cite{song2020scorebased}.
\subsubsection*{Example 2: Flow-based generative modeling}
Next, we explain a method called flow-based generative modeling~\cite{lipman2022flow}, which consists of flow matching and data generation via the ODE.
Flow matching~\cite{lipman2022flow} is a method to estimate the velocity field of the forward process without estimating the score function. In flow matching, we model the velocity field of the forward process $\bm{\nu}_t^{\mathcal{P}}(\bm{x})$ by a neural network $\bm{u}_t^{\theta} \in \mathbb{R}^{n_{\rm d}}$ where $\theta$ is the set of parameters. The objective function,
\begin{align}
    L_{\rm FM}(\theta)=\mathbb{E}_{\mathcal{P}_t,\mathcal{U}}\left[\|\bm{u}_t^{\theta}(\bm{x})-\bm{\nu}^{\mathcal{P}}_t(\bm{x})\|^2\right],\label{eq.FMLoss}
\end{align}
is minimized with respect to $\theta$ using numerical optimization methods, and $\theta^*_{\rm FM}$ is the optimal set of the parameters satisfying
\begin{align}
\theta^*_{\rm FM} \in\operatorname{argmin}_{\theta}L_{\rm FM}(\theta).
\label{opt.FMLoss}
\end{align}
The optimal set of the parameters $\theta^*_{\rm FM}$ gives the velocity field for the estimated reverse process $ \tilde{\bm{\nu}}^{\dagger}_{\tilde{t}}(\bm{x})$ as follows,
\begin{align}
    \tilde{\bm{\nu}}^{\dagger}_{\tilde{t}}(\bm{x})=-{\bm{u}}_{\tau- \tilde{t}}^{\theta^*_{\rm FM}}(\bm{x}).
\end{align}
This expression corresponds to the velocity field of the reverse process $ \bm{\nu}^{\dagger}_{\tilde{t}}(\bm{x})=-\bm{\nu}_{\tau-\tilde{t}}^{\mathcal{P}}(\bm{x})$.

In the flow-based generative modeling, the data generation is done by simulating the ODE~\cite{lipman2022flow,kobyzev2020normalizing},
\begin{align}
    \frac{d\tilde{\bm{x}}_{\tilde{t}}^{\dagger}}{d\tilde{t}}=-{\bm{u}}_{\tau-\tilde{t}}^{\theta^*_{\rm FM}}(\tilde{\bm{x}}_{\tilde{t}}^{\dagger}),
\end{align}
which is similar to the one used in the probability flow ODE method [Eq.~(\ref{eq.deterministicDiffusion})]. This ODE corresponds to the continuity equation for the estimated reverse process [Eq.~(\ref{eq.inferedFP})].

\subsection{Formulations with conditional Gaussian distributions}\label{sec.conditionalGauss}
In a practical implementation of diffusion models, we may consider a process such that the external force $\bm{f}_t(\bm{x})$ is linear,
\begin{align}
    \bm{f}_t(\bm{x})= \mathsf{A}_t \bm{x} + \bm{b}_t,
    \label{linearf}
\end{align}
to reduce the computational complexity~\cite{sohl2015deep,song2020scorebased,ho2020denoising,song2019generative,lipman2022flow}, where $\mathsf{A}_t\in \mathbb{R}^{n_{\rm d} \times n_{\rm d}}$ and $\bm{b}_t \in \mathbb{R}^{n_{\rm d}}$ are the matrix and the vector, respectively. When the initial condition is $\mathcal{P}_{t=0}(\bm{x}) = q(\bm{x})$, the solution $\mathcal{P}_t(\bm{x})$ for the process can be given by
\begin{align}
     &\mathcal{P}_t(\bm{x})=\int d\bm{y}\mathcal{P}^{\rm c}_t(\bm{x}|\bm{y})q(\bm{y}).\label{eq.marginalization}
\end{align}
Here, the transition probability $\mathcal{P}^{\rm c}_t(\bm{x}|\bm{y})$ is a Gaussian distribution $\mathcal{P}^{\rm c}_t(\bm{x}|\bm{y})=\mathcal{N}(\bm{x}|\bm{\mu}_t(\bm{y}),\mathsf{\Sigma}_t)$ due to the linearity of the external force~\cite{risken1996fokker}, where $\bm{\mu}_t(\bm{y})$ is the mean, which depends on $\bm{y}$, and $\mathsf{\Sigma}_t$ is the covariance matrix $\mathsf{\Sigma}_t$. We assume that $\mathsf{\Sigma}_t$ does not depend on the state $\bm{y}$. The equation~(\ref{eq.marginalization}) at $t=0$ gives the condition $\mathcal{P}^{\rm c}_0(\bm{x}|\bm{y})= \delta (\bm{x}-\bm{y})$, where $\delta (\bm{x}-\bm{y})$ is the delta function. Thus, the covariance matrix at $t=0$ is $\mathsf{\Sigma}_{t=0} = \mathsf{O}$, and the mean is $\bm{\mu}_{t=0}(\bm{y}) = \bm{y}$, where $\mathsf{O}$ is the zero matrix.

Under the above condition, the optimization problems [Eqs.~(\ref{eq.SMLoss}), (\ref{eq.FMLoss})] are easier to solve~\cite{lipman2022flow,song2019generative}. These training methods using conditional probability in score matching and flow matching are known as denoising score matching~\cite{vincent2011connection} and conditional flow matching~\cite{lipman2022flow}, respectively. To reduce the computational complexity, we consider new objective functions, which correspond to the objective functions [Eqs.~(\ref{eq.SMLoss}), (\ref{eq.FMLoss})] as follows,
\begin{align}
    &L_{\rm SM}^{\rm c}(\theta)=\mathbb{E}_{\mathcal{P}^{\rm c}_t,q, \mathcal{U}}\left[T_t^2\|\bm{s}^{\theta}_t(\bm{x})-\nabla\ln \mathcal{P}^{\rm c}_t(\bm{x}|\bm{y})\|^2\right], \label{eq.cSMLoss}\\\
    &L_{\rm FM}^{\rm c}(\theta)=\mathbb{E}_{\mathcal{P}^{\rm c}_t,q, \mathcal{U}}\left[\|\bm{u}_t^{\theta}(\bm{x})-\bm{\nu}^{\mathcal{P}^{\rm c}}_t(\bm{x}|\bm{y})\|^2\right], \label{eq.cFMLoss}
\end{align}
where $\mathbb{E}_{\mathcal{P}^{\rm c}_t,q, \mathcal{U}} \left[ \cdots \right] = \int_0^{\tau} dt \mathcal{U}(t) \int d\bm{x} d\bm{y} \mathcal{P}^{\rm c}_t(\bm{x}| \bm{y}) q(\bm{y}) 
 \cdots$ is the expected value with respect to $\mathcal{P}^{\rm c}_t(\bm{x}| \bm{y})$, $q(\bm{y})$, $\mathcal{U}(t)$, and $\bm{\nu}_t^{\mathcal{P}^{\rm c}}(\bm{x}|\bm{y})\coloneqq\bm{f}_t(\bm{x})-T_t\nabla\ln \mathcal{P}^{\rm c}_t(\bm{x}|\bm{y})$. As proved in Appendix A, new objective functions satisfy $\nabla_{\theta} L_{\rm SM}^{\rm c}(\theta) =\nabla_{\theta} L_{\rm SM}(\theta)$ and $\nabla_{\theta} L_{\rm FM}^{\rm c}(\theta) =\nabla_{\theta} L_{\rm FM}(\theta)$, where $\nabla_{\theta}$ is the gradient through $\theta$. Since the optimal solutions are given by the conditions $\nabla_{\theta} L_{\rm SM}^{\rm c}(\theta) = \nabla_{\theta} L_{\rm SM}(\theta) =0$ and $\nabla_{\theta} L_{\rm FM}^{\rm c}(\theta) = \nabla_{\theta} L_{\rm FM}(\theta) =0$, we get the optimal solutions,
 \begin{align}
    \theta^{*}_{\rm SM} &\in\operatorname{argmin}_{\theta}L_{\rm SM}^{\rm c}(\theta), \\
    \theta^{*}_{\rm FM} &\in\operatorname{argmin}_{\theta}L_{\rm FM}^{\rm c}(\theta),
\end{align}
which are equivalent to the solutions of the original problems [Eqs.~(\ref{opt.SMLoss}) and (\ref{opt.FMLoss})]. Here, the quantities $\nabla\ln \mathcal{P}^{\rm c}_t(\bm{x}|\bm{y})$ and $\bm{\nu}^{\mathcal{P}^{\rm c}}_t(\bm{x}|\bm{y})$ are linear functions of $\bm{x}$ because of the Gaussian property, and the conditional probability $\mathcal{P}^{\rm c}_t(\bm{x}|\bm{y})$ can be sampled more easily than $\mathcal{P}_t(\bm{x})$ due to the reproductive property of Gaussian distributions. Therefore, these optimization problems [Eqs.~(\ref{eq.cSMLoss}) and (\ref{eq.cFMLoss})] are numerically simpler than the original optimization problems [Eqs.~(\ref{eq.SMLoss}) and (\ref{eq.FMLoss})].

\subsection{Examples of diffusion models}\label{sec.concreteSystem}
We present some representative examples of diffusion models \cite{nichol2021improved,song2020denoising,lipman2022flow,ho2020denoising}. Here we discuss the conditional Gaussian formulations from the previous section. We consider the temperature $T_t$ and the external force $\bm{f}_t(\bm{x})$ as functions of non-negative parameters $\sigma_t (\geq 0)$, $m_t (\geq 0)$ such that
\begin{align}
    &T_t=m_t\sigma_t \partial_t \left(\frac{\sigma_t}{m_t}\right) = \partial_t \left( \frac{\sigma_t^2}
{2} \right) - \sigma_t^2 \partial_t \ln m_t ,
    \label{eq.temperature}\\\
    &\bm{f}_t(\bm{x})=\nabla\left(\frac{\partial_t \ln m_t}{2}\left\|\bm{x}\right\|^2\right) = (\partial_t \ln m_t) \bm{x}.
    \label{eq.externalForce}
\end{align}
The external force $\bm{f}_t(\bm{x})$ is described by the linear function [Eq.~(\ref{linearf})] with $\mathsf{A}_t = (\partial_t \ln m_t) \mathsf{I}$, $\bm{b}_t =\boldsymbol{0}$, where $\mathsf{I}$ is the identity matrix and $\boldsymbol{0}$ is the zero vector. The diffusion dynamics can be determined by the time dependence of $m_t$ and $\sigma_t$ instead of the time dependence of $T_t$ and $\bm{f}_t(\bm{x})$. The time dependence of $m_t$ and $\sigma_t$ is also called the noise schedule~\cite{nichol2021improved,song2020denoising,lipman2022flow,ho2020denoising}. 

We sometimes make the following assumption for the noise schedules. To satisfy $T_t \geq 0$, the monotony condition
\begin{align}
    \begin{split}
        \partial_t\left(\frac{\sigma_t}{m_t}\right)\geq 0,
    \end{split}\label{eq.noiseSceduleCondition}
\end{align}
is assumed. The term $m_t/\sigma_t$ is called the signal-to-noise ratio~\cite{kingma2021variational}, which must be monotonically non-increasing. Furthermore, we assume the initial conditions $m_{t=0}=1$ and $\sigma_{t=0}=0$. Under these initial conditions, the solutions of the mean $\bm{\mu}_t(\boldsymbol{y})$ and the covariance matrix $\mathsf{\Sigma}_t$ in $\mathcal{P}^{\rm c}_t(\bm{x}|\bm{y})$ are given by
\begin{align}
\bm{\mu}_t(\bm{y})=m_t\bm{y}, \: \: \mathsf{\Sigma}_t = \sigma_t^2\mathsf{I},
\end{align}
[see Appendix~\ref{ap.GaussianPath}]. For high-dimensional datasets such as images, we often use this assumption~\cite{ho2020denoising,song2020denoising,lipman2022flow,song2020scorebased}. In the following, we show some examples of the noise schedules.

\subsubsection*{Example 1: Variance exploding diffusion (VE-diffusion)}
We first introduce the variance exploding diffusion (VE-diffusion)~\cite{song2020scorebased}. The VE-diffusion is a method based on the noise conditional score networks (NCSN)~\cite{song2019generative}, which is described by the following Langevin equation~\cite{song2020scorebased} 
\begin{align}
    d\bm{x}_t=\sqrt{2T_t}dB_t.
\end{align}
This equation corresponds to the condition that there is no external force $\bm{f}_t(\bm{x})=\bm{0}$. From Eqs.~(\ref{eq.temperature}) and (\ref{eq.externalForce}), we obtain $\partial_t \ln m_t =0$ and
\begin{align}
T_t = \partial_t \left( \frac{\sigma_{t}^2}{2} \right).
\end{align}
The condition $\partial_t \ln m_t =0$ means that the noise schedule is given by $m_t=1 \:(0 \leq t \leq \tau)$ because the initial condition is given by $m_0=1$. Because of Eq.~(\ref{eq.noiseSceduleCondition}), $\sigma_t$ should be a monotonically non-decreasing function of time. While we explain the VE-diffusion in terms of the Langevin equation, this method can be described in the implementation as the probabilistic flow ODE.

\subsubsection*{Example 2: Conditional optimal transport}
Next, we introduce the conditional optimal transport (cond-OT) schedule used in flow matching. The cond-OT schedule is known as a noise schedule that optimizes the transport of the conditional distribution $\mathcal{P}_t^{\rm c}(\bm{x}|\bm{y})$~\cite{lipman2022flow}, which is described as follows,
\begin{align}
\begin{split}
    &m_t=1-\frac{t}{\tau},\\
    &\sigma_t=\frac{t}{\tau}.
\end{split}\label{eq.CondOTSchedule}
\end{align}
This noise schedule represents the change of the parameters along the geodesic from $(\sigma_t, m_t) = (0,1)$ to $(\sigma_t, m_t) = (1,0)$ on the $2$-dimensional Euclidean space of the standard deviation $\sigma_t$ and the mean $m_t$. The dynamics along the geodesic can be regarded as optimal transport based on the $2$-Wasserstein distance for the conditional probability distributions~\cite{lipman2022flow,tong2023improving}.

\subsubsection*{Example 3: Variance Preserving diffusion (VP-diffusion)}
Finally, we introduce the variance preserving diffusion (VP-diffusion)~\cite{song2020scorebased}. The VP-diffusion is a method based on the denoising diffusion probabilistic model (DDPM)~\cite{ho2020denoising,nichol2021improved}. The DDPM is a diffusion model that uses the diffusion from the input data distribution to the Gaussian distribution $\mathcal{N}(\bm{x}|\bm{0},\mathsf{I})$. The VP-diffusion is described by the following Langevin equation~\cite{song2020scorebased},
\begin{align}
    d\bm{x}_t=-T_t\bm{x}_tdt+\sqrt{2T_t}dB_t.
\end{align}
From Eq.~(\ref{eq.Langevin}) and (\ref{eq.externalForce}) we get $\bm{f}_t (\bm{x})=-T_t\bm{x}$ and $T_t=-\partial_t \ln m_t $. By substituting $T_t=-\partial_t \ln m_t $ into Eq.~(\ref{eq.temperature}), we also obtain the relation
\begin{align}
    \partial_t \ln (m_t^2) = \frac{\partial_t (\sigma_t^2)}{\sigma_t^2 -1} = \partial_t \ln (1- \sigma^2_t).
\end{align}
Since the initial conditions are given by $\sigma_0 =0$ and $m_0 =1$, the condition 
\begin{align}
    m_t^2+\sigma_t^2=1,
\end{align}
holds. This condition implies that $m_t$ and $\sigma_t$ are on the unit circle. Thus, we can consider a noise schedule that changes the parameters along the geodesic on the unit circle from $(\sigma_t, m_t) = (0,1)$ to $(\sigma_t, m_t) = (1,0)$ as follows,
\begin{align}
\begin{split}
    &m_t=\cos\left(\frac{\pi}{2}\frac{t}{\tau}\right),\\
    &\sigma_t=\sin\left(\frac{\pi}{2}\frac{t}{\tau}\right).
    \end{split}\label{eq.cosineSchedule}
\end{align}
This noise schedule is called the cosine schedule~\cite{nichol2021improved}.

Under the condition of the VP-diffusion ($\sigma_t^2+m_t^2=1$), we consider the coordinate transformation to treat the diffusion process without the external force. The method based on the coordinate transformation is called the denoising diffusion implicit model (DDIM)~\cite{song2020denoising}. The coordinate transformation in the DDIM is introduced as follows,
\begin{align}
    \bar{\bm{x}}=\frac{\bm{x}}{m_t}.\label{eq.DDIMConvert}
\end{align}
Under this coordinate transformation, the conditions Eq.~(\ref{eq.temperature}) with $T_t = -\partial_t \ln m_t$, the Fokker-Planck equation is given by
\begin{align}
    \partial_t  \bar{\mathcal{P}}_t(\bar{\bm{x}})=
     -\nabla_{\bar{\bm{x}}} \cdot \left[ \left[ \frac{\partial_t \ln m_t}{(m_t)^2} \nabla_{\bar{\bm{x}}} \ln \bar{\mathcal{P}}_t(\bar{\bm{x}})\right]\bar{\mathcal{P}}_t(\bar{\bm{x}}) \right],
\end{align}
(see Appendix~\ref{sec.fokker-changev}),
where $\bar{\mathcal{P}}_t(\bar{\bm{x}})\coloneqq\mathcal{P}_t(\bm{x})(|d\bar{\bm{x}}/d\bm{x}|)^{-1}$, $\nabla_{\bar{\bm{x}}} \coloneqq m_t \nabla$, and $|d\bar{\bm{x}}/d\bm{x}|$ is the Jacobian. Therefore, the corresponding Langevin equation is 
\begin{align}
    d\bar{\bm{x}}_t=\sqrt{\frac{2T_t}{(m_t)^2}}dB_t.
\end{align}
This Langevin equation describes a situation where the external force is absent. In the DDIM, we generate data using the probability flow ODE [Eq.~(\ref{eq.deterministicDiffusion})]~\cite{song2020denoising}.

\section{Relationships between stochastic thermodynamics and diffusion models}\label{Stochastic-thermo&diffusion}
In this section, we review stochastic thermodynamics~(Sec.~\ref{sec.OverDampedThermodynamics}) and the relationship between stochastic thermodynamics and optimal transport theory (Sec.~\ref{sec.OverDampedSystemAndOptimalTransport}) from the perspective of diffusion models.
\subsection{Review on stochastic thermodynamics}\label{sec.OverDampedThermodynamics}
We introduce nonequilibrium thermodynamics for the overdamped Fokker--Planck equation [Eq.~(\ref{eq.FP})], namely stochastic thermodynamics~\cite{seifert2012stochastic}, and discuss its relation to diffusion models. In stochastic thermodynamics, we mainly consider the quantity called the entropy production rate as a measure of thermodynamic dissipation rate. For the overdamped Fokker--Planck equation [Eq.~(\ref{eq.FP})], the entropy production rate $\dot{S}^{\mathrm{tot}}_t$ is defined as
\begin{align}
    \dot{S}^{\mathrm{tot}}_t 
    &=\frac{1}{T_t}\int d\bm{x} \|\bm{\nu}_t^{\mathcal{P}}(\bm{x})\|^2\mathcal{P}_t(\bm{x}).\label{eq.overdampedEntropyProduction}
\end{align}
Here, the Boltzmann constant $k_{\rm B}$ is regularized to $k_{\rm B}=1$. This entropy production rate can be decomposed into the entropy change rate in the system $\dot{S}^{\mathrm{sys}}_t$ and the entropy change rate in the heat bath $\dot{S}^{\mathrm{bath}}_t$ as follows,
\begin{align}
    \dot{S}^{\mathrm{tot}}_t 
    =& \dot{S}^{\mathrm{sys}}_t +\dot{S}^{\mathrm{bath}}_t, \\
\dot{S}^{\mathrm{sys}}_t =& \partial_t \left[-\int d\bm{x} \mathcal{P}_t(\bm{x}) \ln \mathcal{P}_t(\bm{x}) \right],\\
 \dot{S}^{\mathrm{bath}}_t =&\frac{\int d\bm{x} \bm{f}_t(\bm{x})\cdot \bm{\nu}_t^{\mathcal{P}}(\bm{x}) \mathcal{P}_t(\bm{x})}{T_t},
\end{align}
where we used the partial integration and $\int d\bm{x} \partial_t  \mathcal{P}_t(\bm{x}) =0$ to derive the expression $\dot{S}^{\mathrm{sys}}_t$.
Thus, the entropy production rate $\dot{S}^{\mathrm{tot}}_t$ is regarded as the entropy change rate in the total system. Its non-negativity $\dot{S}^{\mathrm{tot}}_t \geq 0$ means the second law of thermodynamics. As a measure of thermodynamic dissipation, the entropy production from time $t=0$ to $t=\tau$ is also defined as
\begin{align}
    {S}^{\mathrm{tot}}_{\tau} 
    =&\int_0^{\tau} dt \dot{S}^{\mathrm{tot}}_t.
\end{align}

In the original paper of the diffusion model~\cite{sohl2015deep}, the idea of a reverse process in diffusion models may be inspired by the fluctuation theorem~\cite{evans2002fluctuation,crooks1999entropy, chernyak2006path} or the Jarzynski equality~\cite{jarzynski1997nonequilibrium}, which are the relations between the entropy production and the path probabilities of the forward and backward trajectories. 
Based on mathematical techniques in the fluctuation theorem such as dual dynamics~\cite{seifert2012stochastic}, we introduce the following dynamics
\begin{align}
d\bm{x}_{t}=\bm{f}_{t}(\bm{x}_{t}) d t - 2 \boldsymbol{\nu}_{t}^{\mathcal{P}}(\bm{x}_{t})d t + \sqrt{2T_{t}}dB_{t},\label{eq.backwardLangevin}
\end{align}
which corresponds to the estimated reverse process of the score-based generative modeling [Eq.~(\ref{eq.SMLangevin})] and is a special case of the interpolated dynamics~\cite{ito2023geometric} for the velocity field $-\boldsymbol{\nu}_{t}^{\mathcal{P}}(\bm{x}_{t})$.

Here, we discuss two path probabilities of the path $\Gamma= \{\boldsymbol{x}_{0}, \boldsymbol{x}_{\Delta t}, \cdots,  \boldsymbol{x}_{N_{\tau} \Delta t}\}$. Here, $\Delta t (>0)$ is the infinitesimal time interval, and thus we consider the limit $\Delta t \to 0$ with fixed $\tau = N_{\tau} \Delta t$. For the Langevin equation [Eq.~(\ref{eq.Langevin})], the transition probability from the state $\boldsymbol{x}$ at time $t$ to the state $\boldsymbol{y}$ at time $t+ \Delta t$ is given by the expression of the Onsager--Machlup function~\cite{risken1996fokker} as follows,
\begin{align}
\mathcal{T}_t(\boldsymbol{y}|\boldsymbol{x}) =\frac{1}{(4 \pi T_t \Delta t)^{\frac{n_{\rm d}}{2}}} e^{-\frac{\left\|\boldsymbol{y} -\boldsymbol{x}-\boldsymbol{f}_t (\boldsymbol{x})\Delta t \right\|^2 }{4T_t \Delta t}}.
\label{transitionf}
\end{align}
Similarly, the transition probability from the state $\boldsymbol{x}$ at time $t$ to the state $\boldsymbol{y}$ at time $t+ \Delta t$ for the process [Eq.~(\ref{eq.backwardLangevin})] is given by
\begin{align}
\mathcal{T}^{\dagger}_{t}(\boldsymbol{y}|\boldsymbol{x}) =\frac{1}{(4 \pi T_{t} \Delta t)^{\frac{n_{\rm d}}{2}}} e^{-\frac{\left\|\boldsymbol{y} -\boldsymbol{x}-\boldsymbol{f}_{t} (\boldsymbol{x}) \Delta t + 2 \boldsymbol{\nu}^\mathcal{P}_{t}(\boldsymbol{x}) \Delta t \right \|^2 }{4T_{t} \Delta t }}.
\end{align}
Using the transition probability [Eq.~(\ref{transitionf})], we define the path probability for the forward process as
\begin{align}
\mathbb{P}_{\rm F} (\Gamma)= \mathcal{P}_0 (\boldsymbol{x}_{0}) \prod_{N=0}^{N_{\tau}-1}\mathcal{T}_{N \Delta t}(\boldsymbol{x}_{(N+1)\Delta t}|\boldsymbol{x}_{N \Delta t}),
\end{align}
This path probability $\mathbb{P}_{\rm F} (\Gamma)$ is related to the distribution of the forward process $\mathcal{P}_t(\bm{x})$. If we consider the path excluding the state $\{\boldsymbol{x}_{n \Delta t}\}$ defined as $\Gamma_{t \neq n \Delta t}= \Gamma \setminus \{ \boldsymbol{x}_{n \Delta t}\}$, marginalizing the path probability $\mathbb{P}_{\rm F} (\Gamma)$ gives $\mathcal{P}_{n \Delta t} (\bm{x}_{n \Delta t}) = \int d\Gamma_{t \neq n \Delta t} \mathbb{P}_{\rm F} (\Gamma)$.
Similarly, the path probability for the process in Eq.~(\ref{eq.backwardLangevin}) is given by
\begin{align}
\mathbb{P}_{\rm B} (\Gamma)= \mathcal{P}_0 (\boldsymbol{x}_{0}) \prod_{N=0}^{N_{\tau}-1}\mathcal{T}^{\dagger}_{N \Delta t}(\boldsymbol{x}_{(N+1)\Delta t}|\boldsymbol{x}_{N \Delta t}).
\label{backward1}
\end{align}
We remark that this path probability $\mathbb{P}_{\rm B}$ is not related to the reverse process or the estimated reverse process, and marginalizing the path probability $\mathbb{P}_{\rm B} (\Gamma)$ does not yield $\mathcal{P}_{\tilde{t}}^{\dagger} (\bm{x})$ or $\tilde{\mathcal{P}}^{\dagger}_{\tilde{t}} (\bm{x})$ because this path probability $\mathbb{P}_{\rm B}$ is not introduced as the time-reversed dynamics from the probability distribution function ${\mathcal{P}}^{\dagger}_{0}$ or $\tilde{\mathcal{P}}^{\dagger}_{0}$.

The entropy production can be interpreted as the statistical difference between two path probabilities $\mathbb{P}_{\rm F}$ and $\mathbb{P}_{\rm B}$.
If we consider the Kullback--Leibler divergence between $\mathbb{P}_{\rm F} (\Gamma)$ and $\mathbb{P}_{\rm B} (\Gamma)$ defined as
\begin{align}
D_{\rm KL} (\mathbb{P}_{\rm F} \| \mathbb{P}_{\rm B} ) = \int d\Gamma \mathbb{P}_{\rm F}(\Gamma) \ln \frac{\mathbb{P}_{\rm F}(\Gamma) }{\mathbb{P}_{\rm B}(\Gamma)},
\end{align}
we obtain the following relation between the Kullback-Leibler divergence and the entropy production~\cite{seifert2012stochastic, dechant2021continuous, ito2023geometric}
\begin{align}
{S}^{\mathrm{tot}}_{\tau}  = D_{\rm KL} (\mathbb{P}_{\rm F} \| \mathbb{P}_{\rm B} ) \label{eq.KLEntropyProduction},
\end{align}
(see also Appendix~\ref{ap.KLandEntropyProduction}). 

We remark that there are several expressions of the entropy production as the Kullback-Leibler divergence. For example, the entropy production can be formulated as the projection in information geometry, which is a minimization problem of the Kullback-Leibler divergence~\cite{ito2020unified, ito2023geometric}. In the context of the fluctuation theorem, the expression based on the path probability for the backward trajectory is well discussed. The path probability for the backward trajectory is defined as
\begin{align}
\mathbb{P}_{\rm B}' (\Gamma)= \mathcal{P}_{\tau} (\boldsymbol{x}_{N_{\tau} \Delta t}) \prod_{N=0}^{N_{\tau}-1}\mathcal{T}_{ N \Delta t}(\boldsymbol{x}_{N\Delta t}|\boldsymbol{x}_{(N+1) \Delta t}).
\end{align}
We remark that this path probability $\mathbb{P}'_{\rm B}$ is not related to the reverse process or the estimated reverse process because this path probability $\mathbb{P}'_{\rm B}$ uses the transition probability $\mathcal{T}_t$, which is not the transition probability of the reverse process or the estimated reverse process, in the time-reversed dynamics from the initial condition ${\mathcal{P}}_{\tau}= {\mathcal{P}}^{\dagger}_{0}$. We can also obtain the similar relation between the Kullback-Leibler divergence and the entropy production~\cite{kawai2007dissipation, seifert2012stochastic, ito2023geometric}
\begin{align}
{S}^{\mathrm{tot}}_{\tau}  = D_{\rm KL} (\mathbb{P}_{\rm F} \| \mathbb{P}_{\rm B}' ),
\end{align}
 (see also Appendix \ref{ap.KLandEntropyProduction}). 
We here define the stochastic entropy production for the path $\Gamma$ as $s^{\rm tot}_{\tau}(\Gamma)=\ln [\mathbb{P}_{\rm F}(\Gamma) /\mathbb{P}_{\rm B}'(\Gamma)]$. In stochastic thermodynamics, the identity $\exp[ s^{\rm tot}_{\tau}(\Gamma)]=\mathbb{P}_{\rm F}(\Gamma) /\mathbb{P}_{\rm B}'(\Gamma)$ is known as the detailed fluctuation theorem~\cite{evans2002fluctuation,crooks1999entropy, chernyak2006path}. If we introduce the expected value with respect to $\mathbb{P}_{\rm F}$ as $\mathbb{E}_{\mathbb{P}_{\rm F}} [\cdots ] = \int d\Gamma \mathbb{P}_{\rm F}(\Gamma) \cdots $, the entropy production satisfies $S^{\rm tot}_{\tau} = \mathbb{E}_{\mathbb{P}_{\rm F}} [s^{\rm tot}_{\tau}]$. We also can obtain the following identity
\begin{align}
\mathbb{E}_{\mathbb{P}_{\rm F}} [\exp (-s^{\rm tot}_{\tau} ) ]=1,
\end{align}
where we used the normalization $\int d\Gamma \mathbb{P}_{\rm B}'(\Gamma)=1$.
This formula is known as the integral fluctuation theorem~\cite{evans2002fluctuation,crooks1999entropy}, and a special case of the integral fluctuation theorem is known as the Jarnzynski equality~\cite{jarzynski1997nonequilibrium}.

We remark on the analogous connection between the fluctuation theorem and the diffusion model discussed in the original paper on diffusion models~\cite{sohl2015deep}. In Ref.~\cite{sohl2015deep}, the path probability for the estimated reverse process $\mathbb{P}_{\rm E}$ is introduced as
\begin{align}
\mathbb{P}_{\rm E} (\Gamma)= \tilde{\mathcal{P}}^\dagger_{0} (\boldsymbol{x}_{N_{\tau} \Delta t}) \prod_{N=0}^{N_{\tau}-1}\mathcal{T}^{\rm E}_{ N \Delta t}(\boldsymbol{x}_{N\Delta t}|\boldsymbol{x}_{(N+1) \Delta t}),
\end{align}
and $p(\boldsymbol{x}_0)$ is obtained from the marginalization $p(\boldsymbol{x}_0) = \int d\Gamma_{t \neq 0} \mathbb{P}_{\rm E} (\Gamma)$, where $\mathcal{T}^{\rm E}_t$ is the transition probability for the estimated reverse process. The expression of $\mathbb{P}_{\rm E}$ is analogous to the expression of $\mathbb{P}'_{\rm B}$. While we consider the statistical difference between $\mathbb{P}_{\rm F}$ and $\mathbb{P}'_{\rm B}$ in stochastic thermodynamics, we consider the situation where $\mathbb{P}_{\rm E}$ statistically mimics $\mathbb{P}_{\rm F}$ in diffusion models.
In Ref.~\cite{sohl2015deep}, the transition probability $\mathcal{T}^{\rm E}_t$ is estimated from the problem of maximizing the lower bound $K$ on the model log-likelihood $L(q\| p)= \int d\boldsymbol{x} q(\boldsymbol{x}) \ln p(\boldsymbol{x}) (\geq K)$. The lower bound $K$ includes the terms of the Kullback-Leibler divergence and the differential entropy, and the derivation of $K$ is analogous to the derivation of the Jarzynski equality. The maximization of $K$ is solved as a partial minimization of the terms of the Kullback-Leibler divergence. 

As described above, there are several analogous connections between stochastic thermodynamics and diffusion models. Although not trivial, it may be possible to apply stochastic thermodynamics techniques to diffusion models due to similarities in mathematical structure. In particular, interesting techniques in stochastic thermodynamics are optimal transport theoretic techniques for the minimum entropy production problem. We next discuss optimal transport theory~\cite{villani2009optimal, villani2021topics}. 

\subsection{Optimal transport theory and minimum entropy production}\label{sec.OverDampedSystemAndOptimalTransport}
In this section, we introduce optimal transport theory and its application to stochastic thermodynamics.  
We start with the distance in optimal transport theory, namely the $p$-Wasserstein distance~\cite{villani2009optimal}. The $p$-Wasserstein distance $\mathcal{W}_p(P,Q)$ between probability density functions $P(\bm{x})$ and $Q(\bm{y})$ satisfying  $P(\bm{x})\geq 0$, $\int d\boldsymbol{x} P(\bm{x}) =1$, $Q(\bm{y})\geq 0$ and $\int d\boldsymbol{y} Q(\bm{y}) =1$, is defined as
\begin{align}
\mathcal{W}_p(P,Q) \coloneqq \left( \inf_{\pi\in \Pi(P,Q)} \int d\bm{x}d\bm{y} \| \bm{x}-\bm{y}\| ^p\pi(\bm{x},\bm{y}) \right)^{\frac{1}{p}}.\label{eq.defineWasserstein}
\end{align}
Here $\Pi(P,Q)$ is a set of joint probability density functions defined as
\begin{align}
\Pi(P,Q) =\{ \pi(\boldsymbol{x},\boldsymbol{y})|& \pi(\boldsymbol{x},\boldsymbol{y})\geq 0, \int d\boldsymbol{y}\pi(\boldsymbol{x},\boldsymbol{y}) = P(\boldsymbol{x}), . \nonumber \\
  &\int d\boldsymbol{x}\pi(\boldsymbol{x},\boldsymbol{y}) = Q(\boldsymbol{y})\}.\label{eq.optimaltransportSubject}
\end{align}
If the $p$-th order moment is finite, the $p$-Wasserstein distance $\mathcal{W}_p(P,Q)$ remains finite and is well defined. In the space of probability distribution functions which have a finite $p$-th order moment, the metric space axioms are satisfied~\cite{villani2009optimal} as follows,
\begin{align}
\mathcal{W}_p(P,Q) &\geq 0, \\
\quad \mathcal{W}_p(P,Q)=0 &\Leftrightarrow P=Q, \\
\mathcal{W}_p(P,Q)&=\mathcal{W}_p(Q,P), \\
\mathcal{W}_p(P,Q) &\leq \mathcal{W}_p(P,R)+\mathcal{W}_p(R,Q),
\end{align}
where $P,Q,R$ are probability density functions. It is also known that $\mathcal{W}_p(P,Q)\leq \mathcal{W}_q(P,Q)$ for $p\leq q$ because of H\"{o}lder's inequality~\cite{villani2009optimal}. The $1$-Wasserstein distance and the $2$-Wasserstein distance are well used in machine learning and stochastic thermodynamics due to some mathematical properties.

For example, the $1$-Wasserstein distance has a computational advantage because it is computed from expected values. Due to the Kantorovich--Rubinstein duality~\cite{villani2009optimal}, the $1$-Wasserstein distance is given by the following maximization problem,
\begin{align}
    \mathcal{W}_1(P,Q)=\sup_{\psi \in\mathrm{Lip}^1}(\mathbb{E}_{P}[\psi]-\mathbb{E}_{Q}[\psi]),\label{eq.KantorovichDuality}
\end{align}
where $\mathrm{Lip}^1$ is the set of scalar functions satisfying $1$-Lipschitz continuity defined as $\mathrm{Lip}^1=\{\psi(\bm{x})\;|\; \|\nabla \psi(\bm{x})\|\leq 1\}$, and $\mathbb{E}_{P}[\psi]$ is the expected value with respect to the probability density function $P(\bm{x})$ defined as $\mathbb{E}_{P} \left[ \psi \right] = \int d\bm{x} P(\bm{x}) \psi(\bm{x})$. The computational complexity of the $1$-Wasserstein distance via the above maximization problem is relatively simple. Therefore, the $1$-Wasserstein distance is well used as an objective function in generative models such as the Wasserstein generative adversarial network~\cite{arjovsky2017wasserstein}.

In addition, the $2$-Wasserstein distance has good differential geometric properties. If $P$ and $Q$ are $n_{\rm d}$-dimensional Gaussian distributions $\mathcal{N}(\bm{\mu}_P,\mathsf{\Sigma}_P)$ and $\mathcal{N}(\bm{\mu}_Q,\mathsf{\Sigma}_Q)$ respectively, then the $2$-Wasserstein distance can be computed as follows~\cite{clark1984class},
\begin{equation}
    \begin{split}
    \mathcal{W}_2(P,Q)^2&= \|\bm{\mu}_P-\bm{\mu}_Q\|^2\\
    &+\mathrm{tr}\;[\mathsf{\Sigma}_P+\mathsf{\Sigma}_Q-2((\mathsf{\Sigma}_P)^{\frac{1}{2}}\mathsf{\Sigma}_Q(\mathsf{\Sigma}_P)^{\frac{1}{2}})^{\frac{1}{2}}],
    \end{split}
    \label{eq.WD-Gaussian}
\end{equation}
where $\mathrm{tr}$ denotes the trace of the matrix. If the variance-covariance matrices are given by $\mathsf{\Sigma}_P=\sigma_P^2\mathsf{I}$ and $\mathsf{\Sigma}_Q=\sigma_Q^2\mathsf{I}$, Eq.~(\ref{eq.WD-Gaussian}) is rewritten as
\begin{align}
    \mathcal{W}_2(P,Q)^2&= \|\bm{\mu}_P-\bm{\mu}_Q\|^2+\mathrm{tr}\;[(\sigma_P-\sigma_Q)^2\mathsf{I}]\nonumber \\
    &= \|\bm{\mu}_P-\bm{\mu}_Q\|^2+(\sigma_P-\sigma_Q)^2n_{\rm d} \nonumber\\
    &= \|\bm{\mu}_P-\bm{\mu}_Q\|^2 + \| \sigma_P \bm{1} -\sigma_Q \bm{1} \|^2,
    \label{eq.WD-diagonal}
\end{align}
 where $\bm{1}$ is the vector defined as $[\bm{1}]_i=1$, $(i=1, \dots, n_{\rm d})$. In this case, the $2$-Wasserstein distance is equivalent to the distance between $(\bm{\mu}_P, \sigma_P \bm{1})$ and $(\bm{\mu}_Q, \sigma_Q \bm{1})$ in the $2n_{\rm d}$-dimensional Euclidean space of the means and the standard deviations.

Note that the right-hand side of Eq.~(\ref{eq.WD-Gaussian}) is called the Fr\'{e}chet inception distance (FID)~\cite{heusel2017gans} when $P$ and $Q$ represent the distributions of the real-world image data and the generated image data in the latent space of the inception model~\cite{szegedy2015going}. The right-hand side of Eq.~(\ref{eq.WD-Gaussian}) (i.e., the FID) is always less than $\mathcal{W}_2(P,Q)^2$ for the non-Gaussian distributions $P$ and $Q$~\cite{gelbrich1990formula}, where their means and covariances are given by $(\bm{\mu}_P,\mathsf{\Sigma}_P)$ and  $(\bm{\mu}_Q,\mathsf{\Sigma}_Q)$, respectively.

Next, we explain that the $2$-Wasserstein distance provides a lower bound on the entropy production in stochastic thermodynamics. We start with the following Benamou-Brenier formula~\cite{benamou2000computational},
\begin{align}
\mathcal{W}_2(\mathcal{P}_{\tau},\mathcal{P}_{0}) = \sqrt{\:\inf \tau\int^{\tau}_0 \;dt \int\;d\bm{x} \| \bm{v}_{t}(\bm{x})\|^2 q_{t}(\bm{x}) }, 
\label{eq.Benamou-Brenier}
\end{align}
where the infimum is taken  among all paths $(\bm{v}_{t}(\bm{x}),q_{t}(\bm{x}))_{0\leq t \leq \tau}$ satisfying
\begin{align}
&\partial_t q_{t}(\bm{x}) = - \nabla \cdot (\bm{v}_{t}(\bm{x}) q_{t}(\bm{x})), \nonumber\\
& q_{0}(\bm{x}) = \mathcal{P}_{0}(\bm{x}) , \: q_{\tau}(\bm{x}) = \mathcal{P}_{\tau}(\bm{x}).
\end{align}
Because the Fokker--Planck equation [Eq.~(\ref{eq.FP})] is the continuity equation, we obtain
\begin{align}
[\mathcal{W}_2(\mathcal{P}_{0},\mathcal{P}_{\tau})]^2 &\leq \tau \int^{\tau}_0 dt \int d\bm{x}\| \bm{\nu}^{\mathcal{P}}_{t}(\bm{x})\|^2\mathcal{P}_{t}(\bm{x}),
\end{align}
from  Eq.~(\ref{eq.Benamou-Brenier}). If $T_t$ does not depend on $t$, this result can be written as
\begin{align}
S_{\tau}^{\rm tot} \geq \frac{\mathcal{W}_2(\mathcal{P}_{0},\mathcal{P}_{\tau})^2}{\tau T}.
\label{tspeedlimit}
\end{align}
where $T (=T_t)$ is the time-independent temperature. 
Thus, the $2$-Wasserstein distance provides the formula for the minimum entropy production in a finite time $\tau$ under the condition that the initial distribution $\mathcal{P}_{0}$ and the final distribution $\mathcal{P}_{\tau}$ are fixed~\cite{aurell2012refined}. 

This result [Eq.~(\ref{tspeedlimit})] can be interpreted from the perspective of the geodesic in the space of the $2$-Wasserstein distance~\cite{nakazato2021geometrical,ito2023geometric}. If we consider the infinitesimal time evolution from $t$ to $t+\Delta t$, Eq. (\ref{eq.Benamou-Brenier}) indicates
\begin{align}
&[\mathcal{W}_2(\mathcal{P}_{t},\mathcal{P}_{t+\Delta t})]^2 \nonumber\\
&\leq \Delta t
\int^{t+ \Delta t}_{t} dt' \int d\bm{x}\| \bm{\nu}^{\mathcal{P}}_{t'}(\bm{x})\|^2\mathcal{P}_{t'}(\bm{x}) \nonumber \\
&=  (\Delta t)^2\int d\bm{x} \| \bm{\nu}^{\mathcal{P}}_{t}(\bm{x})\|^2\mathcal{P}_{t}(\bm{x}) +O((\Delta t)^3),\label{eq.localBB}
\end{align}
where $O((\Delta t)^3)$ denotes the Landau's big O notation. If we define the speed in the space of the $2$-Wasserstein distance as
\begin{align}
v_2 (t) = \lim_{\Delta t \to +0}\frac{\mathcal{W}_2(\mathcal{P}_{t},\mathcal{P}_{t+\Delta t})}{\Delta t},\label{eq.v2}
\end{align}
we obtain the lower bound of the entropy production rate~\cite{nakazato2021geometrical} from Eq.~(\ref{eq.localBB}) as
\begin{align}
\dot{S}^{\mathrm{tot}}_t \geq \frac{[v_2(t)]^2}{T_t}.
\label{excessepr}
\end{align}
This lower bound $[v_2(t)]^2/T_t$ is called the excess entropy production rate~\cite{dechant2022geometric, dechant2022geometric2, ito2023geometric}. The equality $\dot{S}^{\mathrm{tot}}_t = [v_2(t)]^2/T_t$ holds when the velocity field is given by a gradient of a potential function $\phi_t(\boldsymbol{x})$ as $\boldsymbol{\nu}_t^{\mathcal{P}}(\boldsymbol{x}) = T_t \nabla \phi_t(\boldsymbol{x})$~\cite{benamou1999numerical, nakazato2021geometrical, ito2023geometric}. The equality condition physically means the situation where the external force $\boldsymbol{f}_t(\boldsymbol{x})$ is given by conservative force $\boldsymbol{f}_t(\boldsymbol{x}) = -\nabla U_t(\boldsymbol{x})$ with the potential energy $U_t(\boldsymbol{x})$. For the case where the external force is given by $\boldsymbol{f}_t(\boldsymbol{x}) =\mathsf{A}_t \boldsymbol{x}+\boldsymbol{b}_t$, the equality holds if $\mathsf{A}_t$ is a symmetric matrix~\cite{sekizawa2023decomposing}. 
The equation~(\ref{excessepr}) can be written as $T_t\dot{S}_t^{\rm tot} \geq [v_2(t)]^2$. The upper bound $T_t\dot{S}_t^{\rm tot}$ can be regarded as the instantaneous dissipative work~\cite{crooks2000path} when we consider the transition between equilibrium states.

If $T_t (=T)$ does not depend on $t$, we obtain the following hierarchy of inequalities for the minimum entropy production known as the thermodynamic speed limits~\cite{nakazato2021geometrical},
\begin{align}
S_{\tau}^{\rm tot} = \int_0^{\tau} dt \dot{S}^{\mathrm{tot}}_t   &\geq \frac{\int_0^{\tau} dt [v_2(t)]^2}{T} \nonumber \\
& \geq \frac{\mathcal{L}_\tau^2}{\tau T} \nonumber \\
& \geq \frac{\mathcal{W}_2(\mathcal{P}_{0},\mathcal{P}_{\tau})^2}{\tau T},
\label{hierachy}
\end{align}
where $\mathcal{L}_\tau = \int_0^{\tau} dt v_2(t)$ is the path length in the space of the $2$-Wasserstein distance, and we used the Cauchy-Schwarz inequality $[\int_0^{\tau} dt][\int_0^{\tau} dt [v_2(t)]^2] \geq [\int_0^{\tau} dt v_2(t)]^2$ and the triangle inequality $\mathcal{L}_{\tau} \geq \mathcal{W}_2(\mathcal{P}_{0},\mathcal{P}_{\tau})$. The path length $\mathcal{L}_\tau$ can also be calculated as $\mathcal{L}_{\tau} = \lim_{\Delta t\to +0} \sum^{N_
{\tau} -1}_{n=0}\mathcal{W}_2(\mathcal{P}_{(n+1)\Delta t},\mathcal{P}_{n\Delta t})$ with fixed $\tau=N_
{\tau} \Delta t$. If the probability density function $\mathcal{P}_t$ evolves along the geodesic in the space of the $2$-Wasserstein distance, the following condition
\begin{align}
v_2(t) = \frac{\mathcal{W}_2(\mathcal{P}_{0},\mathcal{P}_{\tau})}{\tau},
\end{align}
holds. This condition provides the equality condition of Eq.~(\ref{hierachy}),
\begin{align}
\frac{\int_0^{\tau} dt [v_2(t)]^2}{T} =\frac{\mathcal{W}_2(\mathcal{P}_{0},\mathcal{P}_{\tau})^2}{\tau T}.
\end{align}
If the system is driven by a conservative force, $\dot{S}^{\mathrm{tot}}_t = [v_2(t)]^2/T$ holds. Thus, the minimum entropy production ${S}^{\mathrm{tot}}_{\tau}= [\mathcal{W}_2(P_0, P_{\tau})]^2/ (\tau T)$ is achieved when the system is driven by a conservative force and the probability density function $\mathcal{P}_t$ evolves along the geodesic in the space of the $2$-Wasserstein distance.

We discuss the thermodynamic uncertainty relations~\cite{horowitz2020thermodynamic,dechant2022geometric, dechant2022geometric2, ito2023geometric} as other lower bounds on the the entropy production rate $\dot{S}_t^{\rm tot}$ and the excess entropy production rate $[v_2(t)]^2/T_t$.
Since the equality in Eq.~(\ref{excessepr}) is achieved when $\boldsymbol{\nu}^{\mathcal{P}}_t (\boldsymbol{x})= T_t \nabla \phi_t(\boldsymbol{x})$, the following equality
\begin{align}
\frac{[v_2(t)]^2}{T_t} = T_t \int d\boldsymbol{x} \|\nabla \phi_t (\boldsymbol{x}) \|^2 \mathcal{P}_t(\boldsymbol{x}),
\label{velocity-potential}
\end{align}
holds. This gives us the following inequality
\begin{align}
&\left[ \int d\boldsymbol{x} \| \nabla \phi_t (\boldsymbol{x}) \|^2 \mathcal{P}_t(\boldsymbol{x})\right] \mathbb{E}_{\mathcal{P}_t} [ \|\nabla r \|^2] \nonumber\\
&\geq \left[ \int d\boldsymbol{x} [\nabla \phi_t (\boldsymbol{x}) ] \cdot [\nabla r (\boldsymbol{x}) ] \mathcal{P}_t(\boldsymbol{x} )\right]^2 \nonumber \\
&=\left[ \frac{1}{T_t} \int d\boldsymbol{x}  r (\boldsymbol{x}) \partial_t \mathcal{P}_t(\boldsymbol{x} )\right]^2 = \left[ \frac{1}{T_t} \partial_t \mathbb{E}_{\mathcal{P}_t}  [r]\right]^2,
\label{inequalitytur}
\end{align}
for any time-independent observable $r(\boldsymbol{x})$, where the expected value $\mathbb{E}_{\mathcal{P}_t}  [r]$ is defined as $ \mathbb{E}_{\mathcal{P}_t}  [r]= \int d\boldsymbol{x} \mathcal{P}_t(\boldsymbol{x} )r(\boldsymbol{x} )$, and we used the Cauchy-Schwarz inequality, the continuity equation $\partial_t \mathcal{P}_{t} (\boldsymbol{x}) = -\nabla \cdot (T_t [\nabla \phi_t(\boldsymbol{x})] \mathcal{P}_t (\boldsymbol{x}) )$ and the partial integration.  From Eqs.~(\ref{excessepr}) and (\ref{inequalitytur}), we obtain the thermodynamic uncertainty relation for the excess entropy production rate~\cite{dechant2022geometric, dechant2022geometric2},
\begin{align}
\dot{S}^{\rm tot}_t \geq \frac{[v_2(t)]^2}{T_t} \geq \frac{\left[ \partial_t \mathbb{E}_{\mathcal{P}_t}  [r]\right]^2}{T_t \mathbb{E}_{\mathcal{P}_t} [ \|\nabla r \|^2]},
\label{tur}
\end{align}
which implies a trade-off relation between the speed of the observable $|\partial_t \mathbb{E}_{\mathcal{P}_t}  [r]|$ and thermodynamic dissipation rate $\dot{S}^{\rm tot}_t$. If we define the normalized speed of the observable $r(\bm{x})$ as $v_r(t)=|\partial_t\mathbb{E}_{\mathcal{P}_t}[r]|/\sqrt{\mathbb{E}_{\mathcal{P}_t}[\|\nabla r\|^2]}$, this inequality [Eq.~(\ref{tur})] can be written as
\begin{align}
    [v_2(t)]^2\geq[v_r(t)]^2.\label{tur2}
\end{align}
Thus, this expression [Eq.~(\ref{tur2})] implies that the speed in the space of the $2$-Wasserstein distance is the upper bound on the speed of any time-independent observable $r(\bm{x})$. We note that this expression [Eq.~(\ref{tur2})] can be regarded as the inequality $[\mathcal{W}_2(P_t, P_{t+\Delta t})/\Delta t ]^2 \geq [\mathcal{W}_1(P_t, P_{t+\Delta t})/\Delta t ]^2$ in the limit $\Delta t \to +0$ if we substitute the $1$-Lipshitz function $\psi \in {\rm Lip}^1$, which gives the $1$-Wasserstein distance, into $r$~\cite{nagayama2023geometric}.

By considering the following maximization problem of this lower bound with respect to the observable $r(\boldsymbol{x})$,
\begin{align}
r^* (\boldsymbol{x})  &\in {\rm argmax}_{r (\boldsymbol{x})} \frac{\left[ \partial_t \mathbb{E}_{\mathcal{P}_t}  [r]\right]^2}{T_t \mathbb{E}_{\mathcal{P}_t} [ \|\nabla r \|^2]}, 
\end{align}
the optimal observable $r^* (\boldsymbol{x})$ gives the speed 
\begin{align}
v_2(t) &= \frac{| \partial_t \mathbb{E}_{\mathcal{P}_t}  [r^*]|}{\sqrt{\mathbb{E}_{\mathcal{P}_t} [ \|\nabla r^* \|^2]}},
\end{align}
and the gradient of the optimal observable $\nabla r^* (\boldsymbol{x})$ is proportional to the velocity field $T_t \nabla \phi_t(\bm{x}) = c \nabla r^* (\boldsymbol{x})$ with the proportional coefficient $c =v_2 (t)/\sqrt{\mathbb{E}_{\mathcal{P}_t} [ \|\nabla r^* \|^2]}$~\cite{dechant2022geometric,dechant2022geometric2}. Even if the external force $\boldsymbol{f}_t(\boldsymbol{x})$ is not conservative, there exists a method to estimate the velocity field from a time series data of the Langevin dynamics by considering the maximization problem based on the other thermodynamic uncertainty relation~\cite{li2019quantifying, otsubo2020estimating, otsubo2022estimating}, which may be useful for
estimating the velocity field in diffusion models. 

\section{Main result}\label{Main-result}
In this section, we explain the main results of the paper. Based on optimal transport theory and stochastic thermodynamics, 
we derive upper bounds on the estimation error in data generation of diffusion models by the entropy production rate, namely the speed-accuracy relations for diffusion models. This upper bound is derived based on an analogy to the thermodynamic speed limit and the thermodynamic uncertainty relation. Based on the upper bound, we can introduce an optimal protocol for the accuracy of data generation in diffusion models given by the geodesic in the space of the $2$-Wasserstein distance. This protocol is analogous to the optimal protocol for minimum entropy production. We discuss the relationship between the optimal protocol and the noise schedules such as the cond-OT schedule and the cosine schedule, and we illustrate the main results numerically.

\subsection{Speed-accuracy relations for diffusion models}~\label{SADM}
\begin{figure}
    \centering
    \includegraphics[width=\linewidth]{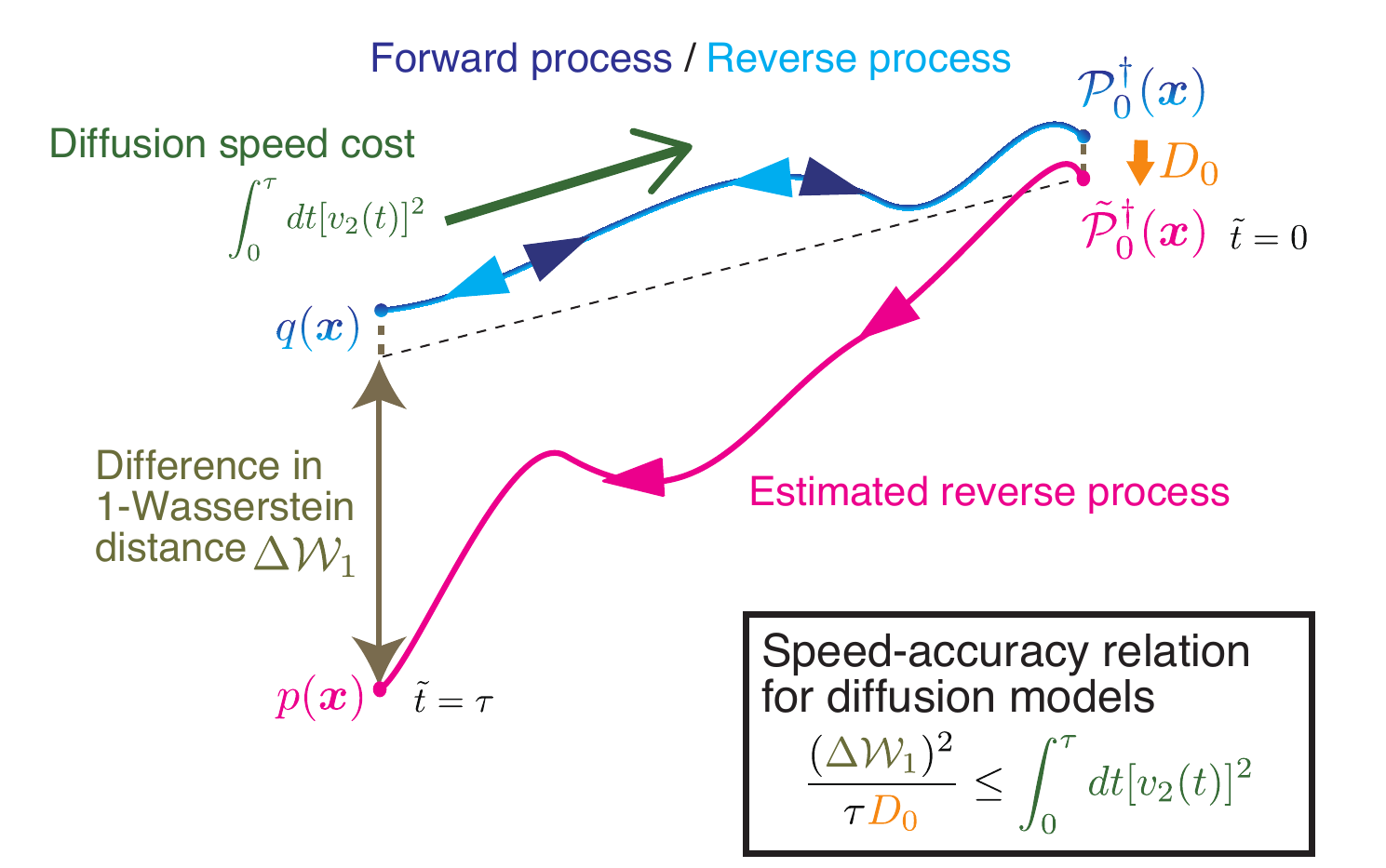}
    \caption{Illustration of the speed-accuracy relation for diffusion models [Eq.~(\ref{eq.SADM1})] in the absence of the non-conservative force}. We evaluate the difference between the initial conditions of the reverse process $\mathcal{P}^{\dagger}_{0}$ and the estimated reverse process $\tilde{\mathcal{P}}^{\dagger}_{0}$ by the Pearson's $\chi^2$-divergence $D_0$, and the estimation error of the generated data by the change of the $1$-Wasserstein distance $\Delta \mathcal{W}_1$.
    The speed-accuracy relation for diffusion models implies that the diffusion speed cost $\int_0^\tau d{t} [v_2(t)]^2$ measured by the $2$-Wasserstein distance in the forward process affects the accuracy of the generated data given by the response function $(\Delta\mathcal{W}_1)^2/D_0$.
    \label{fig:SADMConcept}
\end{figure}

We explain the main result of the paper. If the velocity field of the forward process is accurately reconstructed ($\hat{\bm{\nu}}_t(\bm{x})=\bm{\nu}_t^{\mathcal{P}}(\bm{x})$) in the probability flow ODE or the flow-based generative modeling, we can derive the following inequality
\begin{align}
    \frac{1}{\tau}\frac{(\Delta \mathcal{W}_1)^2}{D_0}
    \leq\int_0^{\tau}dt T_t \dot{S}_t^{\rm tot},\label{DiffusionTUR}
\end{align}
which is a relation between the thermodynamic quantity and the estimation error in data generation. Inspired by the notion of energy-speed-accuracy trade-off relation~\cite{lan2012energy}, we call this result the speed-accuracy relation of diffusion models because $\Delta \mathcal{W}_1$ corresponds to the accuracy, and $T_t \dot{S}_t^{\rm tot}$ can correspond to the speed in the absence of the non-conservative force, as discussed later.
The right-hand side of this inequality is given by thermodynamic quantities such as the entropy production rate $\dot{S}_t^{\rm tot}$ and the temperature $T_t$, and these thermodynamic quantities are computed in the forward process. The quantities on the left-hand side of this inequality are given by the difference between the reverse process and the estimated reverse process. The quantity $D_0$ is the Pearson's $\chi^2$-divergence~\cite{pearson1900x} defined as
\begin{align}
    D_0:=\int d\bm{x} \frac{ (\tilde{\mathcal{P}}^{\dagger}_{0}(\bm{x})-\mathcal{P}^{\dagger}_{0}(\bm{x}))^2 }{\mathcal{P}^{\dagger}_{0}(\bm{x})},
\end{align}
which quantifies the difference between the initial conditions of the reverse process $\mathcal{P}^{\dagger}_0(\bm{x})$ and the estimated reverse process $\tilde{\mathcal{P}}^{\dagger}_0(\bm{x})$ at $\tilde{t}=0$.
This Pearson's $\chi^2$-divergence is the $f$-divergence, which becomes the Fisher information when the difference $\tilde{\mathcal{P}}^{\dagger}_{0}(\bm{x})-\mathcal{P}^{\dagger}_{0}(\bm{x})$ is sufficiently small~\cite{amari2016information}. The quantity $\Delta \mathcal{W}_1:=\mathcal{W}_1(q,p)-\mathcal{W}_1(\mathcal{P}^{\dagger}_{0},\tilde{\mathcal{P}}^{\dagger}_{0})$ is the change of the $1$-Wasserstein distance between reverse and estimated reverse processes from time $\tilde{t}=0$ to $\tilde{t}=\tau$. In generative models, we sometimes consider the estimation error using the $1$-Wasserstein distance $\mathcal{W}_1(q,p)$ for model evaluation~\cite{oko2023diffusion}. If we consider the initial condition satisfying $\mathcal{W}_1(\mathcal{P}^{\dagger}_{0},\tilde{\mathcal{P}}^{\dagger}_{0})\approx 0$, $\Delta \mathcal{W}_1$ becomes the estimation error $\Delta \mathcal{W}_1\approx\mathcal{W}_1(q,p)$.  Therefore, $(\Delta \mathcal{W}_1)^2/D_0$ is considered as a response function, which quantifies how much the perturbation of the initial distribution $D_0$ for the 
estimated reverse process at $\tilde{t}=0$ affects the estimation error of the generated data $\Delta \mathcal{W}_1$ at $\tilde{t}=\tau$ [see also Fig~\ref{fig:SADMConcept}]. If the response function $(\Delta \mathcal{W}_1)^2/D_0$ is smaller, the accuracy of the generated data is better regardless of the initial condition deviations. Thus, a smaller value of $(\Delta \mathcal{W}_1)^2/D_0$ means robustness of diffusion models against the change of the initial distribution $\tilde{\mathcal{P}}^{\dagger}_0 (\boldsymbol{x})$.  In other words, the value of $(\Delta \mathcal{W}_1)^2/D_0$ represents the degree of change in the generated data with respect to the initial perturbation. Therefore, the value of $(\Delta \mathcal{W}_1)^2/D_0$ can also be considered as the sensitivity of the generated data. Our result shows that the thermodynamic quantities corresponding to the thermodynamic dissipation in the forward process $\int_0^\tau dt T_t \dot{S}_t^{\rm tot}$ give the upper bound on the sensitivity of diffusion models $(\Delta \mathcal{W}_1)^2/D_0$.

In particular, when the external force is conservative and thus the velocity field of the forward process is given by the gradient of a potential  ($\bm{\nu}_t^{\mathcal{P}}(\bm{x})=T_t\nabla\phi_t(\bm{x})$), we also obtain
\begin{align}
    \frac{1}{\tau}\frac{(\Delta \mathcal{W}_1)^2}{D_0}\leq\int_0^{\tau} dt [v_2(t)]^2\label{eq.SADM1},
\end{align}
from the equality condition of Eq.~(\ref{excessepr}). This result in the absence of the non-conservative force is also called as the speed-accuracy relation for diffusion models. The right-hand side of this inequality $\int_0^\tau dt [v_2(t)]^2$, which corresponds to the excess entropy production, is given by the diffusion speed measured by the $2$-Wasserstein distance $v_2(t)$ in the forward process. This quantity $\int_0^\tau dt [v_2(t)]^2$ is a purely information-theoretic quantity which is considered as the diffusion speed cost in the forward process [see also Fig~\ref{fig:SADMConcept}].  According to this result [Eq.~(\ref{eq.SADM1})], the response function $(\Delta \mathcal{W}_1)^2/D_0$ becomes larger as the diffusion speed cost $\int_0^\tau dt [v_2(t)]^2$ becomes larger. In other words, the accuracy of the generated data can decrease as the diffusion speed increases in the forward process. Since we usually consider a conservative force in the implementations of the flow-based generative modeling and the probability flow ODE, this result without an explicit expression by thermodynamic dissipation may be more useful and intuitive rather than the general result [Eq.~(\ref{DiffusionTUR})]. 

According to the speed-accuracy relation for diffusion models [Eq.~(\ref{eq.SADM1})], we can consider minimizing the diffusion speed cost $\int_0^\tau d{t} [v_2(t)]^2$ to increase the accuracy of data generation, which is quantified by the smallness of the response function $(\Delta \mathcal{W}_1)^2/D_0$.  Therefore, minimizing the diffusion speed cost $\int_0^\tau d{t} [v_2(t)]^2$ can improve the performance of diffusion models. When the initial state $\mathcal{P}_0$, the final state $\mathcal{P}_\tau$, and the time duration $\tau$ are fixed, the minimization problem can be discussed based on the geodesic in the $2$-Wasserstein metric space. We discuss this minimization problem of the diffusion speed cost in Sec.~\ref{sec.optimal_noise_sce}.

As an instantaneous expression of the speed-accuracy relations for diffusion models, the following detailed inequality can be derived,
\begin{align}
    &\frac{[\partial_{t} \mathcal{W}_1({\mathcal{P}}^{\dagger}_{\tau-t},\tilde{\mathcal{P}}_{\tau-t}^{\dagger})]^2}{{D}_{0}}\leq T_t \dot{S}_t^{\rm tot}.\label{eq.derivativeSADM1}
\end{align}
This equation is also regarded as $[v_{\rm loss}(t)]^2\leq T_t \dot{S}_t^{\rm tot}$, where $v_{\rm loss}(t):= |\partial_{t} \mathcal{W}_1({\mathcal{P}}^{\dagger}_{\tau-t},\tilde{\mathcal{P}}_{\tau-t}^{\dagger})| / \sqrt{{D}_{0}}$ is the change rate of the normalized estimation error. This inequality is analogous to the thermodynamic uncertainty relation [Eq.~(\ref{tur2})]. We also obtain the hierarchy of inequalities corresponding to the speed-accuracy relations for diffusion models as follows,
\begin{align}
    \frac{(\Delta \mathcal{W}_1)^2}{\tau D_0} \leq \int_0^\tau dt [v_{\rm loss} (t)]^
2 \leq  \int_0^\tau dt T_t \dot{S}_t^{\rm tot},
    \label{eq.SADM-loss-nonconservative}
\end{align}
which is analogous to the hierarchy of the thermodynamic speed limits [Eq.~(\ref{hierachy})].

Especially when the external force is conservative, $[v_{2}(t)]^2 = T_t \dot{S}_t^{\rm tot}$ holds. Thus, Eq.~(\ref{eq.derivativeSADM1}) indicates that the speed in the space of the $2$-Wasserstein distance $v_2(t)$ is the upper bound on $v_{\rm loss}(t):= |\partial_{t} \mathcal{W}_1({\mathcal{P}}^{\dagger}_{\tau-t},\tilde{\mathcal{P}}_{\tau-t}^{\dagger})| / \sqrt{{D}_{0}}$ as 
\begin{align}
     [v_2(t)]^2 \geq [v_{\rm loss}(t)]^2.
\end{align}
The equation~(\ref{eq.SADM-loss-nonconservative}) is also rewritten as
\begin{align}
    \frac{(\Delta \mathcal{W}_1)^2}{\tau D_0} \leq \int_0^\tau dt [v_{\rm loss} (t)]^
2 \leq  \int_0^\tau dt [v_2 (t)]^2.
    \label{eq.SADM-loss}
\end{align}

\begin{figure}
    \centering
    \includegraphics[width=\linewidth]{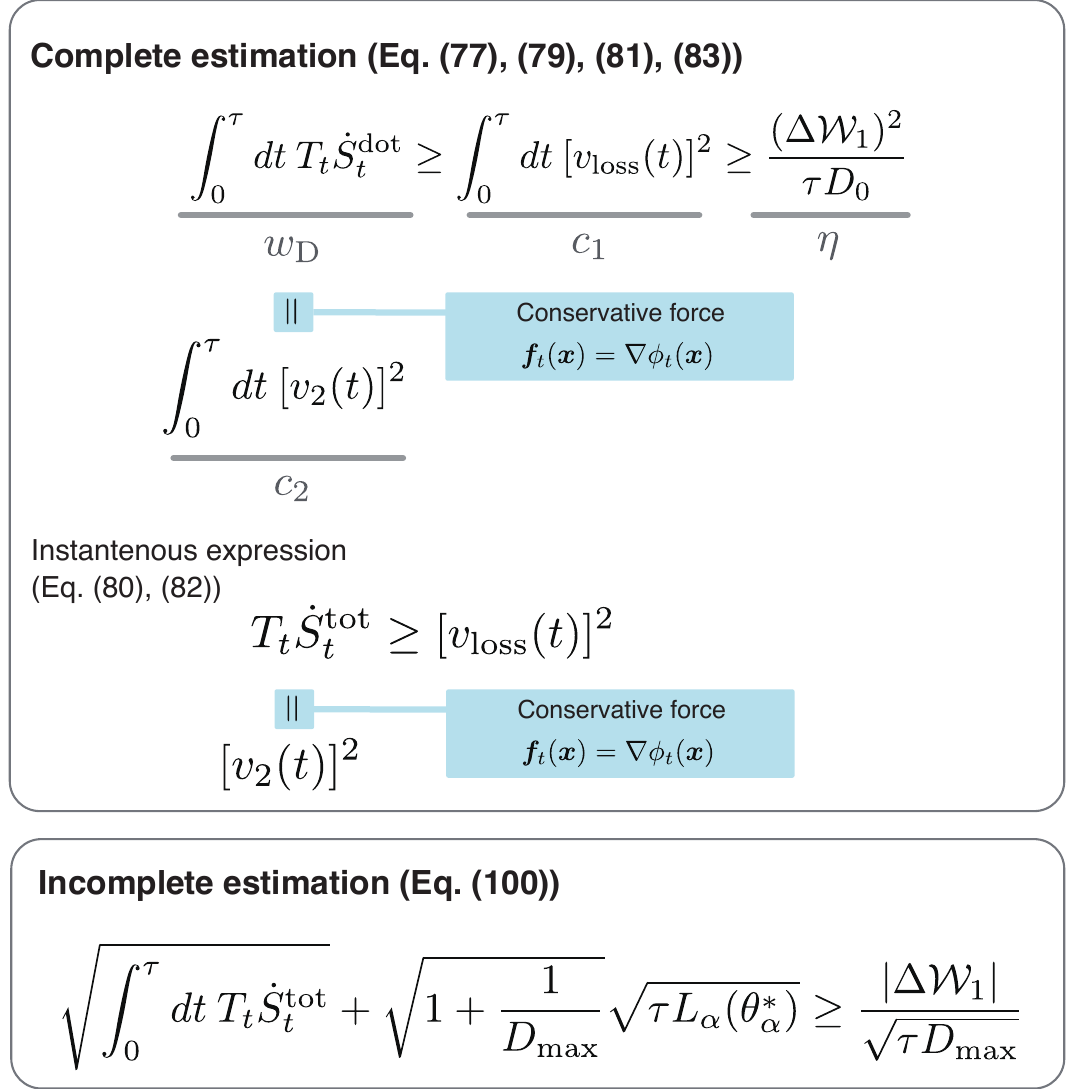}
    \caption{Summary of the main results. We show a hierarchy between the inequalities, and a generalization for incomplete estimation discussed in Sec.~\ref{sec:generalized-sadm}.}
    \label{fig:sadm_logic}
\end{figure}

To summarize, we have illustrated various inequalities in Fig.~\ref{fig:sadm_logic}. As shown in Fig.~\ref{fig:sadm_logic}, we also note that we will later discuss a generalization of the main result for incomplete estimation in Sec.~\ref{sec:generalized-sadm}. In Sec.~\ref{sec.numericalCalculation}, we also use the abbreviated notations $w_{\rm D} = \int_0^{\tau}dt T_t\dot{S}^{\rm tot}_t, c_2 = \int_0^{\tau}dt\:[v_2(t)]^2$, $c_1 = \int_0^{\tau} dt\; [v_{\rm loss}]^2$ and $\eta = (\Delta \mathcal{W}_1)^2/\tau D_0$. Thus, Eqs.~\eqref{eq.SADM-loss-nonconservative} and~\eqref{eq.SADM-loss} can be rewritten as $w_{\rm D} \geq c_1 \geq \eta$ and $c_2 \geq c_1 \geq \eta$, respectively (see also Fig.~\ref{fig:sadm_logic}).

\subsection{Proofs of the main result}\label{sec.derive_sadm}
Here we show the derivation of the main result. We start with the derivation of Eq.~(\ref{eq.derivativeSADM1}). For the sake of simplicity, we introduce a notation for the probability density function of the estimated reverse process $\tilde{\mathcal{P}}_t(\bm{x}):=\tilde{\mathcal{P}}^{\dagger}_{\tau-t}(\bm{x})$ according to the time direction of the forward process. Based on the Kantrovich-Rubinstein duality, the $1$-Wasserstein distance is given by
\begin{align}
\mathcal{W}_1({\mathcal{P}}^{\dagger}_{\tau-t},\tilde{\mathcal{P}}_{\tau-t}^{\dagger})= \mathcal{W}_1(\mathcal{P}_t,\tilde{\mathcal{P}}_t)=\mathbb{E}_{\mathcal{P}_{t}}[\psi^*_t]-\mathbb{E}_{ \tilde{\mathcal{P}}_{t}}[\psi^*_t],
\label{rubinstein-proof}
\end{align}
where $\psi^*_{t}=\underset{\psi\in\mathrm{Lip}^1}{\operatorname{argmax}}\left(\mathbb{E}_{\mathcal{P}_{t}}[\psi]-\mathbb{E}_{ \tilde{\mathcal{P}}_{t}}[\psi]\right)$ is the optimal solution of the maximization problem [Eq.~(\ref{eq.KantorovichDuality})]. Since we assume that the velocity field of the forward process is accurately reconstructed ($\hat{\bm{\nu}}_t(\bm{x})=\bm{\nu}_t^{\mathcal{P}}(\bm{x})$), the continuity equations for the reverse process and the estimated reverse process [Eqs.~(\ref{eq.reverseFP}) and (\ref{eq.inferedFP})] are given by $\partial_{t} \mathcal{P}_{t}(\bm{x})=-\nabla\cdot\left(\bm{\nu}_{t}^{\mathcal{P}}(\bm{x})\mathcal{P}_{t}(\bm{x})\right)$ and $\partial_{t} \tilde{\mathcal{P}}_{t}(\bm{x})=-\nabla\cdot\left(\bm{\nu}_{t}^{\mathcal{P}}(\bm{x})\tilde{\mathcal{P}}_{t}(\bm{x})\right)$. Therefore, the time evolution of the difference between two probability density functions $\delta \mathcal{P}_{t}(\bm{x}):=\mathcal{P}_t(\bm{x})-\tilde{\mathcal{P}}_t(\bm{x})$ is given by
\begin{align}
    \partial_{t}\delta \mathcal{P}_{t}(\bm{x})=-\nabla\cdot\left(\bm{\nu}_{t}^{\mathcal{P}}(\bm{x})\delta \mathcal{P}_{t}(\bm{x})\right).\label{FMdeltaContinuety}
\end{align}
Using Eq.~(\ref{FMdeltaContinuety}), we obtain
\begin{align}
 \partial_t \left(\mathbb{E}_{\mathcal{P}_{t}}[\psi]-\mathbb{E}_{ \tilde{\mathcal{P}}_{t}}[\psi]\right) 
=& \int d\bm{x} \;\psi(\bm{x}) \partial_{t}\delta \mathcal{P}_{t}(\bm{x})\nonumber\\
    =&\int d\bm{x} \;\nabla\psi(\bm{x})\cdot \bm{\nu}_{t}^\mathcal{P}(\bm{x})\delta \mathcal{P}_{t}(\bm{x}), \label{eq.differenceExpectation}
\end{align}
for any time-independent $1$-Lipshitz function $\psi \in\mathrm{Lip}^1$, where we used the partial integration and we assumed that $\delta \mathcal{P}_{t}(\bm{x})$ disappears at infinity. From the Cauchy--Schwarz inequality and the $1$-Lipshitz continuity $\|\nabla \psi (\boldsymbol{x}) \| \leq 1$, we obtain
\begin{align}
&\left(\partial_t \left(\mathbb{E}_{\mathcal{P}_{t}}[\psi]-\mathbb{E}_{ \tilde{\mathcal{P}}_{t}}[\psi]\right) \right)^2\nonumber\\
    &=\left( \int d\bm{x} \;\nabla\psi(\bm{x})\cdot \bm{\nu}_{t}^\mathcal{P}(\bm{x})\delta \mathcal{P}_{t}(\bm{x})\right)^2 \nonumber \\
    &\leq \left(\int d\bm{x}\;\|\bm{\nu}_{t}^\mathcal{P}(\bm{x})\|^2\mathcal{P}_{t}(\bm{x}) \right) \left( \int d\bm{x}\;\|\nabla\psi(\bm{x})\|^2\frac{(\delta \mathcal{P}_{t}(\bm{x}))^2}{\mathcal{P}_{t}(\bm{x})} \right)\nonumber\\
    &\leq \left( \int d\bm{x}\;\|\bm{\nu}_{t}^\mathcal{P}(\bm{x})\|^2\mathcal{P}_{t}(\bm{x}) \right) \left( \int d\bm{x}\; \frac{(\delta \mathcal{P}_{t}(\bm{x}))^2}{\mathcal{P}_{t}(\bm{x})} \right) \nonumber \\
    &= T_t \dot{S}^{\rm tot}_t D_{\tau-t},\label{ap.cauchy}
\end{align}
where ${D}_{\tau-t}$ is the Pearson's $\chi^2$-divergence at time $\tilde{t}=\tau-t$ defined as ${D}_{\tau-t}:=\int d\bm{x} [{\mathcal{P}}^{\dagger}_{\tau-t}(\bm{x})-\tilde{\mathcal{P}}_{\tau-t}^{\dagger}(\bm{x})]^2/\mathcal{P}^{\dagger}_{\tau-t}(\bm{x})$. We can show that the time derivative of $D_{\tau-t}$ is $\partial_t D_{\tau -t } =0$ (see Appendix \ref{sec.proof}). Therefore,  $D_{\tau-t}=D_0\;(=\rm{const.})$ holds, and we obtain the inequality
\begin{align}
&\left|\partial_t \left(\mathbb{E}_{\mathcal{P}_{t}}[\psi]-\mathbb{E}_{ \tilde{\mathcal{P}}_{t}}[\psi]\right) \right|\leq \sqrt{ T_t \dot{S}^{\rm tot}_t D_{0}},\label{eq.diffAbs}
\end{align}
for any $1$-Lipshitz function $\psi\in\mathrm{Lip}^1$. We now implicitly assume the existence of $\partial_{t} \mathcal{W}_1({\mathcal{P}}^{\dagger}_{\tau-t},\tilde{\mathcal{P}}_{\tau-t}^{\dagger})$, which can be justified by the smoothness of $\mathcal{W}_1({\mathcal{P}}^{\dagger}_{\tau-t},\tilde{\mathcal{P}}_{\tau-t}^{\dagger})$ as a function of $t$. In the case of $\partial_{t} \mathcal{W}_1({\mathcal{P}}^{\dagger}_{\tau-t},\tilde{\mathcal{P}}_{\tau-t}^{\dagger}) \geq 0$, we obtain the lower bound on $\sqrt{T_t \dot{S}^{\rm tot}_t D_{0}}$ as follows, 
\begin{align}
0 \leq & \partial_{t}  \mathcal{W}_1({\mathcal{P}}^{\dagger}_{\tau-t},\tilde{\mathcal{P}}_{\tau-t}^{\dagger}) \nonumber \\
=& \lim_{\Delta t \to +0} \frac{\mathcal{W}_1 (\mathcal{P}_{t}, \tilde{\mathcal{P}}_{t}) -\mathcal{W}_1 (\mathcal{P}_{t-\Delta t}, \tilde{\mathcal{P}}_{t-\Delta t}) }{\Delta t}\nonumber\\
\leq& \lim_{\Delta t \to +0} \left[ \frac{\mathbb{E}_{\mathcal{P}_{t}}[\psi^*_{t}]-\mathbb{E}_{ \tilde{\mathcal{P}}_{t}}[\psi^*_{t}]-\mathbb{E}_{\mathcal{P}_{t- \Delta t}}[\psi^*_{t}]+\mathbb{E}_{ \tilde{\mathcal{P}}_{t- \Delta t}}[\psi^*_{t}]}{\Delta t}\right] \nonumber\\
=& \left. \partial_t \left(\mathbb{E}_{\mathcal{P}_{t}}[\psi^*_{s}]-\mathbb{E}_{ \tilde{\mathcal{P}}_{t}}[\psi^*_{s}]\right) \right|_{s=t} \leq \sqrt{T_t \dot{S}^{\rm tot}_t D_{0}},
\label{proofinequality1}
\end{align}
where we used the fact that $\mathcal{W}_1 (\mathcal{P}_{t-\Delta t}, \tilde{\mathcal{P}}_{t-\Delta t}) =  \mathbb{E}_{\mathcal{P}_{t-\Delta t}}[\psi^*_{t-\Delta t}]-\mathbb{E}_{ \tilde{\mathcal{P}}_{t-\Delta t}}[\psi^*_{t-\Delta t}] \geq \mathbb{E}_{\mathcal{P}_{t-\Delta t}}[\psi^*_{t}]-\mathbb{E}_{ \tilde{\mathcal{P}}_{t-\Delta t}}[\psi^*_{t}]$ holds because of the definition of the $1$-Wasserstein distance, and we used the fact that Eq.~(\ref{ap.cauchy}) holds for any $t$-independent $1$-Lipshitz function $\psi^*_{s} \in \mathrm{Lip}^1$. In the case of $\partial_{t}  \mathcal{W}_1({\mathcal{P}}^{\dagger}_{\tau-t},\tilde{\mathcal{P}}_{\tau-t}^{\dagger}) \leq 0$, we obtain the lower bound on $\sqrt{T_t \dot{S}^{\rm tot}_t D_{0}}$ similarly as follows, 
\begin{align}
0 \leq& - \partial_{t}  \mathcal{W}_1({\mathcal{P}}^{\dagger}_{\tau-t},\tilde{\mathcal{P}}_{\tau-t}^{\dagger})  \nonumber \\
=& \lim_{\Delta t \to +0} \frac{\mathcal{W}_1 (\mathcal{P}_{t}, \tilde{\mathcal{P}}_{t}) -\mathcal{W}_1 (\mathcal{P}_{t+\Delta t}, \tilde{\mathcal{P}}_{t+\Delta t}) }{\Delta t}\nonumber\\
\leq& \left.- \partial_t \left(\mathbb{E}_{\mathcal{P}_{t}}[\psi^*_{s}]-\mathbb{E}_{ \tilde{\mathcal{P}}_{t}}[\psi^*_{s}]\right) \right|_{s=t} \leq \sqrt{T_t \dot{S}^{\rm tot}_t D_{0}},
\label{proofinequality2}
\end{align}
where we used the inequality $\mathcal{W}_1 (\mathcal{P}_{t+\Delta t}, \tilde{\mathcal{P}}_{t+\Delta t}) =  \mathbb{E}_{\mathcal{P}_{t+\Delta t}}[\psi^*_{t+\Delta t}]-\mathbb{E}_{ \tilde{\mathcal{P}}_{t+\Delta t}}[\psi^*_{t+\Delta t}] \geq \mathbb{E}_{\mathcal{P}_{t+\Delta t}}[\psi^*_{t}]-\mathbb{E}_{ \tilde{\mathcal{P}}_{t+\Delta t}}[\psi^*_{t}]$ and Eq.~(\ref{ap.cauchy}).
By squaring both sides of the inequalities (\ref{proofinequality1}) and (\ref{proofinequality2}), the resulting inequalities are equivalent to Eq.~(\ref{eq.derivativeSADM1}). Thus, our main result Eq.~(\ref{eq.derivativeSADM1}) is proved regardless of the sign of $\partial_{t}  \mathcal{W}_1({\mathcal{P}}^{\dagger}_{\tau-t},\tilde{\mathcal{P}}_{\tau-t}^{\dagger})$.

We next derive Eq.~(\ref{DiffusionTUR}). By integrating Eq.~(\ref{eq.derivativeSADM1}) with respect to time and applying the Cauchy-Schwartz inequality, we obtain
\begin{align}
\int_0^\tau dtT_t\dot{S}_t^{\rm tot}  &\geq \frac{1}{D_0} \int_0^\tau d{t}[\partial_{t }\mathcal{W}_1(\mathcal{P}_{t},\tilde{\mathcal{P}}_{t})]^2 \nonumber\\
   &\geq \frac{1}{\tau D_0}\left(\int_0^\tau \;d{t}\;\partial_{t} \mathcal{W}_1(\mathcal{P}_{t},\tilde{\mathcal{P}}_{t})\right)^2\nonumber\\
   &=\frac{\left(\Delta \mathcal{W}_1\right)^2}{\tau D_0},\label{eq.cauchy}
\end{align}
which is equivalent to Eq.~(\ref{DiffusionTUR}). This equation is also equivalent to Eq.~(\ref{eq.SADM-loss-nonconservative}) because $v_{\rm loss}(t)=|\partial_{t} \mathcal{W}_1({\mathcal{P}}_{t},\tilde{\mathcal{P}}_{t})| / \sqrt{{D}_{0}}$.

\subsection{Optimality in the noise schedule}\label{sec.optimal_noise_sce}
We discuss the diffusion speed cost $\int_0^{\tau}dt [v_2(t)]^2$, which is the upper bound on the response function in the speed-accuracy relation for diffusion models [Eq.~(\ref{eq.SADM1})] in the absence of the non-conservative force. Lowering this upper bound in Eq.~(\ref{eq.SADM1}) means making the response function $(\Delta \mathcal{W}_1)^2/D_0$ small, and this lowering leads to robust data generation against perturbations of the initial distribution in the estimated reverse process. Since the diffusion speed cost depends only on the dynamics of the forward process, it is possible to discuss the optimality of the noise schedules based on the diffusion speed cost.

We explain why we consider only the upper bound in Eq.~(\ref{eq.SADM1}) and not the upper bound in Eq.~(\ref{DiffusionTUR}). If we start with the upper bound in Eq.~(\ref{DiffusionTUR}), 
the lowering of the upper bound in Eq.~(\ref{DiffusionTUR}), $\int_0^{\tau}dt [v_2(t)]^2$, is achieved only if the non-conservative force does not exist due to the inequality $T_t\dot{S}_t^{\rm tot}\geq [v_2(t)]^2 $ in Eq.~(\ref{tur}).
Thus, we may only need to consider the situation where the non-conservative force is absent, and Eq.~(\ref{DiffusionTUR}) is identical to Eq.~(\ref{eq.SADM1}). Therefore, the absence of the non-conservative force is a necessary condition for an optimal protocol.
 
\begin{figure}
    \centering
    \includegraphics[width=\linewidth]{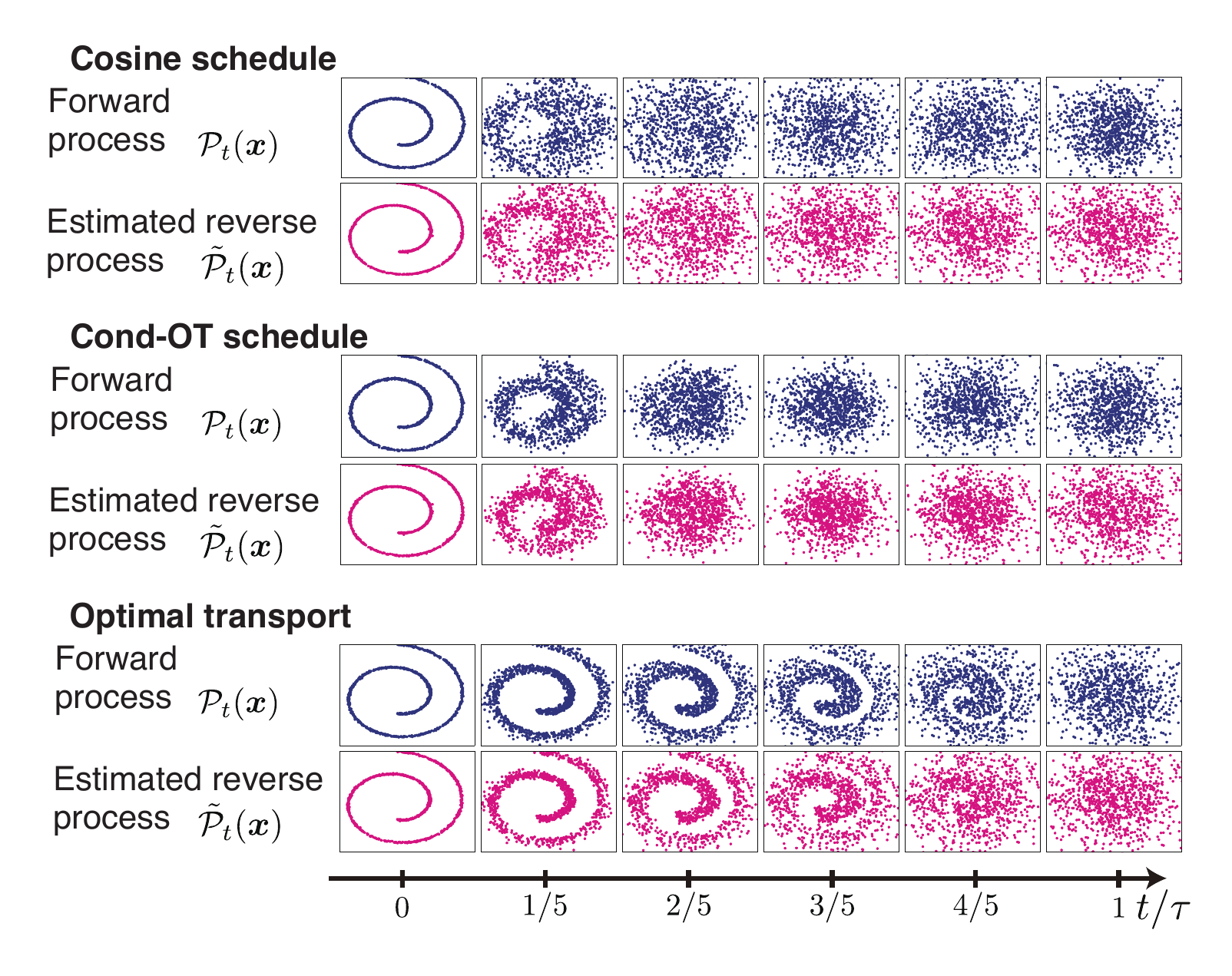}
    \caption{Examples of the forward process and the estimated reverse process according to three noise schedules: the cosine schedule [Eq.~(\ref{eq.cosineSchedule})], the cond-OT schedule [Eq.~(\ref{eq.CondOTSchedule})], and the optimal transport which gives the dynamics along the geodesic in the space of the 2-Wasserstein distance. These figures show the time evolution of samples from a $2$-dimensional Swiss-roll dataset ($n_{\rm{d}}=2$) in the flow-based generative modeling. In all noise schedules,  we used the same sample for the initial distribution of the estimated reverse process $\tilde{\mathcal{P}}_{\tau} (\boldsymbol{x})$. Compared with the cosine schedule and the cond-OT schedule, the noise in the generated figure is relatively small when we consider the optimal transport. This result is consistent with the optimality based on the speed-accuracy relations for diffusion models.}
    \label{fig:swiss_roll}
\end{figure}

We focus on the case where the non-conservative force is absent to discuss an optimal noise schedule.
 Similar to the discussion of the minimum entropy production [Eq.~(\ref{hierachy})], the lower bound of the diffusion speed cost is given by
\begin{align}
    \int_0^{\tau}dt [v_2(t)]^2 \geq  \frac{\mathcal{L}^2_{\tau}}{\tau } \geq \frac{(\mathcal{W}_2(\mathcal{P}_{0},\mathcal{P}_{\tau}))^2}{\tau}.\label{eq.geodesicCondition}
\end{align}
The equality condition is $v_2(t) =\mathcal{W}_2(\mathcal{P}_{0},\mathcal{P}_{\tau})/\tau$, which causes the time evolution of $\mathcal{P}_{t}$ to be driven along the geodesic in the space of the $2$-Wasserstein distance. This time evolution can be discussed in terms of the optimal transport protocol discussed in Ref.~\cite{benamou2000computational}. At least from the perspective of the speed-accuracy relations for diffusion models, an optimal noise schedule is introduced by considering the optimal transport, which causes the time evolution of $\mathcal{P}_{t}$ to be driven along the geodesic in the space of the $2$-Wasserstein distance. We illustrate the example of an optimal noise schedule given by the optimal transport in comparison to other noise schedules such as the cosine schedule and the cond-OT schedule (see Fig.\ref{fig:swiss_roll}).

While optimal transport can lead to accurate data generation, 
it is computationally difficult to construct the diffusion process that achieves optimal transport in high-dimensional data. Therefore, the cond-OT schedule, which can be considered as an approximate optimal transport, is proposed. If the input distribution $q(\bm{x})$ is normalized such that $\mathbb{E}_{q} [\bm{x}] = \boldsymbol{0}$ and $\mathbb{E} _{q} [\|\bm{x}\|^2] = n_{\rm d}$, and the external force and the temperature are given by Eqs.~(\ref{eq.temperature}) and (\ref{eq.externalForce}), the diffusion speed cost is given asymptotically by the conditional kinetic energy $\mathcal{E}_{\rm c}$~\cite{shaul2023kinetic} in the limit $N_{\rm D}/\sqrt{n_{\rm d}} \to 0$,
\begin{align}
    \int_0^{\tau} dt\left[v_2 (t) \right]^2 \simeq \tau n_{\rm d} \mathcal{E}_{\rm c},
\end{align}
 where the dimension of the data $n_{\rm d}$ is sufficiently large compared to the number of data $N_{\rm D}$ [Theorem 4.2 in Ref.~\cite{shaul2023kinetic}]. Here, the conditional kinetic energy $\mathcal{E}_{\rm c}$ for the conditional Gaussian probabilities are given by $\mathcal{E}_{\rm c}=(1/\tau) \int_0^{\tau} dt [(\partial_t\sigma_t)^2+(\partial_tm_t)^2]$~\cite{shaul2023kinetic}.
Thus, if the data dimension is sufficiently larger than the number of data, the optimality based on the speed-accuracy relations for diffusion models can be discussed by considering the minimization of this quantity $n_{\rm d}\int_0^{\tau}dt [(\partial_t\sigma_t)^2+(\partial_t m_t)^2]$ instead of the minimization of the diffusion speed cost $\int_0^{\tau}dt [v_2(t)]^2$.

We explain that minimizing $n_{\rm d}\int_0^{\tau}dt [(\partial_t\sigma_t)^2+(\partial_t m_t)^2]$ leads to the cond-OT schedule [Eq.~(\ref{eq.CondOTSchedule})] and the cosine schedule [Eq.~(\ref{eq.cosineSchedule})]. When $\sigma_t$ and $m_t$ are unconstrained, we obtain an lower bound 
\begin{align}
   &n_{\rm d}\int_0^{\tau}dt [(\partial_t\sigma_t)^2+(\partial_t m_t)^2] \nonumber\\
   &\geq n_{\rm d}\frac{(\sigma_0 - \sigma_{\tau})^2 +(m_0 - m_{\tau})^2}{\tau},
\end{align}
from the Cauchy-Schwarz inequality. 
The equality conditions of this inequality are $\partial_t \sigma_t = ( \sigma_{\tau}-\sigma_0)/\tau$ and $\partial_t m_t = ( m_{\tau}-m_{0})/\tau$, which can be regarded as the cond-OT schedule [Eq.~(\ref{eq.CondOTSchedule})]. Under the constraint of the VP-diffusion, $\sigma_t^2+m_t^2=1$, we can introduce the angle $\theta_t$ which satisfies $(m_t, \sigma_t)=(\cos \theta_t, \sin \theta_t)$. From the Cauchy-Schwarz inequality, we also obtain another lower bound 
\begin{align}
   n_{\rm d}\int_0^{\tau}dt [(\partial_t\sigma_t)^2+(\partial_t m_t)^2]&= n_{\rm d} \int_0^{\tau}dt [(\partial_t\theta_t)^2] \nonumber \\
   &\geq n_{\rm d} \frac{(\theta_0 - \theta_{\tau})^2 }{\tau}.
\end{align}
The equality condition of this inequality is $\partial_t \theta_t = ( \theta_{\tau}-\theta_0)/\tau$, which can be regarded as the cosine schedule [Eq.~(\ref{eq.cosineSchedule})]. The cond-OT schedule or the cosine schedule are considered 
practical suboptimal protocols in terms of the speed-accuracy relations for diffusion models when the dimension of the data is sufficiently larger than the number of data.
\begin{figure*}
    \centering
    \includegraphics[width=\linewidth]{figures/sadm_plot.pdf}
    \caption{Numerical calculations of the speed-accuracy relations for diffusion models [Eqs.~(\ref{eq.SADM1}), (\ref{eq.derivativeSADM1}) and (\ref{eq.SADM-loss})]. (a) The time evolution of the probability density functions in the forward process and the estimated reverse process with the cosine schedule, the cond-OT schedule and the optimal transport (OT). Here $t$ denotes time in the forward direction, and we show the probability density functions at $t=0$, $t=\tau/3$, $t=2\tau/3$ and $t=\tau$. The final state of the forward process is $\mathcal{P}_{\tau}(\bm{x}) = \mathcal{N} (\bm{x}|0,1)$. We used a Gaussian distribution with a different mean $\tilde{\mathcal{P}}_0^{\dagger}(\bm{x}) = \mathcal{N} (\bm{x}|1,1)$ as the initial condition of the estimated reverse process. (b-e) We show the situation where the initial condition of the estimated reverse process $\tilde{\mathcal{P}}_0^{\dagger}(\bm{x})$ is (b) a Gaussian distribution with a different mean, (c) a Gaussian distribution with a different variance, (d) a Gaussian mixture distribution, (e) a uniform distribution. The plot at the top left of each section shows the comparison between $\tilde{\mathcal{P}}_0^{\dagger}(\bm{x})$ and $\mathcal{P}_{\tau}(\bm{x}) = \mathcal{N} (\bm{x}|0,1)$, where $\mathcal{N} (\bm{x}|0,1)$ is shown in gray. In the top right-hand section, we show the bar graph which displays the quantities $\eta=(\Delta \mathcal{W}_1)^2/(\tau D_0)$, $c_{\rm l}=\int_0^{\tau} dt \; [v_{\rm loss}(t)]^2$, $c_2=\int_0^{\tau} dt \;[v_{2}(t)]^2$, and this figure shows the validity of the inequality $\eta \leq c_1 \leq c_2$ [Eq.~(\ref{eq.SADM-loss})]. In the bottom row we show the time evolution of $[v_2 (t)]^2$ and $[v_{\rm loss}(t)]^2$, and this figure shows the validity of the inequality $[v_2 (t)]^2 \geq [v_{\rm loss}(t)]^2$ [Eq.~(\ref{eq.derivativeSADM1})]. Since $[v_2(t)]^2$ and $c_2$ depend only on the forward process, these values are consistent within different initial conditions in (b)-(e). See also Table~\ref{tab.initialConditions} in Appendix~\ref{ap.numericalCalculation} for more detailed explanations, including the probability density function $\tilde{\mathcal{P}}_0^{\dagger}(\bm{x})$ used in (b)-(e).
}
    \label{fig.1dSAGMFM}
\end{figure*}

Finally, we mention about the tightness of the bound in Eq.~(\ref{eq.SADM1}) since lowering the diffusion speed cost $\int_0^{\tau}dt [v_2(t)]^2$ can reduce the response function $(\Delta \mathcal{W}_1)^2/D_0$  if the bound is tight enough. The tightness of the bound in Eq.~(\ref{eq.SADM1}) can be discussed based on the equality condition of the inequalities used in the proof [Eqs.~(\ref{ap.cauchy}), (\ref{proofinequality1}), (\ref{proofinequality2}) and (\ref{eq.cauchy})]. The equality conditions of Eq.~(\ref{ap.cauchy}) are $\bm{\nu}_t^{\mathcal{P}}(\bm{x}) \propto \nabla\psi(\bm{x}) \delta \mathcal{P}_t(\bm{x})/\mathcal{P}_t(\bm{x})$ and $\| \nabla \psi(\boldsymbol{x})\|=1$.
If we consider $ \psi(\boldsymbol{x})=\psi^*_t(\boldsymbol{x})$, the condition $\| \nabla \psi^*_t(\boldsymbol{x})\|=1$ may be satisfied almost everywhere because $\psi^*_t(\boldsymbol{x})$ is the optimal solution of linear programming with the linear inequality constraint $\| \nabla \psi^*_t(\boldsymbol{x})\| \leq 1$~\cite{gulrajani2017improved}. The condition $\bm{\nu}_t^{\mathcal{P}}(\bm{x}) \propto \nabla\psi(\bm{x}) \delta \mathcal{P}_t(\bm{x})/\mathcal{P}_t(\bm{x})$ for $ \psi(\boldsymbol{x})=\psi^*_t(\boldsymbol{x})$ is generally not expected to be satisfied. However, the bound is not so loose if $\bm{\nu}_t^{\mathcal{P}}(\bm{x})$ is not orthogonal to $\nabla\psi(\bm{x}) \delta \mathcal{P}_t(\bm{x})/\mathcal{P}_t(\bm{x})$. If $\left|\partial_t \left(\mathbb{E}_{\mathcal{P}_{t}}[\psi^*_t]-\mathbb{E}_{\tilde{\mathcal{P}}_{t}}[\psi^*_t]\right) \right|$ has a finite value, the non-orthogonality can be guaranteed and the bound may not be loose.
The equality conditions of Eqs.~(\ref{proofinequality1}) and (\ref{proofinequality2}) are $\psi^*_t = \psi^*_{t- \Delta t}$ and $\psi^*_t = \psi^*_{t+ \Delta t}$, respectively. Because $\mathcal{P}_t (\bm{x})\simeq \mathcal{P}_{t\pm \Delta t}(\bm{x})$ and $\tilde{\mathcal{P}}_t (\bm{x})\simeq \tilde{\mathcal{P}}_{t\pm \Delta t}(\bm{x})$ may be satisfied as a consequence of dynamics, $\psi^*_t \simeq \psi^*_{t\pm \Delta t}$ may be expected, and the bound may be tight. The inequalities Eq.~(\ref{eq.cauchy}) become tight when $\partial_t\mathcal{W}_1(\mathcal{P}_t,\tilde{\mathcal{P}}_t)$ is constant. While this condition is generally not expected to be satisfied, the numerical result in Sec.~\ref{sec.numericalCalculation} supports that this condition may be achievable if we use the optimal transport schedule. In Sec.~\ref{sec.numericalCalculation}, we also confirm numerically that both sides of Eq.~(\ref{eq.SADM1}) do not differ by an order of magnitude in simple examples and realistic image generation, and the bound in Eq.~(\ref{eq.SADM1}) can become relatively tight. This fact implies that the discussion in this section may be valid.
\subsection{Numerical experiments}\label{sec.numericalCalculation}
In this section, we first illustrate the validity of our main result [Eq.~(\ref{eq.SADM1})] in a $1$-dimensional situation where all variables of Eq.~(\ref{eq.SADM1}) can be computed numerically by density estimation [Sec.~\ref{sec.1d-gaussian}]. Second, we illustrate a $2$-dimensional situation with a non-conservative force, where we have to consider the bound in Eq.~(\ref{DiffusionTUR}) instead of the bound in Eq.~(\ref{eq.SADM1}). We show the importance of considering the bound given by the entropy production rate in diffusion models [Sec.~\ref{sec.2d-swissroll}]. Finally, we show that our main result [Eq.~(\ref{eq.SADM1})] provides meaningful bounds in practical situations using the real image dataset and trained flow matching models [Sec.~\ref{sec.image-dataset}].
\subsubsection{One-dimensional Gaussian mixture distribution}\label{sec.1d-gaussian}
We illustrate the validity of the speed-accuracy relations for diffusion models by simple numerical experiments. We consider the $1$-dimensional flow-based generative modeling ($\bm{x} \in \mathbb{R}$). We now consider the situation where the velocity field of the forward process is accurately reconstructed and $q(\bm{x})$ is given by the Gaussian mixture distribution
\begin{align}
    &q(\bm{x})=\frac{1}{2}\mathcal{N}(\bm{x}|m^{\rm a},\sigma^2)+\frac{1}{2}\mathcal{N}(\bm{x}|m^{\rm b},\sigma^2).\label{eq.targetDistribution}
\end{align}
Here, $\mathcal{N}(\bm{x}|m,\sigma^2)$ is the $1$-dimensional Gaussian distribution with the mean $m$ and the variance $\sigma^2$.
The velocity field $\bm{\nu}_t^{\mathcal{P}}(\bm{x}) (\in \mathbb{R})$ can be calculated analytically as follows (see Appendix~\ref{ap.numericalCalculation} for details of the numerical calculation),
 \begin{align}
      &\bm{\nu}_t^{\mathcal{P}}(\bm{x}) =\frac{\partial_t\sigma_t}{\sigma_t}\bm{x}-\frac{m_t}{\mathcal{P}_t(\bm{x})}\left(\frac{\partial_t\sigma_t}{\sigma_t}-\frac{\partial_tm_t}{m_t}\right) \mathbb{E}_{\mathcal{P}^c_t, q} [\bm{y}], \\
 &\mathbb{E}_{\mathcal{P}^c_t, q} [\bm{y}] =\frac{1}{2}\frac{\sigma^2 m_t\bm{x}+\sigma_t^2m^{\rm a} \boldsymbol{1}}{\sigma_t^2+\sigma^2m_t^2} \mathcal{N}^{\rm a} + \frac{1}{2}\frac{\sigma^2 m_t\bm{x}+\sigma_t^2m^{\rm b}\boldsymbol{1}}{\sigma_t^2+\sigma^2m_t^2} \mathcal{N}^{\rm b},\label{eq.analyticalVel}
\end{align}
where $\mathcal{N}^{\rm a}$ and $\mathcal{N}^{\rm b}$ are the Gaussian distributions defined as $\mathcal{N}^{\rm x}=\mathcal{N}\left(\bm{x}|m_t m^{\rm x},\sigma_t^2+\sigma^2 m_t^2\right)\;({\rm x}= {\rm a}, {\rm b})$. We set the parameters $m^{\rm a}=-m^{\rm b}=\sqrt{0.35}$ and $\sigma=0.3$. We consider various perturbations of the initial condition in the estimated reverse process $\tilde{ \mathcal{P}}^{\dagger}_{0}(\bm{x})$ at $\tilde{t}=0$ (see also Appendix~\ref{ap.numericalCalculation} for details). We numerically compute the quantities in the speed-accuracy relations for diffusion models [Eqs.~(\ref{eq.SADM1}), (\ref{eq.SADM-loss}) and (\ref{eq.derivativeSADM1})] in Fig. \ref{fig.1dSAGMFM}.

We discuss the interpretations of the numerical results in Fig.~\ref{fig.1dSAGMFM}. In Fig.~\ref{fig.1dSAGMFM}(a), we show the time evolution of the probability density functions in the forward process $\mathcal{P}_t(\bm{x})$ and the time evolution of the probability density functions in the estimated reverse process $\tilde{\mathcal{P}}_t(\bm{x})$ with three noise schedules: the cosine schedule, the cond-OT schedule and the optimal transport. Note that the cosine and cond-OT schedules are not considered the optimal transport because are considering the situation where $N_{\rm D}/\sqrt{n_{\rm d}}$ is not sufficiently small. The initial condition in the estimated reverse process $\mathcal{P}_0^{\dagger}(\bm{x})$ is fixed to the Gaussian distribution $\mathcal{N}(\boldsymbol{x}|1,1)$ while the final state in the forward process is given by the Gaussian distribution $\mathcal{N}(\boldsymbol{x}|0,1)$. In Appendix~\ref{ap.numericalCalculation}, we also show the cases with other initial conditions $\mathcal{P}_0^{\dagger}(\bm{x})$ (see also Fig.~\ref{fig.initialConditions}).
In Fig.~\ref{fig.1dSAGMFM}(a), the probability density functions in the forward process with the cosine schedule and the cond-OT schedule are significantly changed between time $t=0$ and $t=(1/3) \tau$ compared to the optimal transport. Thus, the data structure corresponding to the two peaks of the probability density function is not well recovered in the estimated reverse process for the cosine and cond-OT schedules between $t=(1/3) \tau$ and $t=\tau$ while the two peaks can be seen between $t=(1/3) \tau$ and $t=(2/3) \tau$ in the estimated reverse process for the optimal transport.

Next, we numerically confirm the speed-accuracy relations for diffusion models [Eqs.~(\ref{eq.SADM1}) and (\ref{eq.SADM-loss})] and the detailed inequality [Eq.~(\ref{eq.derivativeSADM1})] hold under the various conditions of $\mathcal{P}_0^{\dagger}(\bm{x})$ in Fig.~\ref{fig.1dSAGMFM}(b)-(e): (b) a Gaussian distribution with a different mean, (c) a Gaussian
distribution with a different variance, (d) a Gaussian mixture distribution and (e) a uniform distribution. Here, we used the notation $\eta \leq c_1 \leq c_2$ for Eq.~(\ref{eq.SADM-loss}), where $\eta=(\Delta \mathcal{W}_1)^2/D_0$, $c_1=\int_0^{\tau}dt [v_{\rm loss}(t)]^2 $ and $c_2=\int_0^{\tau}dt [v_2(t)]^2$. We confirm that the speed-accuracy relations for diffusion models are valid under all conditions. We also observe that the bounds in Eq.~(\ref{eq.derivativeSADM1}) are tighter in the case of optimal transport (OT) than the bounds in the cases of the cosine and cond-OT schedules. There is a tendency for the bound in the detailed inequality [Eq.~(\ref{eq.derivativeSADM1})] to become loose during the time evolution for the cosine schedule, and the bound in the detailed inequality [Eq.~(\ref{eq.derivativeSADM1})] to become loose at the beginning and at the end of the time evolution for the cond-OT schedule. The upper bounds of the response function $\eta$, namely $c_1$ and $c_2$, are tighter for the optimal transport compared to the cosine and cond-OT schedules. The value of the response function $\eta$ for the optimal transport is also the smallest for any schedule (see Tab.~\ref{tab.ValuesOfEta}), which supports our conclusion that the optimal transport provides the most accurate data generation. We also confirm that $[v_{\rm loss}(t)]^2$ is constant, and thus $|\partial_t \mathcal{W}_1(\mathcal{P}_t, \tilde{\mathcal{P}}_t)|$ is also constant in the case of OT. Moreover, our bounds become relatively tight in the case of OT for conditions (b) and (c).

\begin{table}
    \centering
    \begin{tabular}{cc| c}
         Initial conditions&Noise schedules& Values of $\eta$  \\
        \hline\hline
         \begin{tabular}{c}Gaussian distribution\\ with a different mean\end{tabular}&\begin{tabular}{c}
              Cosine\\Cond-OT\\OT
         \end{tabular} &\begin{tabular}{c}
              $9.188 \times 10^{-2}$\\$8.781\times 10^{-2}$\\ $8.537\times 10^{-2}$
         \end{tabular}\\\hline \begin{tabular}{c}
         Gaussian distribution\\ with a different variance
         \end{tabular}&\begin{tabular}{c}
              Cosine\\Cond-OT\\OT
         \end{tabular} &\begin{tabular}{c}
             $1.078 \times 10^{-1}$\\$1.047\times 10^{-1}$\\$1.022\times 10^{-1}$
         \end{tabular}\\ \hline
         Gaussian mixture distribution&\begin{tabular}{c}
              Cosine\\Cond-OT\\OT
         \end{tabular} &\begin{tabular}{c}
              $2.764 \times 10^{-2}$\\$2.674 \times 10^{-2}$\\$2.611\times 10^{-2}$
         \end{tabular}\\ \hline
         Uniform distribution&\begin{tabular}{c}
              Cosine\\Cond-OT\\OT
         \end{tabular} &\begin{tabular}{c}
              $2.142 \times 10^{-2}$\\$2.068 \times 10^{-2}$\\$2.033 \times 10^{-2}$
         \end{tabular}
    \end{tabular}
    \caption{The values of the response function $\eta=(\Delta\mathcal{W}_1)^2/(\tau D_0)$ in the numerical calculations [in Fig.~\ref{fig.1dSAGMFM}]. Since we computed $3 \times 10^8$ samples, we display the value with four significant digits.}
    \label{tab.ValuesOfEta}
\end{table}

Interestingly, although the values of the upper bounds $c_1$ and $c_2$ for the optimal transport are so small compared to the cosine and cond-OT schedules, the value of the response function $\eta$ for the optimal transport is not so small compared to the cosine and cond-OT schedules. In fact, the probability density functions corresponding to the generated data $p(\bm{x})$ in Fig. \ref{fig.1dSAGMFM}(a) are not significantly different under different conditions. The reason for this fact may be that we are only considering a simple data structure of $1$-dimensional Gaussian mixtures, so larger differences in $\eta$ between different noise schedules may occur when the data structure is more complex. Reducing the time duration $\tau$ and rapidly changing the time evolution in the forward process may also cause large $\eta$ differences between different noise schedules. 

Our discussion based on the speed-accuracy relations for diffusion models may not be sufficient to explain why the cosine and cond-OT schedules work well even when $N/\sqrt{n_{\rm d}}$ is not sufficiently small. Since the value of $\eta$ is indeed approximately equal to the value of $c_1$ for the optimal transport, the value of $\eta$ itself can be considered to be determined by the speed-accuracy relations for diffusion models. To understand why the response function $\eta$ for the optimal transport is not so small compared to the cosine and cond-OT schedules, it may be necessary to consider not only the upper bound on $\eta$ derived in this paper, but also the lower bound on $\eta$.

\begin{figure*}
    \centering
    \includegraphics[width=\linewidth]{figures/sagm_plot_2d.pdf}
    \caption{Numerical experiments of the speed-accuracy relations for diffusion models [Eqs.~(\ref{DiffusionTUR}),~(\ref{eq.derivativeSADM1}) and~(\ref{eq.SADM-loss-nonconservative})] in two-dimensional Swiss-roll situation. (a)-(c) The time evolution of the probability density function in the forward process and the estimated reverse process with the noise schedules using $g=0,3,5$ in Eq.~(\ref{eq:2dns}). The value $g$ stands for the degree of non-conservativity and $g=0$ denotes a situation with conservative external force $\bm{f}_t(\bm{x})$. (d) The time evolution of the probability density function with optimal transport noise schedule. In (a)-(d) $t$ denotes time in the forward direction, and we show the probability density functions at $t=0$, $t=\tau/3$, $t=2\tau/3$ and $t=\tau$. Each axis of the color map denotes the axis of the data space $\mathbb{R}^{n_{\rm d}}$. Throughout this experiment, we used a Gaussian distribution with different variance $\tilde{\mathcal{P}}_0^{\dagger}(\bm{x})=\mathcal{N}(\bm{0},(0.9)^2\mathsf{I}$) as the initial condition of the estimated reverse process. (e) Temporal evolution of  $T_t\dot{S}_t^{\rm tot}$ and $[v_{\rm loss}(t)]^2$ proving the validity of $[v_{\rm loss}(t)]^2\leq T_t\dot{S}_t^{\rm tot}$[Eq.~(\ref{eq.derivativeSADM1})].  (f) A bar graph showing the quantities $\eta = (\Delta \mathcal{W}_1)^2/(\tau D_0)$, $c_{\rm l} = \int_0^{\tau}dt\:[v_{\rm loss}(t)]^2, w_{\rm D}=\int_0^{\tau}dt\: T_t\dot{S}_t^{\rm tot}$. This figure shows the relationship $\eta\leq c_{\rm l}\leq w_{\rm D}$  [Eq.~(\ref{eq.SADM-loss-nonconservative})].   For detailed explanations see Appendix~\ref{ap.numericalCalculation}.}
    \label{fig:2dsadm}
\end{figure*}

\subsubsection{Two-dimensional Swiss-roll dataset}\label{sec.2d-swissroll}

Second, we illustrate a situation where the non-conservative force is present and we need to consider the bound given by the entropy production rate $\dot{S}_t^{\rm tot}$ [Eq.~(\ref{DiffusionTUR})] instead of the bound given by $v_2(t)$ [Eq.~(\ref{eq.SADM1})]. Specifically, we consider a $2$-dimensional Swiss roll dataset and use a noise schedule with a non-conservative external force as
\begin{align}
\begin{split}
    &\bm{f}_t(\bm{x})=\begin{pmatrix}
        H_t&-G_t\\G_t&H_t
    \end{pmatrix}\bm{x},\\
    &H_t = -h \left(1-\frac{t}{\tau}\right) - T_t,\\
    &G_t = g \left(1-\frac{t}{\tau}\right),\\
    &T_t = T_{\rm fin}\left(\frac{t}{\tau}\right)^2,
\end{split}\label{eq:2dns}
\end{align}
where $h\in\mathbb{R}_{>0}$, $T_{\rm fin}\in\mathbb{R}_{>0}$ and $g\in \mathbb{R}_{\geq 0}$ are parameters to control the noise schedule. In case of $g=0$, this external force $\bm{f}_t(\bm{x})$ is conservative. In the case of $g >0$, this external force $\bm{f}_t(\bm{x})$ is non-conservative. Since it is difficult to derive a closed form of the velocity field in this situation, we adopted a denoising score matching method with sufficiently large artificial neural networks and simulated the forward and estimated reverse processes using the estimated score function to evaluate Eq.~(\ref{DiffusionTUR}). We set the parameters $h=1.1,T_{\rm fin}=0.5$ and changed the value of $g$ in $\{0, 3, 5\}$. We set the initial condition of
the estimated reverse process $\tilde{\mathcal{P}}_{\tau}(\boldsymbol{x})$ as the Gaussian distribution with different variance $\mathcal{N}(\bm{0},(0.9)^2\mathsf{I})$ because the probability distribution ${\mathcal{P}}_{\tau}(\boldsymbol{x})$ becomes the Gaussian distribution $\mathcal{N}(\bm{0}, \mathsf{I})$ in the limit $\tau \to \infty$ (see also Appendix~\ref{ap.numericalCalculation}). For comparison, we also numerically estimated the optimal transport noise schedule using a method introduced in~\cite{liu2022flow} (see also Appendix~\ref{ap.numericalCalculation}). We numerically computed the quantities in the speed accuracy relations for diffusion models [Eqs.~(\ref{DiffusionTUR}) and (\ref{eq.derivativeSADM1})] in Fig.~\ref{fig:2dsadm}.

Through the illustration in Fig.~\ref{fig:2dsadm}, we can confirm the validity of minimizing the upper bound of our results [Eqs.~(\ref{DiffusionTUR}),~(\ref{eq.derivativeSADM1}) and~(\ref{eq.SADM-loss-nonconservative})] against improving the robustness of the diffusion model. In Fig.~\ref{fig:2dsadm}(a)-(d), we can see that the generated data $\tilde{\mathcal{P}}_0(\bm{x}) =p(\bm{x})$ becomes significantly different from the training data $\mathcal{P}_0(\bm{x})=q(\bm{x})$ in the case of $g=0,3,5$ [(a)-(c)] compared to the OT noise schedule [(d)]. We confirm that the bounds [Eqs.~(\ref{DiffusionTUR}),~(\ref{eq.derivativeSADM1}) and~(\ref{eq.SADM-loss-nonconservative})] are valid for each noise schedule in Fig.~\ref{fig:2dsadm}(e) and Fig.~\ref{fig:2dsadm}(f).
We also confirm that minimizing the thermodynamic cost $w_{\rm D}=\int_0^{\tau} dt\: T_t\dot{S}_t^{\rm tot}$ qualitatively improves the robustness of the generation in Fig.~\ref{fig:2dsadm}(f). The values of $\eta$ in $g=0,3,5$ differ by orders of magnitude compared to other noise schedules, while the value of $\eta$ in the OT noise schedule remains in an acceptable range. We confirm that minimizing the upper bound $w_{\rm D}$ in Eq.~(\ref{DiffusionTUR}) and considering the OT noise schedule are effective in ensuring that the generated data does not differ from the training data.

We discuss the effects of non-conservative forces and thermodynamic dissipation, which could not be dealt with in the $1$-dimensional Gaussian mixture distribution. The numerical results also show that the strength of the non-conservative forces $g$ affects the sensitivity of the data generation. The sensitivity $(\Delta \mathcal{W}_1)^2/(\tau D_0)$ increases with increasing $g$ in the region up to $g=5$. This trend is similar for the upper bound $w_D$, and it can be said that the sensitivity of data generation in the diffusion model is higher in situations of strong non-conservative forces, where more thermodynamic dissipation occurs. Although non-conservative forces have not been considered in conventional diffusion models using cond-OT and cosine schedules, it can be assumed that non-conservative forces have not been considered to ensure that data generation is robust. On the other hand, if one wants to deliberately create a generated result that is different from the training data, it is necessary to increase the sensitivity $(\Delta \mathcal{W}_1)^2/(\tau D_0)$. Therefore, one could consider a situation where such non-conservative forces are included in diffusion models and thermodynamic dissipation is dared to occur in the forward processes.

In addition, Figs.~\ref{fig:2dsadm}(a) and~\ref{fig:2dsadm}(d) show a situation where OT provides superior generation results even in a conservative situation as discussed in Fig.~\ref{fig.1dSAGMFM}. As mentioned earlier, the noise schedule in Eq.~(\ref{eq:2dns}) with $g=0$ provides a conservative external force $\bm{f}_t(\bm{x})$ and in this case $[v_2(t)]^2=T_t\dot{S}_t^{\rm tot}, w_{\rm D}=\int_0^{\tau}dt\:[v_2(t)]^2$ holds. From this perspective, the comparison between the schedule of $g=0$ and the OT schedule can be seen as a confirmation of Eqs.~(\ref{eq.SADM1}) and (\ref{eq.SADM-loss}), similar to the situation we dealt with in Fig.~\ref{fig.1dSAGMFM}. Looking at Fig.~\ref{fig:2dsadm}(f), we can see that the $g=0$ schedule has a higher $\eta$ than the OT schedule by a sufficient margin [Tab.~\ref{tab.ValuesOfEta2}] compared to the cosine, cond-OT and OT schedules in the previous setup of a Gaussian mixture in $1$-dimension [Tab.~\ref{tab.ValuesOfEta}]. This is because the schedule of $g=0$ is far from the optimal schedule,
while the cosine and cond-OT schedules are suboptimal as discussed in Sec.~\ref{sec.optimal_noise_sce}.

\begin{table}
    \centering
    \begin{tabular}{cc| c}
         Initial conditions&Noise schedules& Values of $\eta$  \\
        \hline\hline
         \begin{tabular}{c}
         Gaussian distribution\\ with a different variance
         \end{tabular}&\begin{tabular}{c}
              $g=0$\\$g=3$\\$g=5$\\OT
         \end{tabular} &\begin{tabular}{c}
              $9.301 \times 10^{-2}$\\
              $3.430\times 10^{-1}$\\ 
              $5.615\times 10^{-1}$\\
              $7.221\times 10^{-6}$
         \end{tabular}
    \end{tabular}
    \caption{The values of the response function $\eta=(\Delta\mathcal{W}_1)^2/(\tau D_0)$ in the numerical calculations [Fig.~\ref{fig:2dsadm}]. Since we computed $1 \times 10^8$ samples, we display the value with four significant digits.}
    \label{tab.ValuesOfEta2}
\end{table}

\begin{figure*}
    \centering
    \includegraphics[width=\linewidth]{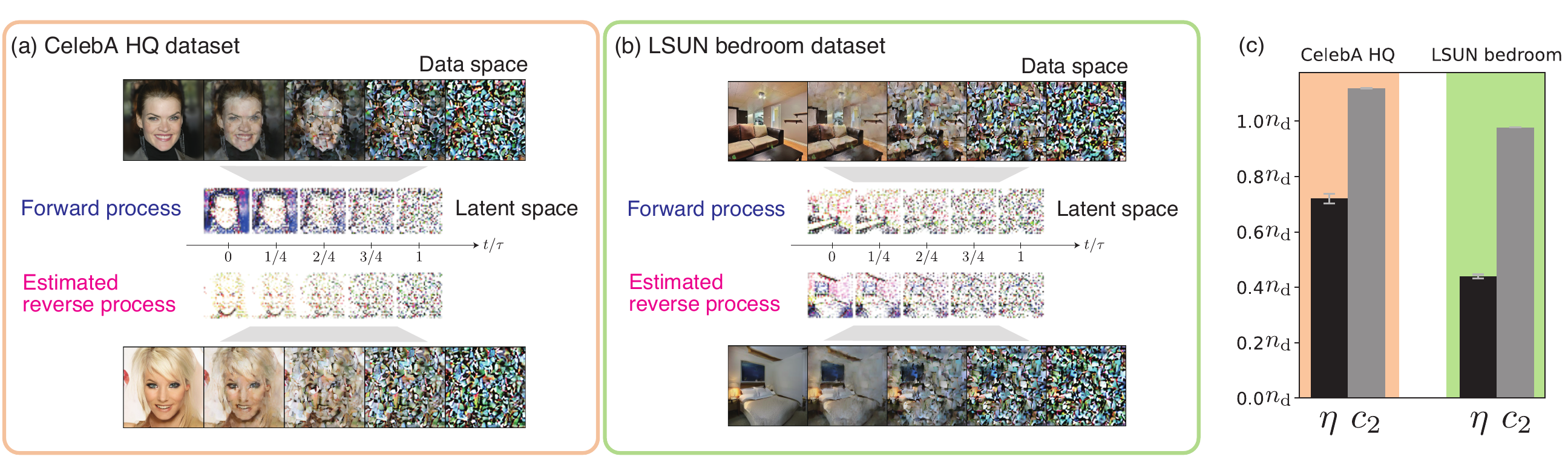}
    \caption{The application of diffusion models using the real-world image datasets and the validity of Eq.~(\ref{eq.SADM1}) on the latent space. (a) The forward process and the estimated reverse process for the downscaled version of the CelebA HQ dataset~\cite{karras2018progressive}.
    (b) The forward process and the estimated reverse process for the bedrooms subset of the LSUN dataset~\cite{yu2015lsun}. 
    (c) The estimated values of $\eta = (\Delta \mathcal{W}_1)^2/(\tau D_0),\; c_2=\int_0^{\tau}dt [v_2(t)]^2$ computed on the latent space. The error bars shows the standard deviation among 5 calculations using $3\times10^4$ samples for each calculations. $n_{\rm d}=4096$ denotes the dimension of the latent space. The inequality $c_2 \geq \eta$ [Eq.~(\ref{eq.SADM1})] on the latent space is still valid for realistic applications of diffusion models using the real-world image datasets.}
    \label{fig:latent-sadm}
\end{figure*}

\subsubsection{Real-world image datasets}\label{sec.image-dataset}

We show that the analysis using our main result [Eq.~(\ref{eq.SADM1})] is feasible in a more realistic situation for the application of diffusion models using the real-world image datasets. We estimated the values of Eq.~(\ref{eq.SADM1}) on the latent space using trained models and benchmark image datasets. Here, we used a trained flow matching model in Ref.~\cite{dao2023flow}, which creates a diffusion process on a latent space~\cite{rombach2022high,dao2023flow}. For the datasets, we used a $256\times 256$ downscaled version of the CelebA HQ dataset~\cite{karras2018progressive} and the bedroom subset of the LSUN dataset~\cite{yu2015lsun}.

In Fig.~\ref{fig:latent-sadm} we have illustrated diffusion models for the real-world image datasets, and the validity of Eq.~(\ref{eq.SADM1}) on the latent space is shown in Fig.~\ref{fig:latent-sadm} (see also the details in Appendix~\ref{ap.numericalCalculation}). These trained models were trained with the cond-OT schedule [Eq.~(\ref{eq.CondOTSchedule})] on the latent space, and we used $\mathcal{N}(\bm{x}|\bm{0},\mathsf{I})$ for the initial distribution of the estimated reverse process on the latent space. In Figs.~\ref{fig:latent-sadm}(a) and \ref{fig:latent-sadm}(b), we have illustrated the samples from the forward process and the estimated reverse process. In this setting, the dimension of the latent space is $n_{\rm d}=4096$, and it may be difficult to estimate the time evolution of the probability density functions in the forward and estimated reverse processes with a finite number of samples. However, we may estimate the quantities in our main result [Eq.~(\ref{eq.SADM1})], as shown in Fig.~\ref{fig:latent-sadm}(c), because we need the velocity field in the forward process and the initial and final probability distributions in the forward and estimated reverse processes to estimate the quantities. We find that our result [Eq.~(\ref{eq.SADM1})] is still valid. Moreover, the response function $(\Delta \mathcal{W}_1)^2/(\tau D_0)$ has the same order of magnitude as the upper bound $\int_0^{\tau}dt [v_2(t)]^2$. Thus, the numerical analysis based on [Eq.~(\ref{eq.SADM1})] may be reasonable even if we discuss diffusion models on the latent space for the real-world image datasets with data of large dimension $n_{\rm d}\gg 1$.

We explain that our treatment based on the response function may be reasonable in realistic situations such as this experiment. In this experiment, the starting point of the estimated reverse process $\tilde{\mathcal{P}}^{\dagger}_0(\bm{x})$ is chosen to be similar to the final state of the forward process $\mathcal{P}_{\tau}(\bm{x})$, so that the estimated reverse process will be close to the genuine reverse process. One of the typical choices of $\tilde{\mathcal{P}}^{\dagger}_0(\bm{x})$ is the standard Gaussian distribution $\mathcal{N}(\bm{x}|\bm{0},\mathsf{I})$, since $\mathcal{P}_{\tau}(\bm{x})$ will converge to this distribution for sufficiently large $\tau$ in the formulation of Eqs.~(\ref{eq.temperature}) and (\ref{eq.externalForce}). However, the discrepancy between $\tilde{\mathcal{P}}^{\dagger}_0(\bm{x})$ and $\mathcal{P}_{\tau}(\bm{x})$ usually exists, since $\mathcal{P}_{\tau}(\bm{x})=\int d\bm{y} \:\mathcal{P}^{\rm c}_{\tau}(\bm{x}|\bm{y})q(\bm{y})$ [Eq.~(\ref{eq.marginalization})] is not generally equal to $\tilde{\mathcal{P}}^{\dagger}_0(\bm{x})=\mathcal{N}(\bm{x}|\bm{0},\mathsf{I})$ with a finite $\tau$. Due to the large dimension of the data, it is also difficult to estimate $\mathcal{P}_{\tau}(\bm{x})$ from the finite number of samples. Therefore, the difference between $\tilde{\mathcal{P}}^{\dagger}_0(\bm{x})$ and $\mathcal{P}_{\tau}(\bm{x})$ may be unavoidable, and our main result based on the response function [Eq.~(\ref{eq.SADM1})] may be useful to consider the effect of the difference between $\tilde{\mathcal{P}}^{\dagger}_0(\bm{x})$ and $\mathcal{P}_{\tau}(\bm{x})$.

\section{Discussions}
\label{discussion}
In this paper, we summarize the relationship between diffusion models and nonequilibrium thermodynamics, and derive the speed-accuracy relations for diffusion models based on techniques from stochastic thermodynamics and optimal transport theory.  The speed-accuracy relations for diffusion models explain that the robustness of data generation to perturbations is generally limited by the thermodynamic dissipation in the forward diffusion process. The speed-accuracy relations for diffusion models also quantitatively explain the validity of using the cosine and cond-OT schedules as the noise schedules in the diffusion model and the importance of using optimal transport in the forward diffusion process. We also derive the speed-accuracy relations for diffusion models as analogues of the thermodynamic trade-off relations such as the thermodynamic uncertainty relations~\cite{horowitz2020thermodynamic,barato2015thermodynamic,dechant2022geometric} and the thermodynamic speed limits~\cite{aurell2012refined,nakazato2021geometrical}.
Unlike the conventional thermodynamic trade-off relations, which consider thermodynamic limits on the speed of the observable in a stochastic process, the speed-accuracy relations for diffusion models are conceptually different because they give limits on how the difference between two different processes, the forward process and the estimated reverse process, changes. Our conclusion for the noise schedule is also justified in light of recent developments in the diffusion model. In our conclusion, the optimal forward process for the data generation should be given by the geodesic in the space of the $2$-Wasserstein distance. In recent years, several methods have been proposed to realize the dynamics along the geodesic approximately in the flow-based generative modeling and the probability flow ODE~\cite{liu2022flow,pooladian2023multisample,tong2023improving,shaul2023kinetic,kornilov2024optimal}, which is becoming the mainstream methods.

As below, we further discuss the generalization of the main result for the incomplete estimation of the velocity field, a comparison with other similar bounds, a comparison with the conventional understanding of noise schedules, the importance of the entropy production rate in diffusion models, and possible future studies.

\subsection{Generalization of Eq.~(\ref{DiffusionTUR}) for incomplete estimation}\label{sec:generalized-sadm}
The main result is due to the error-free estimation of the velocity field in the forward process and its use in the estimated reverse process. If the velocity field estimation is incomplete, then corrections to the speed-accuracy relations for the diffusion model would have to be made due to the incompleteness. 

A possible generalization of our main result [Eq.~(\ref{DiffusionTUR})] for incomplete estimation can be obtained as
\begin{align}
    \frac{|\Delta \mathcal{W}_1|}{\sqrt{\tau D_{\rm max}}}\leq \sqrt{\int_0^{\tau}dt\: T_t\dot{S}_t^{\mathrm{tot}}} +    \sqrt{1+\frac{1}{D_{\rm max}}}\sqrt{\tau L_{\alpha}(\theta_{\alpha}^*)}\label{eq:underestimateSADM},
\end{align}
where $L_{\alpha}(\theta_{\alpha}^*)$ ($\alpha \in \{ {\rm SM}, {\rm FM} \}$) is the objective function after training defined in Eqs.~(\ref{eq.SMLoss}), (\ref{opt.SMLoss}) and Eqs.~(\ref{eq.FMLoss}), (\ref{opt.FMLoss}), and $D_{\rm max}:=\max_{t\in [0,\tau]}D_t$ is the maximum value of the $\chi^2$-divergence. The derivation of this bound is given in Appendix~\ref{sec:proofIncompleteEstimate}. We can see that  Eq.~(\ref{eq:underestimateSADM}) becomes Eq.~(\ref{DiffusionTUR}) when the objective function is zero ($L_{\alpha}(\theta_{\alpha}^*)=0$) since $D_{\rm max}=D_0$ holds in the error-free estimation of the velocity field (see Appendix~\ref{sec.proof}).

If the estimation of the velocity field is satisfactorily achieved and $L_{\alpha}(\theta_{\alpha}^*)\ll 
 (1+1/D_{\rm max})^{-1}\tau^{-1}\int_0^{\tau}dt\: T_t\dot{S}_t^{\mathrm{tot}}$ holds, we obtain approximately a similar form of our main result [Eq.~(\ref{DiffusionTUR})] by replacing $D_0$ with $D_{\rm max}$, and our understanding based on Eq.~(\ref{DiffusionTUR}) may be useful even for the incomplete estimation of the velocity field. This is due to the fact that the second term of Eq.~(\ref{eq:underestimateSADM}) may be negligible if $ L_{\alpha}(\theta_{\alpha}^*)\ll 
 (1+1/D_{\rm max})^{-1}\tau^{-1}\int_0^{\tau}dt\: T_t\dot{S}_t^{\rm{tot}}$. Thus, noise schedules close to the optimal transport may be a better option to use in this case.

If the estimation of the velocity field is not well achieved, the second term $\sqrt{(1+1/D_{\rm max})\tau L_{\alpha}(\theta_{\alpha}^*)}$ is not negligible. If we want to reduce the left hand side $|\Delta \mathcal{W}_1|/\sqrt{\tau D_{\rm max}}$ by minimizing the upper bound, the noise schedule in the forward process corresponding to $\sqrt{\int_0^{\tau}dt\: T_t\dot{S}_t^{\mathrm{tot}}}$, and the minimization of the objective function corresponding to  $\sqrt{(1+1/D_{\rm max})\tau L_{\alpha}(\theta_{\alpha}^*)}$, are both important. Compared to noise schedules that are close to the optimal transport, we may rather use noise schedules 
which is advantageous for training efficiency at lower values of the objective function.
In fact, it is known that one of the viable options for an approximate optimal transport noise schedule~\cite{tong2023improving} may not achieve the minimum value of the generation error in a particular situation~\cite{fukumizu2024flow}. Therefore, the optimal transport noise schedule may not be the best choice for a particular situation where the objective function is not negligible.

\subsection{Comparison with bounds similar to Eq.~(\ref{DiffusionTUR}) }
We compare our main result [Eq.~(\ref{DiffusionTUR})] with other similar known bounds. As we mentioned in Sec.~\ref{Main-result}, our bound is inspired by the thermodynamic uncertainty relation and the final form [Eq.~(\ref{DiffusionTUR})] resembles the thermodynamic uncertainty relation. However, our result cannot be derived from the thermodynamic uncertainty relation by changing the observable [Eq.~(\ref{tur})]. For example, a derivation of the thermodynamic uncertainty relation in Ref.~\cite{dechant2022geometric} is based on the Cauchy--Schwartz inequality 
\begin{align}
&\left(\int d\bm{x} \mathcal{P}_t(\bm{x})\:\bm{\nu}^{\mathcal{P}}_t(\bm{x})\cdot \nabla r(\bm{x})\right)^2\nonumber\\
&\leq \left( \int d\bm{x} \mathcal{P}_t(\bm{x})\: \|\bm{\nu}_t^{\mathcal{P}}(\bm{x})\|^2 \right)\left( \int d\bm{x} \mathcal{P}_t(\bm{x})\: \|\nabla r(\bm{x})\|^2 \right),
\end{align} 
where $r(\bm{x})$ denotes an arbitrary time-independent scalar observable [see also Eq.~(\ref{tur})]. On the other hand, the derivation of our main result [Eq.~(\ref{DiffusionTUR})] is based on a different Cauchy--Schwartz inequality,
\begin{align}
    &\left(\int d\bm{x} \delta \mathcal{P}_t(\bm{x}) \:\bm{\nu}^{\mathcal{P}}_t(\bm{x})\cdot \nabla \psi(\bm{x})\right)^2\nonumber\\
    &\leq \left( \int d\bm{x} \mathcal{P}_t(\bm{x}) \:\|\bm{\nu}^{\mathcal{P}}_t(\bm{x})\|^2 \right) \left( \int d\bm{x}\frac{(\delta\mathcal{P}_t(\bm{x}) )^2 \|\nabla \psi(\bm{x})\|^2}{\mathcal{P}_t(\bm{x})} \right).
\end{align}
and the $1$-Lipshitz continuity $\|\nabla \psi\| \leq 1$.
Therefore, our bound uses the Cauchy--Schwartz inequality in a different way. In addition, our bound requires the discussion of the Kantorovich potential, which is used in Eqs.~(\ref{proofinequality1}) and (\ref{proofinequality2}). Therefore, our result is not regarded as a simple application of the thermodynamic uncertainty relation.

We also compare our main results [Eqs.~(\ref{DiffusionTUR}) and (\ref{eq.SADM-loss})] with the trivial inequality in optimal transport theory. We have a trivial relation between the 1-Wasserstein distance and the 2-Wasserstein distance $\mathcal{W}_1(P,Q)\leq\mathcal{W}_2(P,Q)$ for arbitrary probabilities $P$ and $Q$, and this trivial relation resembles our result [Eq.~(\ref{DiffusionTUR})]. However, our result is the non-trivial relation between the $1$-Wasserstein distance and $2$-Wasserstein distance because our result compares the time derivative of the $1$-Wasserstein distance $\partial_t \mathcal{W}_1(\mathcal{P}^{\dagger}_{\tau-t}, \tilde{\mathcal{P}}^{\dagger}_{\tau-t})$ for the two processes $\mathcal{P}^{\dagger}_{\tau-t}(\boldsymbol{x})$ and $\tilde{\mathcal{P}}^{\dagger}_{\tau-t}(\boldsymbol{x})$ with the speed in the space of the $2$-Wasserstein distance $v_2(t)=\lim_{\Delta t  \to 0}\mathcal{W}_2(\mathcal{P}_t,\mathcal{P}_{t+\Delta t})/\Delta t$ for the forward process $\mathcal{P}_t (\boldsymbol{x})$. This inequality is not straightforward consequence of the trivial relation between the $1$-Wasserstein distance and the $2$-Wasserstein distance. We also note that the 
trivial inequality $\mathcal{W}_1(\mathcal{P}_t,\mathcal{P}_{t+\Delta t})\leq \mathcal{W}_2(\mathcal{P}_t,\mathcal{P}_{t+\Delta t})$ leads to the thermodynamic uncertainty relation [see Sec.~\ref{sec.OverDampedSystemAndOptimalTransport}], and our result [Eq.~(\ref{DiffusionTUR})] differs from the thermodynamic uncertainty relation as discussed above.

In the study of machine learning, there are some studies on a similar problem that derive some bounds in a similar form~\cite{kwon2022score,oko2023diffusion,fukumizu2024flow}. However, these bounds are conceptually different from ours. In Eq.~(23) of Ref.~\cite{kwon2022score}, the authors consider a setup with perturbation on the initial training data and evaluate the effect of the perturbation on the generated data. The authors used the contraction property of the $2$-Wasserstein distance, which is
\begin{align}
    \mathcal{W}_2(\mathcal{P}_t,\mathcal{P}^{\rm p}_t)\leq \exp\left(\int_0^tdt'\: K_{\rm L}(t')\right)\mathcal{W}_2(\mathcal{P}_0,\mathcal{P}^{\rm p}_0),
\end{align}
where $K_{\rm L}(t)$ is the Lipshitz constant of the external force $\bm{f}_t(\bm{x})$ satisfying $\|\bm{f}_t(\bm{x})-\bm{f}_t(\bm{y})\|\leq K_{\rm L}(t)\|\bm{x}-\bm{y}\|$ for any $\bm{x}, \bm{y}\in\mathbb{R}^{n_{\rm{d}}}$ and $\mathcal{P}^{\rm p}_t(\bm{x})$ is the density function of the forward process with perturbation in the initial data distribution $\mathcal{P}_0(\bm{x})$. The author also discussed the upper bound on the $2$-Wasserstein distance $\mathcal{W}_2(\tilde{\mathcal{P}}_0, \tilde{\mathcal{P}}^{\rm p}_0)$ based on the triangle inequality, where $\tilde{\mathcal{P}}^{\rm p}_t (\bm{x})$ is the density function of the estimated reverse process with perturbation in the initial data distribution $\mathcal{P}_0(\bm{x})$.
This bound differs from ours because the paper did not focus on the perturbation in the estimated reverse process and the time evolution of the forward process. Indeed, the result in Ref.~\cite{kwon2022score} does not provide insight into the optimality of the noise schedule in the forward process. In Refs.~\cite{oko2023diffusion, fukumizu2024flow}, they consider the bound on the estimation error when the initial perturbation is negligible $D_{\rm max} \simeq 0$ and the objective function is not negligible $L_{\alpha} (\theta^*_{\alpha}) \neq 0$. In Ref.~\cite{oko2023diffusion}, the authors considered score-based generative modeling and focused on the specific situation where the forward process is given by the Ornstein-Uhlenbeck process and the objective function is optimized with respect to specific parameters. The result is therefore not applicable to the general case where, for example, optimal transport is achieved. In Ref.~\cite{fukumizu2024flow}, the authors considered flow-based generative modeling and obtained the similar bound obtained in Ref.~\cite{oko2023diffusion}, and thus the result is not applicable to the general case either. These analyses are examples of the convergence analysis of generative models~\cite{song2021maximum,lee2022convergence,chen2023sampling,kwon2022score,oko2023diffusion,de2022convergence}, which provide bounds on the estimation errors with respect to the objective function. Our generalized result [Eq.~(\ref{eq:underestimateSADM})] can be considered as the convergence analysis. Compared with the results in Refs.~\cite{oko2023diffusion, fukumizu2024flow}, our results are generally applicable to diffusion models where the forward process is arbitrarily chosen and the objective function is not well optimized. In fact, optimal transport is generally not achieved with the Ornstein-Uhlenbeck process. Thus, our general inequality allows us to discuss an optimality of data generation via optimal transport, while the results in Refs.~\cite{oko2023diffusion, fukumizu2024flow} do not. This fact may be a possible merit of our result based on the universal thermodynamic framework.

Finally, we discuss the possibility of generalizing our result based on other metrics of estimation error instead of the 1-Wasserstein distance. In practical diffusion models, metrics such as the FID~\cite{heusel2017gans} and the inception score (IS)~\cite{salimans2016improved} are widely used as an index of generation quality. Adapting our methodology to these metrics may not be straightforward and some other ideas may be needed. First, both metrics are heavily based on the pre-trained inception model~\cite{szegedy2015going}, which is a black-box classification model on images. Therefore, it may be difficult to provide general bounds on these metrics without adding specific assumptions to this black-box model. Although it may be difficult to apply our methodology, the FID may be more appropriate than the SI for considering a similar upper bound. As discussed in Sec.~\ref{sec.OverDampedSystemAndOptimalTransport}, the FID is a lower bound on the $2$-Wasserstein distance $\mathcal{W}_2(p,q)$, and it may be possible to evaluate it by providing an upper bound on $\mathcal{W}_2(p,q)$. Since our bounds are on the $1$-Wasserstein distance $\mathcal{W}_1(p,q)$ and the trivial relation $\mathcal{W}_1(p,q)\leq\mathcal{W}_2(p,q)$ does not lead to an upper bound on the FID, a simple modification of our proof may not give an upper bound on the FID. Thus, another idea such as the triangle inequality for the $2$-Wasserstein distance may be needed to consider a similar thermodynamic bound on the FID.

\subsection{Comparison with conventional understanding of noise schedules}\label{dis:diffusion}

We discuss our result with the conventional understanding on noise schedules in diffusion models. In studies of diffusion models, there have been many numerical attempts to investigate the effect of noise schedules on the quality of data generation~\cite{nichol2021improved,albergo2022building,tong2023improving,lipman2022flow,liu2022flow,esser2024scaling}. For example, the cosine schedule has been shown experimentally to be an efficient protocol for image generation tasks compared to other conventional noise schedules under the $m_t^2 + \sigma_t^2=1$ constraint~\cite{nichol2021improved}, and optimal transport based flow matching methods have also been shown experimentally to provide better sampling quality and efficiency~\cite{tong2023improving,albergo2022building,liu2022flow,esser2024scaling}. However, there are almost no theoretical studies that explain their efficiency in terms of the universal upper bound on the estimation error. For example, Ref.~\cite{tong2023improving} introduced a method for using an approximation of optimal transport for noise schedules in flow matching, and the authors did not provide any reasons for using optimal transport as noise schedules except for numerical illustration of the generated quality and efficient sampling.

The speed-accuracy relations for diffusion models derived in this study may provide insight into how to determine this noise schedule for accurate data generation. In fact, we have shown that the existing methods, such as the cosine and cond-OT schedules, are suboptimal, and that the genuine optimal transport should be the best noise schedules from the perspective of the speed-accuracy relations for diffusion models. Thus, the room for improvement in existing methods can be discussed quantitatively based on the tightness of the inequalities. Our treatment would be a new direction in the theoretical understanding of the optimality of noise schedules for diffusion models.

\subsection{Possible importance of the entropy production rate in diffusion models}
We discuss the possible importance of the entropy production rate in diffusion models in diffusion models by comparing our approach with other physically inspired machine learning methods, so-called energy-based methods. In our main result [Eq.~(\ref{DiffusionTUR})], we used the entropy production rate for the upper bound on the estimation error. As an analogy to the entropy production minimization problem, the reduction of the thermodynamic quantity, $\int_0^{\tau} dt T_t \dot{S}_t^{\rm tot}$ for the forward dynamics, is related to a smallness of the response function $(\Delta \mathcal{W}_1)^2/D_0$ against arbitrary initial perturbations. 

In machine learning, there are energy-based methods that minimize physical quantities, such as the energy and the free energy, similar to the entropy production. For example, Amari-Hopfield networks~\cite{amari1972learning, hopfield1982neural} modeled the brain's memory function using a weighted graph called a neural network, which formed the basis of current machine learning methods using artificial neural networks. Amari-Hopfield networks are physically viewed as spin-grass systems, and the association and retrieval of given memories is characterized by the energy function in spin-grass systems. Mathematically, the energy function in Amari-Hopfield networks is just the objective function, and minimizing the energy function is understood in terms of optimizing the objective function. Boltzmann machines~\cite{hinton1984boltzmann, salakhutdinov2009deep, hinton2010practical} are machine learning models deeply inspired by Amari-Hopfield networks, which formed the basis of modern machine learning models. Boltzmann machines treat training data as memories in Amari-Hopfield networks, and slightly modify  the framework to treat it in a probabilistic way. Specifically, using the same neural network and the energy function as Amari-Hopfield networks, the data in Boltzmann machines is expressed as samples of the Gibbs measure with respect to the energy function. In the training process of Boltzmann machines, we minimize the Kullback-Leibler divergence between the training dataset distribution and the Gibbs measure by changing the weights of the neural network.  In this way, Boltzmann machines estimates the probability distribution behind the input data, and Boltzmann machines can be seen as a variant of generative models. In stochastic thermodynamics, the Kullback-Leibler divergence between the current probability and the Gibbs measure is the Lyapunov function for the system with the detailed balance condition and is considered as a nonequilibrium generalization of the free energy~\cite{schnakenberg1976network, qian2001relative}, and thus this learning is also translated in terms of the free energy minimization.

From the perspective of stochastic thermodynamics, the application of the energy function and the free energy is limited to the stochastic process that satisfies the detailed balance condition. For example, it is often the case that the non-conservative force exists in Langevin systems or the Fokker-Planck systems, and the detailed balance condition is violated. In this case, the energy may not be physically defined, but the entropy production rate is still well defined. Moreover, if the detailed balance condition holds, the entropy production rate is rewritten in terms of the potential energy or the free energy by the first law of thermodynamics~\cite{sekimoto2010stochastic}. Thus, the entropy production rate can be a more general thermodynamic quantity than the energy for stochastic processes such as Langevin dynamics and Fokker-Planck dynamics.

The entropy production rate in diffusion models may be discussed in parallel with the energy and the free energy in Amari-Hopfield networks and Boltzmann machines.
Amari-Hopfield networks and Boltzmann machines may be based on the systems where the energy function can be defined, and it means that the detailed balance condition may hold. It makes sense that the energy function would be useful in such a system. Since diffusion models are based on the Langevin dynamics or the Fokker-Planck dynamics, our framework based on the entropy production rate may make sense because the entropy production rate is more informative rather than the energy function, especially when the detailed balance condition is violated by the non-conservative force. It is surprising that the idea of minimizing a physically inspired cost $\int_0^{\tau} dt T_t \dot{S}_t^{\rm tot}$ is one of the criteria for model selection in diffusion models as in the traditional energy-based methods such as Amari-Hopfield networks and Boltzmann machines. Interestingly, our proposed optimal protocol is related to the minimization of the entropy production when the temperature of the system does not depend on time, analogous to the minimization of the energy and the minimization of the free energy. In fact, the entropy production can also be related to the potential energy $U_t(\boldsymbol{x})$ when the non-conservative force is absent, i.e., $\boldsymbol{f}_t(\boldsymbol{x})=-\nabla U_t(\boldsymbol{x})$. Thus, it may be interesting to consider the connection between our framework and the energy-based treatment, where the detailed equilibrium condition holds and the potential energy is defined. In Ref.~\cite{biroli2024dynamical}, the symmetry breaking of the phase transition is discussed in diffusion models based on the energy-based spin glass theory. Such an energy-based treatment of the diffusion model based on statistical physics may also be promising.

\subsection{Future studies}
In this paper, we have demonstrated only one aspect of the usefulness of the analogy between diffusion models and stochastic thermodynamics. We believe that the conventional thermodynamic uncertainty relations~\cite{barato2015thermodynamic,horowitz2020thermodynamic,otsubo2020estimating, koyuk2020thermodynamic, dechant2022geometric} are also useful because we can consider the speed of any observable in the forward process, for example, the speed of data structure breakage in the diffusion process. It is also noteworthy that the short-time thermodynamic uncertainty relation is used to estimate the time-varying velocity field~\cite{otsubo2020estimating, otsubo2022estimating}, which is important in the flow-based generative modeling, and it is worth considering whether there are any aspects in which such a thermodynamic-based method is superior to conventional flow matching methods.

It is also interesting to reconsider the path probability based method as discussed in the original paper on the diffusion model from a thermodynamic point of view. This is because the method discussed in the original paper can handle not only the simple diffusion processes described by the Langevin and Fokker-Planck equations, but also the diffusion process on the graph described by the Markov jump processes, which may provide a more scalable method than the current one. In such a case, the thermodynamic trade-off relations and optimal transport for the Markov jump process~\cite{dechant2022minimum, yoshimura2023housekeeping,van2023thermodynamic,kolchinsky2022information} may be useful to consider the optimality of the diffusion model.
In such a case, analogies to an information geometric structure of the Kullback-Leibler divergence in stochastic thermodynamics~\cite{ito2020unified,kolchinsky2022information, ito2023geometric} may also be important, since the objective function is introduced by the Kullback-Leibler divergence and its minimization is mathematically well discussed as the projection theorem in information geometry~\cite{amari2016information}.
In fact, there are several stochastic methods based on the Schr\"odinger bridge in the diffusion model~\cite{wang2021deep,de2021diffusion,chen2021likelihood,shi2022conditional}, which is given by the minimization of the Kullback-Leibler divergence. We may be able to obtain some trade-off relations for such a system by considering an analogy to stochastic thermodynamics based on path probability.

It would also be interesting to examine our results from the perspective of increasing the variety of data generated by raising the upper bound of the response function, which is considered to be the sensitivity. Indeed, numerical experiments with thermodynamic non-conservative forces show that as the upper bound becomes looser, it becomes easier to generate data that are different from the original data. Increasing the strength of the non-conservative force may be a way to generate new data that is difficult to obtain using conventional methods. It is also possible to deliberately account for non-optimal transport or increase the error in estimating the velocity field to increase such sensitivity, and our main result [Eq.~(\ref{DiffusionTUR})] and the generalized result [Eq.~(\ref{eq:underestimateSADM})]  may also be useful in such cases.

\section*{ACKNOWLEDGMENTS}
S.I. thanks Sachinori Watanabe for discussions on diffusion models. K.I. and S.I. thank Taiji Suzuki for insightful discussions. S.I. is supported by JSPS KAKENHI Grants No.~21H01560, No.~22H01141, No.~23H00467, and No.~24H00834, JST ERATO Grant No.~JPMJER2302, 
and UTEC-UTokyo FSI Research Grant Program.

\begin{widetext}
\appendix
\section{Derivation of the relations between the two gradients of the objective functions}
We show $\nabla_{\theta} L_{\rm SM}^{\rm c}(\theta) =\nabla_{\theta} L_{\rm SM}(\theta)$ for solving the optimization problem by score matching, and that $\nabla_{\theta} L_{\rm FM}^{\rm c}(\theta) =\nabla_{\theta} L_{\rm FM}(\theta)$ is satisfied for solving the optimization problem by flow matching.

\subsubsection{Score matching}
We show that two objective functions
\begin{align}
    L_{\rm{SM}}(\theta)=\mathbb{E}_{\mathcal{P}_t,\mathcal{U}}\left[T_t^2\|\bm{s}^{\theta}_t(\bm{x})-\nabla\ln \mathcal{P}_t(\bm{x})\|^2\right] ,
\end{align}
and
\begin{align}
    L^{\rm c}_{\rm{SM}}(\theta)=\mathbb{E}_{\mathcal{P}^{\rm c}_t, q,\mathcal{U}}\left[T_t^2\|\bm{s}^{\theta}_t(\bm{x})-\nabla\ln \mathcal{P}^c_t(\bm{x}|\bm{y})\|^2\right],
\end{align}
give the same gradient for the parameter $\theta$. The gradient $\nabla_{\theta}L^{\rm c}_{\rm{SM}}(\theta)$ can be calculated as follows,
\begin{align}
    \nabla_{\theta}L^{\rm c}_{\rm{SM}}(\theta) 
    =&\frac{2}{\tau}\int_0^\tau dt \int d\bm{x}  \int d\bm{y} \mathcal{P}^{\rm c}_t(\bm{x}|\bm{y})q(\bm{y})T_t^2[\nabla_{\theta}\bm{s}^{\theta}_t(\bm{x})]\cdot\left(\bm{s}^{\theta}_t(\bm{x})-\frac{\nabla\mathcal{P}^{\rm c}_t(\bm{x}|\bm{y})}{\mathcal{P}^{\rm c}_t(\bm{x}|\bm{y})}\right)\nonumber\\
    =& \frac{2}{\tau}\int_0^\tau dt \int d\bm{x} \mathcal{P}_t(\bm{x}) T_t^2[\nabla_{\theta}\bm{s}^{\theta}_t(\bm{x})]\cdot\left(\bm{s}^{\theta}_t(\bm{x})-\frac{\nabla\mathcal{P}_t(\bm{x})}{\mathcal{P}_t(\bm{x})}\right)\nonumber\\
    =& \nabla_{\theta}  L_{\rm{SM}}(\theta),
\end{align}
which means that $\nabla_{\theta}L^{\rm c}_{\rm{SM}}(\theta)$ coincides with $\nabla_{\theta}  L_{\rm{SM}}(\theta)$.
Here, we used $\int d\boldsymbol{y} \mathcal{P}^{\rm c}_t(\bm{x}|\bm{y}) q(\bm{y}) = \mathcal{P}_t (\bm{x})$ and $q(\bm{y}) \nabla\mathcal{P}^{\rm c}_t(\bm{x}|\bm{y}) = \nabla[ \mathcal{P}^{\rm c}_t(\bm{x}|\bm{y})q(\bm{y})]$.
\subsubsection{Flow matching}
Similarly, in flow matching, we show that two objective functions
\begin{align}
    L_{\rm{FM}}(\theta)=\mathbb{E}_{\mathcal{P}_t, \mathcal{U}}\left[\|\bm{u}_t^{\theta}(\bm{x})-\bm{\nu}^{\mathcal{P}}_t(\bm{x})\|^2\right],
\end{align}
and
\begin{align}
    L^c_{\rm{FM}}(\theta)=\mathbb{E}_{\mathcal{P}_t,q, \mathcal{U}}\left[\|\bm{u}_t^{\theta}(\bm{x})-\bm{\nu}^{\mathcal{P}^c}_t(\bm{x}|\bm{y})\|^2\right],
\end{align}
gives the same gradient for the parameter $\theta$. The gradient $\nabla_{\theta}L^{\rm c}_{\rm{FM}}(\theta)$ can be calculated as follows,
\begin{align}
      \nabla_{\theta}  L^{\rm c}_{\rm{FM}}(\theta) =& \frac{2}{\tau}\int dt\int d\bm{y} \int d\bm{x}\mathcal{P}^{\rm c}_t(\bm{x}|\bm{y})q( \bm{y})[\nabla_{\theta} \bm{u}_t^{\theta}(\bm{x})]\cdot(\bm{u}_t^{\theta}(\bm{x})-\bm{\nu}^{\mathcal{P}^{\rm c}}_t(\bm{x}|\bm{y}))\nonumber\\
      =& \frac{2}{\tau}\int dt \int d\bm{x} \mathcal{P}_t(\bm{x})[\nabla_{\theta} \bm{u}_t^{\theta}(\bm{x})]\cdot(\bm{u}_t^{\theta}(\bm{x})-\bm{\nu}^{\mathcal{P}}_t(\bm{x})) \nonumber\\
      =&\nabla_{\theta}  L_{\rm{FM}}(\theta),
\end{align}
which means that $\nabla_{\theta}L^{\rm c}_{\rm{FM}}(\theta)$ coincides with $\nabla_{\theta}  L_{\rm{FM}}(\theta)$.
Here, we used $\int d\boldsymbol{y} \mathcal{P}^{\rm c}_t(\bm{x}|\bm{y}) q(\bm{y}) = \mathcal{P}_t (\bm{x})$ and
\begin{align}
    \bm{\nu}_t^{\mathcal{P}}(\bm{x}) \mathcal{P}_t(\bm{x})=\int d\bm{y} \bm{\nu}_t^{\mathcal{P}^c}(\bm{x}|\bm{y})\mathcal{P}_t^c(\bm{x}|\bm{y})q(\bm{y}).
\end{align} 
\section{Parameters of the conditional probability density function}\label{ap.GaussianPath}
Here, we explain the specific expressions for the parameters $\bm{\mu}_t(\bm{y})$ and $\mathsf{\Sigma}_t$ of the conditional probability density function $\mathcal{P}^{\rm c}_t(\bm{x}|\bm{y})= \mathcal{N}(\bm{x}|\bm{\mu}_t(\bm{y}),\mathsf{\Sigma}_t)$.The Fokker--Planck equation for the conditional probability density function $\mathcal{P}^{\rm c}_t(\bm{x}|\bm{y})$ is given as follows,
\begin{align}
    \partial_t \mathcal{P}^{\rm c}_t(\bm{x}|\bm{y}) &= -\nabla\cdot \left(\bm{\nu}^{\mathcal{P}^{\rm c}}_t(\bm{x}|\bm{y})\mathcal{P}^{\rm c}_t(\bm{x}|\bm{y})\right),\\
    \bm{\nu}_t^{\mathcal{P}^{\rm c}}(\bm{x}|\bm{y})&\coloneqq \mathsf{A}_t\bm{x}+\bm{b}_t-T_t\nabla\ln \mathcal{P}^{\rm c}_t(\bm{x}|\bm{y}) = (\mathsf{A}_t + T_t\mathsf{\Sigma}_t^{-1} ) (\bm{x} - \bm{\mu}_t(\bm{y}))+ \mathsf{A}_t \bm{\mu}_t(\bm{y}) +\bm{b}_t.
\end{align}
Thus, the time evolution of the parameter $\bm{\mu}_t(\bm{y})$ can be calculated as follows,
\begin{align}
\partial_t \bm{\mu}_t(\bm{y})&=\partial_t \int d\bm{x} \bm{x} \mathcal{P}
^{\rm{c}}_t(\bm{x}|\bm{y}) \nonumber\\
    &=\int d\bm{x} \bm{x}\left(-\nabla\cdot \left(\bm{\nu}^{\mathcal{P}^{\rm c}}_t(\bm{x}|\bm{y})\mathcal{P}^{\rm c}_t(\bm{x}|\bm{y})\right)\right) \nonumber\\
    &=\int d\bm{x} \bm{\nu}^{\mathcal{P}^{\rm c}}_t(\bm{x}|\bm{y})\mathcal{P}^{\rm c}_t(\bm{x}|\bm{y}) \nonumber\\
    &=\int d\bm{x} [A_t\bm{x}+\bm{b}_t]\mathcal{P}^{\rm c}_t(\bm{x}|\bm{y}) - \int d\bm{x} T_t\nabla  \mathcal{P}^{\rm c}_t(\bm{x}|\bm{y}) \nonumber\\
    &= \mathsf{A}_t\bm{\mu}_t(\bm{y}) +\bm{b}_t,
    \label{difeqmean}
\end{align}
where we assumed that the contribution vanishes at infinity in the partial integration, and we used $\bm{\mu}_t(\bm{y})=\int d\bm{x} \bm{x} \mathcal{P}
^{\rm{c}}_t(\bm{x}|\bm{y})$ and $\int d\bm{x}\nabla \mathcal{P}^{\rm c}_t(\bm{x}|\bm{y}) =0$. The time evolution of the parameter $\mathsf{\Sigma}_t$ is also calculated as follows,
\begin{align}
\partial_t \mathsf{\Sigma}_t&=\partial_t \int d\bm{x} (\bm{x} -\bm{\mu}_t(\bm{y}) ) (\bm{x} -\bm{\mu}_t(\bm{y}) )^{\top}\mathcal{P}
^{\rm{c}}_t(\bm{x}|\bm{y}) \nonumber\\
    &=\int d\bm{x}  (\bm{x} -\bm{\mu}_t(\bm{y}) ) (\bm{x} -\bm{\mu}_t(\bm{y}) )^{\top}\left(-\nabla\cdot \left(\bm{\nu}^{\mathcal{P}^{\rm c}}_t(\bm{x}|\bm{y})\mathcal{P}^{\rm c}_t(\bm{x}|\bm{y})\right)\right) \nonumber\\
    &=\int d\bm{x} (\mathsf{A}_t + T_t\mathsf{\Sigma}_t^{-1} ) (\bm{x} -\bm{\mu}_t(\bm{y}) ) (\bm{x} -\bm{\mu}_t(\bm{y}) )^{\top} \mathcal{P}^{\rm c}_t(\bm{x}|\bm{y})  +\int d\bm{x} (\bm{x} -\bm{\mu}_t(\bm{y}) ) \left[(\mathsf{A}_t + T_t\mathsf{\Sigma}_t^{-1} )  (\bm{x} -\bm{\mu}_t(\bm{y}))  \right]^{\top} \mathcal{P}^{\rm c}_t(\bm{x}|\bm{y})   \nonumber\\
    &=\mathsf{A}_t\mathsf{\Sigma}_t + \mathsf{\Sigma}_t \mathsf{A}_t^{\top} + 2 T_t \mathsf{I},
    \label{difeqcov}
\end{align}
where $\top$ stands for the transpose, and we used $\mathsf{\Sigma}_t =\int d\bm{x} (\bm{x} -\bm{\mu}_t(\bm{y}) ) (\bm{x} -\bm{\mu}_t(\bm{y}) )^{\top}\mathcal{P}
^{\rm{c}}_t(\bm{x}|\bm{y})$, $\int d\bm{x} (\bm{x} -\bm{\mu}_t(\bm{y}) ) \mathcal{P}^{\rm{c}}_t(\bm{x}|\bm{y}) = \boldsymbol{0}$ and $(\mathsf{\Sigma}_t^{-1})^{\top}= \mathsf{\Sigma}_t^{-1}$.

Now, we consider a situation where $\mathsf{A}_t$, $\bm{b}_t$ and $T_t$ are expressed as $\mathsf{A}_t = (\partial_t \ln m_t) \mathsf{I}$, $\bm{b}_t =\boldsymbol{0}$, $T_t =  \partial_t  (\sigma_t^2/2) - \sigma_t^2 \partial_t \ln m_t $ with non-negative parameters $\sigma_t (\geq 0)$, $m_t (\geq 0)$, and the initial conditions of the parameters are given by $m_{t=0}=1$ and $\sigma_{t=0}=0$. In this situation, Eqs.~(\ref{difeqmean}) and (\ref{difeqcov}) are given as follows,
\begin{align}
\partial_t \bm{\mu}_t(\bm{y})&= (\partial_t \ln m_t)  \bm{\mu}_t(\bm{y}), \\
\partial_t \mathsf{\Sigma}_t(\bm{y}) &= (2 \partial_t \ln m_t) \mathsf{\Sigma}_t(\bm{y})+ (\partial_t  (\sigma_t^2) - 2\sigma_t^2 \partial_t \ln m_t )\mathsf{I}.
\end{align}
Therefore, the $k$-th element of the vector $[\bm{\mu}_t(\bm{y})]_k$ and, the $(k,l)$-element of the matrix $[ \mathsf{\Sigma}_t(\bm{y})]_{kl}$ satisfy the following equations,
\begin{align}
\partial_t [\bm{\mu}_t(\bm{y})]_k &= (\partial_t \ln m_t) [\bm{\mu}_t(\bm{y})]_k, \\
\partial_t ([ \mathsf{\Sigma}_t(\bm{y})]_{kl} - \sigma_t^2 \delta_{kl} )&= (2 \partial_t \ln m_t) ([ \mathsf{\Sigma}_t(\bm{y})]_{kl} - \sigma_t^2 \delta_{kl} ),
\end{align}
where $\delta_{kl}= [ \mathsf{I}]_{kl}$ is the Kronecker delta. Since $\bm{\mu}_{t=0} (\bm{y})=\bm{y}$ and $\mathsf{\Sigma}_{t=0}=\mathsf{O}$, performing a time integral from $t=0$ to $t$ yields
\begin{align}
[\bm{\mu}_t(\bm{y})]_k &= m_t [\bm{y}]_k, \\
[\mathsf{\Sigma}_t(\bm{y})]_{kl} - \sigma_t^2 \delta_{kl} &=0,
\end{align}
and thus the following specific expressions are obtained,
\begin{align}
\bm{\mu}_t(\bm{y})=m_t\bm{y}, \: \: \mathsf{\Sigma}_t = \mathsf{\sigma}_t^2\mathsf{I}.
\end{align}

\section{Coordinate transformation in the DDIM}\label{sec.fokker-changev}
We show the detailed calculation of the coordinate transformation in the DDIM. The Fokker--Planck equation [Eq.~(\ref{eq.FP})] with Eq.~(\ref{eq.externalForce}) is given as follows,
\begin{align}
    \partial_t \mathcal{P}_t(\bm{x})=-\nabla\cdot \left[ \left[ (\partial_t \ln m_t) \boldsymbol{x} - T_t \nabla\ln \mathcal{P}_t(\bm{x})\right]\mathcal{P}_t(\bm{x})\right].  \label{appendix.fokker}
\end{align}
From the coordinate transform
\begin{align}
    \bar{\bm{x}}=\frac{\bm{x}}{m_t},
\end{align}
we obtain 
$\mathcal{P}_t(\bm{x})=  \bar{\mathcal{P}}_t(\bar{\bm{x}}) (|d\bar{\bm{x}}/d\bm{x}|)= \bar{\mathcal{P}}_t(\bar{\bm{x}}) (m_t)^{-{n_{\rm d}}}$. Substituting this into Eq.~(\ref{appendix.fokker}), the left-hand side is calculated as 
\begin{align}
    \partial_t \mathcal{P}_t(\bm{x})= (m_t)^{-{n_{\rm d}}}\partial_t  \bar{\mathcal{P}}_t(\bar{\bm{x}}) -{n_{\rm d}}  (\partial_t \ln m_t) (m_t)^{-{n_{\rm d}}} \bar{\mathcal{P}}_t(\bar{\bm{x}}) - (\partial_t \ln m_t) (m_t)^{-(n_{\rm d} +1)} \bm{x} \cdot \nabla_{\bar{\bm{x}}}  \bar{\mathcal{P}}_t(\bar{\bm{x}}),\label{eq.fokkerplancka}
\end{align}
where $\nabla_{\bar{\bm{x}}}$ is defined as $\nabla_{\bar{\bm{x}}} = m_t \nabla$. The right hand side of Eq.~(\ref{appendix.fokker}) is calculated as 
\begin{align}
    &-\nabla\cdot \left[ \left[ (\partial_t \ln m_t) \boldsymbol{x} - T_t \nabla\ln \mathcal{P}_t(\bm{x})\right]\mathcal{P}_t(\bm{x})\right]    \nonumber \\
    =&  -{n_{\rm d}}(\partial_t \ln m_t) (m_t)^{-{n_{\rm d}}}\bar{\mathcal{P}}_t(\bar{\bm{x}}) - (\partial_t \ln m_t )(m_t)^{-(n_{\rm d} +1)} \bm{x} \cdot \nabla_{\bar{\bm{x}}}  \bar{\mathcal{P}}_t(\bar{\bm{x}})
    + \nabla_{\bar{\bm{x}}} \cdot \left[ \left[ T_t  \nabla_{\bar{\bm{x}}} \ln \bar{\mathcal{P}}_t(\bar{\bm{x}})\right]\bar{\mathcal{P}}_t(\bar{\bm{x}}) (m_t)^{-(n_{\rm d}+2)} \right]. \label{eq.fokkerplanckb}
\end{align}
From Eqs.~(\ref{eq.fokkerplancka}) and (\ref{eq.fokkerplanckb}), we obtain
\begin{align}
    \partial_t  \bar{\mathcal{P}}_t(\bar{\bm{x}})=
     -\nabla_{\bar{\bm{x}}} \cdot \left[ \left[ -\frac{T_t}{(m_t)^2} \nabla_{\bar{\bm{x}}} \ln \bar{\mathcal{P}}_t(\bar{\bm{x}})\right]\bar{\mathcal{P}}_t(\bar{\bm{x}}) \right].
\end{align}
By substituting $T_t=-\partial_t\ln m_t$ into this equation, 
\begin{align}
    \partial_t  \bar{\mathcal{P}}_t(\bar{\bm{x}})=
     -\nabla_{\bar{\bm{x}}} \cdot \left[ \left[ \frac{\partial_t \ln m_t}{(m_t)^2} \nabla_{\bar{\bm{x}}} \ln \bar{\mathcal{P}}_t(\bar{\bm{x}})\right]\bar{\mathcal{P}}_t(\bar{\bm{x}}) \right],
\end{align}
is derived.

\section{The relations between the Kullback--Leibler divergence and the entropy production}\label{ap.KLandEntropyProduction}
First, we prove that the Kullback--Leibler divergence
\begin{align}
D_{\rm KL} (\mathbb{P}_{\rm F} \| \mathbb{P}_{\rm B})&=\sum_{N=0}^{N_{\tau}-1} \int d\Gamma \mathbb{P}_{\rm F}(\Gamma) \ln \frac{\mathcal{T}_{N \Delta t}(\boldsymbol{x}_{(N+1)\Delta t}|\boldsymbol{x}_{N \Delta t})}{\mathcal{T}^{\dagger}_{N \Delta t}(\boldsymbol{x}_{(N+1)\Delta t}|\boldsymbol{x}_{N \Delta t})},
\label{supp:kldivergence}
\end{align}
coincides with the entropy production
\begin{align}
S^{\rm tot}_{\tau} = \int_0^{\tau} dt \left[ \frac{1}{T_t}\int d\boldsymbol{x} \| \boldsymbol{\nu}^{\mathcal{P}}_t (\boldsymbol{x})\|^2  \mathcal{P}_t (\boldsymbol{x})\right],
\end{align}
in the limit $\Delta t \to 0$ with fixed $\tau$. By calculating $\ln [\mathcal{T}_{N \Delta t}(\boldsymbol{x}_{(N+1)\Delta t}|\boldsymbol{x}_{N \Delta t})/ \mathcal{T}^{\dagger}_{N \Delta t}(\boldsymbol{x}_{(N+1)\Delta t}|\boldsymbol{x}_{N \Delta t})]$, we obtain
\begin{align}
& \ln \frac{\mathcal{T}_{N \Delta t}(\boldsymbol{x}_{(N+1)\Delta t}|\boldsymbol{x}_{N \Delta t})}{ \mathcal{T}^{\dagger}_{N \Delta t}(\boldsymbol{x}_{(N+1)\Delta t}|\boldsymbol{x}_{N \Delta t})} \nonumber \\
=& \frac{\left\|\boldsymbol{x}_{(N+1) \Delta t} -\boldsymbol{x}_{N  \Delta t} -\boldsymbol{f}_{N \Delta t} (\boldsymbol{
x}_{N  \Delta t})\Delta t + 2 \boldsymbol{\nu}_{N \Delta t}^{\mathcal{P}} (\boldsymbol{x}_{N  \Delta t}) \Delta t \right\|^2 }{4 T_{N \Delta t} \Delta t} - \frac{\left\|\boldsymbol{x}_{(N+1) \Delta t} -\boldsymbol{x}_{N  \Delta t} -\boldsymbol{f}_{N \Delta t} (\boldsymbol{
x}_{N  \Delta t})\Delta t \right \|^2 }{4 T_{N \Delta t} \Delta t}
\nonumber \\
=& \frac{\left\|\boldsymbol{\nu}_{N \Delta t}^{\mathcal{P}} (\boldsymbol{x}_{N  \Delta t}) \right \|^2 }{T_{N \Delta t} } \Delta t + \frac{(\boldsymbol{x}_{(N+1) \Delta t} -\boldsymbol{x}_{N  \Delta t} -\boldsymbol{f}_{N \Delta t} (\boldsymbol{
x}_{N  \Delta t})\Delta t  ) \cdot  \boldsymbol{\nu}_{N \Delta t}^{\mathcal{P}} (\boldsymbol{x}_{N  \Delta t}) }{ T_{N \Delta t} }.
\end{align}
By using $\int d\boldsymbol{x}_{(N+1) \Delta t} \mathcal{T}_{N \Delta t}(\boldsymbol{x}_{(N+1)\Delta t}|\boldsymbol{x}_{N \Delta t})[\boldsymbol{x}_{(N+1) \Delta t} -\boldsymbol{x}_{N  \Delta t} -\boldsymbol{f}_{N \Delta t} (\boldsymbol{
x}_{N  \Delta t})\Delta t  ]= \boldsymbol{0}$ and the normalization of the probability distributions, we obtain  
\begin{align}
&\int d\Gamma \mathbb{P}^{\rm F}(\Gamma) \ln \frac{\mathcal{T}_{N \Delta t}(\boldsymbol{x}_{(N+1)\Delta t}|\boldsymbol{x}_{N \Delta t})}{\mathcal{T}^{\dagger}_{N \Delta t}(\boldsymbol{x}_{(N+1)\Delta t}|\boldsymbol{x}_{N \Delta t})}  \nonumber \\
&= \int d \boldsymbol{x}_{N \Delta t} \int d\boldsymbol{x}_{(N+1) \Delta t} \mathcal{T}_{N \Delta t}(\boldsymbol{x}_{(N+1)\Delta t}|\boldsymbol{x}_{N \Delta t}) \mathcal{P}_{N \Delta t} (\boldsymbol{x}_{N \Delta t})\ln \frac{\mathcal{T}_{N \Delta t}(\boldsymbol{x}_{(N+1)\Delta t}|\boldsymbol{x}_{N \Delta t})}{\mathcal{T}^{\dagger}_{N \Delta t}(\boldsymbol{x}_{(N+1)\Delta t}|\boldsymbol{x}_{N \Delta t})}\nonumber \\
&= \frac{\Delta t}{T_{N \Delta t}}\int d \boldsymbol{x}_{N \Delta t}   \left\|\boldsymbol{\nu}_{N \Delta t}^{\mathcal{P}} (\boldsymbol{x}_{N  \Delta t}) \right \|^2 \mathcal{P}_{N \Delta t} (\boldsymbol{x}_{N \Delta t}).
\end{align}
By taking the limit $\Delta t \to 0$ with fixed $\tau$, we obtain
\begin{align}
D_{\rm KL} (\mathbb{P}_{\rm F} \| \mathbb{P}_{\rm B})&=\sum_{N=0}^{N_{\tau}-1} \frac{\Delta t}{T_{N \Delta t}}\int d \boldsymbol{x}  \left\|\boldsymbol{\nu}_{N \Delta t}^{\mathcal{P}} (\boldsymbol{x}) \right \|^2 \mathcal{P}_{N \Delta t} (\boldsymbol{x}) \nonumber\\
&\to \int_0^{\tau} dt \left[ \frac{1}{T_t}\int d\boldsymbol{x} \| \boldsymbol{\nu}^{\mathcal{P}}_t (\boldsymbol{x})\|^2  \mathcal{P}_t (\boldsymbol{x})\right] =S^{\rm tot}_{\tau}.
\end{align}

Next, we prove that Kullback-Leibler divergence $D_{\rm KL} (\mathbb{P}_{\rm F} \| \mathbb{P}_{\rm B}')= \mathbb{E}_{\mathbb{P}_{\rm F}} [s^{\rm tot}_{\tau} ]$ coincides with $S^{\rm tot}_{\tau} = \int_0^{\tau} dt \dot{S}_{t}^{\rm sys} +\int_0^{\tau} dt \dot{S}_{t}^{\rm bath}$ in the limit $\Delta t \to 0$ with fixed $\tau$. Since the stochastic entropy production is given by
\begin{align}
s^{\rm tot}_{\tau}=\ln \frac{\mathcal{P}_{0}(\boldsymbol{x}_{0})}{\mathcal{P}_{N_{\tau} \Delta t}(\boldsymbol{x}_{N_{\tau} \Delta t})}+ \sum_{N=0}^{N_{\tau}-1} \ln \frac{\mathcal{T}_{N \Delta t}(\boldsymbol{x}_{(N+1)\Delta t}|\boldsymbol{x}_{N \Delta t})}{\mathcal{T}_{N \Delta t}(\boldsymbol{x}_{N \Delta t}|\boldsymbol{x}_{(N+1) \Delta t})},
\end{align}
$D_{\rm KL} (\mathbb{P}_{\rm F} \| \mathbb{P}_{\rm B}') $ is calculated as follows,
\begin{align}
D_{\rm KL} (\mathbb{P}_{\rm F} \| \mathbb{P}_{\rm B}') &= \mathbb{E}_{\mathbb{P}_{\rm F}} [s^{\rm tot}_{\tau} ] \nonumber \\
&= \int d\Gamma \mathbb{P}_{\rm F}(\Gamma) \ln \frac{\mathcal{P}_{0}(\boldsymbol{x}_{0})}{\mathcal{P}_{N_{\tau} \Delta t}(\boldsymbol{x}_{N_{\tau} \Delta t})}+ \sum_{N=0}^{N_{\tau}-1} \int d\Gamma \mathbb{P}_{\rm F}(\Gamma) \ln \frac{\mathcal{T}_{N \Delta t}(\boldsymbol{x}_{(N+1)\Delta t}|\boldsymbol{x}_{N \Delta t})}{\mathcal{T}_{N \Delta t}(\boldsymbol{x}_{N \Delta t}|\boldsymbol{x}_{(N+1) \Delta t})}.
\label{sup:calc-a}
\end{align}
From the marginalization of the path probability, the first term of Eq.~(\ref{sup:calc-a}) is rewritten as follows,
\begin{align}
\int d\Gamma \mathbb{P}_{\rm F}(\Gamma) \ln \frac{\mathcal{P}_{0}(\boldsymbol{x}_{0})}{\mathcal{P}_{N_{\tau} \Delta t}(\boldsymbol{x}_{N_{\tau} \Delta t})}&= \int d\boldsymbol{x}_{0}  \mathcal{P}_{0}(\boldsymbol{x}_{0})\ln \mathcal{P}_{0}(\boldsymbol{x}_{0}) - \int d\boldsymbol{x}_{N_{\tau} \Delta t} \mathcal{P}_{N_{\tau} \Delta t}(\boldsymbol{x}_{N_{\tau} \Delta t}) \ln \mathcal{P}_{N_{\tau} \Delta t}(\boldsymbol{x}_{N_{\tau} \Delta t}) \nonumber \\
&=  \int_0^{\tau} dt \partial_t \left[ -\int d \boldsymbol{x} \mathcal{P}_{t}(\boldsymbol{x}) \ln \mathcal{P}_{t}(\boldsymbol{x})\right] \nonumber\\
&= \int_0^{\tau} dt \dot{S}_{t}^{\rm sys}.
\label{sup:calc-b}
\end{align}
The second term of Eq.~(\ref{sup:calc-a}) is also calculated as
\begin{align}
&\int d\Gamma \mathbb{P}_{\rm F}(\Gamma) \ln \frac{\mathcal{T}_{N \Delta t}(\boldsymbol{x}_{(N+1)\Delta t}|\boldsymbol{x}_{N \Delta t})}{\mathcal{T}_{N \Delta t}(\boldsymbol{x}_{N \Delta t}|\boldsymbol{x}_{(N+1) \Delta t})} \nonumber\\
=& \int d\Gamma \mathbb{P}_{\rm F}(\Gamma)\left[ -\frac{\left\|\boldsymbol{x}_{(N +1)\Delta t} -\boldsymbol{x}_{N \Delta t}-\boldsymbol{f}_{N \Delta t} (\boldsymbol{x}_{N \Delta t})\Delta t \right\|^2 }{4T_{N \Delta t} \Delta t}  +\frac{\left\|\boldsymbol{x}_{N \Delta t} -\boldsymbol{x}_{(N +1)\Delta t} -\boldsymbol{f}_{N \Delta t} (\boldsymbol{x}_{(N +1)\Delta t} )\Delta t \right\|^2 }{4T_{N \Delta t} \Delta t} \right] \nonumber \\
=& \int d\Gamma \mathbb{P}_{\rm F}(\Gamma)\left[ \frac{ (\boldsymbol{x}_{(N +1)\Delta t} -\boldsymbol{x}_{N \Delta t})\cdot \frac{ \boldsymbol{f}_{N \Delta t} (\boldsymbol{x}_{N \Delta t}) + \boldsymbol{f}_{N \Delta t} (\boldsymbol{x}_{(N +1)\Delta t})}{2}}{T_{N \Delta t}}  + \frac{ \| \boldsymbol{f}_{N \Delta t} (\boldsymbol{x}_{(N +1)\Delta t})\|^2 -\|\boldsymbol{f}_{N \Delta t} (\boldsymbol{x}_{N \Delta t}) \|^2}{4T_{N\Delta t}}\Delta t \right] \nonumber \\
=&\int d\boldsymbol{x}\mathcal{P}_{N \Delta t}(\boldsymbol{x})  \int d\boldsymbol{y} \mathcal{T}_{N \Delta t}(\boldsymbol{y}|\boldsymbol{x})\left[ \frac{ (\boldsymbol{y} -\boldsymbol{x})\cdot \frac{ \boldsymbol{f}_{N \Delta t} (\boldsymbol{x}) + \boldsymbol{f}_{N \Delta t} (\boldsymbol{y})}{2}}{T_{N \Delta t}} \right]\nonumber\\
&+\int d\boldsymbol{x}\mathcal{P}_{N \Delta t}(\boldsymbol{x})  \int d\boldsymbol{y} \mathcal{T}_{N \Delta t}(\boldsymbol{y}|\boldsymbol{x})\left[  \frac{ \| \boldsymbol{f}_{N \Delta t} (\boldsymbol{y})\|^2 -\|\boldsymbol{f}_{N \Delta t} (\boldsymbol{x}) \|^2}{4T_{N\Delta t}}\Delta t \right]\nonumber\\
=& \frac{1}{T_{N\Delta t}}\int d\boldsymbol{x}\mathcal{P}_{N \Delta t}(\boldsymbol{x})  \int d\boldsymbol{y} \mathcal{T}_{N \Delta t}(\boldsymbol{y}|\boldsymbol{x})\left[ (\boldsymbol{y} -\boldsymbol{x})\cdot \boldsymbol{f}_{N \Delta t} (\boldsymbol{x}) \right] \nonumber\\
&+\frac{1}{T_{N\Delta t}}\int d\boldsymbol{x}\mathcal{P}_{N \Delta t}(\boldsymbol{x})  \int d\boldsymbol{y} \mathcal{T}_{N \Delta t}(\boldsymbol{y}|\boldsymbol{x})\left[ \sum_{i,j} \frac{[ y_i -x_i ][y_j -x_j] }{2}\nabla_{x_j} [\boldsymbol{f}_{N \Delta t} (\boldsymbol{x})]_i \right]+  O((\Delta t)^2) \nonumber\\
=&\frac{\Delta t}{T_{N\Delta t}} \int d\boldsymbol{x}\mathcal{P}_{N \Delta t}(\boldsymbol{x}) \| \boldsymbol{f}_{N \Delta t} (\boldsymbol{x}) \|^2 + \Delta t  \int d\boldsymbol{x}\mathcal{P}_{N \Delta t}(\boldsymbol{x})  \nabla \cdot \boldsymbol{f}_{N \Delta t} (\boldsymbol{x})+  O((\Delta t)^2)  \nonumber\\
=&\frac{\Delta t}{T_{N\Delta t}} \int d\boldsymbol{x}\mathcal{P}_{N \Delta t}(\boldsymbol{x}) \boldsymbol{f}_{N \Delta t} (\boldsymbol{x}) \cdot \boldsymbol{\nu}_{N \Delta t}^{\mathcal{P}} (\boldsymbol{x})+  O((\Delta t)^2).
\end{align}
Therefore, in the limit $\Delta t \to 0$ with fixed $\tau$, we obtain
\begin{align}
\sum_{N=0}^{N_{\tau}-1} \int d\Gamma \mathbb{P}_{\rm F}(\Gamma) \ln \frac{\mathcal{T}_{N \Delta t}(\boldsymbol{x}_{(N+1)\Delta t}|\boldsymbol{x}_{N \Delta t})}{\mathcal{T}_{N \Delta t}(\boldsymbol{x}_{N \Delta t}|\boldsymbol{x}_{(N+1) \Delta t})} = \int_0^{\tau}dt\frac{\int d\boldsymbol{x} \mathcal{P}_{t}(\boldsymbol{x}) \boldsymbol{f}_{t} (\boldsymbol{x}) \cdot \boldsymbol{\nu}_{t}^{\mathcal{P}} (\boldsymbol{x})}{T_{t}} = \int_0^{\tau} dt \dot{S}_{t}^{\rm bath}.
\label{sup:calc-c}
\end{align}
 From Eqs.~(\ref{sup:calc-a}), (\ref{sup:calc-b}) and (\ref{sup:calc-c}), we obtain
\begin{align}
D_{\rm KL} (\mathbb{P}_{\rm F} \| \mathbb{P}_{\rm B}')=\int_0^{\tau} dt \dot{S}_{t}^{\rm sys}+ \int_0^{\tau} dt \dot{S}_{t}^{\rm bath} = {S}_{\tau}^{\rm tot}.
\end{align}
We note that $\mathbb{E}_{\mathbb{P}_{\rm F}}[\ln \mathbb{P}_{\rm B}]=\mathbb{E}_{\mathbb{P}_{\rm F}}[\ln \mathbb{P}^{\prime}_{\rm B}]$ is satisfied because ${S}_{t}^{\rm tot}= D_{\rm KL} (\mathbb{P}_{\rm F} \| \mathbb{P}_{\rm B})=D_{\rm KL} (\mathbb{P}_{\rm F} \| \mathbb{P}_{\rm B}')$, while $\mathbb{P}_{\rm B}(\Gamma) \neq \mathbb{P}_{\rm B}'(\Gamma)$.

\section{Time-independence of the Pearson's $\chi^2$-divergence}\label{sec.proof}
We show that $D_{\tau-t}$ is independent of time if the velocity field of the forward process is accurately reconstructed ($\hat{\bm{\nu}}_t(\bm{x})=\bm{\nu}_t^{\mathcal{P}}(\bm{x})$) in the flow-based generative modeling or the probability flow ODE. From Eq.~(\ref{FMdeltaContinuety}) and the continuity equation $\partial_{t} \mathcal{P}_{t}(\bm{x})=-\nabla\cdot\left(\bm{\nu}_{t}^{\mathcal{P}}(\bm{x})\mathcal{P}_{t}(\bm{x})\right)$, the time derivative of $D_{\tau-t}$ is calculated as
\begin{align}
    \partial_t D_{\tau-t}&=\int d\bm{x} \partial_{t} \frac{(\delta \mathcal{P}_{t}(\bm{x}))^2}{\mathcal{P}_{t}(\bm{x})}\nonumber\\
    &=\int d\bm{x} \left[\frac{2\delta \mathcal{P}_{t}(\bm{x})}{\mathcal{P}_{t}(\bm{x})}\partial_{t}\delta \mathcal{P}_{t}(\bm{x}) -\left(\frac{\delta \mathcal{P}_{t}(\bm{x})}{\mathcal{P}_{t}(\bm{x})}\right)^2\partial_{t}\mathcal{P}_{t}(\bm{x})\right]\nonumber\\
    &=-\int d\bm{x} \left[\frac{2\delta \mathcal{P}_{t}(\bm{x})}{\mathcal{P}_{t}(\bm{x})}(\nabla\cdot(\bm{\nu}^{\mathcal{P}}_{t}(\bm{x})\delta\mathcal{P}_{t}(\bm{x})))\right] +\int d\bm{x}\left[\left(\frac{\delta \mathcal{P}_{t}(\bm{x})}{\mathcal{P}_{t}(\bm{x})}\right)^2(\nabla\cdot(\bm{\nu}^{\mathcal{P}}_{t}(\bm{x})\mathcal{P}_{t}(\bm{x})))\right]\nonumber\\
    &=\int d\bm{x} \left[2\nabla\left(\frac{\delta \mathcal{P}_{t}(\bm{x})}{\mathcal{P}_{t}(\bm{x})}\right)\cdot\bm{\nu}^{\mathcal{P}}_{t}(\bm{x})\delta\mathcal{P}_{t}(\bm{x})\right]-\int d\bm{x} \left[2\nabla\left(\frac{\delta \mathcal{P}_{t}(\bm{x})}{\mathcal{P}_{t}(\bm{x})}\right)\cdot\bm{\nu}^{\mathcal{P}}_{t}(\bm{x})\frac{\delta\mathcal{P}_{t}(\bm{x})}{\mathcal{P}_{t}(\bm{x})}\mathcal{P}_{t}(\bm{x})\right]\nonumber\\
    &=0,
\end{align}
where we used the partial integration and we assumed that $\mathcal{P}_{t}(\bm{x})$ and $\delta\mathcal{P}_{t}(\bm{x})$ vanish at infinity. Therefore, we obtain $\partial_t D_{\tau-t}=0$, which means that $D_{\tau-t}$ is independent of time.  

\section{The derivation of Eq.~(\ref{eq:underestimateSADM})}\label{sec:proofIncompleteEstimate}
Here we derive Eq.~(\ref{eq:underestimateSADM}), which is a generalization of the main result for incomplete estimation. In the case of flow-based generative modeling or the probability flow ODE, the time evolution of the probability densities for the forward process is given by $\partial_{t} \mathcal{P}_{t}(\bm{x})=-\nabla\cdot\left(\bm{\nu}_{t}^{\mathcal{P}}(\bm{x})\mathcal{P}_{t}(\bm{x})\right)$. For the estimated reverse process, the time evolution is also given by $\partial_{t} \tilde{\mathcal{P}}_t = -\nabla\cdot (\tilde{\bm{\nu}}_{t}(\bm{x}) \tilde{\mathcal{P}}_t(\bm{x}))$ with the incomplete estimated velocity field $\tilde{\bm{\nu}}_t(\bm{x})\: (\neq \bm{\nu}_{t}^{\mathcal{P}}(\bm{x}))$. Analogous to Eq.~(\ref{FMdeltaContinuety}), the time evolution of $\delta \mathcal{P}_{t}(\bm{x}) = {\mathcal{P}}_t (\bm{x})-\tilde{\mathcal{P}}_t(\bm{x})$ is calculated as 
\begin{align}
    \partial_{t} \delta\mathcal{P}_{t}(\bm{x}) &= -\nabla\cdot (\bm{\nu}^{\mathcal{P}}_t(\bm{x})\mathcal{P}_t(\bm{x}) - \tilde{\bm{\nu}}_{t}(\bm{x})\tilde{\mathcal{P}}_{t}(\bm{x}) ) \nonumber\\
    &= -\nabla\cdot (\bm{\nu}^{\mathcal{P}}_t(\bm{x})\delta\mathcal{P}_t(\bm{x}) + \delta\bm{\nu}_t(\bm{x})\tilde{\mathcal{P}}_{t}(\bm{x}) ), \label{eq.witherror}
\end{align}
where $\delta\bm{\nu}_t(\bm{x})$ is the difference of the velocity fields defined as $\delta\bm{\nu}_t(\bm{x})\coloneqq \bm{\nu}^{\mathcal{P}}_t(\bm{x}) - \tilde{\bm{\nu}}_{t}(\bm{x})$. From Eq.~(\ref{eq.witherror}), we obtain
\begin{align}
     \partial_t \left(\mathbb{E}_{\mathcal{P}_{t}}[\psi]-\mathbb{E}_{ \tilde{\mathcal{P}}_{t}}[\psi]\right) 
    &= \int d\bm{x} \;\psi(\bm{x}) \partial_{t}\delta \mathcal{P}_{t}(\bm{x})\nonumber\\
    =&\int d\bm{x} \;\nabla\psi(\bm{x})\cdot (\bm{\nu}_{t}^\mathcal{P}(\bm{x})\delta \mathcal{P}_{t}(\bm{x})+\delta\bm{\nu}_t(\bm{x})\tilde{\mathcal{P}}_t(\bm{x}) ),\label{eq.differenceExpectationWitherror}
\end{align}
for any time-independent $1$-Lipshitz function $\psi \in\mathrm{Lip}^1$, using partial integration under the assumption that the surface term is negligible, similar to the proof of our main result [Eqs.~(\ref{eq.differenceExpectation}) and (\ref{ap.cauchy})]. From the Cauchy-Schwarz inequality, the 1-Lipshitz continuity $\| \nabla\psi(\bm{x})\|\leq 1$, the triangle inequality and Eq.~(\ref{eq.differenceExpectationWitherror}), and $\int d\bm{x}\:\delta\mathcal{P}_t(\bm{x})=0$, we obtain 
\begin{align}
    &|\partial_t \left(\mathbb{E}_{\mathcal{P}_{t}}[\psi]-\mathbb{E}_{ \tilde{\mathcal{P}}_{t}}[\psi]\right)|\nonumber\\
    &= \left|\int d\bm{x} \;\nabla\psi(\bm{x})\cdot (\bm{\nu}_{t}^\mathcal{P}(\bm{x})\delta \mathcal{P}_{t}(\bm{x})+\delta\bm{\nu}_t(\bm{x})\tilde{\mathcal{P}}_t(\bm{x}) )\right| \nonumber \\
    &\leq \left| \int d\bm{x} \;\nabla\psi(\bm{x})\cdot\bm{\nu}^{\mathcal{P}}_t(\bm{x})\delta\mathcal{P}_t(\bm{x})\right| + \left| \int d\bm{x} \;\nabla\psi(\bm{x})\cdot\delta\bm{\nu}_t(\bm{x})\tilde{\mathcal{P}}_t(\bm{x})\right| \nonumber\\
    &\leq \sqrt{\left( \int d\bm{x}\;\| \nabla\psi(\bm{x})\|^2\frac{(\delta\mathcal{P}_t(\bm{x}))^2}{\mathcal{P}_t(\bm{x})}\right)\left( \int d\bm{x}\;\| \bm{\nu}^{\mathcal{P}}_t(\bm{x})\|^2\mathcal{P}_t(\bm{x})\right)}+\sqrt{\left( \int d\bm{x}\;\| \nabla\psi(\bm{x})\|^2\dfrac{\tilde{\mathcal{P}}_t(\bm{x})^2}{\mathcal{P}_t(\bm{x})}\right)\left( \int d\bm{x}\;\| \delta\bm{\nu}_t(\bm{x})\|^2\mathcal{P}_t(\bm{x})\right)}\nonumber\\
    &\leq \sqrt{\left( \int d\bm{x}\;\frac{(\delta\mathcal{P}_t(\bm{x}))^2}{\mathcal{P}_t(\bm{x})}\right)\left( \int d\bm{x}\;\| \bm{\nu}^{\mathcal{P}}_t(\bm{x})\|^2\mathcal{P}_t(\bm{x})\right)} + \sqrt{\left( \int d\bm{x}\;\dfrac{(\mathcal{P}_t(\bm{x})-\delta\mathcal{P}_t(\bm{x}))^2}{\mathcal{P}_t(\bm{x})}\right)\left( \int d\bm{x}\;\| \delta\bm{\nu}_t(\bm{x})\|^2\mathcal{P}_t(\bm{x})\right)}\nonumber\\
    &\leq \sqrt{T_t\dot{S}_t^{\mathrm{tot}}D_{\tau-t}}+\sqrt{1+D_{\tau-t}}\sqrt{\int d\bm{x}\;\| \delta\bm{\nu}_t(\bm{x})\|^2\mathcal{P}_t(\bm{x})}.
\end{align}
 From $|\partial_t\mathcal{W}_1(\mathcal{P}_t(\bm{x}),\tilde{\mathcal{P}}_t(\bm{x}))| \leq |\partial_t \left(\mathbb{E}_{\mathcal{P}_{t}}[\psi]-\mathbb{E}_{ \tilde{\mathcal{P}}_{t}}[\psi]\right)|$ in Eqs.~(\ref{proofinequality1}) and (\ref{proofinequality2}), we obtain a generalization of the instantaneous bound for incomplete estimation as
\begin{align}
    |\partial_t\mathcal{W}_1(\mathcal{P}_t(\bm{x}),\tilde{\mathcal{P}}_t(\bm{x}))|\leq \sqrt{D_{\tau-t}}\sqrt{T_t\dot{S}_t^{\mathrm{tot}}}+\sqrt{1+D_{\tau-t}}\sqrt{\int d\bm{x}\;\| \delta\bm{\nu}_t(\bm{x})\|^2\mathcal{P}_t(\bm{x})}.\label{eq:sadm-underestimate}
\end{align}
Compared to Eq.~(\ref{eq.derivativeSADM1}), this result contains an additional term $\sqrt{1+D_{\tau-t}}\sqrt{\int d\bm{x}\;\| \delta\bm{\nu}_t(\bm{x})\|^2\mathcal{P}_t(\bm{x})}$.

From Eq.~(\ref{eq:sadm-underestimate}), the triangle inequality, Cauchy--Schwarz inequality, $\Delta\mathcal{W}_1=\mathcal{W}_1(\mathcal{P}_{\tau},\tilde{\mathcal{P}}_{\tau})-\mathcal{W}_1(\mathcal{P}_{0},\tilde{\mathcal{P}}_0)$ and $D_{\rm max}:= \max_{t\in[0,\tau]} D_t$, we can also obtain an upper bound on $|\Delta\mathcal{W}_1|$ as
\begin{align}
    |\Delta\mathcal{W}_1|&\leq\int_0^{\tau}dt^{\prime}\: |\partial_{t^{\prime}}\mathcal{W}_1(\mathcal{P}_{t^{\prime}}(\bm{x}),\tilde{\mathcal{P}}_{t^{\prime}}(\bm{x}))|\nonumber\\
    &\leq \int_0^{\tau}dt^{\prime}\;\sqrt{D_{\tau-t^{\prime}}}\sqrt{T_{t^{\prime}}\dot{S}_{t^{\prime}}^{\mathrm{tot}}}+\int_0^{\tau}dt^{\prime}\: \sqrt{1+D_{\tau-t^{\prime}}}\sqrt{\int d\bm{x}\;\| \delta\bm{\nu}_{t^{\prime}}(\bm{x})\|^2\mathcal{P}_{t^{\prime}}(\bm{x})}\nonumber\\
    &\leq \sqrt{D_{\rm max}}\int_0^{\tau}dt^{\prime}\: \sqrt{T_{t^{\prime}}\dot{S}_{t^{\prime}}^{\mathrm{tot}}} +  \sqrt{1+D_{\rm max}}\int_0^{\tau}dt^{\prime}\sqrt{\int d\bm{x}\;\| \delta\bm{\nu}_{t^{\prime}}(\bm{x})\|^2\mathcal{P}_{t^{\prime}}(\bm{x})}\nonumber\\
    &\leq \sqrt{D_{\rm max}}\sqrt{\tau\int_0^{\tau}dt^{\prime}\: T_{t^{\prime}}\dot{S}_{t^{\prime}}^{\mathrm{tot}}} +  \sqrt{1+D_{\rm max}}\sqrt{\tau\int_0^{\tau}dt^{\prime}\int d\bm{x}\;\| \delta\bm{\nu}_{t^{\prime}}(\bm{x})\|^2\mathcal{P}_{t^{\prime}}(\bm{x})}.
\end{align}
The term $\int_0^{\tau}dt^{\prime}\int d\bm{x}\;\| \delta\bm{\nu}_{t^{\prime}}(\bm{x})\|^2\mathcal{P}_{t^{\prime}}(\bm{x})$ is considered as the objective function since $L_{\rm SM}(\theta_{\rm SM}^*)=(1/{\tau}) \int_0^{\tau}dt^{\prime}\int d\bm{x}\;\| \delta\bm{\nu}_{t^{\prime}}(\bm{x}) \|^2\mathcal{P}_{t^{\prime}}(\bm{x})$ and $L_{\rm FM}(\theta_{\rm FM}^*)=(1/{\tau}) \int_0^{\tau}dt^{\prime}\int d\bm{x}\;\| \delta\bm{\nu}_{t^{\prime}}(\bm{x}) \|^2\mathcal{P}_{t^{\prime}}(\bm{x})$ hold in the probability flow ODE and in the flow matching, respectively. Taking this into account, we can rewrite this bound as
\begin{align}
    \frac{|\Delta \mathcal{W}_1|}{\sqrt{\tau D_{\rm max}}}\leq \sqrt{\int_0^{\tau}dt\: T_t\dot{S}_t^{\mathrm{tot}}} +    \sqrt{1+\frac{1}{D_{\rm max}}}\sqrt{\tau L_{\alpha}(\theta_{\alpha}^*)}\quad (\alpha=\text{SM, FM}).
\end{align}

\section{Details of the numerical experiments}\label{ap.numericalCalculation}

\subsection{One-dimensional Gaussian mixture distribution}
We explain the details of the numerical calculations in Fig.~\ref{fig.1dSAGMFM}. In the numerical calculation, the input distribution $q(\bm{x})$ is given by the $1$-dimensional situation Gaussian mixture distribution 
\begin{align}
     q(\bm{x})=\frac{1}{2}\mathcal{N}(\bm{x}|m^{\rm a},\sigma^2)+\frac{1}{2}\mathcal{N}(\bm{x}|m^{\rm b},\sigma^2),
\end{align}
with $\sigma=0.3$ and $m^{\rm a}=-m^{\rm b}=\sqrt{0.35}$. Here, we can analytically obtain $\bm{\nu}_t^{\mathcal{P}}(\bm{x})$ as follows. By considering the Fokker--Planck equation for the conditional distribution, $\partial_t\mathcal{P}_t^{\rm c}(\bm{x}|\bm{y})=-\nabla\cdot(\bm{\nu}_t^{\mathcal{P}^{\rm{c}}}(\bm{x}|\bm{y})\mathcal{P}_t^{\rm c}(\bm{x}|\bm{y}))$, the time evolution of $\mathcal{P}_t (\bm{x})= \int d\bm{y} \mathcal{P}_t^{\rm c}(\bm{x}|\bm{y})q(\bm{y})$ is given by
\begin{align}
\partial_t\mathcal{P}_t(\bm{x}) &= \partial_t \left[\int d\bm{y} \mathcal{P}_t^{\rm c}(\bm{x}|\bm{y})q(\bm{y}) \right] \nonumber\\
&= 
-\nabla\cdot \left( \frac{\int d\bm{y} \bm{\nu}_t^{\mathcal{P}^{\rm{c}}}(\bm{x}|\bm{y})\mathcal{P}_t^{\rm c}(\bm{x}|\bm{y})q(\bm{y})}{\mathcal{P}_t(\bm{x})} \mathcal{P}_t(\bm{x})\right).
\end{align}
Comparing this equation with Eq.~(\ref{eq.FP}), we obtain 
\begin{align}
\bm{\nu}_t^{\mathcal{P}}(\bm{x})=\int d\bm{y} \frac{\bm{\nu}_t^{\mathcal{P}^{\rm c}}(\bm{x}|\bm{y}) \mathcal{P}_t^{\rm c}(\bm{x}|\bm{y})q(\bm{y}) }{\mathcal{P}_t(\bm{x}) }.
\end{align}
From $\bm{\nu}_t^{\mathcal{P}^{\rm c}}(\bm{x}|\bm{y}) =\bm{f}_t(\bm{x})-T_t\nabla\ln \mathcal{P}^{\rm c}_t(\bm{x}|\bm{y})$ and Eqs.~(\ref{eq.temperature}), (\ref{eq.externalForce}), we obtain
\begin{align}
\bm{\nu}_t^{\mathcal{P}^{\rm c}}(\bm{x}|\bm{y}) &= (\partial_t \ln m_t) \bm{x} - \left[\partial_t \left( \frac{\sigma_t^2}
{2} \right) - \sigma_t^2 \partial_t \ln m_t \right]\nabla\ln \mathcal{P}^{\rm c}_t(\bm{x}|\bm{y})\nonumber\\
&=\frac{\partial_t m_t}{m_t}\bm{x} -\sigma_t^2 \left(\frac{\partial_t\sigma_t}{\sigma_t}-\frac{\partial_tm_t}{m_t}\right) \nabla\ln \mathcal{P}^{\rm c}_t(\bm{x}|\bm{y}).
\end{align}
Therefore, $\bm{\nu}_t^{\mathcal{P}}(\bm{x})$ is calculated as,
\begin{align}
\bm{\nu}_t^{\mathcal{P}}(\bm{x}) &=\int d\bm{y} \left[ \frac{\partial_t m_t}{m_t}\bm{x} -\sigma_t^2 \left(\frac{\partial_t\sigma_t}{\sigma_t}-\frac{\partial_tm_t}{m_t}\right) \nabla \ln \mathcal{P}^{\rm c}_t(\bm{x}|\bm{y}) \right] \frac{\mathcal{P}_t^{\rm c}(\bm{x}|\bm{y})q(\bm{y})} {\mathcal{P}_t(\bm{x}) } \nonumber \\
&=\int d\bm{y} \left[\frac{\partial_tm_t}{m_t}\bm{x}+\left(\frac{\partial_t\sigma_t}{\sigma_t}-\frac{\partial_tm_t}{m_t}\right)(\bm{x}-m_t\bm{y})\right]\frac{\mathcal{P}_t^{\rm c}(\bm{x}|\bm{y})q(\bm{y})}{\mathcal{P}_t(\bm{x})} \nonumber \\
&=\frac{\partial_t\sigma_t}{\sigma_t}\bm{x}-\frac{m_t}{\mathcal{P}_t(\bm{x})}\left(\frac{\partial_t\sigma_t}{\sigma_t}-\frac{\partial_tm_t}{m_t}\right) \mathbb{E}_{\mathcal{P}^c_t, q} [\bm{y}],
\end{align}
where we used $\mathcal{P}_t^{\rm c}(\bm{x}|\bm{y})=\mathcal{N}(\bm{x}|m_t\bm{y},\sigma_t^2)$ and the expected value $\mathbb{E}_{\mathcal{P}^c_t, q} [\bm{y}]$ is defined as $\mathbb{E}_{\mathcal{P}^c_t, q} [\bm{y}]=\int d\bm{y} \bm{y}  \mathcal{P}^c_t(\bm{x}|\bm{y}) q(\bm{y})$. Here, $\mathbb{E}_{\mathcal{P}^c_t, q} [\bm{y}]$ is calculated as
\begin{align}
 \mathbb{E}_{\mathcal{P}^c_t, q} [\bm{y}]&= \frac{1}{2} \int d\bm{y} \bm{y} \mathcal{N}(\bm{x}|m_t\bm{y},\sigma_t^2) \mathcal{N}(\bm{y}|m^{\rm a},\sigma^2)+\frac{1}{2}\int d\bm{y} \bm{y} \mathcal{N}(\bm{x}|m_t\bm{y},\sigma_t^2)\mathcal{N}(\bm{y}|m^{\rm b},\sigma^2)\nonumber \\
&=\frac{1}{2}\frac{\sigma^2 m_t\bm{x}+\sigma_t^2m^{\rm a}\boldsymbol{1}}{\sigma_t^2+\sigma^2m_t^2}\mathcal{N}\left(\bm{x}|m_tm^{\rm a},\sigma_t^2+\sigma^2m_t^2\right) + \frac{1}{2}\frac{\sigma^2 m_t\bm{x}+\sigma_t^2m^{\rm b}\boldsymbol{1}}{\sigma_t^2+\sigma^2m_t^2}\mathcal{N}\left(\bm{x}|m_tm^{\rm b},\sigma_t^2+\sigma^2m_t^2\right) \nonumber\\
&=\frac{1}{2}\frac{\sigma^2 m_t\bm{x}+\sigma_t^2m^{\rm a}\boldsymbol{1}}{\sigma_t^2+\sigma^2m_t^2} \mathcal{N}^{\rm a} + \frac{1}{2}\frac{\sigma^2 m_t\bm{x}+\sigma_t^2m^{\rm b}\boldsymbol{1}}{\sigma_t^2+\sigma^2m_t^2} \mathcal{N}^{\rm b}.
\end{align}
We use this analytical expression of the velocity field $\bm{\nu}_t^{\mathcal{P}}(\bm{x})$ [Eq.~(\ref{eq.analyticalVel})] for the cosine and cond-OT schedules.

To compute the velocity field for the optimal transport, we adopted the optimal transport conditional flow matching (OTCFM) method~\cite{tong2023improving}. This method is a conditional flow matching method that estimates the velocity field for the optimal transport by using samples generated from the optimal transport plan. To compute the optimal transport plan, we used the Python optimal transport (POT) library~\cite{flamary2021pot}. As an estimation architecture, we used a fully connected multi-layer perceptron with 5 layers consisting of 1024 neurons in each layer. The hyperbolic tangent function was used as the activation function. For training, we used the Adam optimizer~\cite{kingma2014adam} with a learning rate of $3\times 10^{-4}$ for learning $10^{7}$ data points with a batch size of $4056$.

To obtain the probability density function for the estimated reverse process, we numerically computed the ODE $d\tilde{\bm{x}}^{\dagger}_{\tilde{t}}/d\tilde{t} = \tilde{\bm{\nu}}^{\dagger}_{\tilde{t}}(\tilde{\bm{x}}^{\dagger}_{\tilde{t}})=-\bm{\nu}_{\tau-\tilde{t}}^{\mathcal{P}}(\tilde{\bm{x}}^{\dagger}_{\tilde{t}})$ for the velocity fields of the cosine schedule, the cond-OT schedule and the optimal transport with various initial conditions, using the $4$-th order explicit Runge--Kutta method. Specifically, we computed $3\times10^8$ samples and used a cubic spline interpolation for the histogram of the samples to approximate the probability density function.
To numerically compute the each term in the inequality [Eq.~(\ref{eq.derivativeSADM1})], we used numerical integration methods for $D_{\tau-t}=\int d\bm{x}\; (\delta \mathcal{P}_t(\bm{x}))^2/\mathcal{P}_t(\bm{x})$ and $[v_2(t)]^2=\int  d\bm{x}\;\|\bm{\nu}_t(\bm{x})\|^2\mathcal{P}_t(\bm{x})$, and we used the POT library for the $1$-Wasserstein distance. We used $\tau/\Delta t = 100$ for the time step $\Delta t$, and created a uniform grid for the spatial step $\Delta x$ as $(4.5-(-4.5))/\Delta x = 5\times10^2$ in $x\in[-4.5,4.5]$. When the initial condition is given by the uniform distribution, we set the spatial step to $(4.5-(-4.5))/\Delta x = 5\times10^3$  because the computation can be numerically unstable for a large spatial step due to the discontinuity of the distribution. We show the initial conditions in Tab.~\ref{tab.initialConditions}. We also show the time evolution of the probability density functions in the forward and estimated reverse process for each condition in Fig.~\ref{fig.initialConditions}.

\begin{table*}
    \centering
    \begin{tabular}{c|c}
        Initial conditions &Probability density functions\\
        \hline
        \hline
              Gaussian distribution 
              with a different mean
         
          & $q(\bm{x})=\mathcal{N}(\bm{x}|1.0,1.0)$\\
          \hline
         
             Gaussian distribution
              with a different variance

                  &$q(\bm{x})=\mathcal{N}(\bm{x}|0.0,0.8^2)$\\
          \hline

             Gaussian mixture
             distribution
 &
            $q(\bm{x})=\frac{1}{3}[\mathcal{N}(\bm{x}|\frac{-1}{0.94},\left(\frac{0.3}{0.94}\right)^2)+\mathcal{N}(\bm{x}|0,\left(\frac{0.8}{0.94}\right)^2)+\mathcal{N}(\bm{x}|\frac{1}{0.94},\left(\frac{0.3}{0.94}\right)^2)$
\\
        \hline

             Uniform distribution
     & $q(\bm{x})=\frac{1}{2\sqrt{3}}\;(\bm{x}\in[-\sqrt{3},\sqrt{3}])$
    \end{tabular}
    \caption{The initial conditions used in the numerical calculations [Fig.~\ref{fig.1dSAGMFM}]. Here $\mathcal{N}(\bm{x}| \mu, \sigma^2)$ denotes the $1$-dimensional Gaussian distribution with mean $\mu$ and variance $\sigma^2$.}
    \label{tab.initialConditions}
\end{table*}

\begin{figure*}
    \centering
    \includegraphics[width=\linewidth]{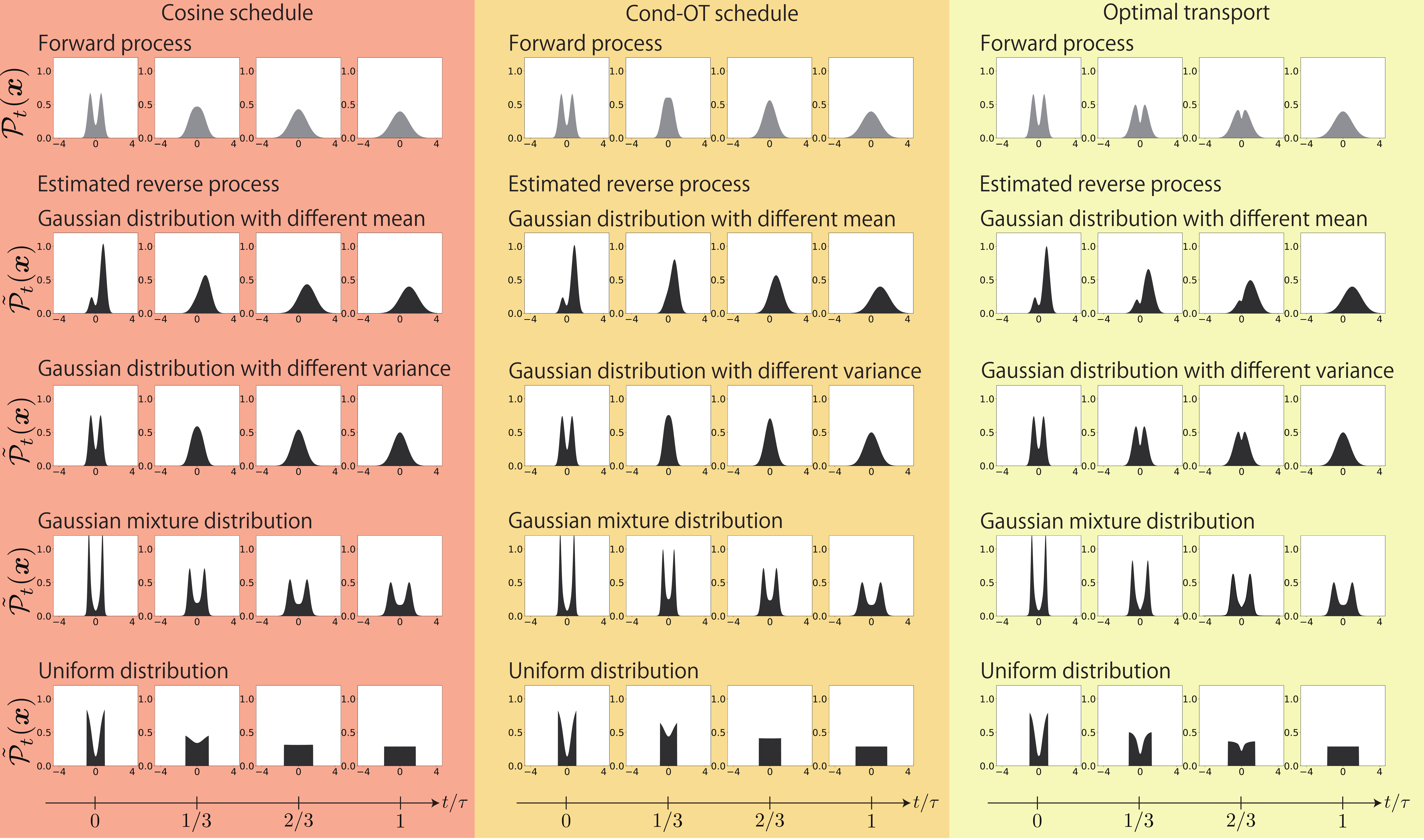}
    \caption{The time evolution of the probability density functions in the forward and estimated reverse processes, for each condition used in the numerical calculations [Fig.~\ref{fig.1dSAGMFM}].}
    \label{fig.initialConditions}
\end{figure*}

\subsection{Two-dimensional Swiss-roll dataset}
We explain the details of the numerical calculations in Fig.~\ref{fig:2dsadm}. 
First, we provide the details of the score estimation method under Eq.~(\ref{eq:2dns}). We adopted a slightly modified version of the denosing score matching method introduced in Sec.~\ref{sec.conditionalGauss} due to the non-conservativity of $\bm{f}_t(\bm{x})$. Similar to  the derivation in Appendix~\ref{ap.GaussianPath}, we solve the differential equations (\ref{difeqmean}) and (\ref{difeqcov}) to obtain a closed-form solution of $\mathcal{P}_t^{\rm c}(\bm{x}|\bm{y})=\mathcal{N}(\bm{x}| \bm{\mu}_t(\bm{y}),\mathsf{\Sigma}_t)$ under the initial conditions $\bm{\mu}_0(\bm{y})=\bm{y}$ and $\mathsf{\Sigma}_0 =\mathsf{O}$. Substituting Eq.~(\ref{eq:2dns}) into Eq.~(\ref{difeqmean}), we obtain
\begin{align}
     \partial_t \bm{\mu}_t(\bm{y}) &= \begin{pmatrix}
        H_t&-G_t\\
        G_t &H_t
    \end{pmatrix}\bm{\mu}_t(\bm{y}).
\end{align}
This differential equation is solvable, and the solution is given by
\begin{align}
    &\bm{\mu}_t(\bm{y}) = \mathsf{P}\exp\left(\int_0^{t} dt^{\prime} \begin{pmatrix}
        H_{t^{\prime}}+iG_{t^{\prime}} & 0 \\
        0 &  H_{t^{\prime}}-iG_{t^{\prime}}
    \end{pmatrix}\right)\mathsf{P}^{*} \mu_0(\bm{y}),\nonumber\\
    &=\exp\left(\int_0^{t}dt^{\prime} H_{t^{\prime}}\right)\begin{pmatrix}
        \cos\int_0^t dt^{\prime} G_{t^{\prime}} &-\sin\int_0^t dt^{\prime} G_{t^{\prime}}\\\sin\int_0^t dt^{\prime} G_{t^{\prime}}&\cos\int_0^t dt^{\prime} G_{t^{\prime}}
    \end{pmatrix}\mu_0(\bm{y}),
    \label{solution1}
\end{align}
where $i=\sqrt{-1}$ is an imaginary unit, $\mathsf{P}$ is an unitary matrix defined as
\begin{align}
    \mathsf{P} = \frac{1}{\sqrt{2}}\begin{pmatrix}
        i&-i\\1&1
    \end{pmatrix},
\end{align}
and $\mathsf{P}^*$ denotes an adjoint matrix of $\mathsf{P}$.

Next, we solve the differential equation for the covariance matrix [Eq.~(\ref{difeqcov})]. First we define the matrix $\mathsf{A}_t$ as
\begin{align}
    \mathsf{A}_t=\begin{pmatrix}
        H_t&-G_t\\
        G_t &H_t
    \end{pmatrix}=\begin{pmatrix}
        -h\left(1-\frac{t}{\tau}\right) -  T_{\rm fin}\left(\frac{t}{\tau}\right)^2,&-g\left(1-\frac{t}{\tau}\right)\\
        g\left(1-\frac{t}{\tau}\right) & -h\left(1-\frac{t}{\tau}\right) -  T_{\rm fin}\left(\frac{t}{\tau}\right)^2
    \end{pmatrix}.
\end{align} Here, Eq.~(\ref{difeqcov}) will be
\begin{align}
     \mathsf{A}_t\mathsf{\Sigma}_t + \mathsf{\Sigma}_t\mathsf{A}_t^{*}+2T_t\mathsf{I}-\partial_t \mathsf{\Sigma}_t =0
\end{align}
The real part of the eigenvalue of $\mathsf{A}_t$, that is $H_t$, is negative due to $h>0$ and $T_{\rm fin}>0$. Thus, an analytic expression of $\mathsf{\Sigma}_t$ is formally obtained as
\begin{align}
    \mathsf{\Sigma}_t &= \int_0^{\infty}ds\: e^{\mathsf{A}_ts}(2T_t\mathsf{I} -\partial_t\mathsf{\Sigma}_t)e^{\mathsf{A}^*_ts}\nonumber\\
    &=2T_t\int_0^{\infty}ds\: e^{(\mathsf{A}_t+\mathsf{A}_t^*)s}-\int_0^{\infty}ds\: e^{\mathsf{A}_t s}\partial_t\mathsf{\Sigma}_te^{\mathsf{A}^*_ts}\nonumber\\
    &=2T_t\int_0^{\infty}ds\: e^{2 H_ts}\mathsf{I}-\int_0^{\infty}ds\: e^{\mathsf{A}_t s}\partial_t\mathsf{\Sigma}_te^{\mathsf{A}_t^{*} s}\nonumber\\
    &=-\frac{T_t}{H_t}\mathsf{I}-\int_0^{\infty}ds\: e^{\mathsf{A}_t s}\partial_t\mathsf{\Sigma}_te^{\mathsf{A}_t^{*} s},\label{eq:lyapunov-solution}
\end{align}
because we can obtain $\mathsf{A}_t\mathsf{\Sigma}_t + \mathsf{\Sigma}_t\mathsf{A}_t^{*} = \int_0^{\infty}ds (d/ds)\left[ e^{\mathsf{A}_ts}(2T_t\mathsf{I} -\partial_t\mathsf{\Sigma}_t)e^{\mathsf{A}^*_ts} \right] = -2T_t\mathsf{I}+\partial_t\mathsf{\Sigma}_t$ if the real part of the eigenvalue of $\mathsf{A}_t$ is negative.
In this case, it is possible to derive a closed-form solution. Let $\tilde{\mathsf{A}}_t$ and $\tilde{\mathsf{\Sigma}}_t$ be $\tilde{\mathsf{A}}_t = \mathsf{P}^*\mathsf{A}_t\mathsf{P}$ and $ \tilde{\mathsf{\Sigma}}_t=\mathsf{P^*}\mathsf{\Sigma}_t\mathsf{P}$, respectively. The equation~(\ref{eq:lyapunov-solution}) can be rewritten as
\begin{align}
    &\tilde{\mathsf{\Sigma}}_t=-\frac{T_t}{H_t}\mathsf{I}-\int_0^{\infty}ds e^{\tilde{\mathsf{A}}_ts}\partial_t(\mathsf{P^*}\mathsf{\Sigma}_t\mathsf{P})e^{\tilde{\mathsf{A}}^*_ts}\nonumber\\
    &=-\frac{T_t}{H_t}\mathsf{I} -\int_0^{\infty}ds \begin{pmatrix}e^{(H_t+iG_t)s}&0\\  0 & e^{(H_t-iG_t)s}\end{pmatrix}\partial_t\tilde{\mathsf{\Sigma}}_t\begin{pmatrix}e^{(H_t-iG_t)s}&0\\  0 & e^{(H_t+iG_t)s}\end{pmatrix}\nonumber\\
    &=-\frac{T_t}{H_t}\mathsf{I}-\int_0^{\infty}ds\begin{pmatrix}
        e^{2H_ts}[\partial_t\tilde{\mathsf{\Sigma}}_t]_{11}&e^{2(H_t+iG_t)s}[\partial_t\tilde{\mathsf{\Sigma}}_t]_{12}\\
        e^{2(H_t-iG_t)s}[\partial_t\tilde{\mathsf{\Sigma}}_t]_{21}& e^{2H_ts}[\partial_t\tilde{\mathsf{\Sigma}}_t]_{22}
    \end{pmatrix}\nonumber\\
    &=-\frac{T_t}{H_t}\mathsf{I}+\frac{1}{2}\begin{pmatrix}
        [\partial_t\tilde{\mathsf{\Sigma}}_t]_{11}/H_t&[\partial_t\tilde{\mathsf{\Sigma}}_t]_{12}/(H_t+iG_t)\\
        [\partial_t\tilde{\mathsf{\Sigma}}_t]_{21}/(H_t-iG_t)& [\partial_t\tilde{\mathsf{\Sigma}}_t]_{22}/H_t
    \end{pmatrix}.
\end{align}
Considering this equation for each component of the matrix $\partial_t\tilde{\mathsf{\Sigma}}_t$, we obtain
\begin{align}
    \partial_t [\tilde{\mathsf{\Sigma}}_t]_{ii} &= 2T_t + 2H_t[\tilde{\mathsf{\Sigma}}_t]_{ii} \quad(i=1,2),\label{eq:diagcov}\\
    \partial_t  [\tilde{\mathsf{\Sigma}}_t]_{12} &= 2(H_t+iG_t)[\tilde{\mathsf{\Sigma}}_t]_{12},\\
     \partial_t  [\tilde{\mathsf{\Sigma}}_t]_{21} &= 2(H_t-iG_t)[\tilde{\mathsf{\Sigma}}_t]_{21},
\end{align}
which is solvable under the initial condition $\mathsf{\Sigma}_0=\tilde{\mathsf{\Sigma}}_0=\mathsf{O}$. The solution is given by
\begin{align}
\tilde{\mathsf{\Sigma}}_t = \left[e^{2\int_0^tdt^{\prime}H_{t^{\prime}}}\int_0^t ds e^{-2\int_0^{s}dt^{\prime}H_{t^{\prime}}}T_s \right]\mathsf{I}.
\end{align}
Since this is a scaled identity matrix $\mathsf{I}$, 
$\mathsf{\Sigma}_t = \mathsf{P}\tilde{\mathsf{\Sigma}}_t\mathsf{P}^*$ is also obtained as
\begin{align}
    \mathsf{\Sigma}_t= \left[e^{2\int_0^tdt^{\prime}H_{t^{\prime}}}\int_0^t ds e^{-2\int_0^{s}dt^{\prime}H_{t^{\prime}}}T_s \right]\mathsf{I}.
    \label{solution2}
\end{align}

 Using the solutions [Eqs.~(\ref{solution1}) and~(\ref{solution2})], we empirically minimized the conditional objective function Eq.~(\ref{eq.cSMLoss}) and estimated the score function $\bm{s}_t^{\theta}(\bm{x})$. We used a fully connected multi-layer perceptron with 8 layers consisting of 1024 neurons in each layer for $\bm{s}_t^{\theta}(\bm{x})$. The hyperbolic tangent function was used as the activation function. For training, we used the Adam optimizer~\cite{kingma2014adam} with a learning rate of $3\times 10^{-4}$ to learn $10^{7}$ data points with a batch size of $1024$. The number of samples for stochastic processes was set to $1\times10^8$. For the sampling procedure, we used the Euler--Maruyama method for the SDE in the forward process and the 4th order Runge--Kutta method for the ODE in the estimated reverse process.

To estimate the optimal transport, we used the Nadaraya-Watson style model described in Ref.~\cite{liu2022flow}. Here, we consider the $2$-Wasserstein distance between $\mathcal{P}_0(\boldsymbol{x}_0)$ and $\mathcal{P}_{\tau}(\boldsymbol{x}_{\tau})$, and obtain the optimal transport plan $\pi^* (\boldsymbol{x}_0, \boldsymbol{x}_{\tau})$ for the $2$-Wasserstein distance $\mathcal{W}_2 (\mathcal{P}_0, \mathcal{P}_{\tau}) = \sqrt{\mathbb{E}_{\pi^*}  [\|\boldsymbol{x}_0- \boldsymbol{x}_{\tau} \|^2]}$, where $\mathbb{E}_{\pi^*}[\cdots]$ is defined as $\mathbb{E}_{\pi^*}[\cdots]= \int d\boldsymbol{x}_0 \int d\boldsymbol{x}_{\tau} \pi^*(\boldsymbol{x}_0, \boldsymbol{x}_{\tau}) \cdots$. We consider the interpolation $\boldsymbol{x}_t =(1-t/\tau) \boldsymbol{x}_0 +(t/\tau) \boldsymbol{x}_{\tau}$ and velocity $(\boldsymbol{x}_{\tau}-\boldsymbol{x}_0)/\tau =(\boldsymbol{x}_{\tau}-\boldsymbol{x}_{t})/(\tau-t)$. We consider the expression of the optimal velocity field $\boldsymbol{\nu}^{\mathcal{P}}_t(\boldsymbol{x}) = \mathbb{E} [(\boldsymbol{x}_{\tau}-\boldsymbol{x}_0)/\tau | \boldsymbol{x}_{t}=\boldsymbol{x}]$, which is the conditional expected value of $(\boldsymbol{x}_{\tau}-\boldsymbol{x}_0)/\tau $ under the condition $\boldsymbol{x}_{t}=\boldsymbol{x}$. We introduce the normalized Gaussian kernel $\kappa(\boldsymbol{x}_{t}, \bm{x})=\exp[-\|\bm{x}_t-\bm{x}\|^2/(2\sigma_{\rm K}^2) ]/\mathbb{E}_{\pi^*}[\exp[-\|\bm{x}_t-\bm{x}\|^2/(2\sigma_{\rm K}^2)]$. If $\sigma_{\rm K}$ is relatively small, $\pi^* (\boldsymbol{x}_0, \boldsymbol{x}_{\tau}) \kappa(\boldsymbol{x}_{t}, \bm{x})$ can be regarded as the conditional probability of $(\boldsymbol{x}_0, \boldsymbol{x}_{\tau})$ under the condition $\boldsymbol{x}_t=\boldsymbol{x}$. Therefore, the optimal velocity field may be obtained from the expression
\begin{align}
\mathbb{E}_{\pi^*}\left[\frac{\bm{x}_\tau-\bm{x}}{\tau-t}\kappa(\bm{x}_t,\bm{x})\right],
\end{align}
and we used $1\times10^7$ samples and $\sigma_{\rm K}=0.08$ for this expression to estimate the optimal velocity field.

To numerically compute each term in the inequality [Eq.~(\ref{eq.derivativeSADM1})], we used numerical integration methods for $D_{\tau-t}=\int d\bm{x}\; (\delta \mathcal{P}_t(\bm{x}))^2/\mathcal{P}_t(\bm{x})$ and $T_t\dot{S}_t^{\rm tot}=\int  d\bm{x}\;\|\bm{\nu}_t(\bm{x})\|^2\mathcal{P}_t(\bm{x})$, and we used the POT library for the $1$-Wasserstein distance. We used $\tau/\Delta t = 100$ for the time step $\Delta t$, and created a uniform two-dimensional grid for the spatial step $\Delta x=5\times10^{-1} $ in the region $x_1\in[-5,5]$ and $x_2 \in[-5,5]$ for $\boldsymbol{x}=(x_1, x_2)$.

\subsection{Real-world image datasets}
Here, we describe the details of the experiment in Fig.~\ref{fig:latent-sadm}. In this experiment, we used the trained flow matching model for the $256\times 256$ downscaled CelebA HQ dataset~\cite{karras2018progressive} and the LSUN bedrooms subset~\cite{yu2015lsun} provided in Ref.~\cite{dao2023flow}. From each dataset, we sampled $3\times 10^4$ points in the latent space and computed the forward and the estimated reverse process. We assume that the estimation of the velocity field in the trained model is accurate. Let $\bm{u}_t^{\theta^*}(\bm{x})$ be the velocity field for the forward process in the trained model, and we computed $\partial_t \bm{x}_t = \bm{u}_t^{\theta^*}(\bm{x}_t)$ and $\partial_{\tilde{t}} \tilde{\bm{x}}^{\dagger}_{\tilde{t}} = - \bm{u}_{\tau - \tilde{t}}^{\theta^*}(\tilde{\bm{x}}^{\dagger}_{\tilde{t}} )$ for forward and estimated reverse processes, respectively. For the initial state of the estimated reverse process $\tilde{\bm{x}}^{\dagger}_{0}$, we used $3\times 10^4$ samples from the standard Gaussian distribution $\mathcal{N}(\bm{x}|\bm{0},\mathsf{I})$.

Using the generated trajectories of the processes $\bm{x}_t$ and $\tilde{\bm{x}}^{\dagger}_{\tilde{t}}$, we estimated $c_2 =\int_0^{\tau} dt [v_2(t)]^2$ by empirically calculating the expectation as follows, 
\begin{align}
    c_2 = \int_0^{\tau} dt\: \int d\bm{x} \|\bm{u}_t^{\theta^*}(\bm{x}) \|^2 \mathcal{P}_t(\bm{x}) = \tau \mathbb{E}_{\mathcal{U},\mathcal{P}_t}\left[\|\bm{u}_t^{\theta^*}(\bm{x})\|^2\right].
\end{align}
For the response function $\eta = (\Delta \mathcal{W}_1)^2/D_0$, we estimated $D_0 = \int d\bm{x}\: (\mathcal{P}_0^{\dagger}(\bm{x})-\tilde{\mathcal{P}}_0^{\dagger}(\bm{x}))^2/\mathcal{P}_0^{\dagger}(\bm{x})$ by assuming that $\mathcal{P}_0^{\dagger}(\bm{x})$ and $\tilde{\mathcal{P}}_0^{\dagger}(\bm{x})$ are Gaussian distributions, $\mathcal{P}_0^{\dagger}(\bm{x})$ is close to $\tilde{\mathcal{P}}_0^{\dagger}(\bm{x})$, and the higher order of the difference $\delta \mathcal{P}_0^{\dagger}(\bm{x}) = \tilde{\mathcal{P}}_0^{\dagger}(\bm{x})-\mathcal{P}_0^{\dagger}(\bm{x})$ is negligible in this experiment. In this situation, $D_0$ may be estimated using the Kullback-Leibler divergence $D_0 /2\approx  D_{\rm KL} ( \mathcal{P}_0^{\dagger}\|\tilde{\mathcal{P}}_0^{\dagger})= \int d\bm{x}\: \mathcal{P}_0^{\dagger}(\bm{x})\ln [\mathcal{P}_0^{\dagger}(\bm{x})/\tilde{\mathcal{P}}_0^{\dagger}(\bm{x})]$ because $D_{\rm KL} ( \mathcal{P}_0^{\dagger}\|\tilde{\mathcal{P}}_0^{\dagger})$ is given by
\begin{align}
    &D_{\rm KL} ( \mathcal{P}_0^{\dagger}\|\tilde{\mathcal{P}}_0^{\dagger}) = \int d\bm{x}\: \mathcal{P}_0^{\dagger}(\bm{x})\ln\frac{\mathcal{P}_0^{\dagger}(\bm{x})}{\tilde{\mathcal{P}}_0^{\dagger}(\bm{x})} = \frac{1}{2} D_0 + O\left(\left( \delta\mathcal{P}_0^{\dagger}(\bm{x})\right)^3\right).
\end{align}
The advantage of using the Kullback-Leibler divergence instead of the $\chi^2$-divergence is that the Kullback-Leibler divergence between two Gaussian distributions can be computed using only the mean and covariance matrix. In a realistic situation of the diffusion model for the real-world image dataset, such as the one we are dealing with here, we can assume that $\mathcal{P}_0^{\dagger}(\bm{x})$ and $\tilde{\mathcal{P}}_0^{\dagger}(\bm{x})$ are given by high dimensional Gaussian distributions and the difference between $\mathcal{P}_0^{\dagger}$ and $\tilde{\mathcal{P}}_0^{\dagger}$ is relatively small. In this case, it is numerically more stable and feasible to compute $D_0$ using the Kullback-Leibler divergence than using the conventional definition of the $\chi^2$-divergence. We also assume that the covariance matrices of the Gaussian distributions are proportional to the identity matrix to ensure that the estimated value of the Kullback-Leibler divergence is finite for the high-dimensional systems. For $\Delta \mathcal{W}_1 = \mathcal{W}_1(p,q)-\mathcal{W}_1(\mathcal{P}_0^{\dagger},\tilde{\mathcal{P}}_0^{\dagger})$, we computed the $1$-Wasserstein distances empirically using the POT library~\cite{flamary2021pot}.

\end{widetext}

\bibliography{ref}

\begin{thebibliography}{145}%
\makeatletter
\providecommand \@ifxundefined [1]{%
 \@ifx{#1\undefined}
}%
\providecommand \@ifnum [1]{%
 \ifnum #1\expandafter \@firstoftwo
 \else \expandafter \@secondoftwo
 \fi
}%
\providecommand \@ifx [1]{%
 \ifx #1\expandafter \@firstoftwo
 \else \expandafter \@secondoftwo
 \fi
}%
\providecommand \natexlab [1]{#1}%
\providecommand \enquote  [1]{``#1''}%
\providecommand \bibnamefont  [1]{#1}%
\providecommand \bibfnamefont [1]{#1}%
\providecommand \citenamefont [1]{#1}%
\providecommand \href@noop [0]{\@secondoftwo}%
\providecommand \href [0]{\begingroup \@sanitize@url \@href}%
\providecommand \@href[1]{\@@startlink{#1}\@@href}%
\providecommand \@@href[1]{\endgroup#1\@@endlink}%
\providecommand \@sanitize@url [0]{\catcode `\\12\catcode `\$12\catcode `\&12\catcode `\#12\catcode `\^12\catcode `\_12\catcode `\%12\relax}%
\providecommand \@@startlink[1]{}%
\providecommand \@@endlink[0]{}%
\providecommand \url  [0]{\begingroup\@sanitize@url \@url }%
\providecommand \@url [1]{\endgroup\@href {#1}{\urlprefix }}%
\providecommand \urlprefix  [0]{URL }%
\providecommand \Eprint [0]{\href }%
\providecommand \doibase [0]{https://doi.org/}%
\providecommand \selectlanguage [0]{\@gobble}%
\providecommand \bibinfo  [0]{\@secondoftwo}%
\providecommand \bibfield  [0]{\@secondoftwo}%
\providecommand \translation [1]{[#1]}%
\providecommand \BibitemOpen [0]{}%
\providecommand \bibitemStop [0]{}%
\providecommand \bibitemNoStop [0]{.\EOS\space}%
\providecommand \EOS [0]{\spacefactor3000\relax}%
\providecommand \BibitemShut  [1]{\csname bibitem#1\endcsname}%
\let\auto@bib@innerbib\@empty
\bibitem [{\citenamefont {Van~Kampen}(1992)}]{van1992stochastic}%
  \BibitemOpen
  \bibfield  {author} {\bibinfo {author} {\bibfnamefont {N.~G.}\ \bibnamefont {Van~Kampen}},\ }\href@noop {} {\emph {\bibinfo {title} {Stochastic processes in physics and chemistry}}},\ Vol.~\bibinfo {volume} {1}\ (\bibinfo  {publisher} {Elsevier},\ \bibinfo {year} {1992})\BibitemShut {NoStop}%
\bibitem [{\citenamefont {Sekimoto}(2010)}]{sekimoto2010stochastic}%
  \BibitemOpen
  \bibfield  {author} {\bibinfo {author} {\bibfnamefont {K.}~\bibnamefont {Sekimoto}},\ }\href@noop {} {\emph {\bibinfo {title} {Stochastic energetics}}},\ \bibinfo {edition} {1st}\ ed.,\ \bibinfo {series} {Lecture Notes in Physics}, Vol.\ \bibinfo {volume} {799}\ (\bibinfo  {publisher} {Springer},\ \bibinfo {year} {2010})\BibitemShut {NoStop}%
\bibitem [{\citenamefont {Seifert}(2012)}]{seifert2012stochastic}%
  \BibitemOpen
  \bibfield  {author} {\bibinfo {author} {\bibfnamefont {U.}~\bibnamefont {Seifert}},\ }\bibfield  {title} {\bibinfo {title} {Stochastic thermodynamics, fluctuation theorems and molecular machines},\ }\href@noop {} {\bibfield  {journal} {\bibinfo  {journal} {Reports on progress in physics}\ }\textbf {\bibinfo {volume} {75}},\ \bibinfo {pages} {126001} (\bibinfo {year} {2012})}\BibitemShut {NoStop}%
\bibitem [{\citenamefont {Sekimoto}(1998)}]{sekimoto1998langevin}%
  \BibitemOpen
  \bibfield  {author} {\bibinfo {author} {\bibfnamefont {K.}~\bibnamefont {Sekimoto}},\ }\bibfield  {title} {\bibinfo {title} {Langevin equation and thermodynamics},\ }\href@noop {} {\bibfield  {journal} {\bibinfo  {journal} {Progress of Theoretical Physics Supplement}\ }\textbf {\bibinfo {volume} {130}},\ \bibinfo {pages} {17} (\bibinfo {year} {1998})}\BibitemShut {NoStop}%
\bibitem [{\citenamefont {Kurchan}(1998)}]{kurchan1998fluctuation}%
  \BibitemOpen
  \bibfield  {author} {\bibinfo {author} {\bibfnamefont {J.}~\bibnamefont {Kurchan}},\ }\bibfield  {title} {\bibinfo {title} {Fluctuation theorem for stochastic dynamics},\ }\href@noop {} {\bibfield  {journal} {\bibinfo  {journal} {Journal of Physics A: Mathematical and General}\ }\textbf {\bibinfo {volume} {31}},\ \bibinfo {pages} {3719} (\bibinfo {year} {1998})}\BibitemShut {NoStop}%
\bibitem [{\citenamefont {Hatano}\ and\ \citenamefont {Sasa}(2001)}]{hatano2001steady}%
  \BibitemOpen
  \bibfield  {author} {\bibinfo {author} {\bibfnamefont {T.}~\bibnamefont {Hatano}}\ and\ \bibinfo {author} {\bibfnamefont {S.-i.}\ \bibnamefont {Sasa}},\ }\bibfield  {title} {\bibinfo {title} {Steady-state thermodynamics of langevin systems},\ }\href@noop {} {\bibfield  {journal} {\bibinfo  {journal} {Physical review letters}\ }\textbf {\bibinfo {volume} {86}},\ \bibinfo {pages} {3463} (\bibinfo {year} {2001})}\BibitemShut {NoStop}%
\bibitem [{\citenamefont {Seifert}(2005)}]{seifert2005entropy}%
  \BibitemOpen
  \bibfield  {author} {\bibinfo {author} {\bibfnamefont {U.}~\bibnamefont {Seifert}},\ }\bibfield  {title} {\bibinfo {title} {Entropy production along a stochastic trajectory and an integral fluctuation theorem},\ }\href@noop {} {\bibfield  {journal} {\bibinfo  {journal} {Physical review letters}\ }\textbf {\bibinfo {volume} {95}},\ \bibinfo {pages} {040602} (\bibinfo {year} {2005})}\BibitemShut {NoStop}%
\bibitem [{\citenamefont {Chernyak}\ \emph {et~al.}(2006)\citenamefont {Chernyak}, \citenamefont {Chertkov},\ and\ \citenamefont {Jarzynski}}]{chernyak2006path}%
  \BibitemOpen
  \bibfield  {author} {\bibinfo {author} {\bibfnamefont {V.~Y.}\ \bibnamefont {Chernyak}}, \bibinfo {author} {\bibfnamefont {M.}~\bibnamefont {Chertkov}},\ and\ \bibinfo {author} {\bibfnamefont {C.}~\bibnamefont {Jarzynski}},\ }\bibfield  {title} {\bibinfo {title} {Path-integral analysis of fluctuation theorems for general langevin processes},\ }\href@noop {} {\bibfield  {journal} {\bibinfo  {journal} {Journal of Statistical Mechanics: Theory and Experiment}\ }\textbf {\bibinfo {volume} {2006}},\ \bibinfo {pages} {P08001} (\bibinfo {year} {2006})}\BibitemShut {NoStop}%
\bibitem [{\citenamefont {Schmiedl}\ and\ \citenamefont {Seifert}(2007)}]{schmiedl2007efficiency}%
  \BibitemOpen
  \bibfield  {author} {\bibinfo {author} {\bibfnamefont {T.}~\bibnamefont {Schmiedl}}\ and\ \bibinfo {author} {\bibfnamefont {U.}~\bibnamefont {Seifert}},\ }\bibfield  {title} {\bibinfo {title} {Efficiency at maximum power: An analytically solvable model for stochastic heat engines},\ }\href@noop {} {\bibfield  {journal} {\bibinfo  {journal} {Europhysics letters}\ }\textbf {\bibinfo {volume} {81}},\ \bibinfo {pages} {20003} (\bibinfo {year} {2007})}\BibitemShut {NoStop}%
\bibitem [{\citenamefont {Allahverdyan}\ \emph {et~al.}(2009)\citenamefont {Allahverdyan}, \citenamefont {Janzing},\ and\ \citenamefont {Mahler}}]{allahverdyan2009thermodynamic}%
  \BibitemOpen
  \bibfield  {author} {\bibinfo {author} {\bibfnamefont {A.~E.}\ \bibnamefont {Allahverdyan}}, \bibinfo {author} {\bibfnamefont {D.}~\bibnamefont {Janzing}},\ and\ \bibinfo {author} {\bibfnamefont {G.}~\bibnamefont {Mahler}},\ }\bibfield  {title} {\bibinfo {title} {Thermodynamic efficiency of information and heat flow},\ }\href@noop {} {\bibfield  {journal} {\bibinfo  {journal} {Journal of Statistical Mechanics: Theory and Experiment}\ }\textbf {\bibinfo {volume} {2009}},\ \bibinfo {pages} {P09011} (\bibinfo {year} {2009})}\BibitemShut {NoStop}%
\bibitem [{\citenamefont {Van~den Broeck}\ and\ \citenamefont {Esposito}(2010)}]{van2010three}%
  \BibitemOpen
  \bibfield  {author} {\bibinfo {author} {\bibfnamefont {C.}~\bibnamefont {Van~den Broeck}}\ and\ \bibinfo {author} {\bibfnamefont {M.}~\bibnamefont {Esposito}},\ }\bibfield  {title} {\bibinfo {title} {Three faces of the second law. ii. fokker-planck formulation},\ }\href@noop {} {\bibfield  {journal} {\bibinfo  {journal} {Physical Review E}\ }\textbf {\bibinfo {volume} {82}},\ \bibinfo {pages} {011144} (\bibinfo {year} {2010})}\BibitemShut {NoStop}%
\bibitem [{\citenamefont {Sagawa}\ and\ \citenamefont {Ueda}(2012)}]{sagawa2012nonequilibrium}%
  \BibitemOpen
  \bibfield  {author} {\bibinfo {author} {\bibfnamefont {T.}~\bibnamefont {Sagawa}}\ and\ \bibinfo {author} {\bibfnamefont {M.}~\bibnamefont {Ueda}},\ }\bibfield  {title} {\bibinfo {title} {Nonequilibrium thermodynamics of feedback control},\ }\href@noop {} {\bibfield  {journal} {\bibinfo  {journal} {Physical Review E}\ }\textbf {\bibinfo {volume} {85}},\ \bibinfo {pages} {021104} (\bibinfo {year} {2012})}\BibitemShut {NoStop}%
\bibitem [{\citenamefont {Ito}\ and\ \citenamefont {Sagawa}(2013)}]{ito2013information}%
  \BibitemOpen
  \bibfield  {author} {\bibinfo {author} {\bibfnamefont {S.}~\bibnamefont {Ito}}\ and\ \bibinfo {author} {\bibfnamefont {T.}~\bibnamefont {Sagawa}},\ }\bibfield  {title} {\bibinfo {title} {Information thermodynamics on causal networks},\ }\href@noop {} {\bibfield  {journal} {\bibinfo  {journal} {Physical review letters}\ }\textbf {\bibinfo {volume} {111}},\ \bibinfo {pages} {180603} (\bibinfo {year} {2013})}\BibitemShut {NoStop}%
\bibitem [{\citenamefont {Horowitz}\ and\ \citenamefont {Sandberg}(2014)}]{horowitz2014second}%
  \BibitemOpen
  \bibfield  {author} {\bibinfo {author} {\bibfnamefont {J.~M.}\ \bibnamefont {Horowitz}}\ and\ \bibinfo {author} {\bibfnamefont {H.}~\bibnamefont {Sandberg}},\ }\bibfield  {title} {\bibinfo {title} {Second-law-like inequalities with information and their interpretations},\ }\href@noop {} {\bibfield  {journal} {\bibinfo  {journal} {New Journal of Physics}\ }\textbf {\bibinfo {volume} {16}},\ \bibinfo {pages} {125007} (\bibinfo {year} {2014})}\BibitemShut {NoStop}%
\bibitem [{\citenamefont {Ito}\ and\ \citenamefont {Sagawa}(2015)}]{ito2015maxwell}%
  \BibitemOpen
  \bibfield  {author} {\bibinfo {author} {\bibfnamefont {S.}~\bibnamefont {Ito}}\ and\ \bibinfo {author} {\bibfnamefont {T.}~\bibnamefont {Sagawa}},\ }\bibfield  {title} {\bibinfo {title} {Maxwell’s demon in biochemical signal transduction with feedback loop},\ }\href@noop {} {\bibfield  {journal} {\bibinfo  {journal} {Nature communications}\ }\textbf {\bibinfo {volume} {6}},\ \bibinfo {pages} {1} (\bibinfo {year} {2015})}\BibitemShut {NoStop}%
\bibitem [{\citenamefont {Gingrich}\ \emph {et~al.}(2017)\citenamefont {Gingrich}, \citenamefont {Rotskoff},\ and\ \citenamefont {Horowitz}}]{gingrich2017inferring}%
  \BibitemOpen
  \bibfield  {author} {\bibinfo {author} {\bibfnamefont {T.~R.}\ \bibnamefont {Gingrich}}, \bibinfo {author} {\bibfnamefont {G.~M.}\ \bibnamefont {Rotskoff}},\ and\ \bibinfo {author} {\bibfnamefont {J.~M.}\ \bibnamefont {Horowitz}},\ }\bibfield  {title} {\bibinfo {title} {Inferring dissipation from current fluctuations},\ }\href@noop {} {\bibfield  {journal} {\bibinfo  {journal} {Journal of Physics A: Mathematical and Theoretical}\ }\textbf {\bibinfo {volume} {50}},\ \bibinfo {pages} {184004} (\bibinfo {year} {2017})}\BibitemShut {NoStop}%
\bibitem [{\citenamefont {Dechant}\ and\ \citenamefont {Sasa}(2018)}]{dechant2018entropic}%
  \BibitemOpen
  \bibfield  {author} {\bibinfo {author} {\bibfnamefont {A.}~\bibnamefont {Dechant}}\ and\ \bibinfo {author} {\bibfnamefont {S.-i.}\ \bibnamefont {Sasa}},\ }\bibfield  {title} {\bibinfo {title} {Entropic bounds on currents in langevin systems},\ }\href@noop {} {\bibfield  {journal} {\bibinfo  {journal} {Physical Review E}\ }\textbf {\bibinfo {volume} {97}},\ \bibinfo {pages} {062101} (\bibinfo {year} {2018})}\BibitemShut {NoStop}%
\bibitem [{\citenamefont {Li}\ \emph {et~al.}(2019)\citenamefont {Li}, \citenamefont {Horowitz}, \citenamefont {Gingrich},\ and\ \citenamefont {Fakhri}}]{li2019quantifying}%
  \BibitemOpen
  \bibfield  {author} {\bibinfo {author} {\bibfnamefont {J.}~\bibnamefont {Li}}, \bibinfo {author} {\bibfnamefont {J.~M.}\ \bibnamefont {Horowitz}}, \bibinfo {author} {\bibfnamefont {T.~R.}\ \bibnamefont {Gingrich}},\ and\ \bibinfo {author} {\bibfnamefont {N.}~\bibnamefont {Fakhri}},\ }\bibfield  {title} {\bibinfo {title} {Quantifying dissipation using fluctuating currents},\ }\href@noop {} {\bibfield  {journal} {\bibinfo  {journal} {Nature communications}\ }\textbf {\bibinfo {volume} {10}},\ \bibinfo {pages} {1666} (\bibinfo {year} {2019})}\BibitemShut {NoStop}%
\bibitem [{\citenamefont {Hasegawa}\ and\ \citenamefont {Van~Vu}(2019)}]{hasegawa2019uncertainty}%
  \BibitemOpen
  \bibfield  {author} {\bibinfo {author} {\bibfnamefont {Y.}~\bibnamefont {Hasegawa}}\ and\ \bibinfo {author} {\bibfnamefont {T.}~\bibnamefont {Van~Vu}},\ }\bibfield  {title} {\bibinfo {title} {Uncertainty relations in stochastic processes: An information inequality approach},\ }\href@noop {} {\bibfield  {journal} {\bibinfo  {journal} {Physical Review E}\ }\textbf {\bibinfo {volume} {99}},\ \bibinfo {pages} {062126} (\bibinfo {year} {2019})}\BibitemShut {NoStop}%
\bibitem [{\citenamefont {Ito}\ and\ \citenamefont {Dechant}(2020)}]{ito2020stochastic}%
  \BibitemOpen
  \bibfield  {author} {\bibinfo {author} {\bibfnamefont {S.}~\bibnamefont {Ito}}\ and\ \bibinfo {author} {\bibfnamefont {A.}~\bibnamefont {Dechant}},\ }\bibfield  {title} {\bibinfo {title} {Stochastic time evolution, information geometry, and the cram{\'e}r-rao bound},\ }\href@noop {} {\bibfield  {journal} {\bibinfo  {journal} {Physical Review X}\ }\textbf {\bibinfo {volume} {10}},\ \bibinfo {pages} {021056} (\bibinfo {year} {2020})}\BibitemShut {NoStop}%
\bibitem [{\citenamefont {Otsubo}\ \emph {et~al.}(2020)\citenamefont {Otsubo}, \citenamefont {Ito}, \citenamefont {Dechant},\ and\ \citenamefont {Sagawa}}]{otsubo2020estimating}%
  \BibitemOpen
  \bibfield  {author} {\bibinfo {author} {\bibfnamefont {S.}~\bibnamefont {Otsubo}}, \bibinfo {author} {\bibfnamefont {S.}~\bibnamefont {Ito}}, \bibinfo {author} {\bibfnamefont {A.}~\bibnamefont {Dechant}},\ and\ \bibinfo {author} {\bibfnamefont {T.}~\bibnamefont {Sagawa}},\ }\bibfield  {title} {\bibinfo {title} {Estimating entropy production by machine learning of short-time fluctuating currents},\ }\href@noop {} {\bibfield  {journal} {\bibinfo  {journal} {Physical Review E}\ }\textbf {\bibinfo {volume} {101}},\ \bibinfo {pages} {062106} (\bibinfo {year} {2020})}\BibitemShut {NoStop}%
\bibitem [{\citenamefont {Dechant}\ and\ \citenamefont {Sasa}(2021)}]{dechant2021continuous}%
  \BibitemOpen
  \bibfield  {author} {\bibinfo {author} {\bibfnamefont {A.}~\bibnamefont {Dechant}}\ and\ \bibinfo {author} {\bibfnamefont {S.-i.}\ \bibnamefont {Sasa}},\ }\bibfield  {title} {\bibinfo {title} {Continuous time reversal and equality in the thermodynamic uncertainty relation},\ }\href@noop {} {\bibfield  {journal} {\bibinfo  {journal} {Physical Review Research}\ }\textbf {\bibinfo {volume} {3}},\ \bibinfo {pages} {L042012} (\bibinfo {year} {2021})}\BibitemShut {NoStop}%
\bibitem [{\citenamefont {Otsubo}\ \emph {et~al.}(2022)\citenamefont {Otsubo}, \citenamefont {Manikandan}, \citenamefont {Sagawa},\ and\ \citenamefont {Krishnamurthy}}]{otsubo2022estimating}%
  \BibitemOpen
  \bibfield  {author} {\bibinfo {author} {\bibfnamefont {S.}~\bibnamefont {Otsubo}}, \bibinfo {author} {\bibfnamefont {S.~K.}\ \bibnamefont {Manikandan}}, \bibinfo {author} {\bibfnamefont {T.}~\bibnamefont {Sagawa}},\ and\ \bibinfo {author} {\bibfnamefont {S.}~\bibnamefont {Krishnamurthy}},\ }\bibfield  {title} {\bibinfo {title} {Estimating time-dependent entropy production from non-equilibrium trajectories},\ }\href@noop {} {\bibfield  {journal} {\bibinfo  {journal} {Communications Physics}\ }\textbf {\bibinfo {volume} {5}},\ \bibinfo {pages} {11} (\bibinfo {year} {2022})}\BibitemShut {NoStop}%
\bibitem [{\citenamefont {Koyuk}\ and\ \citenamefont {Seifert}(2020)}]{koyuk2020thermodynamic}%
  \BibitemOpen
  \bibfield  {author} {\bibinfo {author} {\bibfnamefont {T.}~\bibnamefont {Koyuk}}\ and\ \bibinfo {author} {\bibfnamefont {U.}~\bibnamefont {Seifert}},\ }\bibfield  {title} {\bibinfo {title} {Thermodynamic uncertainty relation for time-dependent driving},\ }\href@noop {} {\bibfield  {journal} {\bibinfo  {journal} {Physical Review Letters}\ }\textbf {\bibinfo {volume} {125}},\ \bibinfo {pages} {260604} (\bibinfo {year} {2020})}\BibitemShut {NoStop}%
\bibitem [{\citenamefont {Villani}\ \emph {et~al.}(2009)\citenamefont {Villani} \emph {et~al.}}]{villani2009optimal}%
  \BibitemOpen
  \bibfield  {author} {\bibinfo {author} {\bibfnamefont {C.}~\bibnamefont {Villani}} \emph {et~al.},\ }\href@noop {} {\emph {\bibinfo {title} {Optimal transport: old and new}}},\ Vol.\ \bibinfo {volume} {338}\ (\bibinfo  {publisher} {Springer},\ \bibinfo {year} {2009})\BibitemShut {NoStop}%
\bibitem [{\citenamefont {Jordan}\ \emph {et~al.}(1998)\citenamefont {Jordan}, \citenamefont {Kinderlehrer},\ and\ \citenamefont {Otto}}]{jordan1998variational}%
  \BibitemOpen
  \bibfield  {author} {\bibinfo {author} {\bibfnamefont {R.}~\bibnamefont {Jordan}}, \bibinfo {author} {\bibfnamefont {D.}~\bibnamefont {Kinderlehrer}},\ and\ \bibinfo {author} {\bibfnamefont {F.}~\bibnamefont {Otto}},\ }\bibfield  {title} {\bibinfo {title} {The variational formulation of the fokker--planck equation},\ }\href@noop {} {\bibfield  {journal} {\bibinfo  {journal} {SIAM journal on mathematical analysis}\ }\textbf {\bibinfo {volume} {29}},\ \bibinfo {pages} {1} (\bibinfo {year} {1998})}\BibitemShut {NoStop}%
\bibitem [{\citenamefont {Aurell}\ \emph {et~al.}(2011)\citenamefont {Aurell}, \citenamefont {Mej{\'\i}a-Monasterio},\ and\ \citenamefont {Muratore-Ginanneschi}}]{aurell2011optimal}%
  \BibitemOpen
  \bibfield  {author} {\bibinfo {author} {\bibfnamefont {E.}~\bibnamefont {Aurell}}, \bibinfo {author} {\bibfnamefont {C.}~\bibnamefont {Mej{\'\i}a-Monasterio}},\ and\ \bibinfo {author} {\bibfnamefont {P.}~\bibnamefont {Muratore-Ginanneschi}},\ }\bibfield  {title} {\bibinfo {title} {Optimal protocols and optimal transport in stochastic thermodynamics},\ }\href@noop {} {\bibfield  {journal} {\bibinfo  {journal} {Physical review letters}\ }\textbf {\bibinfo {volume} {106}},\ \bibinfo {pages} {250601} (\bibinfo {year} {2011})}\BibitemShut {NoStop}%
\bibitem [{\citenamefont {Chen}\ \emph {et~al.}(2019)\citenamefont {Chen}, \citenamefont {Georgiou},\ and\ \citenamefont {Tannenbaum}}]{chen2019stochastic}%
  \BibitemOpen
  \bibfield  {author} {\bibinfo {author} {\bibfnamefont {Y.}~\bibnamefont {Chen}}, \bibinfo {author} {\bibfnamefont {T.~T.}\ \bibnamefont {Georgiou}},\ and\ \bibinfo {author} {\bibfnamefont {A.}~\bibnamefont {Tannenbaum}},\ }\bibfield  {title} {\bibinfo {title} {Stochastic control and nonequilibrium thermodynamics: Fundamental limits},\ }\href@noop {} {\bibfield  {journal} {\bibinfo  {journal} {IEEE transactions on automatic control}\ }\textbf {\bibinfo {volume} {65}},\ \bibinfo {pages} {2979} (\bibinfo {year} {2019})}\BibitemShut {NoStop}%
\bibitem [{\citenamefont {Nakazato}\ and\ \citenamefont {Ito}(2021)}]{nakazato2021geometrical}%
  \BibitemOpen
  \bibfield  {author} {\bibinfo {author} {\bibfnamefont {M.}~\bibnamefont {Nakazato}}\ and\ \bibinfo {author} {\bibfnamefont {S.}~\bibnamefont {Ito}},\ }\bibfield  {title} {\bibinfo {title} {Geometrical aspects of entropy production in stochastic thermodynamics based on wasserstein distance},\ }\href@noop {} {\bibfield  {journal} {\bibinfo  {journal} {Physical Review Research}\ }\textbf {\bibinfo {volume} {3}},\ \bibinfo {pages} {043093} (\bibinfo {year} {2021})}\BibitemShut {NoStop}%
\bibitem [{\citenamefont {Aurell}\ \emph {et~al.}(2012)\citenamefont {Aurell}, \citenamefont {Gaw{\c e}dzki}, \citenamefont {Mej{\'\i}a-Monasterio}, \citenamefont {Mohayaee},\ and\ \citenamefont {Muratore-Ginanneschi}}]{aurell2012refined}%
  \BibitemOpen
  \bibfield  {author} {\bibinfo {author} {\bibfnamefont {E.}~\bibnamefont {Aurell}}, \bibinfo {author} {\bibfnamefont {K.}~\bibnamefont {Gaw{\c e}dzki}}, \bibinfo {author} {\bibfnamefont {C.}~\bibnamefont {Mej{\'\i}a-Monasterio}}, \bibinfo {author} {\bibfnamefont {R.}~\bibnamefont {Mohayaee}},\ and\ \bibinfo {author} {\bibfnamefont {P.}~\bibnamefont {Muratore-Ginanneschi}},\ }\bibfield  {title} {\bibinfo {title} {Refined second law of thermodynamics for fast random processes},\ }\href {https://doi.org/10.1007/s10955-012-0478-x} {\bibfield  {journal} {\bibinfo  {journal} {Journal of Statistical Physics}\ }\textbf {\bibinfo {volume} {147}},\ \bibinfo {pages} {487} (\bibinfo {year} {2012})}\BibitemShut {NoStop}%
\bibitem [{\citenamefont {Dechant}\ \emph {et~al.}(2022{\natexlab{a}})\citenamefont {Dechant}, \citenamefont {Sasa},\ and\ \citenamefont {Ito}}]{dechant2022geometric2}%
  \BibitemOpen
  \bibfield  {author} {\bibinfo {author} {\bibfnamefont {A.}~\bibnamefont {Dechant}}, \bibinfo {author} {\bibfnamefont {S.-i.}\ \bibnamefont {Sasa}},\ and\ \bibinfo {author} {\bibfnamefont {S.}~\bibnamefont {Ito}},\ }\bibfield  {title} {\bibinfo {title} {Geometric decomposition of entropy production into excess, housekeeping, and coupling parts},\ }\href@noop {} {\bibfield  {journal} {\bibinfo  {journal} {Physical Review E}\ }\textbf {\bibinfo {volume} {106}},\ \bibinfo {pages} {024125} (\bibinfo {year} {2022}{\natexlab{a}})}\BibitemShut {NoStop}%
\bibitem [{\citenamefont {Ito}(2024)}]{ito2023geometric}%
  \BibitemOpen
  \bibfield  {author} {\bibinfo {author} {\bibfnamefont {S.}~\bibnamefont {Ito}},\ }\bibfield  {title} {\bibinfo {title} {Geometric thermodynamics for the fokker--planck equation: stochastic thermodynamic links between information geometry and optimal transport},\ }\href@noop {} {\bibfield  {journal} {\bibinfo  {journal} {Information Geometry}\ }\textbf {\bibinfo {volume} {7}},\ \bibinfo {pages} {441} (\bibinfo {year} {2024})}\BibitemShut {NoStop}%
\bibitem [{\citenamefont {Nagayama}\ \emph {et~al.}(2023)\citenamefont {Nagayama}, \citenamefont {Yoshimura}, \citenamefont {Kolchinsky},\ and\ \citenamefont {Ito}}]{nagayama2023geometric}%
  \BibitemOpen
  \bibfield  {author} {\bibinfo {author} {\bibfnamefont {R.}~\bibnamefont {Nagayama}}, \bibinfo {author} {\bibfnamefont {K.}~\bibnamefont {Yoshimura}}, \bibinfo {author} {\bibfnamefont {A.}~\bibnamefont {Kolchinsky}},\ and\ \bibinfo {author} {\bibfnamefont {S.}~\bibnamefont {Ito}},\ }\bibfield  {title} {\bibinfo {title} {Geometric thermodynamics of reaction-diffusion systems: Thermodynamic trade-off relations and optimal transport for pattern formation},\ }\href@noop {} {\bibfield  {journal} {\bibinfo  {journal} {arXiv preprint arXiv:2311.16569}\ } (\bibinfo {year} {2023})}\BibitemShut {NoStop}%
\bibitem [{\citenamefont {Tomczak}(2022)}]{tomczak2022deep}%
  \BibitemOpen
  \bibfield  {author} {\bibinfo {author} {\bibfnamefont {J.~M.}\ \bibnamefont {Tomczak}},\ }\href@noop {} {\emph {\bibinfo {title} {Deep generative modeling}}}\ (\bibinfo  {publisher} {Springer},\ \bibinfo {year} {2022})\BibitemShut {NoStop}%
\bibitem [{\citenamefont {Sohl-Dickstein}\ \emph {et~al.}(2015)\citenamefont {Sohl-Dickstein}, \citenamefont {Weiss}, \citenamefont {Maheswaranathan},\ and\ \citenamefont {Ganguli}}]{sohl2015deep}%
  \BibitemOpen
  \bibfield  {author} {\bibinfo {author} {\bibfnamefont {J.}~\bibnamefont {Sohl-Dickstein}}, \bibinfo {author} {\bibfnamefont {E.}~\bibnamefont {Weiss}}, \bibinfo {author} {\bibfnamefont {N.}~\bibnamefont {Maheswaranathan}},\ and\ \bibinfo {author} {\bibfnamefont {S.}~\bibnamefont {Ganguli}},\ }\bibfield  {title} {\bibinfo {title} {Deep unsupervised learning using nonequilibrium thermodynamics},\ }in\ \href@noop {} {\emph {\bibinfo {booktitle} {International conference on machine learning}}}\ (\bibinfo {organization} {PMLR},\ \bibinfo {year} {2015})\ pp.\ \bibinfo {pages} {2256--2265}\BibitemShut {NoStop}%
\bibitem [{\citenamefont {Song}\ \emph {et~al.}(2020{\natexlab{a}})\citenamefont {Song}, \citenamefont {Sohl-Dickstein}, \citenamefont {Kingma}, \citenamefont {Kumar}, \citenamefont {Ermon},\ and\ \citenamefont {Poole}}]{song2020scorebased}%
  \BibitemOpen
  \bibfield  {author} {\bibinfo {author} {\bibfnamefont {Y.}~\bibnamefont {Song}}, \bibinfo {author} {\bibfnamefont {J.}~\bibnamefont {Sohl-Dickstein}}, \bibinfo {author} {\bibfnamefont {D.~P.}\ \bibnamefont {Kingma}}, \bibinfo {author} {\bibfnamefont {A.}~\bibnamefont {Kumar}}, \bibinfo {author} {\bibfnamefont {S.}~\bibnamefont {Ermon}},\ and\ \bibinfo {author} {\bibfnamefont {B.}~\bibnamefont {Poole}},\ }\bibfield  {title} {\bibinfo {title} {Score-based generative modeling through stochastic differential equations},\ }in\ \href@noop {} {\emph {\bibinfo {booktitle} {International Conference on Learning Representations}}}\ (\bibinfo {year} {2020})\BibitemShut {NoStop}%
\bibitem [{\citenamefont {Evans}\ and\ \citenamefont {Searles}(2002)}]{evans2002fluctuation}%
  \BibitemOpen
  \bibfield  {author} {\bibinfo {author} {\bibfnamefont {D.~J.}\ \bibnamefont {Evans}}\ and\ \bibinfo {author} {\bibfnamefont {D.~J.}\ \bibnamefont {Searles}},\ }\bibfield  {title} {\bibinfo {title} {The fluctuation theorem},\ }\href@noop {} {\bibfield  {journal} {\bibinfo  {journal} {Advances in Physics}\ }\textbf {\bibinfo {volume} {51}},\ \bibinfo {pages} {1529} (\bibinfo {year} {2002})}\BibitemShut {NoStop}%
\bibitem [{\citenamefont {Crooks}(1999)}]{crooks1999entropy}%
  \BibitemOpen
  \bibfield  {author} {\bibinfo {author} {\bibfnamefont {G.~E.}\ \bibnamefont {Crooks}},\ }\bibfield  {title} {\bibinfo {title} {Entropy production fluctuation theorem and the nonequilibrium work relation for free energy differences},\ }\href@noop {} {\bibfield  {journal} {\bibinfo  {journal} {Physical Review E}\ }\textbf {\bibinfo {volume} {60}},\ \bibinfo {pages} {2721} (\bibinfo {year} {1999})}\BibitemShut {NoStop}%
\bibitem [{\citenamefont {Jarzynski}(1997)}]{jarzynski1997nonequilibrium}%
  \BibitemOpen
  \bibfield  {author} {\bibinfo {author} {\bibfnamefont {C.}~\bibnamefont {Jarzynski}},\ }\bibfield  {title} {\bibinfo {title} {Nonequilibrium equality for free energy differences},\ }\href@noop {} {\bibfield  {journal} {\bibinfo  {journal} {Physical Review Letters}\ }\textbf {\bibinfo {volume} {78}},\ \bibinfo {pages} {2690} (\bibinfo {year} {1997})}\BibitemShut {NoStop}%
\bibitem [{\citenamefont {Ho}\ \emph {et~al.}(2020)\citenamefont {Ho}, \citenamefont {Jain},\ and\ \citenamefont {Abbeel}}]{ho2020denoising}%
  \BibitemOpen
  \bibfield  {author} {\bibinfo {author} {\bibfnamefont {J.}~\bibnamefont {Ho}}, \bibinfo {author} {\bibfnamefont {A.}~\bibnamefont {Jain}},\ and\ \bibinfo {author} {\bibfnamefont {P.}~\bibnamefont {Abbeel}},\ }\bibfield  {title} {\bibinfo {title} {Denoising diffusion probabilistic models},\ }\href@noop {} {\bibfield  {journal} {\bibinfo  {journal} {Advances in neural information processing systems}\ }\textbf {\bibinfo {volume} {33}},\ \bibinfo {pages} {6840} (\bibinfo {year} {2020})}\BibitemShut {NoStop}%
\bibitem [{\citenamefont {Kingma}\ \emph {et~al.}(2021)\citenamefont {Kingma}, \citenamefont {Salimans}, \citenamefont {Poole},\ and\ \citenamefont {Ho}}]{kingma2021variational}%
  \BibitemOpen
  \bibfield  {author} {\bibinfo {author} {\bibfnamefont {D.}~\bibnamefont {Kingma}}, \bibinfo {author} {\bibfnamefont {T.}~\bibnamefont {Salimans}}, \bibinfo {author} {\bibfnamefont {B.}~\bibnamefont {Poole}},\ and\ \bibinfo {author} {\bibfnamefont {J.}~\bibnamefont {Ho}},\ }\bibfield  {title} {\bibinfo {title} {Variational diffusion models},\ }\href@noop {} {\bibfield  {journal} {\bibinfo  {journal} {Advances in neural information processing systems}\ }\textbf {\bibinfo {volume} {34}},\ \bibinfo {pages} {21696} (\bibinfo {year} {2021})}\BibitemShut {NoStop}%
\bibitem [{\citenamefont {Song}\ and\ \citenamefont {Ermon}(2019)}]{song2019generative}%
  \BibitemOpen
  \bibfield  {author} {\bibinfo {author} {\bibfnamefont {Y.}~\bibnamefont {Song}}\ and\ \bibinfo {author} {\bibfnamefont {S.}~\bibnamefont {Ermon}},\ }\bibfield  {title} {\bibinfo {title} {Generative modeling by estimating gradients of the data distribution},\ }\href@noop {} {\bibfield  {journal} {\bibinfo  {journal} {Advances in neural information processing systems}\ }\textbf {\bibinfo {volume} {32}} (\bibinfo {year} {2019})}\BibitemShut {NoStop}%
\bibitem [{\citenamefont {Nichol}\ and\ \citenamefont {Dhariwal}(2021)}]{nichol2021improved}%
  \BibitemOpen
  \bibfield  {author} {\bibinfo {author} {\bibfnamefont {A.~Q.}\ \bibnamefont {Nichol}}\ and\ \bibinfo {author} {\bibfnamefont {P.}~\bibnamefont {Dhariwal}},\ }\bibfield  {title} {\bibinfo {title} {Improved denoising diffusion probabilistic models},\ }in\ \href@noop {} {\emph {\bibinfo {booktitle} {International Conference on Machine Learning}}}\ (\bibinfo {organization} {PMLR},\ \bibinfo {year} {2021})\ pp.\ \bibinfo {pages} {8162--8171}\BibitemShut {NoStop}%
\bibitem [{\citenamefont {Rombach}\ \emph {et~al.}(2022)\citenamefont {Rombach}, \citenamefont {Blattmann}, \citenamefont {Lorenz}, \citenamefont {Esser},\ and\ \citenamefont {Ommer}}]{rombach2022high}%
  \BibitemOpen
  \bibfield  {author} {\bibinfo {author} {\bibfnamefont {R.}~\bibnamefont {Rombach}}, \bibinfo {author} {\bibfnamefont {A.}~\bibnamefont {Blattmann}}, \bibinfo {author} {\bibfnamefont {D.}~\bibnamefont {Lorenz}}, \bibinfo {author} {\bibfnamefont {P.}~\bibnamefont {Esser}},\ and\ \bibinfo {author} {\bibfnamefont {B.}~\bibnamefont {Ommer}},\ }\bibfield  {title} {\bibinfo {title} {High-resolution image synthesis with latent diffusion models},\ }in\ \href@noop {} {\emph {\bibinfo {booktitle} {Proceedings of the IEEE/CVF conference on computer vision and pattern recognition}}}\ (\bibinfo {year} {2022})\ pp.\ \bibinfo {pages} {10684--10695}\BibitemShut {NoStop}%
\bibitem [{\citenamefont {Song}\ and\ \citenamefont {Ermon}(2020)}]{song2020improved}%
  \BibitemOpen
  \bibfield  {author} {\bibinfo {author} {\bibfnamefont {Y.}~\bibnamefont {Song}}\ and\ \bibinfo {author} {\bibfnamefont {S.}~\bibnamefont {Ermon}},\ }\bibfield  {title} {\bibinfo {title} {Improved techniques for training score-based generative models},\ }\href@noop {} {\bibfield  {journal} {\bibinfo  {journal} {Advances in neural information processing systems}\ }\textbf {\bibinfo {volume} {33}},\ \bibinfo {pages} {12438} (\bibinfo {year} {2020})}\BibitemShut {NoStop}%
\bibitem [{\citenamefont {Song}\ \emph {et~al.}(2023)\citenamefont {Song}, \citenamefont {Dhariwal}, \citenamefont {Chen},\ and\ \citenamefont {Sutskever}}]{song2023consistency}%
  \BibitemOpen
  \bibfield  {author} {\bibinfo {author} {\bibfnamefont {Y.}~\bibnamefont {Song}}, \bibinfo {author} {\bibfnamefont {P.}~\bibnamefont {Dhariwal}}, \bibinfo {author} {\bibfnamefont {M.}~\bibnamefont {Chen}},\ and\ \bibinfo {author} {\bibfnamefont {I.}~\bibnamefont {Sutskever}},\ }\bibfield  {title} {\bibinfo {title} {Consistency models},\ }\href@noop {} {\bibfield  {journal} {\bibinfo  {journal} {arXiv preprint arXiv:2303.01469}\ } (\bibinfo {year} {2023})}\BibitemShut {NoStop}%
\bibitem [{\citenamefont {Dhariwal}\ and\ \citenamefont {Nichol}(2021)}]{dhariwal2021diffusion}%
  \BibitemOpen
  \bibfield  {author} {\bibinfo {author} {\bibfnamefont {P.}~\bibnamefont {Dhariwal}}\ and\ \bibinfo {author} {\bibfnamefont {A.}~\bibnamefont {Nichol}},\ }\bibfield  {title} {\bibinfo {title} {Diffusion models beat gans on image synthesis},\ }\href@noop {} {\bibfield  {journal} {\bibinfo  {journal} {Advances in neural information processing systems}\ }\textbf {\bibinfo {volume} {34}},\ \bibinfo {pages} {8780} (\bibinfo {year} {2021})}\BibitemShut {NoStop}%
\bibitem [{\citenamefont {Song}\ \emph {et~al.}(2020{\natexlab{b}})\citenamefont {Song}, \citenamefont {Meng},\ and\ \citenamefont {Ermon}}]{song2020denoising}%
  \BibitemOpen
  \bibfield  {author} {\bibinfo {author} {\bibfnamefont {J.}~\bibnamefont {Song}}, \bibinfo {author} {\bibfnamefont {C.}~\bibnamefont {Meng}},\ and\ \bibinfo {author} {\bibfnamefont {S.}~\bibnamefont {Ermon}},\ }\bibfield  {title} {\bibinfo {title} {Denoising diffusion implicit models},\ }in\ \href@noop {} {\emph {\bibinfo {booktitle} {International Conference on Learning Representations}}}\ (\bibinfo {year} {2020})\BibitemShut {NoStop}%
\bibitem [{\citenamefont {Karras}\ \emph {et~al.}(2022)\citenamefont {Karras}, \citenamefont {Aittala}, \citenamefont {Aila},\ and\ \citenamefont {Laine}}]{karras2022elucidating}%
  \BibitemOpen
  \bibfield  {author} {\bibinfo {author} {\bibfnamefont {T.}~\bibnamefont {Karras}}, \bibinfo {author} {\bibfnamefont {M.}~\bibnamefont {Aittala}}, \bibinfo {author} {\bibfnamefont {T.}~\bibnamefont {Aila}},\ and\ \bibinfo {author} {\bibfnamefont {S.}~\bibnamefont {Laine}},\ }\bibfield  {title} {\bibinfo {title} {Elucidating the design space of diffusion-based generative models},\ }\href@noop {} {\bibfield  {journal} {\bibinfo  {journal} {Advances in Neural Information Processing Systems}\ }\textbf {\bibinfo {volume} {35}},\ \bibinfo {pages} {26565} (\bibinfo {year} {2022})}\BibitemShut {NoStop}%
\bibitem [{\citenamefont {Chen}\ \emph {et~al.}(2021)\citenamefont {Chen}, \citenamefont {Liu},\ and\ \citenamefont {Theodorou}}]{chen2021likelihood}%
  \BibitemOpen
  \bibfield  {author} {\bibinfo {author} {\bibfnamefont {T.}~\bibnamefont {Chen}}, \bibinfo {author} {\bibfnamefont {G.-H.}\ \bibnamefont {Liu}},\ and\ \bibinfo {author} {\bibfnamefont {E.}~\bibnamefont {Theodorou}},\ }\bibfield  {title} {\bibinfo {title} {Likelihood training of schr{\"o}dinger bridge using forward-backward sdes theory},\ }in\ \href@noop {} {\emph {\bibinfo {booktitle} {International Conference on Learning Representations}}}\ (\bibinfo {year} {2021})\BibitemShut {NoStop}%
\bibitem [{\citenamefont {Ramesh}\ \emph {et~al.}(2022)\citenamefont {Ramesh}, \citenamefont {Dhariwal}, \citenamefont {Nichol}, \citenamefont {Chu},\ and\ \citenamefont {Chen}}]{ramesh2022hierarchical}%
  \BibitemOpen
  \bibfield  {author} {\bibinfo {author} {\bibfnamefont {A.}~\bibnamefont {Ramesh}}, \bibinfo {author} {\bibfnamefont {P.}~\bibnamefont {Dhariwal}}, \bibinfo {author} {\bibfnamefont {A.}~\bibnamefont {Nichol}}, \bibinfo {author} {\bibfnamefont {C.}~\bibnamefont {Chu}},\ and\ \bibinfo {author} {\bibfnamefont {M.}~\bibnamefont {Chen}},\ }\bibfield  {title} {\bibinfo {title} {Hierarchical text-conditional image generation with clip latents},\ }\href@noop {} {\bibfield  {journal} {\bibinfo  {journal} {arXiv preprint arXiv:2204.06125}\ }\textbf {\bibinfo {volume} {1}},\ \bibinfo {pages} {3} (\bibinfo {year} {2022})}\BibitemShut {NoStop}%
\bibitem [{\citenamefont {Ho}\ and\ \citenamefont {Salimans}(2022)}]{ho2022classifier}%
  \BibitemOpen
  \bibfield  {author} {\bibinfo {author} {\bibfnamefont {J.}~\bibnamefont {Ho}}\ and\ \bibinfo {author} {\bibfnamefont {T.}~\bibnamefont {Salimans}},\ }\bibfield  {title} {\bibinfo {title} {Classifier-free diffusion guidance},\ }\href@noop {} {\bibfield  {journal} {\bibinfo  {journal} {arXiv preprint arXiv:2207.12598}\ } (\bibinfo {year} {2022})}\BibitemShut {NoStop}%
\bibitem [{\citenamefont {Nichol}\ \emph {et~al.}(2022)\citenamefont {Nichol}, \citenamefont {Dhariwal}, \citenamefont {Ramesh}, \citenamefont {Shyam}, \citenamefont {Mishkin}, \citenamefont {Mcgrew}, \citenamefont {Sutskever},\ and\ \citenamefont {Chen}}]{nichol2021glide}%
  \BibitemOpen
  \bibfield  {author} {\bibinfo {author} {\bibfnamefont {A.~Q.}\ \bibnamefont {Nichol}}, \bibinfo {author} {\bibfnamefont {P.}~\bibnamefont {Dhariwal}}, \bibinfo {author} {\bibfnamefont {A.}~\bibnamefont {Ramesh}}, \bibinfo {author} {\bibfnamefont {P.}~\bibnamefont {Shyam}}, \bibinfo {author} {\bibfnamefont {P.}~\bibnamefont {Mishkin}}, \bibinfo {author} {\bibfnamefont {B.}~\bibnamefont {Mcgrew}}, \bibinfo {author} {\bibfnamefont {I.}~\bibnamefont {Sutskever}},\ and\ \bibinfo {author} {\bibfnamefont {M.}~\bibnamefont {Chen}},\ }\bibfield  {title} {\bibinfo {title} {Glide: Towards photorealistic image generation and editing with text-guided diffusion models},\ }in\ \href@noop {} {\emph {\bibinfo {booktitle} {International Conference on Machine Learning}}}\ (\bibinfo {organization} {PMLR},\ \bibinfo {year} {2022})\ pp.\ \bibinfo {pages} {16784--16804}\BibitemShut {NoStop}%
\bibitem [{\citenamefont {Sauer}\ \emph {et~al.}(2024)\citenamefont {Sauer}, \citenamefont {Boesel}, \citenamefont {Dockhorn}, \citenamefont {Blattmann}, \citenamefont {Esser},\ and\ \citenamefont {Rombach}}]{sauer2024fast}%
  \BibitemOpen
  \bibfield  {author} {\bibinfo {author} {\bibfnamefont {A.}~\bibnamefont {Sauer}}, \bibinfo {author} {\bibfnamefont {F.}~\bibnamefont {Boesel}}, \bibinfo {author} {\bibfnamefont {T.}~\bibnamefont {Dockhorn}}, \bibinfo {author} {\bibfnamefont {A.}~\bibnamefont {Blattmann}}, \bibinfo {author} {\bibfnamefont {P.}~\bibnamefont {Esser}},\ and\ \bibinfo {author} {\bibfnamefont {R.}~\bibnamefont {Rombach}},\ }\bibfield  {title} {\bibinfo {title} {Fast high-resolution image synthesis with latent adversarial diffusion distillation},\ }\href@noop {} {\bibfield  {journal} {\bibinfo  {journal} {arXiv preprint arXiv:2403.12015}\ } (\bibinfo {year} {2024})}\BibitemShut {NoStop}%
\bibitem [{\citenamefont {Podell}\ \emph {et~al.}(2024)\citenamefont {Podell}, \citenamefont {English}, \citenamefont {Lacey}, \citenamefont {Blattmann}, \citenamefont {Dockhorn}, \citenamefont {M{\"u}ller}, \citenamefont {Penna},\ and\ \citenamefont {Rombach}}]{podell2023sdxl}%
  \BibitemOpen
  \bibfield  {author} {\bibinfo {author} {\bibfnamefont {D.}~\bibnamefont {Podell}}, \bibinfo {author} {\bibfnamefont {Z.}~\bibnamefont {English}}, \bibinfo {author} {\bibfnamefont {K.}~\bibnamefont {Lacey}}, \bibinfo {author} {\bibfnamefont {A.}~\bibnamefont {Blattmann}}, \bibinfo {author} {\bibfnamefont {T.}~\bibnamefont {Dockhorn}}, \bibinfo {author} {\bibfnamefont {J.}~\bibnamefont {M{\"u}ller}}, \bibinfo {author} {\bibfnamefont {J.}~\bibnamefont {Penna}},\ and\ \bibinfo {author} {\bibfnamefont {R.}~\bibnamefont {Rombach}},\ }\bibfield  {title} {\bibinfo {title} {{SDXL}: Improving latent diffusion models for high-resolution image synthesis},\ }in\ \href@noop {} {\emph {\bibinfo {booktitle} {{The Twelfth International Conference on Learning Representations}}}}\ (\bibinfo {year} {2024})\BibitemShut {NoStop}%
\bibitem [{\citenamefont {Hyv{\"a}rinen}\ and\ \citenamefont {Dayan}(2005)}]{hyvarinen2005estimation}%
  \BibitemOpen
  \bibfield  {author} {\bibinfo {author} {\bibfnamefont {A.}~\bibnamefont {Hyv{\"a}rinen}}\ and\ \bibinfo {author} {\bibfnamefont {P.}~\bibnamefont {Dayan}},\ }\bibfield  {title} {\bibinfo {title} {Estimation of non-normalized statistical models by score matching.},\ }\href@noop {} {\bibfield  {journal} {\bibinfo  {journal} {Journal of Machine Learning Research}\ }\textbf {\bibinfo {volume} {6}} (\bibinfo {year} {2005})}\BibitemShut {NoStop}%
\bibitem [{\citenamefont {Vincent}(2011)}]{vincent2011connection}%
  \BibitemOpen
  \bibfield  {author} {\bibinfo {author} {\bibfnamefont {P.}~\bibnamefont {Vincent}},\ }\bibfield  {title} {\bibinfo {title} {A connection between score matching and denoising autoencoders},\ }\href@noop {} {\bibfield  {journal} {\bibinfo  {journal} {Neural computation}\ }\textbf {\bibinfo {volume} {23}},\ \bibinfo {pages} {1661} (\bibinfo {year} {2011})}\BibitemShut {NoStop}%
\bibitem [{\citenamefont {Kingma}\ and\ \citenamefont {Cun}(2010)}]{kingma2010regularized}%
  \BibitemOpen
  \bibfield  {author} {\bibinfo {author} {\bibfnamefont {D.~P.}\ \bibnamefont {Kingma}}\ and\ \bibinfo {author} {\bibfnamefont {Y.}~\bibnamefont {Cun}},\ }\bibfield  {title} {\bibinfo {title} {Regularized estimation of image statistics by score matching},\ }\href@noop {} {\bibfield  {journal} {\bibinfo  {journal} {Advances in neural information processing systems}\ }\textbf {\bibinfo {volume} {23}} (\bibinfo {year} {2010})}\BibitemShut {NoStop}%
\bibitem [{\citenamefont {Dinh}\ \emph {et~al.}(2015)\citenamefont {Dinh}, \citenamefont {Krueger},\ and\ \citenamefont {Bengio}}]{dinh2014nice}%
  \BibitemOpen
  \bibfield  {author} {\bibinfo {author} {\bibfnamefont {L.}~\bibnamefont {Dinh}}, \bibinfo {author} {\bibfnamefont {D.}~\bibnamefont {Krueger}},\ and\ \bibinfo {author} {\bibfnamefont {Y.}~\bibnamefont {Bengio}},\ }\bibfield  {title} {\bibinfo {title} {Nice: Non-linear independent components estimation},\ }in\ \href@noop {} {\emph {\bibinfo {booktitle} {ICLR (Workshop)}}}\ (\bibinfo {year} {2015})\BibitemShut {NoStop}%
\bibitem [{\citenamefont {Rezende}\ and\ \citenamefont {Mohamed}(2015)}]{rezende2015variational}%
  \BibitemOpen
  \bibfield  {author} {\bibinfo {author} {\bibfnamefont {D.}~\bibnamefont {Rezende}}\ and\ \bibinfo {author} {\bibfnamefont {S.}~\bibnamefont {Mohamed}},\ }\bibfield  {title} {\bibinfo {title} {Variational inference with normalizing flows},\ }in\ \href@noop {} {\emph {\bibinfo {booktitle} {International conference on machine learning}}}\ (\bibinfo {organization} {PMLR},\ \bibinfo {year} {2015})\ pp.\ \bibinfo {pages} {1530--1538}\BibitemShut {NoStop}%
\bibitem [{\citenamefont {Chen}\ \emph {et~al.}(2018)\citenamefont {Chen}, \citenamefont {Rubanova}, \citenamefont {Bettencourt},\ and\ \citenamefont {Duvenaud}}]{chen2018neural}%
  \BibitemOpen
  \bibfield  {author} {\bibinfo {author} {\bibfnamefont {R.~T.}\ \bibnamefont {Chen}}, \bibinfo {author} {\bibfnamefont {Y.}~\bibnamefont {Rubanova}}, \bibinfo {author} {\bibfnamefont {J.}~\bibnamefont {Bettencourt}},\ and\ \bibinfo {author} {\bibfnamefont {D.~K.}\ \bibnamefont {Duvenaud}},\ }\bibfield  {title} {\bibinfo {title} {Neural ordinary differential equations},\ }\href@noop {} {\bibfield  {journal} {\bibinfo  {journal} {Advances in neural information processing systems}\ }\textbf {\bibinfo {volume} {31}} (\bibinfo {year} {2018})}\BibitemShut {NoStop}%
\bibitem [{\citenamefont {Song}(2022)}]{song2022applying}%
  \BibitemOpen
  \bibfield  {author} {\bibinfo {author} {\bibfnamefont {K.-U.}\ \bibnamefont {Song}},\ }\bibfield  {title} {\bibinfo {title} {Applying regularized schr\"odinger-bridge-based stochastic process in generative modeling},\ }\href@noop {} {\bibfield  {journal} {\bibinfo  {journal} {arXiv preprint arXiv:2208.07131}\ } (\bibinfo {year} {2022})}\BibitemShut {NoStop}%
\bibitem [{\citenamefont {Lipman}\ \emph {et~al.}(2022)\citenamefont {Lipman}, \citenamefont {Chen}, \citenamefont {Ben-Hamu}, \citenamefont {Nickel},\ and\ \citenamefont {Le}}]{lipman2022flow}%
  \BibitemOpen
  \bibfield  {author} {\bibinfo {author} {\bibfnamefont {Y.}~\bibnamefont {Lipman}}, \bibinfo {author} {\bibfnamefont {R.~T.}\ \bibnamefont {Chen}}, \bibinfo {author} {\bibfnamefont {H.}~\bibnamefont {Ben-Hamu}}, \bibinfo {author} {\bibfnamefont {M.}~\bibnamefont {Nickel}},\ and\ \bibinfo {author} {\bibfnamefont {M.}~\bibnamefont {Le}},\ }\bibfield  {title} {\bibinfo {title} {Flow matching for generative modeling},\ }in\ \href@noop {} {\emph {\bibinfo {booktitle} {The Eleventh International Conference on Learning Representations}}}\ (\bibinfo {year} {2022})\BibitemShut {NoStop}%
\bibitem [{\citenamefont {Arjovsky}\ \emph {et~al.}(2017)\citenamefont {Arjovsky}, \citenamefont {Chintala},\ and\ \citenamefont {Bottou}}]{arjovsky2017wasserstein}%
  \BibitemOpen
  \bibfield  {author} {\bibinfo {author} {\bibfnamefont {M.}~\bibnamefont {Arjovsky}}, \bibinfo {author} {\bibfnamefont {S.}~\bibnamefont {Chintala}},\ and\ \bibinfo {author} {\bibfnamefont {L.}~\bibnamefont {Bottou}},\ }\bibfield  {title} {\bibinfo {title} {Wasserstein generative adversarial networks},\ }in\ \href@noop {} {\emph {\bibinfo {booktitle} {International conference on machine learning}}}\ (\bibinfo {organization} {PMLR},\ \bibinfo {year} {2017})\ pp.\ \bibinfo {pages} {214--223}\BibitemShut {NoStop}%
\bibitem [{\citenamefont {Kwon}\ \emph {et~al.}(2022)\citenamefont {Kwon}, \citenamefont {Fan},\ and\ \citenamefont {Lee}}]{kwon2022score}%
  \BibitemOpen
  \bibfield  {author} {\bibinfo {author} {\bibfnamefont {D.}~\bibnamefont {Kwon}}, \bibinfo {author} {\bibfnamefont {Y.}~\bibnamefont {Fan}},\ and\ \bibinfo {author} {\bibfnamefont {K.}~\bibnamefont {Lee}},\ }\bibfield  {title} {\bibinfo {title} {Score-based generative modeling secretly minimizes the wasserstein distance},\ }\href@noop {} {\bibfield  {journal} {\bibinfo  {journal} {Advances in Neural Information Processing Systems}\ }\textbf {\bibinfo {volume} {35}},\ \bibinfo {pages} {20205} (\bibinfo {year} {2022})}\BibitemShut {NoStop}%
\bibitem [{\citenamefont {Oko}\ \emph {et~al.}(2023)\citenamefont {Oko}, \citenamefont {Akiyama},\ and\ \citenamefont {Suzuki}}]{oko2023diffusion}%
  \BibitemOpen
  \bibfield  {author} {\bibinfo {author} {\bibfnamefont {K.}~\bibnamefont {Oko}}, \bibinfo {author} {\bibfnamefont {S.}~\bibnamefont {Akiyama}},\ and\ \bibinfo {author} {\bibfnamefont {T.}~\bibnamefont {Suzuki}},\ }\bibfield  {title} {\bibinfo {title} {Diffusion models are minimax optimal distribution estimators},\ }in\ \href@noop {} {\emph {\bibinfo {booktitle} {International Conference on Machine Learning}}}\ (\bibinfo {organization} {PMLR},\ \bibinfo {year} {2023})\ pp.\ \bibinfo {pages} {26517--26582}\BibitemShut {NoStop}%
\bibitem [{\citenamefont {Bortoli}(2022)}]{de2022convergence}%
  \BibitemOpen
  \bibfield  {author} {\bibinfo {author} {\bibfnamefont {V.~D.}\ \bibnamefont {Bortoli}},\ }\bibfield  {title} {\bibinfo {title} {Convergence of denoising diffusion models under the manifold hypothesis},\ }\href@noop {} {\bibfield  {journal} {\bibinfo  {journal} {Transactions on Machine Learning Research}\ } (\bibinfo {year} {2022})},\ \bibinfo {note} {expert Certification}\BibitemShut {NoStop}%
\bibitem [{\citenamefont {Kornilov}\ \emph {et~al.}(2024)\citenamefont {Kornilov}, \citenamefont {Mokrov}, \citenamefont {Gasnikov},\ and\ \citenamefont {Korotin}}]{kornilov2024optimal}%
  \BibitemOpen
  \bibfield  {author} {\bibinfo {author} {\bibfnamefont {N.~M.}\ \bibnamefont {Kornilov}}, \bibinfo {author} {\bibfnamefont {P.}~\bibnamefont {Mokrov}}, \bibinfo {author} {\bibfnamefont {A.}~\bibnamefont {Gasnikov}},\ and\ \bibinfo {author} {\bibfnamefont {A.}~\bibnamefont {Korotin}},\ }\bibfield  {title} {\bibinfo {title} {Optimal flow matching: Learning straight trajectories in just one step},\ }in\ \href@noop {} {\emph {\bibinfo {booktitle} {The Thirty-eighth Annual Conference on Neural Information Processing Systems}}}\ (\bibinfo {year} {2024})\BibitemShut {NoStop}%
\bibitem [{\citenamefont {Shaul}\ \emph {et~al.}(2023)\citenamefont {Shaul}, \citenamefont {Chen}, \citenamefont {Nickel}, \citenamefont {Le},\ and\ \citenamefont {Lipman}}]{shaul2023kinetic}%
  \BibitemOpen
  \bibfield  {author} {\bibinfo {author} {\bibfnamefont {N.}~\bibnamefont {Shaul}}, \bibinfo {author} {\bibfnamefont {R.~T.}\ \bibnamefont {Chen}}, \bibinfo {author} {\bibfnamefont {M.}~\bibnamefont {Nickel}}, \bibinfo {author} {\bibfnamefont {M.}~\bibnamefont {Le}},\ and\ \bibinfo {author} {\bibfnamefont {Y.}~\bibnamefont {Lipman}},\ }\bibfield  {title} {\bibinfo {title} {On kinetic optimal probability paths for generative models},\ }in\ \href@noop {} {\emph {\bibinfo {booktitle} {International Conference on Machine Learning}}}\ (\bibinfo {organization} {PMLR},\ \bibinfo {year} {2023})\ pp.\ \bibinfo {pages} {30883--30907}\BibitemShut {NoStop}%
\bibitem [{\citenamefont {Tong}\ \emph {et~al.}(2023)\citenamefont {Tong}, \citenamefont {Malkin}, \citenamefont {Huguet}, \citenamefont {Zhang}, \citenamefont {Rector-Brooks}, \citenamefont {Fatras}, \citenamefont {Wolf},\ and\ \citenamefont {Bengio}}]{tong2023improving}%
  \BibitemOpen
  \bibfield  {author} {\bibinfo {author} {\bibfnamefont {A.}~\bibnamefont {Tong}}, \bibinfo {author} {\bibfnamefont {N.}~\bibnamefont {Malkin}}, \bibinfo {author} {\bibfnamefont {G.}~\bibnamefont {Huguet}}, \bibinfo {author} {\bibfnamefont {Y.}~\bibnamefont {Zhang}}, \bibinfo {author} {\bibfnamefont {J.}~\bibnamefont {Rector-Brooks}}, \bibinfo {author} {\bibfnamefont {K.}~\bibnamefont {Fatras}}, \bibinfo {author} {\bibfnamefont {G.}~\bibnamefont {Wolf}},\ and\ \bibinfo {author} {\bibfnamefont {Y.}~\bibnamefont {Bengio}},\ }\bibfield  {title} {\bibinfo {title} {Improving and generalizing flow-based generative models with minibatch optimal transport},\ }in\ \href@noop {} {\emph {\bibinfo {booktitle} {ICML Workshop on New Frontiers in Learning, Control, and Dynamical Systems}}}\ (\bibinfo {year} {2023})\BibitemShut {NoStop}%
\bibitem [{\citenamefont {Shi}\ \emph {et~al.}(2022)\citenamefont {Shi}, \citenamefont {De~Bortoli}, \citenamefont {Deligiannidis},\ and\ \citenamefont {Doucet}}]{shi2022conditional}%
  \BibitemOpen
  \bibfield  {author} {\bibinfo {author} {\bibfnamefont {Y.}~\bibnamefont {Shi}}, \bibinfo {author} {\bibfnamefont {V.}~\bibnamefont {De~Bortoli}}, \bibinfo {author} {\bibfnamefont {G.}~\bibnamefont {Deligiannidis}},\ and\ \bibinfo {author} {\bibfnamefont {A.}~\bibnamefont {Doucet}},\ }\bibfield  {title} {\bibinfo {title} {Conditional simulation using diffusion schr{\"o}dinger bridges},\ }in\ \href@noop {} {\emph {\bibinfo {booktitle} {Uncertainty in Artificial Intelligence}}}\ (\bibinfo {organization} {PMLR},\ \bibinfo {year} {2022})\ pp.\ \bibinfo {pages} {1792--1802}\BibitemShut {NoStop}%
\bibitem [{\citenamefont {Liu}\ \emph {et~al.}(2023)\citenamefont {Liu}, \citenamefont {Gong},\ and\ \citenamefont {qiang liu}}]{liu2022flow}%
  \BibitemOpen
  \bibfield  {author} {\bibinfo {author} {\bibfnamefont {X.}~\bibnamefont {Liu}}, \bibinfo {author} {\bibfnamefont {C.}~\bibnamefont {Gong}},\ and\ \bibinfo {author} {\bibnamefont {qiang liu}},\ }\bibfield  {title} {\bibinfo {title} {Flow straight and fast: Learning to generate and transfer data with rectified flow},\ }in\ \href@noop {} {\emph {\bibinfo {booktitle} {The Eleventh International Conference on Learning Representations}}}\ (\bibinfo {year} {2023})\BibitemShut {NoStop}%
\bibitem [{\citenamefont {Esser}\ \emph {et~al.}(2024)\citenamefont {Esser}, \citenamefont {Kulal}, \citenamefont {Blattmann}, \citenamefont {Entezari}, \citenamefont {M{\"u}ller}, \citenamefont {Saini}, \citenamefont {Levi}, \citenamefont {Lorenz}, \citenamefont {Sauer}, \citenamefont {Boesel}, \citenamefont {Podell}, \citenamefont {Dockhorn}, \citenamefont {English},\ and\ \citenamefont {Rombach}}]{esser2024scaling}%
  \BibitemOpen
  \bibfield  {author} {\bibinfo {author} {\bibfnamefont {P.}~\bibnamefont {Esser}}, \bibinfo {author} {\bibfnamefont {S.}~\bibnamefont {Kulal}}, \bibinfo {author} {\bibfnamefont {A.}~\bibnamefont {Blattmann}}, \bibinfo {author} {\bibfnamefont {R.}~\bibnamefont {Entezari}}, \bibinfo {author} {\bibfnamefont {J.}~\bibnamefont {M{\"u}ller}}, \bibinfo {author} {\bibfnamefont {H.}~\bibnamefont {Saini}}, \bibinfo {author} {\bibfnamefont {Y.}~\bibnamefont {Levi}}, \bibinfo {author} {\bibfnamefont {D.}~\bibnamefont {Lorenz}}, \bibinfo {author} {\bibfnamefont {A.}~\bibnamefont {Sauer}}, \bibinfo {author} {\bibfnamefont {F.}~\bibnamefont {Boesel}}, \bibinfo {author} {\bibfnamefont {D.}~\bibnamefont {Podell}}, \bibinfo {author} {\bibfnamefont {T.}~\bibnamefont {Dockhorn}}, \bibinfo {author} {\bibfnamefont {Z.}~\bibnamefont {English}},\ and\ \bibinfo {author} {\bibfnamefont {R.}~\bibnamefont {Rombach}},\ }\bibfield  {title} {\bibinfo {title} {Scaling rectified flow transformers for high-resolution image synthesis},\ }in\
  \href@noop {} {\emph {\bibinfo {booktitle} {Forty-first International Conference on Machine Learning}}}\ (\bibinfo {year} {2024})\BibitemShut {NoStop}%
\bibitem [{\citenamefont {Klinger}\ and\ \citenamefont {Rotskoff}(2025)}]{klinger2024universal}%
  \BibitemOpen
  \bibfield  {author} {\bibinfo {author} {\bibfnamefont {J.}~\bibnamefont {Klinger}}\ and\ \bibinfo {author} {\bibfnamefont {G.~M.}\ \bibnamefont {Rotskoff}},\ }\bibfield  {title} {\bibinfo {title} {Universal energy-speed-accuracy trade-offs in driven nonequilibrium systems},\ }\href@noop {} {\bibfield  {journal} {\bibinfo  {journal} {Physical Review E}\ }\textbf {\bibinfo {volume} {111}},\ \bibinfo {pages} {014114} (\bibinfo {year} {2025})}\BibitemShut {NoStop}%
\bibitem [{\citenamefont {Brock}\ \emph {et~al.}(2019)\citenamefont {Brock}, \citenamefont {Donahue},\ and\ \citenamefont {Simonyan}}]{brock2018large}%
  \BibitemOpen
  \bibfield  {author} {\bibinfo {author} {\bibfnamefont {A.}~\bibnamefont {Brock}}, \bibinfo {author} {\bibfnamefont {J.}~\bibnamefont {Donahue}},\ and\ \bibinfo {author} {\bibfnamefont {K.}~\bibnamefont {Simonyan}},\ }\bibfield  {title} {\bibinfo {title} {Large scale {GAN} training for high fidelity natural image synthesis},\ }in\ \href@noop {} {\emph {\bibinfo {booktitle} {International Conference on Learning Representations}}}\ (\bibinfo {year} {2019})\BibitemShut {NoStop}%
\bibitem [{\citenamefont {Brown}\ \emph {et~al.}(2020)\citenamefont {Brown}, \citenamefont {Mann}, \citenamefont {Ryder}, \citenamefont {Subbiah}, \citenamefont {Kaplan}, \citenamefont {Dhariwal}, \citenamefont {Neelakantan}, \citenamefont {Shyam}, \citenamefont {Sastry}, \citenamefont {Askell} \emph {et~al.}}]{brown2020language}%
  \BibitemOpen
  \bibfield  {author} {\bibinfo {author} {\bibfnamefont {T.}~\bibnamefont {Brown}}, \bibinfo {author} {\bibfnamefont {B.}~\bibnamefont {Mann}}, \bibinfo {author} {\bibfnamefont {N.}~\bibnamefont {Ryder}}, \bibinfo {author} {\bibfnamefont {M.}~\bibnamefont {Subbiah}}, \bibinfo {author} {\bibfnamefont {J.~D.}\ \bibnamefont {Kaplan}}, \bibinfo {author} {\bibfnamefont {P.}~\bibnamefont {Dhariwal}}, \bibinfo {author} {\bibfnamefont {A.}~\bibnamefont {Neelakantan}}, \bibinfo {author} {\bibfnamefont {P.}~\bibnamefont {Shyam}}, \bibinfo {author} {\bibfnamefont {G.}~\bibnamefont {Sastry}}, \bibinfo {author} {\bibfnamefont {A.}~\bibnamefont {Askell}}, \emph {et~al.},\ }\bibfield  {title} {\bibinfo {title} {Language models are few-shot learners},\ }\href@noop {} {\bibfield  {journal} {\bibinfo  {journal} {Advances in neural information processing systems}\ }\textbf {\bibinfo {volume} {33}},\ \bibinfo {pages} {1877} (\bibinfo {year} {2020})}\BibitemShut {NoStop}%
\bibitem [{\citenamefont {Oord}\ \emph {et~al.}(2016)\citenamefont {Oord}, \citenamefont {Dieleman}, \citenamefont {Zen}, \citenamefont {Simonyan}, \citenamefont {Vinyals}, \citenamefont {Graves}, \citenamefont {Kalchbrenner}, \citenamefont {Senior},\ and\ \citenamefont {Kavukcuoglu}}]{oord2016wavenet}%
  \BibitemOpen
  \bibfield  {author} {\bibinfo {author} {\bibfnamefont {A.~v.~d.}\ \bibnamefont {Oord}}, \bibinfo {author} {\bibfnamefont {S.}~\bibnamefont {Dieleman}}, \bibinfo {author} {\bibfnamefont {H.}~\bibnamefont {Zen}}, \bibinfo {author} {\bibfnamefont {K.}~\bibnamefont {Simonyan}}, \bibinfo {author} {\bibfnamefont {O.}~\bibnamefont {Vinyals}}, \bibinfo {author} {\bibfnamefont {A.}~\bibnamefont {Graves}}, \bibinfo {author} {\bibfnamefont {N.}~\bibnamefont {Kalchbrenner}}, \bibinfo {author} {\bibfnamefont {A.}~\bibnamefont {Senior}},\ and\ \bibinfo {author} {\bibfnamefont {K.}~\bibnamefont {Kavukcuoglu}},\ }\bibfield  {title} {\bibinfo {title} {Wavenet: A generative model for raw audio},\ }\href@noop {} {\bibfield  {journal} {\bibinfo  {journal} {arXiv preprint arXiv:1609.03499}\ } (\bibinfo {year} {2016})}\BibitemShut {NoStop}%
\bibitem [{\citenamefont {Goodfellow}\ \emph {et~al.}(2016)\citenamefont {Goodfellow}, \citenamefont {Bengio},\ and\ \citenamefont {Courville}}]{goodfellow2016deep}%
  \BibitemOpen
  \bibfield  {author} {\bibinfo {author} {\bibfnamefont {I.}~\bibnamefont {Goodfellow}}, \bibinfo {author} {\bibfnamefont {Y.}~\bibnamefont {Bengio}},\ and\ \bibinfo {author} {\bibfnamefont {A.}~\bibnamefont {Courville}},\ }\href@noop {} {\emph {\bibinfo {title} {Deep Learning}}}\ (\bibinfo  {publisher} {MIT Press},\ \bibinfo {year} {2016})\BibitemShut {NoStop}%
\bibitem [{\citenamefont {Theis}\ \emph {et~al.}(2015)\citenamefont {Theis}, \citenamefont {Oord},\ and\ \citenamefont {Bethge}}]{theis2015note}%
  \BibitemOpen
  \bibfield  {author} {\bibinfo {author} {\bibfnamefont {L.}~\bibnamefont {Theis}}, \bibinfo {author} {\bibfnamefont {A.~v.~d.}\ \bibnamefont {Oord}},\ and\ \bibinfo {author} {\bibfnamefont {M.}~\bibnamefont {Bethge}},\ }\bibfield  {title} {\bibinfo {title} {A note on the evaluation of generative models},\ }\href@noop {} {\bibfield  {journal} {\bibinfo  {journal} {arXiv preprint arXiv:1511.01844}\ } (\bibinfo {year} {2015})}\BibitemShut {NoStop}%
\bibitem [{\citenamefont {Betzalel}\ \emph {et~al.}(2022)\citenamefont {Betzalel}, \citenamefont {Penso}, \citenamefont {Navon},\ and\ \citenamefont {Fetaya}}]{betzalel2022study}%
  \BibitemOpen
  \bibfield  {author} {\bibinfo {author} {\bibfnamefont {E.}~\bibnamefont {Betzalel}}, \bibinfo {author} {\bibfnamefont {C.}~\bibnamefont {Penso}}, \bibinfo {author} {\bibfnamefont {A.}~\bibnamefont {Navon}},\ and\ \bibinfo {author} {\bibfnamefont {E.}~\bibnamefont {Fetaya}},\ }\bibfield  {title} {\bibinfo {title} {A study on the evaluation of generative models},\ }\href@noop {} {\bibfield  {journal} {\bibinfo  {journal} {arXiv preprint arXiv:2206.10935}\ } (\bibinfo {year} {2022})}\BibitemShut {NoStop}%
\bibitem [{\citenamefont {Bischoff}\ \emph {et~al.}(2024)\citenamefont {Bischoff}, \citenamefont {Darcher}, \citenamefont {Deistler}, \citenamefont {Gao}, \citenamefont {Gerken}, \citenamefont {Gloeckler}, \citenamefont {Haxel}, \citenamefont {Kapoor}, \citenamefont {Lappalainen}, \citenamefont {Macke} \emph {et~al.}}]{bischoff2024practical}%
  \BibitemOpen
  \bibfield  {author} {\bibinfo {author} {\bibfnamefont {S.}~\bibnamefont {Bischoff}}, \bibinfo {author} {\bibfnamefont {A.}~\bibnamefont {Darcher}}, \bibinfo {author} {\bibfnamefont {M.}~\bibnamefont {Deistler}}, \bibinfo {author} {\bibfnamefont {R.}~\bibnamefont {Gao}}, \bibinfo {author} {\bibfnamefont {F.}~\bibnamefont {Gerken}}, \bibinfo {author} {\bibfnamefont {M.}~\bibnamefont {Gloeckler}}, \bibinfo {author} {\bibfnamefont {L.}~\bibnamefont {Haxel}}, \bibinfo {author} {\bibfnamefont {J.}~\bibnamefont {Kapoor}}, \bibinfo {author} {\bibfnamefont {J.~K.}\ \bibnamefont {Lappalainen}}, \bibinfo {author} {\bibfnamefont {J.~H.}\ \bibnamefont {Macke}}, \emph {et~al.},\ }\bibfield  {title} {\bibinfo {title} {A practical guide to statistical distances for evaluating generative models in science},\ }\href@noop {} {\bibfield  {journal} {\bibinfo  {journal} {arXiv preprint arXiv:2403.12636}\ } (\bibinfo {year} {2024})}\BibitemShut {NoStop}%
\bibitem [{\citenamefont {Heusel}\ \emph {et~al.}(2017)\citenamefont {Heusel}, \citenamefont {Ramsauer}, \citenamefont {Unterthiner}, \citenamefont {Nessler},\ and\ \citenamefont {Hochreiter}}]{heusel2017gans}%
  \BibitemOpen
  \bibfield  {author} {\bibinfo {author} {\bibfnamefont {M.}~\bibnamefont {Heusel}}, \bibinfo {author} {\bibfnamefont {H.}~\bibnamefont {Ramsauer}}, \bibinfo {author} {\bibfnamefont {T.}~\bibnamefont {Unterthiner}}, \bibinfo {author} {\bibfnamefont {B.}~\bibnamefont {Nessler}},\ and\ \bibinfo {author} {\bibfnamefont {S.}~\bibnamefont {Hochreiter}},\ }\bibfield  {title} {\bibinfo {title} {Gans trained by a two time-scale update rule converge to a local nash equilibrium},\ }\href@noop {} {\bibfield  {journal} {\bibinfo  {journal} {Advances in neural information processing systems}\ }\textbf {\bibinfo {volume} {30}} (\bibinfo {year} {2017})}\BibitemShut {NoStop}%
\bibitem [{\citenamefont {Amari}(2016)}]{amari2016information}%
  \BibitemOpen
  \bibfield  {author} {\bibinfo {author} {\bibfnamefont {S.-i.}\ \bibnamefont {Amari}},\ }\href@noop {} {\emph {\bibinfo {title} {Information geometry and its applications}}},\ Vol.\ \bibinfo {volume} {194}\ (\bibinfo  {publisher} {Springer},\ \bibinfo {year} {2016})\BibitemShut {NoStop}%
\bibitem [{\citenamefont {Poole}\ \emph {et~al.}(2022)\citenamefont {Poole}, \citenamefont {Jain}, \citenamefont {Barron},\ and\ \citenamefont {Mildenhall}}]{poole2022dreamfusion}%
  \BibitemOpen
  \bibfield  {author} {\bibinfo {author} {\bibfnamefont {B.}~\bibnamefont {Poole}}, \bibinfo {author} {\bibfnamefont {A.}~\bibnamefont {Jain}}, \bibinfo {author} {\bibfnamefont {J.~T.}\ \bibnamefont {Barron}},\ and\ \bibinfo {author} {\bibfnamefont {B.}~\bibnamefont {Mildenhall}},\ }\bibfield  {title} {\bibinfo {title} {Dreamfusion: Text-to-3d using 2d diffusion},\ }\href@noop {} {\bibfield  {journal} {\bibinfo  {journal} {arXiv preprint arXiv:2209.14988}\ } (\bibinfo {year} {2022})}\BibitemShut {NoStop}%
\bibitem [{\citenamefont {Xu}\ \emph {et~al.}(2022)\citenamefont {Xu}, \citenamefont {Yu}, \citenamefont {Song}, \citenamefont {Shi}, \citenamefont {Ermon},\ and\ \citenamefont {Tang}}]{xu2022geodiff}%
  \BibitemOpen
  \bibfield  {author} {\bibinfo {author} {\bibfnamefont {M.}~\bibnamefont {Xu}}, \bibinfo {author} {\bibfnamefont {L.}~\bibnamefont {Yu}}, \bibinfo {author} {\bibfnamefont {Y.}~\bibnamefont {Song}}, \bibinfo {author} {\bibfnamefont {C.}~\bibnamefont {Shi}}, \bibinfo {author} {\bibfnamefont {S.}~\bibnamefont {Ermon}},\ and\ \bibinfo {author} {\bibfnamefont {J.}~\bibnamefont {Tang}},\ }\bibfield  {title} {\bibinfo {title} {Geodiff: A geometric diffusion model for molecular conformation generation},\ }\href@noop {} {\bibfield  {journal} {\bibinfo  {journal} {arXiv preprint arXiv:2203.02923}\ } (\bibinfo {year} {2022})}\BibitemShut {NoStop}%
\bibitem [{\citenamefont {Abramson}\ \emph {et~al.}(2024)\citenamefont {Abramson}, \citenamefont {Adler}, \citenamefont {Dunger}, \citenamefont {Evans}, \citenamefont {Green}, \citenamefont {Pritzel}, \citenamefont {Ronneberger}, \citenamefont {Willmore}, \citenamefont {Ballard}, \citenamefont {Bambrick} \emph {et~al.}}]{abramson2024accurate}%
  \BibitemOpen
  \bibfield  {author} {\bibinfo {author} {\bibfnamefont {J.}~\bibnamefont {Abramson}}, \bibinfo {author} {\bibfnamefont {J.}~\bibnamefont {Adler}}, \bibinfo {author} {\bibfnamefont {J.}~\bibnamefont {Dunger}}, \bibinfo {author} {\bibfnamefont {R.}~\bibnamefont {Evans}}, \bibinfo {author} {\bibfnamefont {T.}~\bibnamefont {Green}}, \bibinfo {author} {\bibfnamefont {A.}~\bibnamefont {Pritzel}}, \bibinfo {author} {\bibfnamefont {O.}~\bibnamefont {Ronneberger}}, \bibinfo {author} {\bibfnamefont {L.}~\bibnamefont {Willmore}}, \bibinfo {author} {\bibfnamefont {A.~J.}\ \bibnamefont {Ballard}}, \bibinfo {author} {\bibfnamefont {J.}~\bibnamefont {Bambrick}}, \emph {et~al.},\ }\bibfield  {title} {\bibinfo {title} {Accurate structure prediction of biomolecular interactions with alphafold 3},\ }\href@noop {} {\bibfield  {journal} {\bibinfo  {journal} {Nature}\ ,\ \bibinfo {pages} {1}} (\bibinfo {year} {2024})}\BibitemShut {NoStop}%
\bibitem [{\citenamefont {Ho}\ \emph {et~al.}(2022)\citenamefont {Ho}, \citenamefont {Chan}, \citenamefont {Saharia}, \citenamefont {Whang}, \citenamefont {Gao}, \citenamefont {Gritsenko}, \citenamefont {Kingma}, \citenamefont {Poole}, \citenamefont {Norouzi}, \citenamefont {Fleet} \emph {et~al.}}]{ho2022imagen}%
  \BibitemOpen
  \bibfield  {author} {\bibinfo {author} {\bibfnamefont {J.}~\bibnamefont {Ho}}, \bibinfo {author} {\bibfnamefont {W.}~\bibnamefont {Chan}}, \bibinfo {author} {\bibfnamefont {C.}~\bibnamefont {Saharia}}, \bibinfo {author} {\bibfnamefont {J.}~\bibnamefont {Whang}}, \bibinfo {author} {\bibfnamefont {R.}~\bibnamefont {Gao}}, \bibinfo {author} {\bibfnamefont {A.}~\bibnamefont {Gritsenko}}, \bibinfo {author} {\bibfnamefont {D.~P.}\ \bibnamefont {Kingma}}, \bibinfo {author} {\bibfnamefont {B.}~\bibnamefont {Poole}}, \bibinfo {author} {\bibfnamefont {M.}~\bibnamefont {Norouzi}}, \bibinfo {author} {\bibfnamefont {D.~J.}\ \bibnamefont {Fleet}}, \emph {et~al.},\ }\bibfield  {title} {\bibinfo {title} {Imagen video: High definition video generation with diffusion models},\ }\href@noop {} {\bibfield  {journal} {\bibinfo  {journal} {arXiv preprint arXiv:2210.02303}\ } (\bibinfo {year} {2022})}\BibitemShut {NoStop}%
\bibitem [{\citenamefont {Brooks}\ \emph {et~al.}(2024)\citenamefont {Brooks}, \citenamefont {Peebles}, \citenamefont {Holmes}, \citenamefont {DePue}, \citenamefont {Guo}, \citenamefont {Jing}, \citenamefont {Schnurr}, \citenamefont {Taylor}, \citenamefont {Luhman}, \citenamefont {Luhman}, \citenamefont {Ng}, \citenamefont {Wang},\ and\ \citenamefont {Ramesh}}]{videoworldsimulators2024}%
  \BibitemOpen
  \bibfield  {author} {\bibinfo {author} {\bibfnamefont {T.}~\bibnamefont {Brooks}}, \bibinfo {author} {\bibfnamefont {B.}~\bibnamefont {Peebles}}, \bibinfo {author} {\bibfnamefont {C.}~\bibnamefont {Holmes}}, \bibinfo {author} {\bibfnamefont {W.}~\bibnamefont {DePue}}, \bibinfo {author} {\bibfnamefont {Y.}~\bibnamefont {Guo}}, \bibinfo {author} {\bibfnamefont {L.}~\bibnamefont {Jing}}, \bibinfo {author} {\bibfnamefont {D.}~\bibnamefont {Schnurr}}, \bibinfo {author} {\bibfnamefont {J.}~\bibnamefont {Taylor}}, \bibinfo {author} {\bibfnamefont {T.}~\bibnamefont {Luhman}}, \bibinfo {author} {\bibfnamefont {E.}~\bibnamefont {Luhman}}, \bibinfo {author} {\bibfnamefont {C.}~\bibnamefont {Ng}}, \bibinfo {author} {\bibfnamefont {R.}~\bibnamefont {Wang}},\ and\ \bibinfo {author} {\bibfnamefont {A.}~\bibnamefont {Ramesh}},\ }\href {https://openai.com/research/video-generation-models-as-world-simulators} {\bibinfo {title} {Video generation models as world simulators}} (\bibinfo {year} {2024})\BibitemShut {NoStop}%
\bibitem [{\citenamefont {Chen}\ \emph {et~al.}(2020)\citenamefont {Chen}, \citenamefont {Zhang}, \citenamefont {Zen}, \citenamefont {Weiss}, \citenamefont {Norouzi},\ and\ \citenamefont {Chan}}]{chen2020wavegrad}%
  \BibitemOpen
  \bibfield  {author} {\bibinfo {author} {\bibfnamefont {N.}~\bibnamefont {Chen}}, \bibinfo {author} {\bibfnamefont {Y.}~\bibnamefont {Zhang}}, \bibinfo {author} {\bibfnamefont {H.}~\bibnamefont {Zen}}, \bibinfo {author} {\bibfnamefont {R.~J.}\ \bibnamefont {Weiss}}, \bibinfo {author} {\bibfnamefont {M.}~\bibnamefont {Norouzi}},\ and\ \bibinfo {author} {\bibfnamefont {W.}~\bibnamefont {Chan}},\ }\bibfield  {title} {\bibinfo {title} {Wavegrad: Estimating gradients for waveform generation},\ }in\ \href@noop {} {\emph {\bibinfo {booktitle} {International Conference on Learning Representations}}}\ (\bibinfo {year} {2020})\BibitemShut {NoStop}%
\bibitem [{\citenamefont {Kong}\ \emph {et~al.}(2020)\citenamefont {Kong}, \citenamefont {Ping}, \citenamefont {Huang}, \citenamefont {Zhao},\ and\ \citenamefont {Catanzaro}}]{kong2020diffwave}%
  \BibitemOpen
  \bibfield  {author} {\bibinfo {author} {\bibfnamefont {Z.}~\bibnamefont {Kong}}, \bibinfo {author} {\bibfnamefont {W.}~\bibnamefont {Ping}}, \bibinfo {author} {\bibfnamefont {J.}~\bibnamefont {Huang}}, \bibinfo {author} {\bibfnamefont {K.}~\bibnamefont {Zhao}},\ and\ \bibinfo {author} {\bibfnamefont {B.}~\bibnamefont {Catanzaro}},\ }\bibfield  {title} {\bibinfo {title} {Diffwave: A versatile diffusion model for audio synthesis},\ }in\ \href@noop {} {\emph {\bibinfo {booktitle} {International Conference on Learning Representations}}}\ (\bibinfo {year} {2020})\BibitemShut {NoStop}%
\bibitem [{hug()}]{huggan/smithsonian_butterflies_subset}%
  \BibitemOpen
  \href@noop {} {\bibinfo {title} {Smithsonian butterflies subset}},\ \bibinfo {note} {accessed: 2024-02-09}\BibitemShut {NoStop}%
\bibitem [{\citenamefont {Risken}\ and\ \citenamefont {Risken}(1996)}]{risken1996fokker}%
  \BibitemOpen
  \bibfield  {author} {\bibinfo {author} {\bibfnamefont {H.}~\bibnamefont {Risken}}\ and\ \bibinfo {author} {\bibfnamefont {H.}~\bibnamefont {Risken}},\ }\href@noop {} {\emph {\bibinfo {title} {Fokker-planck equation}}}\ (\bibinfo  {publisher} {Springer},\ \bibinfo {year} {1996})\BibitemShut {NoStop}%
\bibitem [{\citenamefont {Van~Kampen}(1976)}]{van1976stochastic}%
  \BibitemOpen
  \bibfield  {author} {\bibinfo {author} {\bibfnamefont {N.~G.}\ \bibnamefont {Van~Kampen}},\ }\bibfield  {title} {\bibinfo {title} {Stochastic differential equations},\ }\href@noop {} {\bibfield  {journal} {\bibinfo  {journal} {Physics reports}\ }\textbf {\bibinfo {volume} {24}},\ \bibinfo {pages} {171} (\bibinfo {year} {1976})}\BibitemShut {NoStop}%
\bibitem [{\citenamefont {Brooks}(1998)}]{brooks1998markov}%
  \BibitemOpen
  \bibfield  {author} {\bibinfo {author} {\bibfnamefont {S.}~\bibnamefont {Brooks}},\ }\bibfield  {title} {\bibinfo {title} {Markov chain monte carlo method and its application},\ }\href@noop {} {\bibfield  {journal} {\bibinfo  {journal} {Journal of the royal statistical society: series D (the Statistician)}\ }\textbf {\bibinfo {volume} {47}},\ \bibinfo {pages} {69} (\bibinfo {year} {1998})}\BibitemShut {NoStop}%
\bibitem [{\citenamefont {Andrieu}\ \emph {et~al.}(2003)\citenamefont {Andrieu}, \citenamefont {De~Freitas}, \citenamefont {Doucet},\ and\ \citenamefont {Jordan}}]{andrieu2003introduction}%
  \BibitemOpen
  \bibfield  {author} {\bibinfo {author} {\bibfnamefont {C.}~\bibnamefont {Andrieu}}, \bibinfo {author} {\bibfnamefont {N.}~\bibnamefont {De~Freitas}}, \bibinfo {author} {\bibfnamefont {A.}~\bibnamefont {Doucet}},\ and\ \bibinfo {author} {\bibfnamefont {M.~I.}\ \bibnamefont {Jordan}},\ }\bibfield  {title} {\bibinfo {title} {An introduction to mcmc for machine learning},\ }\href@noop {} {\bibfield  {journal} {\bibinfo  {journal} {Machine learning}\ }\textbf {\bibinfo {volume} {50}},\ \bibinfo {pages} {5} (\bibinfo {year} {2003})}\BibitemShut {NoStop}%
\bibitem [{\citenamefont {Welling}\ and\ \citenamefont {Teh}(2011)}]{welling2011bayesian}%
  \BibitemOpen
  \bibfield  {author} {\bibinfo {author} {\bibfnamefont {M.}~\bibnamefont {Welling}}\ and\ \bibinfo {author} {\bibfnamefont {Y.~W.}\ \bibnamefont {Teh}},\ }\bibfield  {title} {\bibinfo {title} {Bayesian learning via stochastic gradient langevin dynamics},\ }in\ \href@noop {} {\emph {\bibinfo {booktitle} {Proceedings of the 28th international conference on machine learning (ICML-11)}}}\ (\bibinfo {organization} {Citeseer},\ \bibinfo {year} {2011})\ pp.\ \bibinfo {pages} {681--688}\BibitemShut {NoStop}%
\bibitem [{\citenamefont {Chen}(2023)}]{chen2023importance}%
  \BibitemOpen
  \bibfield  {author} {\bibinfo {author} {\bibfnamefont {T.}~\bibnamefont {Chen}},\ }\bibfield  {title} {\bibinfo {title} {On the importance of noise scheduling for diffusion models},\ }\href@noop {} {\bibfield  {journal} {\bibinfo  {journal} {arXiv preprint arXiv:2301.10972}\ } (\bibinfo {year} {2023})}\BibitemShut {NoStop}%
\bibitem [{Note1()}]{Note1}%
  \BibitemOpen
  \bibinfo {note} {The score matching objective function allows for arbitrary time-dependent weights~\cite {song2020scorebased}; in this study, we have adopted $T_t^2$ for the consistency with Eq.~\protect \eqref {eq:underestimateSADM}.}\BibitemShut {Stop}%
\bibitem [{\citenamefont {Anderson}(1982)}]{anderson1982reverse}%
  \BibitemOpen
  \bibfield  {author} {\bibinfo {author} {\bibfnamefont {B.~D.}\ \bibnamefont {Anderson}},\ }\bibfield  {title} {\bibinfo {title} {Reverse-time diffusion equation models},\ }\href@noop {} {\bibfield  {journal} {\bibinfo  {journal} {Stochastic Processes and their Applications}\ }\textbf {\bibinfo {volume} {12}},\ \bibinfo {pages} {313} (\bibinfo {year} {1982})}\BibitemShut {NoStop}%
\bibitem [{\citenamefont {Lu}\ \emph {et~al.}(2022{\natexlab{a}})\citenamefont {Lu}, \citenamefont {Zhou}, \citenamefont {Bao}, \citenamefont {Chen}, \citenamefont {Li},\ and\ \citenamefont {Zhu}}]{lu2022dpm}%
  \BibitemOpen
  \bibfield  {author} {\bibinfo {author} {\bibfnamefont {C.}~\bibnamefont {Lu}}, \bibinfo {author} {\bibfnamefont {Y.}~\bibnamefont {Zhou}}, \bibinfo {author} {\bibfnamefont {F.}~\bibnamefont {Bao}}, \bibinfo {author} {\bibfnamefont {J.}~\bibnamefont {Chen}}, \bibinfo {author} {\bibfnamefont {C.}~\bibnamefont {Li}},\ and\ \bibinfo {author} {\bibfnamefont {J.}~\bibnamefont {Zhu}},\ }\bibfield  {title} {\bibinfo {title} {Dpm-solver: A fast ode solver for diffusion probabilistic model sampling in around 10 steps},\ }\href@noop {} {\bibfield  {journal} {\bibinfo  {journal} {Advances in Neural Information Processing Systems}\ }\textbf {\bibinfo {volume} {35}},\ \bibinfo {pages} {5775} (\bibinfo {year} {2022}{\natexlab{a}})}\BibitemShut {NoStop}%
\bibitem [{\citenamefont {Lu}\ \emph {et~al.}(2022{\natexlab{b}})\citenamefont {Lu}, \citenamefont {Zheng}, \citenamefont {Bao}, \citenamefont {Chen}, \citenamefont {Li},\ and\ \citenamefont {Zhu}}]{lu2022maximum}%
  \BibitemOpen
  \bibfield  {author} {\bibinfo {author} {\bibfnamefont {C.}~\bibnamefont {Lu}}, \bibinfo {author} {\bibfnamefont {K.}~\bibnamefont {Zheng}}, \bibinfo {author} {\bibfnamefont {F.}~\bibnamefont {Bao}}, \bibinfo {author} {\bibfnamefont {J.}~\bibnamefont {Chen}}, \bibinfo {author} {\bibfnamefont {C.}~\bibnamefont {Li}},\ and\ \bibinfo {author} {\bibfnamefont {J.}~\bibnamefont {Zhu}},\ }\bibfield  {title} {\bibinfo {title} {Maximum likelihood training for score-based diffusion odes by high order denoising score matching},\ }in\ \href@noop {} {\emph {\bibinfo {booktitle} {International Conference on Machine Learning}}}\ (\bibinfo {organization} {PMLR},\ \bibinfo {year} {2022})\ pp.\ \bibinfo {pages} {14429--14460}\BibitemShut {NoStop}%
\bibitem [{\citenamefont {Zhang}\ and\ \citenamefont {Chen}(2023)}]{zhang2022fast}%
  \BibitemOpen
  \bibfield  {author} {\bibinfo {author} {\bibfnamefont {Q.}~\bibnamefont {Zhang}}\ and\ \bibinfo {author} {\bibfnamefont {Y.}~\bibnamefont {Chen}},\ }\bibfield  {title} {\bibinfo {title} {Fast sampling of diffusion models with exponential integrator},\ }in\ \href@noop {} {\emph {\bibinfo {booktitle} {The Eleventh International Conference on Learning Representations}}}\ (\bibinfo {year} {2023})\BibitemShut {NoStop}%
\bibitem [{\citenamefont {Kobyzev}\ \emph {et~al.}(2020)\citenamefont {Kobyzev}, \citenamefont {Prince},\ and\ \citenamefont {Brubaker}}]{kobyzev2020normalizing}%
  \BibitemOpen
  \bibfield  {author} {\bibinfo {author} {\bibfnamefont {I.}~\bibnamefont {Kobyzev}}, \bibinfo {author} {\bibfnamefont {S.~J.}\ \bibnamefont {Prince}},\ and\ \bibinfo {author} {\bibfnamefont {M.~A.}\ \bibnamefont {Brubaker}},\ }\bibfield  {title} {\bibinfo {title} {Normalizing flows: An introduction and review of current methods},\ }\href@noop {} {\bibfield  {journal} {\bibinfo  {journal} {IEEE transactions on pattern analysis and machine intelligence}\ }\textbf {\bibinfo {volume} {43}},\ \bibinfo {pages} {3964} (\bibinfo {year} {2020})}\BibitemShut {NoStop}%
\bibitem [{\citenamefont {Ito}\ \emph {et~al.}(2020)\citenamefont {Ito}, \citenamefont {Oizumi},\ and\ \citenamefont {Amari}}]{ito2020unified}%
  \BibitemOpen
  \bibfield  {author} {\bibinfo {author} {\bibfnamefont {S.}~\bibnamefont {Ito}}, \bibinfo {author} {\bibfnamefont {M.}~\bibnamefont {Oizumi}},\ and\ \bibinfo {author} {\bibfnamefont {S.-i.}\ \bibnamefont {Amari}},\ }\bibfield  {title} {\bibinfo {title} {Unified framework for the entropy production and the stochastic interaction based on information geometry},\ }\href@noop {} {\bibfield  {journal} {\bibinfo  {journal} {Physical Review Research}\ }\textbf {\bibinfo {volume} {2}},\ \bibinfo {pages} {033048} (\bibinfo {year} {2020})}\BibitemShut {NoStop}%
\bibitem [{\citenamefont {Kawai}\ \emph {et~al.}(2007)\citenamefont {Kawai}, \citenamefont {Parrondo},\ and\ \citenamefont {Van~den Broeck}}]{kawai2007dissipation}%
  \BibitemOpen
  \bibfield  {author} {\bibinfo {author} {\bibfnamefont {R.}~\bibnamefont {Kawai}}, \bibinfo {author} {\bibfnamefont {J.~M.}\ \bibnamefont {Parrondo}},\ and\ \bibinfo {author} {\bibfnamefont {C.}~\bibnamefont {Van~den Broeck}},\ }\bibfield  {title} {\bibinfo {title} {Dissipation: The phase-space perspective},\ }\href@noop {} {\bibfield  {journal} {\bibinfo  {journal} {Physical review letters}\ }\textbf {\bibinfo {volume} {98}},\ \bibinfo {pages} {080602} (\bibinfo {year} {2007})}\BibitemShut {NoStop}%
\bibitem [{\citenamefont {Villani}(2021)}]{villani2021topics}%
  \BibitemOpen
  \bibfield  {author} {\bibinfo {author} {\bibfnamefont {C.}~\bibnamefont {Villani}},\ }\href@noop {} {\emph {\bibinfo {title} {Topics in optimal transportation}}},\ Vol.~\bibinfo {volume} {58}\ (\bibinfo  {publisher} {American Mathematical Soc.},\ \bibinfo {year} {2021})\BibitemShut {NoStop}%
\bibitem [{\citenamefont {Givens}\ and\ \citenamefont {Shortt}(1984)}]{clark1984class}%
  \BibitemOpen
  \bibfield  {author} {\bibinfo {author} {\bibfnamefont {C.~R.}\ \bibnamefont {Givens}}\ and\ \bibinfo {author} {\bibfnamefont {R.~M.}\ \bibnamefont {Shortt}},\ }\bibfield  {title} {\bibinfo {title} {A class of wasserstein metrics for probability distributions},\ }\href {https://doi.org/10.1307/mmj/1029003026} {\bibfield  {journal} {\bibinfo  {journal} {Michigan Mathematical Journal}\ }\textbf {\bibinfo {volume} {31}},\ \bibinfo {pages} {231 } (\bibinfo {year} {1984})}\BibitemShut {NoStop}%
\bibitem [{\citenamefont {Szegedy}\ \emph {et~al.}(2015)\citenamefont {Szegedy}, \citenamefont {Liu}, \citenamefont {Jia}, \citenamefont {Sermanet}, \citenamefont {Reed}, \citenamefont {Anguelov}, \citenamefont {Erhan}, \citenamefont {Vanhoucke},\ and\ \citenamefont {Rabinovich}}]{szegedy2015going}%
  \BibitemOpen
  \bibfield  {author} {\bibinfo {author} {\bibfnamefont {C.}~\bibnamefont {Szegedy}}, \bibinfo {author} {\bibfnamefont {W.}~\bibnamefont {Liu}}, \bibinfo {author} {\bibfnamefont {Y.}~\bibnamefont {Jia}}, \bibinfo {author} {\bibfnamefont {P.}~\bibnamefont {Sermanet}}, \bibinfo {author} {\bibfnamefont {S.}~\bibnamefont {Reed}}, \bibinfo {author} {\bibfnamefont {D.}~\bibnamefont {Anguelov}}, \bibinfo {author} {\bibfnamefont {D.}~\bibnamefont {Erhan}}, \bibinfo {author} {\bibfnamefont {V.}~\bibnamefont {Vanhoucke}},\ and\ \bibinfo {author} {\bibfnamefont {A.}~\bibnamefont {Rabinovich}},\ }\bibfield  {title} {\bibinfo {title} {Going deeper with convolutions},\ }in\ \href@noop {} {\emph {\bibinfo {booktitle} {Proceedings of the IEEE conference on computer vision and pattern recognition}}}\ (\bibinfo {year} {2015})\ pp.\ \bibinfo {pages} {1--9}\BibitemShut {NoStop}%
\bibitem [{\citenamefont {Gelbrich}(1990)}]{gelbrich1990formula}%
  \BibitemOpen
  \bibfield  {author} {\bibinfo {author} {\bibfnamefont {M.}~\bibnamefont {Gelbrich}},\ }\bibfield  {title} {\bibinfo {title} {On a formula for the l2 wasserstein metric between measures on euclidean and hilbert spaces},\ }\href@noop {} {\bibfield  {journal} {\bibinfo  {journal} {Mathematische Nachrichten}\ }\textbf {\bibinfo {volume} {147}},\ \bibinfo {pages} {185} (\bibinfo {year} {1990})}\BibitemShut {NoStop}%
\bibitem [{\citenamefont {Benamou}\ and\ \citenamefont {Brenier}(2000)}]{benamou2000computational}%
  \BibitemOpen
  \bibfield  {author} {\bibinfo {author} {\bibfnamefont {J.-D.}\ \bibnamefont {Benamou}}\ and\ \bibinfo {author} {\bibfnamefont {Y.}~\bibnamefont {Brenier}},\ }\bibfield  {title} {\bibinfo {title} {A computational fluid mechanics solution to the monge-kantorovich mass transfer problem},\ }\href@noop {} {\bibfield  {journal} {\bibinfo  {journal} {Numerische Mathematik}\ }\textbf {\bibinfo {volume} {84}},\ \bibinfo {pages} {375} (\bibinfo {year} {2000})}\BibitemShut {NoStop}%
\bibitem [{\citenamefont {Dechant}\ \emph {et~al.}(2022{\natexlab{b}})\citenamefont {Dechant}, \citenamefont {Sasa},\ and\ \citenamefont {Ito}}]{dechant2022geometric}%
  \BibitemOpen
  \bibfield  {author} {\bibinfo {author} {\bibfnamefont {A.}~\bibnamefont {Dechant}}, \bibinfo {author} {\bibfnamefont {S.-i.}\ \bibnamefont {Sasa}},\ and\ \bibinfo {author} {\bibfnamefont {S.}~\bibnamefont {Ito}},\ }\bibfield  {title} {\bibinfo {title} {Geometric decomposition of entropy production in out-of-equilibrium systems},\ }\href@noop {} {\bibfield  {journal} {\bibinfo  {journal} {Physical Review Research}\ }\textbf {\bibinfo {volume} {4}},\ \bibinfo {pages} {L012034} (\bibinfo {year} {2022}{\natexlab{b}})}\BibitemShut {NoStop}%
\bibitem [{\citenamefont {Benamou}\ and\ \citenamefont {Brenier}(1999)}]{benamou1999numerical}%
  \BibitemOpen
  \bibfield  {author} {\bibinfo {author} {\bibfnamefont {J.-D.}\ \bibnamefont {Benamou}}\ and\ \bibinfo {author} {\bibfnamefont {Y.}~\bibnamefont {Brenier}},\ }\bibfield  {title} {\bibinfo {title} {A numerical method for the optimal time-continuous mass transport problem and related problems},\ }\href@noop {} {\bibfield  {journal} {\bibinfo  {journal} {Contemporary mathematics}\ }\textbf {\bibinfo {volume} {226}},\ \bibinfo {pages} {1} (\bibinfo {year} {1999})}\BibitemShut {NoStop}%
\bibitem [{\citenamefont {Sekizawa}\ \emph {et~al.}(2023)\citenamefont {Sekizawa}, \citenamefont {Ito},\ and\ \citenamefont {Oizumi}}]{sekizawa2023decomposing}%
  \BibitemOpen
  \bibfield  {author} {\bibinfo {author} {\bibfnamefont {D.}~\bibnamefont {Sekizawa}}, \bibinfo {author} {\bibfnamefont {S.}~\bibnamefont {Ito}},\ and\ \bibinfo {author} {\bibfnamefont {M.}~\bibnamefont {Oizumi}},\ }\bibfield  {title} {\bibinfo {title} {Decomposing thermodynamic dissipation of neural dynamics via spatio-temporal oscillatory modes},\ }\href@noop {} {\bibfield  {journal} {\bibinfo  {journal} {arXiv preprint arXiv:2312.03489}\ } (\bibinfo {year} {2023})}\BibitemShut {NoStop}%
\bibitem [{\citenamefont {Crooks}(2000)}]{crooks2000path}%
  \BibitemOpen
  \bibfield  {author} {\bibinfo {author} {\bibfnamefont {G.~E.}\ \bibnamefont {Crooks}},\ }\bibfield  {title} {\bibinfo {title} {Path-ensemble averages in systems driven far from equilibrium},\ }\href@noop {} {\bibfield  {journal} {\bibinfo  {journal} {Physical review E}\ }\textbf {\bibinfo {volume} {61}},\ \bibinfo {pages} {2361} (\bibinfo {year} {2000})}\BibitemShut {NoStop}%
\bibitem [{\citenamefont {Horowitz}\ and\ \citenamefont {Gingrich}(2020)}]{horowitz2020thermodynamic}%
  \BibitemOpen
  \bibfield  {author} {\bibinfo {author} {\bibfnamefont {J.~M.}\ \bibnamefont {Horowitz}}\ and\ \bibinfo {author} {\bibfnamefont {T.~R.}\ \bibnamefont {Gingrich}},\ }\bibfield  {title} {\bibinfo {title} {Thermodynamic uncertainty relations constrain non-equilibrium fluctuations},\ }\href@noop {} {\bibfield  {journal} {\bibinfo  {journal} {Nature Physics}\ }\textbf {\bibinfo {volume} {16}},\ \bibinfo {pages} {15} (\bibinfo {year} {2020})}\BibitemShut {NoStop}%
\bibitem [{\citenamefont {Lan}\ \emph {et~al.}(2012)\citenamefont {Lan}, \citenamefont {Sartori}, \citenamefont {Neumann}, \citenamefont {Sourjik},\ and\ \citenamefont {Tu}}]{lan2012energy}%
  \BibitemOpen
  \bibfield  {author} {\bibinfo {author} {\bibfnamefont {G.}~\bibnamefont {Lan}}, \bibinfo {author} {\bibfnamefont {P.}~\bibnamefont {Sartori}}, \bibinfo {author} {\bibfnamefont {S.}~\bibnamefont {Neumann}}, \bibinfo {author} {\bibfnamefont {V.}~\bibnamefont {Sourjik}},\ and\ \bibinfo {author} {\bibfnamefont {Y.}~\bibnamefont {Tu}},\ }\bibfield  {title} {\bibinfo {title} {The energy--speed--accuracy trade-off in sensory adaptation},\ }\href@noop {} {\bibfield  {journal} {\bibinfo  {journal} {Nature physics}\ }\textbf {\bibinfo {volume} {8}},\ \bibinfo {pages} {422} (\bibinfo {year} {2012})}\BibitemShut {NoStop}%
\bibitem [{\citenamefont {Pearson}(1900)}]{pearson1900x}%
  \BibitemOpen
  \bibfield  {author} {\bibinfo {author} {\bibfnamefont {K.}~\bibnamefont {Pearson}},\ }\bibfield  {title} {\bibinfo {title} {X. on the criterion that a given system of deviations from the probable in the case of a correlated system of variables is such that it can be reasonably supposed to have arisen from random sampling},\ }\href@noop {} {\bibfield  {journal} {\bibinfo  {journal} {The London, Edinburgh, and Dublin Philosophical Magazine and Journal of Science}\ }\textbf {\bibinfo {volume} {50}},\ \bibinfo {pages} {157} (\bibinfo {year} {1900})}\BibitemShut {NoStop}%
\bibitem [{\citenamefont {Gulrajani}\ \emph {et~al.}(2017)\citenamefont {Gulrajani}, \citenamefont {Ahmed}, \citenamefont {Arjovsky}, \citenamefont {Dumoulin},\ and\ \citenamefont {Courville}}]{gulrajani2017improved}%
  \BibitemOpen
  \bibfield  {author} {\bibinfo {author} {\bibfnamefont {I.}~\bibnamefont {Gulrajani}}, \bibinfo {author} {\bibfnamefont {F.}~\bibnamefont {Ahmed}}, \bibinfo {author} {\bibfnamefont {M.}~\bibnamefont {Arjovsky}}, \bibinfo {author} {\bibfnamefont {V.}~\bibnamefont {Dumoulin}},\ and\ \bibinfo {author} {\bibfnamefont {A.}~\bibnamefont {Courville}},\ }\href@noop {} {\bibinfo {title} {Improved training of wasserstein gans}} (\bibinfo {year} {2017}),\ \Eprint {https://arxiv.org/abs/1704.00028} {arXiv:1704.00028 [cs.LG]} \BibitemShut {NoStop}%
\bibitem [{\citenamefont {Karras}\ \emph {et~al.}(2018)\citenamefont {Karras}, \citenamefont {Aila}, \citenamefont {Laine},\ and\ \citenamefont {Lehtinen}}]{karras2018progressive}%
  \BibitemOpen
  \bibfield  {author} {\bibinfo {author} {\bibfnamefont {T.}~\bibnamefont {Karras}}, \bibinfo {author} {\bibfnamefont {T.}~\bibnamefont {Aila}}, \bibinfo {author} {\bibfnamefont {S.}~\bibnamefont {Laine}},\ and\ \bibinfo {author} {\bibfnamefont {J.}~\bibnamefont {Lehtinen}},\ }\bibfield  {title} {\bibinfo {title} {Progressive growing of {GAN}s for improved quality, stability, and variation},\ }in\ \href@noop {} {\emph {\bibinfo {booktitle} {International Conference on Learning Representations}}}\ (\bibinfo {year} {2018})\BibitemShut {NoStop}%
\bibitem [{\citenamefont {Yu}\ \emph {et~al.}(2015)\citenamefont {Yu}, \citenamefont {Seff}, \citenamefont {Zhang}, \citenamefont {Song}, \citenamefont {Funkhouser},\ and\ \citenamefont {Xiao}}]{yu2015lsun}%
  \BibitemOpen
  \bibfield  {author} {\bibinfo {author} {\bibfnamefont {F.}~\bibnamefont {Yu}}, \bibinfo {author} {\bibfnamefont {A.}~\bibnamefont {Seff}}, \bibinfo {author} {\bibfnamefont {Y.}~\bibnamefont {Zhang}}, \bibinfo {author} {\bibfnamefont {S.}~\bibnamefont {Song}}, \bibinfo {author} {\bibfnamefont {T.}~\bibnamefont {Funkhouser}},\ and\ \bibinfo {author} {\bibfnamefont {J.}~\bibnamefont {Xiao}},\ }\bibfield  {title} {\bibinfo {title} {Lsun: Construction of a large-scale image dataset using deep learning with humans in the loop},\ }\href@noop {} {\bibfield  {journal} {\bibinfo  {journal} {arXiv preprint arXiv:1506.03365}\ } (\bibinfo {year} {2015})}\BibitemShut {NoStop}%
\bibitem [{\citenamefont {Dao}\ \emph {et~al.}(2023)\citenamefont {Dao}, \citenamefont {Phung}, \citenamefont {Nguyen},\ and\ \citenamefont {Tran}}]{dao2023flow}%
  \BibitemOpen
  \bibfield  {author} {\bibinfo {author} {\bibfnamefont {Q.}~\bibnamefont {Dao}}, \bibinfo {author} {\bibfnamefont {H.}~\bibnamefont {Phung}}, \bibinfo {author} {\bibfnamefont {B.}~\bibnamefont {Nguyen}},\ and\ \bibinfo {author} {\bibfnamefont {A.}~\bibnamefont {Tran}},\ }\bibfield  {title} {\bibinfo {title} {Flow matching in latent space},\ }\href@noop {} {\bibfield  {journal} {\bibinfo  {journal} {arXiv preprint arXiv:2307.08698}\ } (\bibinfo {year} {2023})}\BibitemShut {NoStop}%
\bibitem [{\citenamefont {Barato}\ and\ \citenamefont {Seifert}(2015)}]{barato2015thermodynamic}%
  \BibitemOpen
  \bibfield  {author} {\bibinfo {author} {\bibfnamefont {A.~C.}\ \bibnamefont {Barato}}\ and\ \bibinfo {author} {\bibfnamefont {U.}~\bibnamefont {Seifert}},\ }\bibfield  {title} {\bibinfo {title} {Thermodynamic uncertainty relation for biomolecular processes},\ }\href@noop {} {\bibfield  {journal} {\bibinfo  {journal} {Physical review letters}\ }\textbf {\bibinfo {volume} {114}},\ \bibinfo {pages} {158101} (\bibinfo {year} {2015})}\BibitemShut {NoStop}%
\bibitem [{\citenamefont {Pooladian}\ \emph {et~al.}(2023)\citenamefont {Pooladian}, \citenamefont {Ben-Hamu}, \citenamefont {Domingo-Enrich}, \citenamefont {Amos}, \citenamefont {Lipman},\ and\ \citenamefont {Chen}}]{pooladian2023multisample}%
  \BibitemOpen
  \bibfield  {author} {\bibinfo {author} {\bibfnamefont {A.~A.}\ \bibnamefont {Pooladian}}, \bibinfo {author} {\bibfnamefont {H.}~\bibnamefont {Ben-Hamu}}, \bibinfo {author} {\bibfnamefont {C.}~\bibnamefont {Domingo-Enrich}}, \bibinfo {author} {\bibfnamefont {B.}~\bibnamefont {Amos}}, \bibinfo {author} {\bibfnamefont {Y.}~\bibnamefont {Lipman}},\ and\ \bibinfo {author} {\bibfnamefont {R.~T.}\ \bibnamefont {Chen}},\ }\bibfield  {title} {\bibinfo {title} {Multisample flow matching: Straightening flows with minibatch couplings},\ }\href@noop {} {\bibfield  {journal} {\bibinfo  {journal} {Proceedings of Machine Learning Research}\ }\textbf {\bibinfo {volume} {202}},\ \bibinfo {pages} {28100} (\bibinfo {year} {2023})}\BibitemShut {NoStop}%
\bibitem [{\citenamefont {Fukumizu}\ \emph {et~al.}(2025)\citenamefont {Fukumizu}, \citenamefont {Suzuki}, \citenamefont {Isobe}, \citenamefont {Oko},\ and\ \citenamefont {Koyama}}]{fukumizu2024flow}%
  \BibitemOpen
  \bibfield  {author} {\bibinfo {author} {\bibfnamefont {K.}~\bibnamefont {Fukumizu}}, \bibinfo {author} {\bibfnamefont {T.}~\bibnamefont {Suzuki}}, \bibinfo {author} {\bibfnamefont {N.}~\bibnamefont {Isobe}}, \bibinfo {author} {\bibfnamefont {K.}~\bibnamefont {Oko}},\ and\ \bibinfo {author} {\bibfnamefont {M.}~\bibnamefont {Koyama}},\ }\bibfield  {title} {\bibinfo {title} {Flow matching achieves almost minimax optimal convergence},\ }in\ \href@noop {} {\emph {\bibinfo {booktitle} {The Thirteenth International Conference on Learning Representations}}}\ (\bibinfo {year} {2025})\BibitemShut {NoStop}%
\bibitem [{\citenamefont {Song}\ \emph {et~al.}(2021)\citenamefont {Song}, \citenamefont {Durkan}, \citenamefont {Murray},\ and\ \citenamefont {Ermon}}]{song2021maximum}%
  \BibitemOpen
  \bibfield  {author} {\bibinfo {author} {\bibfnamefont {Y.}~\bibnamefont {Song}}, \bibinfo {author} {\bibfnamefont {C.}~\bibnamefont {Durkan}}, \bibinfo {author} {\bibfnamefont {I.}~\bibnamefont {Murray}},\ and\ \bibinfo {author} {\bibfnamefont {S.}~\bibnamefont {Ermon}},\ }\bibfield  {title} {\bibinfo {title} {Maximum likelihood training of score-based diffusion models},\ }\href@noop {} {\bibfield  {journal} {\bibinfo  {journal} {Advances in neural information processing systems}\ }\textbf {\bibinfo {volume} {34}},\ \bibinfo {pages} {1415} (\bibinfo {year} {2021})}\BibitemShut {NoStop}%
\bibitem [{\citenamefont {Lee}\ \emph {et~al.}(2022)\citenamefont {Lee}, \citenamefont {Lu},\ and\ \citenamefont {Tan}}]{lee2022convergence}%
  \BibitemOpen
  \bibfield  {author} {\bibinfo {author} {\bibfnamefont {H.}~\bibnamefont {Lee}}, \bibinfo {author} {\bibfnamefont {J.}~\bibnamefont {Lu}},\ and\ \bibinfo {author} {\bibfnamefont {Y.}~\bibnamefont {Tan}},\ }\bibfield  {title} {\bibinfo {title} {Convergence for score-based generative modeling with polynomial complexity},\ }\href@noop {} {\bibfield  {journal} {\bibinfo  {journal} {Advances in Neural Information Processing Systems}\ }\textbf {\bibinfo {volume} {35}},\ \bibinfo {pages} {22870} (\bibinfo {year} {2022})}\BibitemShut {NoStop}%
\bibitem [{\citenamefont {Chen}\ \emph {et~al.}(2023)\citenamefont {Chen}, \citenamefont {Chewi}, \citenamefont {Li}, \citenamefont {Li}, \citenamefont {Salim},\ and\ \citenamefont {Zhang}}]{chen2023sampling}%
  \BibitemOpen
  \bibfield  {author} {\bibinfo {author} {\bibfnamefont {S.}~\bibnamefont {Chen}}, \bibinfo {author} {\bibfnamefont {S.}~\bibnamefont {Chewi}}, \bibinfo {author} {\bibfnamefont {J.}~\bibnamefont {Li}}, \bibinfo {author} {\bibfnamefont {Y.}~\bibnamefont {Li}}, \bibinfo {author} {\bibfnamefont {A.}~\bibnamefont {Salim}},\ and\ \bibinfo {author} {\bibfnamefont {A.~R.}\ \bibnamefont {Zhang}},\ }\bibfield  {title} {\bibinfo {title} {Sampling is as easy as learning the score: theory for diffusion models with minimal data assumptions},\ }in\ \href@noop {} {\emph {\bibinfo {booktitle} {International Conference on Learning Representations}}}\ (\bibinfo {year} {2023})\BibitemShut {NoStop}%
\bibitem [{\citenamefont {Salimans}\ \emph {et~al.}(2016)\citenamefont {Salimans}, \citenamefont {Goodfellow}, \citenamefont {Zaremba}, \citenamefont {Cheung}, \citenamefont {Radford},\ and\ \citenamefont {Chen}}]{salimans2016improved}%
  \BibitemOpen
  \bibfield  {author} {\bibinfo {author} {\bibfnamefont {T.}~\bibnamefont {Salimans}}, \bibinfo {author} {\bibfnamefont {I.}~\bibnamefont {Goodfellow}}, \bibinfo {author} {\bibfnamefont {W.}~\bibnamefont {Zaremba}}, \bibinfo {author} {\bibfnamefont {V.}~\bibnamefont {Cheung}}, \bibinfo {author} {\bibfnamefont {A.}~\bibnamefont {Radford}},\ and\ \bibinfo {author} {\bibfnamefont {X.}~\bibnamefont {Chen}},\ }\bibfield  {title} {\bibinfo {title} {Improved techniques for training gans},\ }\href@noop {} {\bibfield  {journal} {\bibinfo  {journal} {Advances in neural information processing systems}\ }\textbf {\bibinfo {volume} {29}} (\bibinfo {year} {2016})}\BibitemShut {NoStop}%
\bibitem [{\citenamefont {Albergo}\ and\ \citenamefont {Vanden-Eijnden}(2022)}]{albergo2022building}%
  \BibitemOpen
  \bibfield  {author} {\bibinfo {author} {\bibfnamefont {M.~S.}\ \bibnamefont {Albergo}}\ and\ \bibinfo {author} {\bibfnamefont {E.}~\bibnamefont {Vanden-Eijnden}},\ }\bibfield  {title} {\bibinfo {title} {Building normalizing flows with stochastic interpolants},\ }\href@noop {} {\bibfield  {journal} {\bibinfo  {journal} {arXiv preprint arXiv:2209.15571}\ } (\bibinfo {year} {2022})}\BibitemShut {NoStop}%
\bibitem [{\citenamefont {Amari}(1972)}]{amari1972learning}%
  \BibitemOpen
  \bibfield  {author} {\bibinfo {author} {\bibfnamefont {S.-I.}\ \bibnamefont {Amari}},\ }\bibfield  {title} {\bibinfo {title} {Learning patterns and pattern sequences by self-organizing nets of threshold elements},\ }\href@noop {} {\bibfield  {journal} {\bibinfo  {journal} {IEEE Transactions on computers}\ }\textbf {\bibinfo {volume} {100}},\ \bibinfo {pages} {1197} (\bibinfo {year} {1972})}\BibitemShut {NoStop}%
\bibitem [{\citenamefont {Hopfield}(1982)}]{hopfield1982neural}%
  \BibitemOpen
  \bibfield  {author} {\bibinfo {author} {\bibfnamefont {J.~J.}\ \bibnamefont {Hopfield}},\ }\bibfield  {title} {\bibinfo {title} {Neural networks and physical systems with emergent collective computational abilities.},\ }\href@noop {} {\bibfield  {journal} {\bibinfo  {journal} {Proceedings of the national academy of sciences}\ }\textbf {\bibinfo {volume} {79}},\ \bibinfo {pages} {2554} (\bibinfo {year} {1982})}\BibitemShut {NoStop}%
\bibitem [{\citenamefont {Hinton}\ \emph {et~al.}(1984)\citenamefont {Hinton}, \citenamefont {Sejnowski},\ and\ \citenamefont {Ackley}}]{hinton1984boltzmann}%
  \BibitemOpen
  \bibfield  {author} {\bibinfo {author} {\bibfnamefont {G.~E.}\ \bibnamefont {Hinton}}, \bibinfo {author} {\bibfnamefont {T.~J.}\ \bibnamefont {Sejnowski}},\ and\ \bibinfo {author} {\bibfnamefont {D.~H.}\ \bibnamefont {Ackley}},\ }\href@noop {} {\emph {\bibinfo {title} {Boltzmann machines: Constraint satisfaction networks that learn}}}\ (\bibinfo  {publisher} {Carnegie-Mellon University, Department of Computer Science Pittsburgh, PA},\ \bibinfo {year} {1984})\BibitemShut {NoStop}%
\bibitem [{\citenamefont {Salakhutdinov}\ and\ \citenamefont {Hinton}(2009)}]{salakhutdinov2009deep}%
  \BibitemOpen
  \bibfield  {author} {\bibinfo {author} {\bibfnamefont {R.}~\bibnamefont {Salakhutdinov}}\ and\ \bibinfo {author} {\bibfnamefont {G.}~\bibnamefont {Hinton}},\ }\bibfield  {title} {\bibinfo {title} {Deep boltzmann machines},\ }in\ \href@noop {} {\emph {\bibinfo {booktitle} {Artificial intelligence and statistics}}}\ (\bibinfo {organization} {PMLR},\ \bibinfo {year} {2009})\ pp.\ \bibinfo {pages} {448--455}\BibitemShut {NoStop}%
\bibitem [{\citenamefont {Hinton}(2010)}]{hinton2010practical}%
  \BibitemOpen
  \bibfield  {author} {\bibinfo {author} {\bibfnamefont {G.}~\bibnamefont {Hinton}},\ }\bibfield  {title} {\bibinfo {title} {A practical guide to training restricted boltzmann machines},\ }\href@noop {} {\bibfield  {journal} {\bibinfo  {journal} {Momentum}\ }\textbf {\bibinfo {volume} {9}},\ \bibinfo {pages} {926} (\bibinfo {year} {2010})}\BibitemShut {NoStop}%
\bibitem [{\citenamefont {Schnakenberg}(1976)}]{schnakenberg1976network}%
  \BibitemOpen
  \bibfield  {author} {\bibinfo {author} {\bibfnamefont {J.}~\bibnamefont {Schnakenberg}},\ }\bibfield  {title} {\bibinfo {title} {Network theory of microscopic and macroscopic behavior of master equation systems},\ }\href@noop {} {\bibfield  {journal} {\bibinfo  {journal} {Reviews of Modern physics}\ }\textbf {\bibinfo {volume} {48}},\ \bibinfo {pages} {571} (\bibinfo {year} {1976})}\BibitemShut {NoStop}%
\bibitem [{\citenamefont {Qian}(2001)}]{qian2001relative}%
  \BibitemOpen
  \bibfield  {author} {\bibinfo {author} {\bibfnamefont {H.}~\bibnamefont {Qian}},\ }\bibfield  {title} {\bibinfo {title} {Relative entropy: Free energy associated with equilibrium fluctuations and nonequilibrium deviations},\ }\href@noop {} {\bibfield  {journal} {\bibinfo  {journal} {Physical Review E}\ }\textbf {\bibinfo {volume} {63}},\ \bibinfo {pages} {042103} (\bibinfo {year} {2001})}\BibitemShut {NoStop}%
\bibitem [{\citenamefont {Biroli}\ \emph {et~al.}(2024)\citenamefont {Biroli}, \citenamefont {Bonnaire}, \citenamefont {De~Bortoli},\ and\ \citenamefont {M{\'e}zard}}]{biroli2024dynamical}%
  \BibitemOpen
  \bibfield  {author} {\bibinfo {author} {\bibfnamefont {G.}~\bibnamefont {Biroli}}, \bibinfo {author} {\bibfnamefont {T.}~\bibnamefont {Bonnaire}}, \bibinfo {author} {\bibfnamefont {V.}~\bibnamefont {De~Bortoli}},\ and\ \bibinfo {author} {\bibfnamefont {M.}~\bibnamefont {M{\'e}zard}},\ }\bibfield  {title} {\bibinfo {title} {Dynamical regimes of diffusion models},\ }\href@noop {} {\bibfield  {journal} {\bibinfo  {journal} {Nature Communications}\ }\textbf {\bibinfo {volume} {15}},\ \bibinfo {pages} {9957} (\bibinfo {year} {2024})}\BibitemShut {NoStop}%
\bibitem [{\citenamefont {Dechant}(2022)}]{dechant2022minimum}%
  \BibitemOpen
  \bibfield  {author} {\bibinfo {author} {\bibfnamefont {A.}~\bibnamefont {Dechant}},\ }\bibfield  {title} {\bibinfo {title} {Minimum entropy production, detailed balance and wasserstein distance for continuous-time markov processes},\ }\href@noop {} {\bibfield  {journal} {\bibinfo  {journal} {Journal of Physics A: Mathematical and Theoretical}\ }\textbf {\bibinfo {volume} {55}},\ \bibinfo {pages} {094001} (\bibinfo {year} {2022})}\BibitemShut {NoStop}%
\bibitem [{\citenamefont {Yoshimura}\ \emph {et~al.}(2023)\citenamefont {Yoshimura}, \citenamefont {Kolchinsky}, \citenamefont {Dechant},\ and\ \citenamefont {Ito}}]{yoshimura2023housekeeping}%
  \BibitemOpen
  \bibfield  {author} {\bibinfo {author} {\bibfnamefont {K.}~\bibnamefont {Yoshimura}}, \bibinfo {author} {\bibfnamefont {A.}~\bibnamefont {Kolchinsky}}, \bibinfo {author} {\bibfnamefont {A.}~\bibnamefont {Dechant}},\ and\ \bibinfo {author} {\bibfnamefont {S.}~\bibnamefont {Ito}},\ }\bibfield  {title} {\bibinfo {title} {Housekeeping and excess entropy production for general nonlinear dynamics},\ }\href@noop {} {\bibfield  {journal} {\bibinfo  {journal} {Physical Review Research}\ }\textbf {\bibinfo {volume} {5}},\ \bibinfo {pages} {013017} (\bibinfo {year} {2023})}\BibitemShut {NoStop}%
\bibitem [{\citenamefont {Van~Vu}\ and\ \citenamefont {Saito}(2023)}]{van2023thermodynamic}%
  \BibitemOpen
  \bibfield  {author} {\bibinfo {author} {\bibfnamefont {T.}~\bibnamefont {Van~Vu}}\ and\ \bibinfo {author} {\bibfnamefont {K.}~\bibnamefont {Saito}},\ }\bibfield  {title} {\bibinfo {title} {Thermodynamic unification of optimal transport: Thermodynamic uncertainty relation, minimum dissipation, and thermodynamic speed limits},\ }\href@noop {} {\bibfield  {journal} {\bibinfo  {journal} {Physical Review X}\ }\textbf {\bibinfo {volume} {13}},\ \bibinfo {pages} {011013} (\bibinfo {year} {2023})}\BibitemShut {NoStop}%
\bibitem [{\citenamefont {Kolchinsky}\ \emph {et~al.}(2022)\citenamefont {Kolchinsky}, \citenamefont {Dechant}, \citenamefont {Yoshimura},\ and\ \citenamefont {Ito}}]{kolchinsky2022information}%
  \BibitemOpen
  \bibfield  {author} {\bibinfo {author} {\bibfnamefont {A.}~\bibnamefont {Kolchinsky}}, \bibinfo {author} {\bibfnamefont {A.}~\bibnamefont {Dechant}}, \bibinfo {author} {\bibfnamefont {K.}~\bibnamefont {Yoshimura}},\ and\ \bibinfo {author} {\bibfnamefont {S.}~\bibnamefont {Ito}},\ }\bibfield  {title} {\bibinfo {title} {Information geometry of excess and housekeeping entropy production},\ }\href@noop {} {\bibfield  {journal} {\bibinfo  {journal} {arXiv preprint arXiv:2206.14599}\ } (\bibinfo {year} {2022})}\BibitemShut {NoStop}%
\bibitem [{\citenamefont {Wang}\ \emph {et~al.}(2021)\citenamefont {Wang}, \citenamefont {Jiao}, \citenamefont {Xu}, \citenamefont {Wang},\ and\ \citenamefont {Yang}}]{wang2021deep}%
  \BibitemOpen
  \bibfield  {author} {\bibinfo {author} {\bibfnamefont {G.}~\bibnamefont {Wang}}, \bibinfo {author} {\bibfnamefont {Y.}~\bibnamefont {Jiao}}, \bibinfo {author} {\bibfnamefont {Q.}~\bibnamefont {Xu}}, \bibinfo {author} {\bibfnamefont {Y.}~\bibnamefont {Wang}},\ and\ \bibinfo {author} {\bibfnamefont {C.}~\bibnamefont {Yang}},\ }\bibfield  {title} {\bibinfo {title} {Deep generative learning via schr{\"o}dinger bridge},\ }in\ \href@noop {} {\emph {\bibinfo {booktitle} {International conference on machine learning}}}\ (\bibinfo {organization} {PMLR},\ \bibinfo {year} {2021})\ pp.\ \bibinfo {pages} {10794--10804}\BibitemShut {NoStop}%
\bibitem [{\citenamefont {De~Bortoli}\ \emph {et~al.}(2021)\citenamefont {De~Bortoli}, \citenamefont {Thornton}, \citenamefont {Heng},\ and\ \citenamefont {Doucet}}]{de2021diffusion}%
  \BibitemOpen
  \bibfield  {author} {\bibinfo {author} {\bibfnamefont {V.}~\bibnamefont {De~Bortoli}}, \bibinfo {author} {\bibfnamefont {J.}~\bibnamefont {Thornton}}, \bibinfo {author} {\bibfnamefont {J.}~\bibnamefont {Heng}},\ and\ \bibinfo {author} {\bibfnamefont {A.}~\bibnamefont {Doucet}},\ }\bibfield  {title} {\bibinfo {title} {Diffusion schr{\"o}dinger bridge with applications to score-based generative modeling},\ }\href@noop {} {\bibfield  {journal} {\bibinfo  {journal} {Advances in Neural Information Processing Systems}\ }\textbf {\bibinfo {volume} {34}},\ \bibinfo {pages} {17695} (\bibinfo {year} {2021})}\BibitemShut {NoStop}%
\bibitem [{\citenamefont {Flamary}\ \emph {et~al.}(2021)\citenamefont {Flamary}, \citenamefont {Courty}, \citenamefont {Gramfort}, \citenamefont {Alaya}, \citenamefont {Boisbunon}, \citenamefont {Chambon}, \citenamefont {Chapel}, \citenamefont {Corenflos}, \citenamefont {Fatras}, \citenamefont {Fournier} \emph {et~al.}}]{flamary2021pot}%
  \BibitemOpen
  \bibfield  {author} {\bibinfo {author} {\bibfnamefont {R.}~\bibnamefont {Flamary}}, \bibinfo {author} {\bibfnamefont {N.}~\bibnamefont {Courty}}, \bibinfo {author} {\bibfnamefont {A.}~\bibnamefont {Gramfort}}, \bibinfo {author} {\bibfnamefont {M.~Z.}\ \bibnamefont {Alaya}}, \bibinfo {author} {\bibfnamefont {A.}~\bibnamefont {Boisbunon}}, \bibinfo {author} {\bibfnamefont {S.}~\bibnamefont {Chambon}}, \bibinfo {author} {\bibfnamefont {L.}~\bibnamefont {Chapel}}, \bibinfo {author} {\bibfnamefont {A.}~\bibnamefont {Corenflos}}, \bibinfo {author} {\bibfnamefont {K.}~\bibnamefont {Fatras}}, \bibinfo {author} {\bibfnamefont {N.}~\bibnamefont {Fournier}}, \emph {et~al.},\ }\bibfield  {title} {\bibinfo {title} {Pot: Python optimal transport},\ }\href@noop {} {\bibfield  {journal} {\bibinfo  {journal} {Journal of Machine Learning Research}\ }\textbf {\bibinfo {volume} {22}},\ \bibinfo {pages} {1} (\bibinfo {year} {2021})}\BibitemShut {NoStop}%
\bibitem [{\citenamefont {Kingma}\ and\ \citenamefont {Ba}(2015)}]{kingma2014adam}%
  \BibitemOpen
  \bibfield  {author} {\bibinfo {author} {\bibfnamefont {D.}~\bibnamefont {Kingma}}\ and\ \bibinfo {author} {\bibfnamefont {J.}~\bibnamefont {Ba}},\ }\bibfield  {title} {\bibinfo {title} {Adam: A method for stochastic optimization},\ }in\ \href@noop {} {\emph {\bibinfo {booktitle} {International Conference on Learning Representations}}}\ (\bibinfo {year} {2015})\BibitemShut {NoStop}%
\end{thebibliography}%
\end{document}